\documentclass[10pt]{thesis_mw2}

\usepackage[usenames,dvipsnames,svgnames,table]{xcolor} 
\usepackage[obeyspaces,hyphens,spaces]{url}
\usepackage{thesisfrontmatter}
\usepackage{pdfpages}

\definecolor{Blue}{rgb}{0.25, 0.41, 0.88}
\definecolor{Red}{rgb}{0.92,0.,0.}
\definecolor{darkorange}{rgb}{1.0,0.549,0.}
\definecolor{cobalt}{RGB}{44, 98, 120}
\definecolor{Mathematica1}{rgb}{0.368417, 0.506779, 0.709798}
\definecolor{Mathematica2}{rgb}{0.880722, 0.611041, 0.142051}
\definecolor{Mathematica3}{rgb}{0.560181, 0.691569, 0.194885}
\definecolor{Mathematica4}{rgb}{0.922526, 0.385626, 0.209179}
\definecolor{Mathematica5}{rgb}{0.528488, 0.470624, 0.701351}
\definecolor{Mathematica6}{rgb}{0.772079, 0.431554, 0.102387}
\definecolor{Mathematica7}{rgb}{0.363898, 0.618501, 0.782349}
\definecolor{Mathematica8}{rgb}{1, 0.75, 0}
\definecolor{Mathematica9}{rgb}{0.647624, 0.37816, 0.614037}
\definecolor{plotBlue}{RGB}{94, 130, 181}
\definecolor{plotRed}{RGB}{233, 85, 54}
\definecolor{plotGreen}{RGB}{142, 176, 50}
\definecolor{plotPurple}{RGB}{135, 120, 178}
\definecolor{cornellRed}{HTML}{B31B1B}
\definecolor{cornellBlue}{HTML}{0068AC}
\definecolor{cornellGreen}{HTML}{6EB43F}
\definecolor{dullpurple}{rgb}{0.431,0.188,0.534}
\definecolor{darkgreen}{rgb}{0.075,0.302,0.047}
\definecolor{darkergreen}{rgb}{0,0.196,0.125}
\definecolor{darkergreen2}{rgb}{0,0.294,0.188}
\definecolor{dullred}{rgb}{0.706,0.208,0.192}
\definecolor{darkred}{rgb}{0.545,0,0}
\definecolor{antiquefuchsia}{rgb}{0.57, 0.36, 0.51}
\definecolor{MaroonC}{rgb}{0,0.502,0.502}
\definecolor{dullblue}{rgb}{0,0.298,0.49}
\definecolor{blue3}{RGB}{31, 119, 180}
\definecolor{red3}{RGB}{	214, 39, 40}
\definecolor{orange3}{RGB}{255, 127, 14}
\definecolor{green3}{RGB}{44, 160, 44}

\usepackage{ifthen}
\usepackage{fancyhdr}
\usepackage[colorlinks=true
,urlcolor=dullblue
,anchorcolor=blue3
,citecolor=dullblue
,filecolor=blue3
,linkcolor=darkred
,menucolor=blue3
,pagecolor=blue3
,linktocpage=true
,pdfproducer=medialab
]{hyperref}
\usepackage{booktabs}
\usepackage[english, american, dutch]{babel}
\usepackage{amsmath, amssymb, amsbsy, amstext, amsthm, simplewick}
\usepackage{cleveref}
\usepackage{graphicx}
\usepackage{amsfonts}
\usepackage{enumitem}
\usepackage{upgreek}
\usepackage{framed}
\usepackage{tensor}
\usepackage{pifont}
\usepackage{latexsym, mathrsfs}
\usepackage{array}
\usepackage{xspace}
\usepackage{longtable}
\usepackage{multirow}
\usepackage{cases}
\usepackage{empheq}

\usepackage{bm}
\usepackage{amsmath}
\usepackage{bbm}
\usepackage{appendix}

\usepackage{braket}
\usepackage{yfonts}

\usepackage[load-configurations=astronomy, range-units=brackets, range-phrase=-, per-mode=reciprocal, mode=math]{siunitx}
\usepackage{threeparttable}
\usepackage{subfig}

\usepackage{ragged2e}

\usepackage{hhline}
\usepackage{array}
\newcolumntype{P}[1]{>{\centering\arraybackslash}p{#1}}
\usepackage{booktabs}
\usepackage{tikz}
\usetikzlibrary{calc,shadings,patterns,tikzmark,fadings}

\newcolumntype{C}[1]{>{\centering\let\newline\\\arraybackslash\hspace{0pt}}m{#1}}

\def\r{{\bf r}}

\newcommand{\es}{\hspace{0.5pt}}

\makeatletter
\newlength{\apb@width}
\newcommand{\autoparbox}[2][c]{\settowidth{\apb@width}{#2}\parbox[#1]{\apb@width}{#2}}

\makeatother

\numberwithin{equation}{section}

\def\beq{\begin{equation}}
\def\eeq{\end{equation}}

\def\bea{\begin{eqnarray}}
\def\eea{\end{eqnarray}}

\def\beq{\begin{equation}}
\def\eeq{\end{equation}}
\def\bea{\begin{eqnarray}}
\def\eea{\end{eqnarray}}

\newcommand{\ped}[1]{\textormath{\textsubscript{#1}}{_{\mathrm{#1}}}}
\newcommand{\ap}[1]{\textormath{\textsuperscript{#1}}{^{\mathrm{#1}}}}

\renewcommand{\Im}{\operatorname{Im}}
\renewcommand{\Re}{\operatorname{Re}}

\renewcommand{\vec}[1]{\boldsymbol{\mathbf{#1}}}

\newcommand{\dd}{\mathop{\mathrm{d}\!}{}}

\DeclareMathOperator\sgn{sgn}
\DeclareMathOperator\diag{diag}

\DeclareMathOperator\erf{erf}
\DeclareMathOperator\Li{Li}

\DeclareMathOperator\dilog{dilog}
\DeclareMathOperator\HeunC{HeunC}
\DeclareMathOperator\erfc{erfc}

\DeclarePairedDelimiter{\abs}{\lvert}{\rvert}

\newcommand{\lab}[1]{{\mathrm{#1}}}
\newcommand{\slab}[1]{{\textsc{#1}}}
\newcommand{\mb}[1]{{\mathbf{#1}}}
\newcommand{\minus}{{\scalebox {0.75}[1.0]{$-$}}}
\newcommand{\sminus}{{\scalebox {0.6}[0.85]{$-$}}}
\newcommand{\sperp}{{\scalebox{0.7}{$\perp$}}}

\def\r{{\bf{r}}}

\newcommand{\floq}[1]{{\scalebox{0.65}{$(#1)$}}}

\def\r{{\bf{r}}}

\DeclareRobustCommand{\SkipTocEntry}[4]{}

\usepackage{colortbl}
\definecolor{blue2}{cmyk}{1, 0.1, 0.1, 0}

\definecolor{pyBlue}{RGB}{31, 119, 180}
\definecolor{pyRed}{RGB}{214, 39, 40}
\definecolor{pyGreen}{RGB}{44, 160, 44}
\definecolor{pyBlue2}{RGB}{0, 111, 237}
\definecolor{pyRed2}{RGB}{224, 52, 36}

\makeatletter
\def\Ddots{\mathinner{\mkern1mu\raise\p@
\vbox{\kern7\p@\hbox{.}}\mkern2mu
\raise4\p@\hbox{.}\mkern2mu\raise7\p@\hbox{.}\mkern1mu}}
\makeatother

\usepackage{chngcntr}
\usepackage{etoolbox}

\AtBeginEnvironment{subappendices}{%
}

\thesisfont{cmr} 
\thesissectionstyle{\rmfamily\bfseries}
\thesissectionsizes{\Large}{\large}{}

          \usepackage[explicit]{titlesec}
          \usepackage{lmodern}
          \newlength\chapnumb
          \newlength\chapnumbless
          \titleformat{\chapter}[block]
          {\normalfont\sffamily\scshape}{}{0pt}
          {\parbox[b]{\chapnumb}{
             \fontsize{50}{0}\selectfont\thechapter}
            \parbox[b]{\dimexpr\textwidth-\chapnumb\relax}{
              \raggedleft
				  \hfill{\huge#1 \\[5pt]}
              \rule{\dimexpr\textwidth-\chapnumb\relax}{0.4pt}
              }}      
         \titleformat{name=\chapter,numberless}[block]
          {\normalfont\sffamily\scshape}{}{0pt}
          {\parbox[b]{\chapnumbless}{
             \mbox{}}
            \parbox[b]{\dimexpr\textwidth-\chapnumbless\relax}{
              \raggedleft
              \hfill{\Huge#1}\\
              \rule{\dimexpr\textwidth-\chapnumbless\relax}{0.4pt}}}
              
\usepackage{fix-cm}
\newcommand\HUGE{\@setfontsize\Huge{38}{47}}

\title{{\Huge \sffamily Gravitational Atoms and Black Hole Binaries \\[5pt]}}
\thesiscolofon{{    \noindent This work has been accomplished at Gravitation Astroparticle Physics Amsterdam (GRAPPA) at the University of Amsterdam (UvA). This research is funded by the Netherlands Organisation for Scientific Research (NWO).

\vfill
\par\vskip 12cm

\noindent\copyright{} Giovanni Maria Tomaselli, 2024\\[1em]
   \begin{minipage}[b]{\textwidth}
All rights reserved. Without limiting the rights under copyright reserved above, no part of this book may be reproduced, stored in or introduced into a retrieval system, or transmitted, in any form or by any means (electronic, mechanical, photocopying, recording or otherwise) without the written permission of both the copyright owner and the author of the book.
      \end{minipage}    
\vfill
      }}
\author{Giovanni Maria Tomaselli}
\placeofbirth{Benevento}
\theday{woensdag}
\date{30 oktober 2024}
\thetime{16.00}

\rectormagnificus{prof.~dr.~ir.~P.P.C.C.~Verbeek}
\promotor{
\committeeentry{prof.~dr.~D.~Baumann}{Universiteit van Amsterdam},
\committeeentry{prof.~dr.~G.~Bertone}{Universiteit van Amsterdam}
}
\othercomitteemembers{ 
\committeeentry{prof.~dr.~A.~Buonanno}{Albert-Einstein-Institut},
\committeeentry{prof.~dr.~V.~Cardoso}{Universidade Técnica de Lisboa},
\committeeentry{dr.~B.~Freivogel}{Universiteit van Amsterdam},
\committeeentry{prof.~dr.~L.~Hui}{Columbia University},
\committeeentry{prof.~dr.~S.~Markoff}{Universiteit van Amsterdam},
\committeeentry{dr.~S.~Nissanke}{Universiteit van Amsterdam},
}
\department{\small Faculteit der Natuurwetenschappen, Wiskunde en
Informatica}

\usepackage{pifont}
\usepackage[nottoc]{tocbibind}

\newcommand{\committeeentry}[2]{\begin{tabular*}{0.78\textwidth}{p{0.42\textwidth}l}#1 & #2\end{tabular*}}

\begin{document}
\selectlanguage{english}
{
\includepdf[fitpaper]{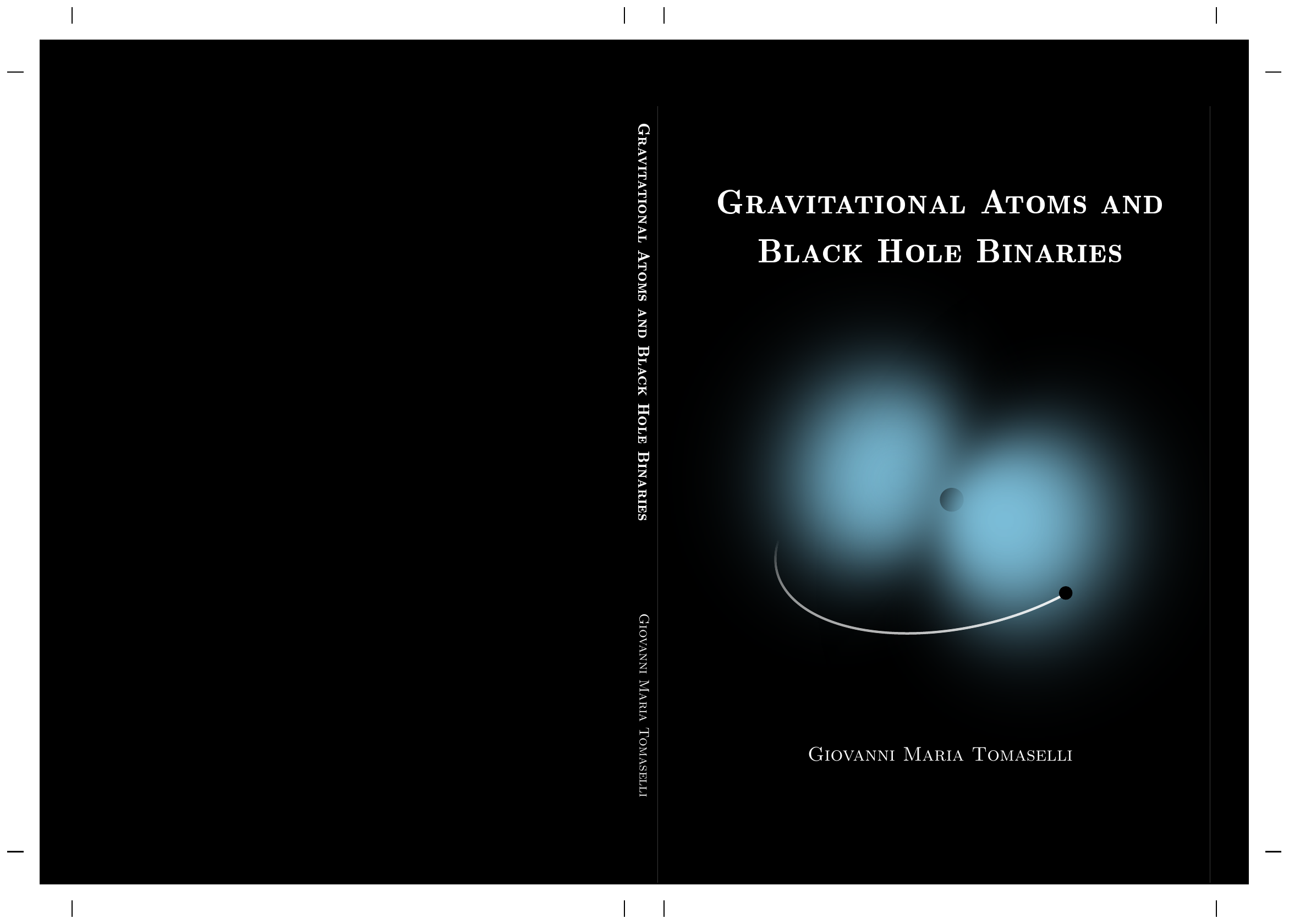}
\let\cleardoublepage\clearpage
\frontmatter
	\maketitle
	}
\basedon{Publications}{
{\scshape This thesis is based on the following publications:} \vskip 20pt

\begin{itemize}[]

{\fontfamily{cmr}\selectfont

\item [\cite{Baumann:2021fkf}]

D.~Baumann, G.~Bertone, J.~Stout and G.~M.~Tomaselli, ``Ionization of gravitational atoms'', \href{http://dx.doi.org/10.1103/PhysRevD.105.115036}{\textit{Phys.\
Rev.\ D} \textbf{105} no.~11, (2022) 115036}, \href{https://arxiv.org/abs/2112.14777}{\ttfamily arXiv:2112.14777}. \vskip 5pt

{\footnotesize G.M.T.~conceived the ionization effect, derived it as in Sec.~\ref{sec:ionization-review} of this thesis, derived all other results and  contributed to writing most of the manuscript. J.S.~derived ionization as presented in Sec.~\ref{sec:thorough-derivation} and App.~\ref{app:approx} to \ref{app:zeroMode} of this thesis and contributed to writing most of the manuscript. D.B.~and G.B.~contributed with discussions and editing.\par}

\item [\cite{Baumann:2022pkl}]

D.~Baumann, G.~Bertone, J.~Stout and G.~M.~Tomaselli, ``Sharp Signals of Boson Clouds in Black Hole Binary Inspirals'', \href{http://dx.doi.org/10.1103/PhysRevD.105.115036}{\textit{Phys.\
Rev.\ Lett.}\ \textbf{128} no.~22, (2022) 221102}, \href{https://arxiv.org/abs/2206.01212}{\ttfamily arXiv:2206.01212}. \vskip 5pt

{\footnotesize G.M.T.~derived all the results presented. All authors contributed to writing the manuscript.\par}

\item [\cite{Tomaselli:2023ysb}]

G.~M.~Tomaselli, T.~F.~M.~Spieksma and G.~Bertone, ``Dynamical friction in gravitational atoms'', \href{http://dx.doi.org/10.1088/1475-7516/2023/07/070}{\textit{JCAP} \textbf{07} (2023) 070}, \href{https://arxiv.org/abs/2305.15460}{\ttfamily arXiv:2305.15460}. \vskip 5pt

{\footnotesize G.M.T.~derived the results, wrote the code and the majority of the manuscript. T.F.M.S.~ran the code and contributed to writing the manuscript. G.B.~contributed with discussions.\par}

\item [\cite{Tomaselli:2024bdd}]

G.~M.~Tomaselli, T.~F.~M.~Spieksma and G.~Bertone, ``Resonant history of gravitational atoms in black hole binaries'', \href{https://doi.org/10.1103/PhysRevD.110.064048}{\textit{Phys.\ Rev.\ D} \textbf{110} no.~6, (2024) 064048}, \href{https://arxiv.org/abs/2403.03147}{\ttfamily arXiv:2403.03147}. \vskip 5pt

{\footnotesize G.M.T.~derived most of the results and wrote the majority of the manuscript. T.F.M.S.~contributed to deriving and checking the results and to writing. G.B.~contributed with discussions.\par}

\item [\cite{Tomaselli:2024dbw}]

G.~M.~Tomaselli, T.~F.~M.~Spieksma and G.~Bertone, ``Legacy of boson clouds on black hole binaries'', \href{https://doi.org/10.1103/PhysRevLett.133.121402}{\textit{Phys.\ Rev.\ Lett.}\ \textbf{133} no.~12, (2024) 121402}, \href{https://arxiv.org/abs/2407.12908}{\ttfamily arXiv:2407.12908}. \vskip 5pt

{\footnotesize G.M.T.~derived most of the results and wrote the majority of the manuscript. T.F.M.S.~contributed to deriving and checking the results and to writing. G.B.~contributed with discussions.\par}

}

\end{itemize}

\newpage

{\scshape During the completion of his PhD, G.~M.~Tomaselli was also an author of the following publications:} \vskip 20pt

\begin{itemize}[]

{\fontfamily{cmr}\selectfont

\item [\cite{Tomaselli:2021pem}]

G.~M.~Tomaselli and A.~Ferrara, ``Lyman-alpha radiation pressure: an analytical exploration'', \href{http://dx.doi.org/10.1093/mnras/stab876}{\textit{Mon.\ Not.\ Roy.\ Astron.\ Soc.}\ \textbf{504} no.~1, (2021) 89--100}, \href{https://arxiv.org/abs/2103.14655}{\ttfamily arXiv:2103.14655 [astro-ph.GA]}. \vskip 10pt

\item [\cite{Hui:2022sri}]

L.~Hui, Y.~T.~A.~Law, L.~Santoni, G.~Sun, G.~M.~Tomaselli and E.~Trincherini, ``Black hole superradiance with dark matter accretion'', \href{http://dx.doi.org/10.1103/PhysRevD.107.104018}{\textit{Phys.\
Rev.\ D} \textbf{107} no.~10, (2023) 104018}, \href{https://arxiv.org/abs/2208.06408}{\ttfamily arXiv:2208.06408 [gr-qc]}. \vskip 10pt

\item [\cite{Cole:2022yzw}]

P.~S.~Cole, G.~Bertone, A.~Coogan, D.~Gaggero, T.~Karydas, B.~J.~Kavanagh, T.~F.~M.~Spieksma and G.~M.~Tomaselli, ``Distinguishing environmental effects on binary black hole gravitational waveforms'', \href{http://dx.doi.org/10.1038/s41550-023-01990-2}{\textit{Nature Astron.}\ \textbf{7} no.~8, (2023) 943--950}, \href{https://arxiv.org/abs/2211.01362}{\ttfamily arXiv:2211.01362 [gr-qc]}. \vskip 10pt

\item [\cite{Reyes:2023fde}]

I.~A.~Reyes and G.~M.~Tomaselli, ``Quantum field theory on compact stars near the Buchdahl limit'', \href{http://dx.doi.org/10.1103/PhysRevD.108.065006}{\textit{Phys.\
Rev.\ D} \textbf{107} no.~10, (2023) 104018}, \href{https://arxiv.org/abs/2301.00826}{\ttfamily arXiv:2301.00826 [gr-qc]}. \vskip 10pt

}

\end{itemize}

\newpage

}

\setcounter{tocdepth}{2}

\addtocontents{toc}{\protect\enlargethispage*{\baselineskip}} 

\tableofcontents

\mainmatter

\chapter{Introduction}

Advances in fundamental physics are inevitably tied to the reach of experiments. The golden age of physics in the early 20\ap{th} century had its roots in the ability to probe, for the first time ever, the microscopic constituents of matter. Decades later, the era of particle accelerators led to the formulation of the Standard Model of particle physics, one of the greatest scientific achievements of modern science. We now live at a time when the future of ground-based experiments in high-energy physics is very uncertain, due to technological limitations and the lack of clear indications of where new physics might hide. Conversely, the quantity and quality of astrophysical and cosmological observations have recently exploded. Through the collection of large amounts of data and the ability to perform precision tests, these fields hold promise to bring about new discoveries in fundamental physics. In fact, many currently unsolved questions within these areas seem to have strong ties to the underlying laws of Nature, and their solutions appear to be within the reach of observations, at least in principle: a prime example is the dark matter problem \cite{Bertone:2004pz,Bertone:2016nfn,Cirelli:2024ssz}.

\vskip 0pt
Among the many fronts that enjoyed dramatic advances, gravitational-wave (GW) astronomy stands out as one of the most exciting ones. Although our current theory of gravity, General Relativity (GR), has been available since 1915, for one century our ability to test it was limited to measuring tiny effects, such as post-Newtonian corrections to the motion of planets, gravitational redshift, and lensing. The access to the strong-field regime of gravity is a true revolution: we can now fully test GR, instead of relying solely on perturbative expansions. Additionally, we can use GWs as messengers of astrophysical events, similar to how light has been used since Galileo first pointed his telescope at the sky. Since the first direct detection of a binary black hole (BH) merger in 2015 \cite{LIGOScientific:2016aoc}, GWs have already led to many discoveries. For example, the first detection of a binary neutron star merger \cite{LIGOScientific:2017vwq}, with associated electromagnetic counterparts \cite{LIGOScientific:2017ync,Cowperthwaite:2017dyu,Troja:2017nqp}, has single-handedly brought evidence for the creation of many of the heaviest elements \cite{Kasen:2017sxr,Pian:2017gtc}, allowed for an unprecedented measure of the speed of gravitational waves \cite{LIGOScientific:2017zic}, provided a new independent measurement of the Hubble constant \cite{LIGOScientific:2017adf} and severely constrained the equation of state of matter at high densities \cite{LIGOScientific:2018cki}. The future will tell what other secrets of Nature will be unveiled by the detection of new events.

\vskip 0pt
The subject of this thesis is a way to discover new particles using GWs. The topic has been explored for the first time in a previous PhD thesis at the University of Amsterdam \cite{Chia:2020dye}, and has since received significant attention from the scientific community. It might not be obvious, at first glance, how gravity can serve as a probe of new fundamental particles: its extreme weakness usually makes it negligible in particle physics. Furthermore, the most successful strategy of discovery has historically been the use of high-energy colliders. This traditional avenue, however, can only be successful if the new degrees of freedom couple strongly enough to known particles, and is thus completely blind to weakly-coupled new physics. Yet undiscovered particles of very small mass could well exist in our Universe, and at the same time be nearly impossible to produce through the collision of Standard Model particles. As it turns out, there exists a natural amplification mechanism that allows for the spontaneous creation of a large amount of very light particles around BHs: \emph{black hole superradiance} \cite{Brito:2015oca}.

\vskip 0pt
Although superradiance has been known since the 1970s \cite{Penrose:1969pc,Penrose:1971uk}, its full potential was only realized after the late 2000s \cite{Dolan:2007mj,Arvanitaki:2009fg,Arvanitaki:2010sy}. Perhaps the biggest reason behind such a delayed blooming is the renewed interest in new ultralight particles. These have been proposed in a variety of different contexts, such as solutions to the strong CP problem \cite{Weinberg:1977ma,Wilczek:1977pj,Peccei:1977hh}, dark matter candidates \cite{Bergstrom:2009ib,Marsh:2015xka,Hui:2016ltb,Ferreira:2020fam}, by-products of string compactifications \cite{Arvanitaki:2009fg,Svrcek:2006yi} and dark photons \cite{Okun:1982xi,Holdom:1985ag,Cicoli:2011yh}. Superradiance necessitates only two ingredients: a light bosonic degree of freedom (such as the ones mentioned above) and a rapidly spinning black hole. Given these, the exponential amplification of field perturbations is a natural consequence, leading to the extraction of a significant amount of mass and angular momentum from the BH. At the same time, a ``boson cloud'' is formed around it, giving rise to a very dense and exotic astrophysical environment, which can serve as a probe for the existence of the light particle. Due to its mathematical similarity to the proton-electron structure of the hydrogen atom, the BH-cloud system is also referred to as a \emph{gravitational atom}.

\vskip 0pt
It should be kept in mind, however, that superradiance is only efficient on astrophysical timescales if the Compton wavelength of the field is comparable to, or slightly larger than, the size of the BH. As such, BHs of different masses can be used to scan through the boson's mass parameter space, according to the relation
\beq
\mu\sim\SI{e-10}{eV}\,\biggl(\frac{M_\odot}{M}\biggr)\,,
\tag{1}
\eeq
where $\mu$ is the boson's mass and $M$ is the BH mass.

\begin{figure}[t]
\centering
\includegraphics[trim=0pt 30pt 0pt 0pt]{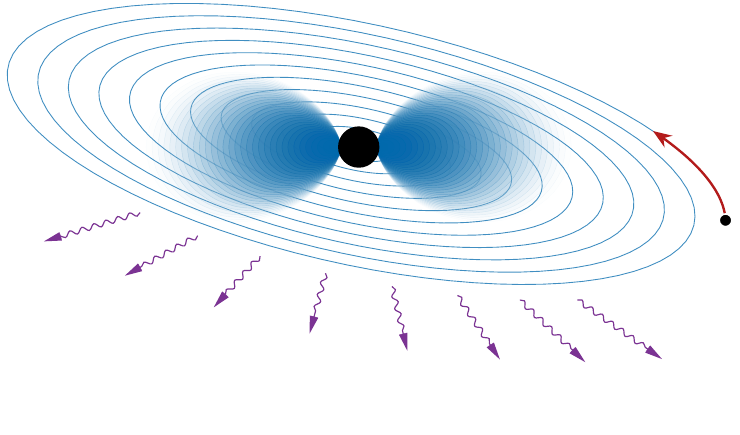}
\caption{Illustration of a gravitational atom in a binary system. As the two black holes inspiral, the smaller of the two goes through the densest regions of the cloud, interacting with it gravitationally. The backreaction on the orbit alters the binary dynamics and the ensuing gravitational waveform, offering clear-cut signatures of the existence of the light boson.}
\label{fig:cover}
\end{figure}

\vskip 0pt
Gravitational atoms lead to a variety of observational signatures, which have been thoroughly explored in the past decade, such as the distribution of BH spins, and GWs emitted by the cloud \cite{Arvanitaki:2010sy,Arvanitaki:2014wva}. While these classical probes remain valid today, a new and richer way of studying gravitational atoms has been discovered more recently~\cite{Chia:2020dye}. The core idea is that the dynamical gravitational interactions of the cloud are directly linked to its specific structure and properties, resulting in extremely distinctive phenomenology. These interactions can be excited by the presence of a binary companion, such as another BH, a setup which also allows for clear observational signatures through modifications of the binary's gravitational waveform. The system is schematically illustrated in Figure~\ref{fig:cover}.

\vskip 0pt
The aforementioned characteristics of gravitational atoms originate in their energy spectrum. Similar to ordinary atoms, a light boson around the BH can occupy many possible discrete states, each coming with its own energy. Superradiance populates one of these states with a macroscopic number of particles, a situation very different from other astrophysical environments where the particles come with a continuous distribution of positions and velocities. Analogous to photons, which can make an electron ``jump'' from one state to another, the gravitational perturbation of the binary companion can mediate similar transitions within the cloud. The discreteness of the spectrum then translates into strict conditions on the binary's frequency, and sharp signals through the emitted GWs.

\vskip 0pt
Before the present thesis, these ``resonant transitions'' had just been discovered~\cite{Baumann:2018vus,Baumann:2019ztm}, together with a few other kinds of binary-cloud interactions. This thesis introduces ionization \cite{Baumann:2021fkf,Baumann:2022pkl}, a new effect with a major impact on the dynamics of the system and with unique observational consequences. Its name is clearly evocative of atomic physics, and indeed ionization can be understood as the analog of the photoeletric effect. Other significant new effects introduced in this thesis include the accretion of the cloud onto the companion BH \cite{Baumann:2021fkf}, and the cloud-mediated dynamical capture of the secondary object~\cite{Tomaselli:2023ysb}. Together, these processes allow for a complete description of the cloud-binary interaction.

\vskip 0pt
After all the different phenomena have been discovered and understood, two additional steps are undertaken to fully describe the observational implications of gravitational atoms in binaries. First, the treatment of resonances and ionization is extended to generic binary's orbits, with nonzero eccentricity and inclination \cite{Tomaselli:2023ysb,Tomaselli:2024bdd}. Second, the ``history'' of the system prior to its observation is studied systematically \cite{Tomaselli:2024bdd}, with the goal to determine the state of the cloud by the time the system enters the band of GW detectors. This step requires to significantly generalize the treatment of resonances, to fully account for their backreaction on the binary. Such a deeper study of resonances unveils new \emph{indirect} observational signatures, in the form of fixed points in the binary's eccentricity and inclination \cite{Tomaselli:2024dbw}.

\vskip 0pt
The completion of this work makes it finally possible to present all the results about the subject in a complete, coherent and self-consistent way. This accomplishment, however, does not imply by any means the end of research on this topic: there is still room for a more accurate description of all the phenomena, and waveform studies are still in their infancy \cite{Cole:2022yzw}. Many of the results presented here are particularly attractive and well-motivated for future GW detectors, which still lie several years ahead. The same can be said for the broader research area of fundamental physics and environmental effects on GWs, which is rapidly evolving and of which I hope the present thesis can constitute a firm cornerstone.

\subsubsection{Roadmap of the thesis}

The outline of the thesis is as follows. Chapter~\ref{chap:background} contains a summary of the theoretical background underlying the rest of the thesis, as well as a short account of the context the research fits into. This includes an overview of black holes (Section~\ref{sec:black-holes}), gravitational-wave theory and astronomy (Section~\ref{sec:gravitational-waves}), and a summary of the motivations and frameworks behind ultralight scalars (Section~\ref{sec:ultralight-scalars}). The chapter concludes with a review of black hole superradiance (Section~\ref{sec:superradiance}) and gravitational atoms (Section~\ref{sec:gravitational-atoms}).

\vskip 0pt
Chapters~\ref{chap:binary-cloud-interaction} to \ref{chap:observational-signatures} present original content, from publications \cite{Baumann:2021fkf,Baumann:2022pkl,Tomaselli:2023ysb,Tomaselli:2024bdd,Tomaselli:2024dbw} (with the exception of Section~\ref{sec:resonances-review}, which serves as a review of \cite{Baumann:2018vus,Baumann:2019ztm}). Due to the interrelatedness of the papers, however, the topics are presented here using a new and more logical order, instead of keeping the contents of the papers separate.

\vskip 0pt
Chapter~\ref{chap:binary-cloud-interaction} contains a rundown of all the phenomena studied in the thesis, namely dynamical capture (Section~\ref{sec:capture}), resonances (Section~\ref{sec:resonances-review}), ionization (Section~\ref{sec:ionization-review}), and accretion (Section~\ref{sec:accretion}). Of these, dynamical capture and accretion are discussed in detail, while resonances and ionization are only introduced, leaving further in-depth discussion to subsequent chapters.

\vskip 0pt
Chapter~\ref{chap:ionization} is entirely devoted to ionization. It starts by discussing its physical interpretation as dynamical friction (Section~\ref{sec:dynamical-friction}). After discussing the scaling of the results with the parameters (Section~\ref{sec:scaling}), a quick look is given at the backreaction of ionization on the orbit (Section~\ref{sec:backreaction-ionization}), postponing the study of its observational signatures to Chapter~\ref{chap:observational-signatures}. The focus is then moved to a different and thorough derivation of ionization that allows a deeper understanding of its features (Section~\ref{sec:thorough-derivation}). The framework is then extended to eccentric (Section~\ref{sec:eccentricity}) and inclined (Section~\ref{sec:inclination}) orbits, including a discussion of the backreaction of ionization on those parameters.

\vskip 0pt
Chapter~\ref{chap:resonances} takes a deeper look at resonances. First, it discusses the novelties introduced by nonzero eccentricity and inclination (Section~\ref{sec:eccentric-inclined-resonances}). Then, it establishes the framework needed to describe the backreaction of resonances on the orbit (Section~\ref{sec:backreaction}), proceeding to explore their phenomenology in the two qualitatively distinct cases of floating (Section~\ref{sec:floating}) and sinking (Section~\ref{sec:sinking}) orbits. Finally, the scalings of all resonance variables with the physical parameters are presented (Section~\ref{sec:types-of-resonances}).

\vskip 0pt
Chapter~\ref{chap:history} explores systematically the sequence of resonances encountered by the system during its evolution. After a general outline of the way resonances impact the history of the system (Section~\ref{sec:generalB}), the evolution is explicitly worked out starting from the two states of the cloud most likely to be populated by superradiance (Sections~\ref{sec:evolution-211} and~\ref{sec:evolution-322}).

\vskip 0pt
Chapter~\ref{chap:observational-signatures} puts together the results of all previous chapters to determine the observational signatures of the cloud. These are of two types: direct signatures of the cloud, mainly due to ionization and resonances (Section~\ref{sec:direct-signatures}), and indirect evidence in case the cloud is destroyed, based on the impact of resonances on the binary parameters (Section~\ref{sec:indirect-evidence}).

\vskip 0pt
Conclusions are given in Chapter~\ref{chap:conclusions}, while Appendices~\ref{app:heunc}, \ref{app:more-on-ionization} and \ref{app:resonance-phenomenology} contain details that were left out of the main text.

\subsubsection{Notations and conventions}

We work in natural units, $G=\hbar=c=1$. The larger BH has mass $M$ and dimensionless spin $\tilde a$, with $0\le\tilde a<1$. The mass and radial distance of the smaller object are denoted by $M_*\equiv qM$ and $R_*$, where $q$ is the mass ratio, while the orbital frequency is $\Omega$ and the mass of the cloud is $M\ped{c}$. The gravitational fine structure constant is $\alpha=\mu M$, where $\mu$ is the mass of the scalar field.

\chapter{Theory and background}

\label{chap:background}

The theoretical foundations of this thesis combine General Relativity with physics beyond the Standard Model. This chapter reviews the concepts needed for later, while also giving an overview of the research context this work fits into, including observational aspects. Each one of the sections contained here is a huge topic on its own. We will thus have no ambition of completeness, focusing instead only on the parts that turn out to be most relevant in the thesis.

\vskip 0pt
We start by reviewing black hole physics in Section~\ref{sec:black-holes}, giving particular emphasis to the Kerr metric and its properties. An overview of gravitational-wave physics and astronomy is provided in Section~\ref{sec:gravitational-waves}. We then move to physics beyond the Standard Model, describing in Section~\ref{sec:ultralight-scalars} the models and motivations behind the hypothetical particles considered in the thesis. In Section~\ref{sec:superradiance}, we give an overview of black hole superradiance, providing a compact derivation of the key results used later. Finally, in Section~\ref{sec:gravitational-atoms}, we describe the properties of gravitational atoms, establishing concepts and notations that are extensively used in the subsequent chapters.

\section{Black holes}

\label{sec:black-holes}

General Relativity is widely considered as one of the most elegant and powerful theories ever developed in physics. It describes gravity, the weakest of the four fundamental forces, in terms of the curvature of four-dimensional spacetime, superseding Newtonian gravity as our most accurate theory of gravitation. So far, GR has passed all experimental and observational tests, which became increasingly more stringent from its formulation in 1915 to today.

\vskip 0pt
The celebrated Einstein field equations of General Relativity read \cite{Einstein:1915ca,Einstein:1916vd}
\beq
R_{\mu\nu}-\frac12g_{\mu\nu}R=8\pi\, T_{\mu\nu}\,,
\eeq
where $g_{\mu\nu}$ is the spacetime metric, $R_{\mu\nu}$ is the Ricci tensor, $R=R^\mu{}_\mu$ is the Ricci scalar and $T_{\mu\nu}$ is the energy-momentum tensor. The first exact solution to the Einstein equations was found by Schwarzschild \cite{Schwarzschild:1916uq}, who used a spherically symmetric ansatz for the metric and looked for vacuum ($T_{\mu\nu}=0$) solutions. In Schwarzschild's coordinates $(t,r,\theta,\phi)$, his metric reads
\beq
ds^2=-\biggl(1-\frac{2M}r\biggr)\dd t^2+\biggl(1-\frac{2M}r\biggr)^{-1}\dd r^2+r^2(\dd\theta^2+\sin^2\theta\dd\phi^2)\,.
\label{eqn:schwarzschild}
\eeq
By matching the Newtonian predictions for the motion of particles at $r\gg2M$, the parameter $M$ can be recognized as the mass of the object sitting at the origin of coordinates.

\vskip 0pt
The leading terms in the $r\gg2M$ expansion of the line element \eqref{eqn:schwarzschild} turned out to accurately predict the motion of planets and light rays in the Solar System. These predictions include small deviations from Newtonian gravity, such as the precession of the perihelion of Mercury and the bending of light, which served as the first observational successes of GR. The structure of the spacetime at (or below) $r=2M$ is instead much harder to probe, as no object in the Solar System is dense enough
to be entirely contained within a radius equal to twice its mass.

\vskip 0pt
The radial distance $r=2M$ in \eqref{eqn:schwarzschild} identifies a three-dimensional null hypersurface known as the \emph{event horizon}. The apparent singularity of the metric at the event horizon can be removed with a suitable change of coordinates: defining the ingoing null coordinate $v$ as
\beq
v=t+r_*,\qquad r_*=r+2M\log\biggl(\frac{r}{2M}-1\biggr)\,,
\label{eqn:v-r_*-finkelstein}
\eeq
we can write the metric in ingoing Eddington-Finkelstein coordinates \cite{Finkelstein:1958zz},
\beq
ds^2=-\biggl(1-\frac{2M}r\biggr)\dd v^2+2\dd v\dd r+r^2(\dd\theta^2+\sin^2\theta\dd\phi^2)\,.
\label{eqn:eddington-finkelstein}
\eeq
Not only is \eqref{eqn:eddington-finkelstein} manifestly regular (and non-degenerate) at $r=2M$, but it shows that outgoing radial null geodesics, which obey
\beq
\frac{\dd v}{\dd r}=2\biggl(1-\frac{2M}r\biggr)^{-1}\,,
\eeq
are actually directed radially inwards for $r<2M$. This property is the key for the general definition of a \emph{black hole} as a region of spacetime that is not contained in the causal past of future null infinity. In other words: a prison that nothing, not even light, can escape.

\vskip 0pt
While the spacetime is still regular at the event horizon, the point $r=0$ is instead a true singularity of the spacetime, where curvature tensors diverge. In the Schwarzschild black hole, the singularity is actually a one-dimensional spacelike region, where spacetime ends, which lies in the causal future of any particle inside the black hole. None of our current theories is able to satisfactorily describe physics in proximity of the singularity.

\vskip 0pt
The properties of black holes, including their very existence, have historically been a source of deep confusion among scientists. Initially, there was a popular belief that some physical mechanism, such as the degeneracy pressure of fermions \cite{Fowler:1926zz}, would eventually kick in and prevent the indefinite collapse of an object. This idea gradually fell apart over time, starting with the works by Chandrasekhar \cite{Chandrasekhar:1931ih} and Oppenheimer and Snyder \cite{Oppenheimer:1939ue}, until the proof by Penrose \cite{Penrose:1964wq} that the creation of a gravitational singularity is inevitable after the formation of a \emph{trapped surface}, which happens at a stage of the collapse where the matter density is not yet very high. The nature of black holes has also been misunderstood for a long time, partly due to the fact that external stationary observers can never see any object cross the event horizon. By the late 1960s, most of these doubts had been cleared up, although observational evidence was still lacking.

\vskip 0pt
If black holes indeed form from the collapse of astrophysical objects, such as degenerate or massive stars, they must retain not just the mass, but also the angular momentum of the original body. The quest for a rotating black hole solution to the Einstein equations came to an end with a landmark discovery by Kerr \cite{Kerr:1963ud}, almost 50 years after the formulation of GR. Remarkably, the metric of a spinning black hole with mass $M$ and angular momentum $J$ can be written in a relatively simple closed form:
\beq
ds^2=-\frac\Delta{\rho^2}(\dd t-a\sin^2\theta\dd\phi)^2 + \frac{\rho^2}\Delta\dd r^2+\rho^2\dd\theta^2+\frac{\sin^2\theta}{\rho^2}(a\dd t-(r^2+a^2)\dd\phi)^2\,,
\label{eqn:Kerr}
\eeq
where $a=J/M$ is the spin parameter, $\Delta=r^2-2Mr+a^2$ and $\rho^2=r^2+a^2\cos^2\theta$. Furthermore, it is useful to define the dimensionless spin parameter $\tilde a=a/M$ and the angular velocity of the event horizon $\Omega_+=a/(2Mr_+)$. The Kerr black hole is significantly more complicated than the Schwarzschild one. First of all, it features two horizons, located at $r_\pm=M\pm\sqrt{M^2-a^2}$, corresponding to the zeros of $\Delta$. The two horizons coincide if $\tilde a=1$, a case known as an \emph{extremal} black hole, and disappear for $\tilde a>1$, which is believed to not be astropyhsically realizable \cite{Penrose:1969pc,Wald:1997wa}. Outside the event horizon $r_+$, the causal structure of the spacetime resembles that of the Schwarzschild black hole. Instead, the inner horizon $r_-$ hides a rich complexity, including a ring-shaped timelike singularity at $\rho=0$, wormholes and so-called ``anti-universes''. For the purposes of this thesis, however, the spacetime at $r<r_+$ is irrelevant and will thus be entirely ignored.

\vskip 0pt
The importance of the Kerr solution is especially highlighted by a classical GR result that goes under the name of \emph{no-hair theorem} \cite{Israel:1967wq,Carter:1971zc,Robinson:1975bv}: \emph{all} neutral black holes are described exclusively by their mass and spin, through the line element \eqref{eqn:Kerr}. There is no other relevant parameter: two neutral BHs with same mass and spin are absolutely identical. This result makes black holes the simplest macroscopic objects known to exists in our Universe, and an exceptionally clean environment to probe the fundamental laws of Nature.

\vskip 0pt
Some of the surprising properties of the Kerr metric are the foundation of the subject of this thesis, so it is worth looking at them in greater detail. The event horizon is not the only interesting place of the spacetime \eqref{eqn:Kerr}: lying just outside it is the \emph{ergosphere}, defined as the region where $g_{00}<0$:
\beq
r<r\ped{erg}=M+\sqrt{M^2-a^2\cos^2\theta}\,.
\eeq
The ergosphere is characterized by several counter-intuitive phenomena. Perhaps the best-known one is that, while objects at $r_+<r<r\ped{erg}$ can still escape the black hole's gravity, they are forced to co-rotate with it. More specifically, this means that any timelike four-velocity of the form $u^\mu=u^t(1,0,0,\Omega)$ must have $\tilde a\Omega>0$.

\vskip 0pt
It is, however, one of the other properties of the ergosphere that turns out to be crucial in our discussion: the Killing vector associated with time translation invariance, $\xi^\mu=\partial_t$, is timelike outside the ergosphere, but becomes spacelike within it. A direct consequence of this fact is that the energy of a particle $E=-\xi^\mu p_\mu$, where $p^\mu$ is the four-momentum, is allowed to be negative inside the ergosphere, according to an observer at infinity. As first noted by Penrose \cite{Penrose:1969pc,Penrose:1971uk}, the black hole can then lose mass by ``eating'' such negative-energy objects, allowing for a mechanism of energy extraction which will be discussed in greater detail in Section~\ref{sec:superradiance}. It is important to realize that such energy extraction cannot happen in an arbitrary way, as the second law of black hole thermodynamics \cite{Bardeen:1973gs} requires the area of the event horizon $A_\slab{bh}=8\pi Mr_+$ to never decrease. The mass of a rotating black hole can thus never become smaller than the mass of a Schwarzschild black hole with equal area: this limit is known as the irreducible mass,
\beq
M\ped{irr}=\sqrt{\frac{M^2+\sqrt{M^4-J^2}}2}\,.
\eeq
The amount of energy extractable from rotating black holes, $M-M\ped{irr}$, is enormous: for an extremal black hole, it is equal to $29\%$ of its rest mass. This phenomenon is at the foundation of some of the most prodigious astrophysical sources (through the Blandford-Znajek process \cite{Blandford:1977ds}), as well as the hypothetical objects studied in this thesis, i.e., gravitational atoms.

\vskip 0pt
Black holes, together with their extraordinary properties, remained a theoretical quirk for a long time. The first indirect observational evidence for black holes came from X-ray binaries \cite{Remillard:2006fc}, which showed the existence of very compact objects with masses of order $\mathcal O(10 M_\odot)$, too high to be neutron stars. Current astronomical observations imply the presence of black holes at the centre of most galaxies, many of them manifesting themselves as quasars: these black holes are typically supermassive, with $M>10^5M_\odot$. In the Milky Way, astronomers have identified a cluster of stars---the S-stars---orbiting an invisible compact object of mass $4.3\times10^6\,M_\odot$, assumed to be a supermassive black hole. The accretion disks around it and around the black hole at the centre of M87 have eventually been directly imaged by the Event Horizon Telescope through a technique known as very-long-baseline interferometry \cite{EventHorizonTelescope:2019dse}.
The most accurate probe of black holes we currently have access to, however, are gravitational waves: gravitational messengers from the reign of strong gravity.

\section{Gravitational waves}

\label{sec:gravitational-waves}

In this section, we review aspects of gravitational-wave physics that are relevant for the thesis. We will not by any means attempt to do justice to this large and fluorishing field, of which a modern and complete overview can be found e.g.~in \cite{Maggiore:2007ulw,Maggiore:2018sht}.

\vskip 0pt
After having introduced GWs (Section~\ref{sec:gw-theory}), we describe their impact on the evolution of binary systems (Section~\ref{sec:binary-inspirals}). Rather than diving into the rich details of the subject, we only lay down explicitly the zeroth-order post-Newtonian results we will use later. We then briefly describe current and future GW detectors (Section~\ref{sec:gw-observations}) and introduce the active research area concerning environmental effects (Section~\ref{sec:environmental-effects}).

\subsection{Linearized gravity}

\label{sec:gw-theory}

Even though Einstein was not able to solve his own equations exactly, he could find approximate solutions in the form of small metric perturbations,
\beq
g_{\mu\nu}=\eta_{\mu\nu}+h_{\mu\nu}\,,
\eeq
where $\eta_{\mu\nu}=\diag(-1,1,1,1)$ is the flat Minkowski metric and $\abs{h_{\mu\nu}}\ll1$. At leading order in $h_{\mu\nu}$, the Einstein equations are
\beq
\partial^\lambda\partial_\mu h_{\nu\lambda}+\partial^\lambda\partial_\nu h_{\mu\lambda}-\Box h_{\mu\nu}-\partial_\mu\partial_\nu h-(\partial^\alpha\partial^\beta h_{\alpha\beta}-\Box h)\eta_{\mu\nu}=16\pi\,T_{\mu\nu}\,,
\label{eqn:linearized-eeq}
\eeq
where we defined $\Box=\partial^\mu\partial_\mu$ and $h=h^\mu{}_\mu$. Equation~\eqref{eqn:linearized-eeq} still appears very complicated, but can be greatly simplified by applying an appropriate infinitesimal coordinate transformation, an operation known as \emph{gauge choice}. In the de Donder gauge, defined as
\beq
\partial^\mu h_{\mu\nu}-\frac12\partial_\nu h=0\,,
\eeq
the field equations reduce to
\beq
\Box\bar h_{\mu\nu}=-16\pi\,T_{\mu\nu}\,,
\label{eqn:sourced-gw}
\eeq
where $\bar h_{\mu\nu}=h_{\mu\nu}-\frac12\eta_{\mu\nu}h$. This is a set of familiar wave equations, with the matter energy-momentum acting as a source for $\bar h_{\mu\nu}$. Where spacetime is vacuum, gravitational waves travel at the speed of light, assuming the form
\beq
\bar h_{\mu\nu}(x)=\Re(H_{\mu\nu}e^{ik_\lambda x^\lambda})\,.
\eeq
The de Donder gauge only requires the polarization tensor $H_{\mu\nu}$ to lie transverse to the wavenumber $k^\mu$, that is, $k^\mu H_{\mu\nu}=0$. We can then exploit the residual gauge freedom to further impose $H_{0\mu}=0$ and $H^\mu{}_\mu=0$, a choice known as the transverse traceless (TT) gauge. For a wave propagating in the $z$ direction, the polarization tensor takes the form
\beq
H_{\mu\nu}=\begin{pmatrix}
0 & 0 & 0 & 0\\
0 & H_+ & H_\times & 0\\
0 & H_\times & -H_+ & 0\\
0 & 0 & 0 & 0
\end{pmatrix}\,,
\eeq
where $H_+$ and $H_\times$ are the amplitudes of the two independent polarizations of the waves.

\vskip 0pt
As for black holes, the interpretation and physical reality of gravitational waves have been the subject of controversies for many decades, partly due to confusion related to the issue of gauge choice in the nonlinear theory. Einstein himself considered GWs to be unphysical~\cite{Einstein:1937qu}. It is now accepted, and experimentally demonstrated, that GWs are physical and carry an energy-momentum density equal to
\beq
\braket{t^{\mu\nu}}=\frac1{32\pi}\braket{\partial^\mu h_{ij}\partial^\nu h_{ij}}.
\label{eqn:gw_tmunu}
\eeq
The average in \eqref{eqn:gw_tmunu} is carried over a suitably large region of spacetime, and is necessary to make $\braket{t^{\mu\nu}}$ gauge-invariant.

\vskip 0pt
In analogy with electromagnetism, gravitational waves are produced by accelerating masses, and their amplitude can be computed by solving \eqref{eqn:sourced-gw} with a retarded Green's function. For sources that move non-relativistically, the spatial components of the metric fluctuation (also known as \emph{strain}) at a large distance $r$ read \cite{Einstein:1918btx}
\beq
\bar h_{ij}\ap{TT}(t,r)=\frac2r\frac{\dd^2Q_{ij}}{\dd t^2}\,,
\label{eqn:htt}
\eeq
where the right-hand side is evaluated at the retarded time $t\ped{ret}=t-r$ and
\beq
Q_{ij}=\int\biggl(x_ix_j-\frac13x^2\delta_{ij}\biggr)T_{00}\dd^3x
\label{eqn:quadrupole-definition}
\eeq
is the mass quadrupole moment of the source. By plugging \eqref{eqn:htt} into \eqref{eqn:gw_tmunu} and averaging over all directions of propagation, we can determine the total power emitted in GWs as
\beq
P=\frac15\Braket{\frac{\dd^3Q_{ij}}{\dd t^3}\frac{\dd^3Q_{ij}}{\dd t^3}}\,.
\label{eqn:quadrupole-general}
\eeq
Equation~\eqref{eqn:quadrupole-general} is popularly known as the quadrupole formula.

\subsection{Binary inspirals}

\label{sec:binary-inspirals}

Any object with a time-varying quadrupole moment generates gravitational waves. Their detection, however, is a formidable challenge due the universal coupling of matter to gravity, and its extreme weakness compared to the other fundamental forces. For this reason, the loudness of a gravitational wave source is among the primary reasons for interest in its study. The loudest and most studied class of astrophysical sources of gravitational waves are binary systems.

\vskip 0pt
Consider two bodies of masses $M$ and $M_*$, with mass ratio $q\equiv M_*/M<1$, orbiting each other circularly at a distance $R_*$. If $R_*\gg M$, relativistic corrections to the orbital motion are small and the orbital frequency can be approximated by Kepler's formula,
\beq
\Omega=\sqrt{\frac{(1+q)M}{R_*^3}}\,.
\eeq
The power emitted by the system in gravitational waves is found by plugging the quadrupole moment \eqref{eqn:quadrupole-definition} of the system into \eqref{eqn:quadrupole-general}:
\beq
P_\slab{gw}=\frac{32}5\frac{q^2M^5(1+q)}{R_*^5}\,.
\label{eqn:p_gw_circular}
\eeq
Assuming that the binary separation changes slowly ($\dd\Omega/\dd t\ll\Omega^2$) and that the orbit remains approximately circular, formula \eqref{eqn:p_gw_circular} predicts that the binary \emph{inspirals} according to
\beq
R_*(t)=R_0\biggl(1-\frac{t}{t_0}\biggr)^{1/4},\qquad t_0=\frac{5R_0^4}{256q(1+q)M^3}\,.
\eeq

\vskip 0pt
As first shown by Peters \cite{Peters:1963ux,Peters:1964zz}, on elliptic orbits with semi-major axis $a$ and eccentricity $\varepsilon$, formula \eqref{eqn:p_gw_circular} generalizes to
\beq
\label{eq:p_gw}
P_\slab{gw}=\frac{32}{5}\frac{q^{2}M^{5}(1+q)}{a^{5}(1-\varepsilon^{2})^{7/2}}\biggl(1 +\frac{73}{24}\varepsilon^{2}+\frac{37}{96}\varepsilon^{4}\biggr)\,,\\
\eeq
where the power is averaged over one full orbit. The average angular momentum per unit time carried away by gravitational waves is instead
\beq
\label{eq:tau_gw}
\tau_\slab{gw}=\frac{32}{5}\frac{q^{2}M^{9/2}\sqrt{1+q}}{a^{7/2}(1-\varepsilon^2)^2}\biggl(1+\frac78\varepsilon^2\biggr)\,.
\eeq
Equations \eqref{eq:p_gw} and \eqref{eq:tau_gw} show that, as the binary inspirals, its eccentricity decreases over time, according to
\beq
\frac1\varepsilon\frac{\dd\varepsilon}{\dd t}=-\frac{304}{15}\frac{q(1+q)M^3}{a^4(1-\varepsilon^2)^{5/2}}\biggl(1+\frac{121}{304}\varepsilon^2\biggr)\,.
\eeq
These predictions have been confirmed for the first time with the discovery by Hulse and Taylor of a binary pulsar system \cite{Hulse:1974eb,Taylor:1982zz}. Its orbital decay has been accurately measured through radio observations, and is in excellent agreement with the predictions of General Relativity.

\vskip 0pt
Formula \eqref{eqn:p_gw_circular} clearly shows that the more compact the binary, the stronger the GW emission. Although the Universe is populated with a large number of binary systems, only those involving small enough objects, such as black holes and/or neutron stars, can be compact enough to emit an amount of gravitational waves that is directly detectable. Smaller separations, however, severely complicate the treatment of the inspiral problem, because they probe the nonlinear nature of General Relativity. The assumptions of slow velocities and weak gravity, that we used above, break down as the inspiral of two compact objects proceeds to to higher frequencies. The time evolution of critical quantities, such as the orbital frequency and phase, can then be expanded in a power series of the velocity $v$.\footnote{The use of a single expansion parameter, such as the orbital velocity, is possible when modelling inspiralling binaries because they are held together by gravitational forces. In more general cases, the expansions powers of $v$ and of Newton's constant $G$ (the latter approach being known as \emph{post-Minkowskian} expansion) can be studied independently.} Terms of order $v^{2n}$ are said to be of $n$-th \emph{post-Newtonian} (PN) order. The phase evolution of binaries involving spinless objects on circular orbits is currently known up to the $4.5$PN order \cite{Blanchet:2023bwj}.

\vskip 0pt
The PN expansion is not the only perturbative approach to the problem of binary inspiral. Instead of organizing the series in powers of the velocity $v$, one may take the mass ratio $q$ as the expansion parameter. In the test particle limit ($q\to0$), the smaller object moves along a geodesic in the spacetime sourced by the mass of the larger body. At finite-$q$ order, the gravitational waves emitted by the smaller object backreact on its motion, causing it to inspiral. The \emph{self-force} program \cite{Barack:2018yvs} aims to compute such a \emph{post-adiabatic} expansion, which is expected to be the approach of choice for extreme mass ratio inspirals (EMRIs), which we can roughly define as $q<10^{-4}$, and intermediate mass ratio inspirals (IMRI), for which $10^{-4}<q<10^{-2}$.

\vskip 0pt
Yet another approach is the Effective One Body (EOB) formalism, where the binary dynamics is mapped to the motion of a test particle in an effective metric~\cite{Buonanno:1998gg,Buonanno:2000ef}. This method requires calibrating free coefficients, which is done by matching to the results of numerical relativity computations.

\vskip 0pt
The inspiral ends with the merger, or \emph{coalescence}, of the two compact objects. At this point, the dynamics of the system can no longer be described by any of the previously mentioned approaches, and numerical relativity techniques are required to model this phase \cite{Pretorius:2005gq}. If a new black hole is created during the merger, it will then undergo a phase of \emph{ringdown}, during which it will shed its asymmetries in GWs, and eventually relax to a stationary, axisymmetric Kerr black hole. Analytical techniques become once again useful, as the frequency of the \emph{quasi-normal modes} excited during the ringdown can be computed within the framework of black hole perturbation theory.

\subsection{Current and future observations}

\label{sec:gw-observations}

The first ever detection of gravitational waves, conducted by the LIGO interferometers \cite{LIGOScientific:2014pky}, occurred in September 2015 \cite{LIGOScientific:2016aoc}. Since then, more than 100 other events have been confirmed, most of them originating from black hole binaries. A minority of them, such as GW170817 \cite{LIGOScientific:2017vwq}, comes instead from binary neutron star (NS) mergers, or BH-NS systems.  Gravitational-wave astronomy has rapidly drawn the attention of the astrophysics and theoretical physics communities, establishing itself as an innovative and promising field.

\begin{figure}[t]
\centering
\includegraphics{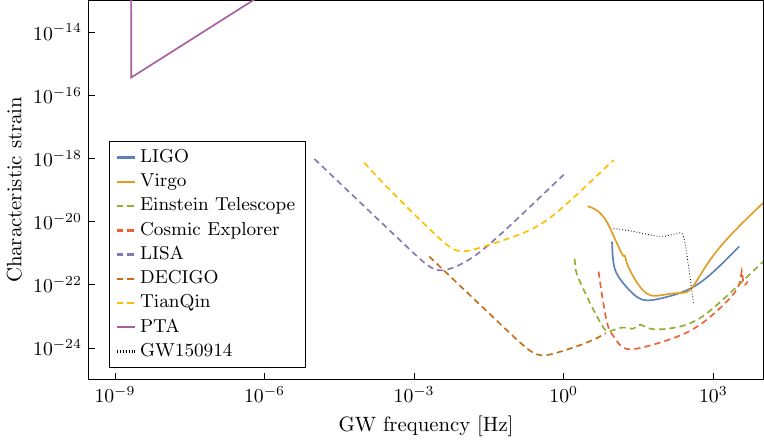}
\caption{Sensitivity curves of current (solid lines) and future (dashed lines) gravitational-wave detectors. The strain of GW150914 is also shown as a thin dotted line. The curves have been generated using data taken from \cite{GWplotter}.}
\label{fig:gw-sensitivities}
\end{figure}

\vskip 0pt
The Livingston and Hanford LIGO observatories have then been joined by Virgo~\cite{VIRGO:2014yos} and KAGRA~\cite{KAGRA:2020tym}. These instruments are sensitive to frequencies between $\SI{10}{Hz}$ and $\SI{e+3}{Hz}$, roughly corresponding to mergers of compact objects of $1$ to $100$ solar masses. Other observatories, currently operational or planned, explore different frequency windows. The Einstein Telescope \cite{Punturo:2010zz} and Cosmic Explorer \cite{Reitze:2019iox} are planned to improve the sensitivity at the same frequencies as LIGO-Virgo-KAGRA, while also expanding the band towards lower frequencies. LISA \cite{LISA:2017pwj} is instead a future space-borne detector with planned sensitivity in the millihertz band, $\SI{e-4}{Hz}$--$\SI{e-1}{Hz}$. Astrophysical sources of gravitational waves at these frequencies include massive black hole binaries, EMRIs, and white dwarf binaries in our galaxy. Proposed interferometers with similar capabilities include TianQin \cite{TianQin:2015yph} and Taiji \cite{Ruan:2018tsw}. Decihertz detectors, such as DECIGO \cite{Kawamura:2011zz} and TianGO \cite{Kuns:2019upi}, are expected to bridge the gap between millihertz and ground-based instruments. Finally, the nanohertz window is currently being scrutinized by various Pulsar Timing Arrays (PTA) such as NANOGrav, which in 2023 announced the first evidence for a stochastic GW background \cite{NANOGrav:2023gor}. The sensitivity curves of many of these detectors as function of the GW frequency are shown in Figure~\ref{fig:gw-sensitivities}.

\vskip 0pt
Such an anormous interest in GW astronomy is justified by its potential to convey information about astrophysics and fundamental physics that has so far been beyond the reach of experiments. Ground-based observations have already allowed to perform tests of General Relativity, such as by measuring the BH quasi-normal mode frequencies \cite{LIGOScientific:2016lio,LIGOScientific:2019fpa} and the speed of gravitational waves \cite{LIGOScientific:2017zic}, and to learn about the internal structure of neutron stars \cite{LIGOScientific:2018cki}, the mass distribution of BHs \cite{LIGOScientific:2018mvr,LIGOScientific:2020kqk,LIGOScientific:2021usb,KAGRA:2021vkt}, and several other important astrophysical implications. This thesis focuses on one avenue through which future GW observations could bring about new discoveries in fundamental physics: black hole environments.

\subsection{Environmental effects}

\label{sec:environmental-effects}

Detection and analysis of gravitational waves has so far been carried out under the assumption that the inspiralling binaries are in vacuum. This hypothesis is well-justified \cite{Macedo:2013qea,Barausse:2014tra,Barausse:2014pra} for the kind of events detected by LIGO-Virgo-KAGRA, which typically involve stellar-mass black holes and comparable mass binaries. The main reasons for this include:
\begin{itemize}
\item the shortness of the signals, which usually only last a few cycles and thus do not leave room for environmental effects build up over time;\footnote{By far the longest-lasting signal so far has been GW170817, which covered approximately 3000 cycles and indeed resulted in the most stringent constraints on the environmental density \cite{CanevaSantoro:2023aol}.}
\item the high GW frequencies, corresponding to the late inspiral and merger phases, during which gravity becomes by far the strongest force at play (environmental effects are instead expected to be relevant earlier in the inspiral);
\item the lack of a strong theoretical case for stellar-mass objects to live in dense astrophysical environments;
\item the near-comparable masses of the objects: only a small-mass secondary object is able to inspiral and merge without sweeping the environment away in the process.
\end{itemize}

\vskip 0pt
All these aspects will however change with future GW detectors. Environments such as accretion disks are expected to be found around intermediate and supermassive BHs, and they can influence the inspiral dynamics in a variety of ways. For example, one of the most prominent effects is \emph{dynamical friction}\footnote{Specifically in accretion disks, gas torques rather than dynamical friction are actually expected to be the leading force \cite{Goldreich:1980wa,Derdzinski:2020wlw}.} \cite{Chandrasekhar:1943ys,1943ApJ....97..263C,1943ApJ....98...54C}, which is a gravitational drag force acting on a body that moves through a dense medium. More specifically, the power lost to dynamical friction is
\beq
P_\slab{df}=\frac{4\pi M_*^2\rho}{v}\log\Lambda\,,
\label{eqn:Pdf-intro}
\eeq
where $M_*$ is the mass of the moving object, $\rho$ is the asymptotic density of the medium, $v$ is the velocity of the object with respect to the medium, and $\Lambda$ is an infrared regulator related to the size of the medium. Even a small effect, $P_\slab{df}\ll P_\slab{gw}$, can become measurable if a long signal is detected, as it induces the system to accumulate a \emph{dephasing} with respect to the vacuum waveform. This kind of precision tests is expected to be achievable with millihertz detectors: as an example, LISA is supposed to deliver year-long waveforms.

\vskip 0pt
Environmental effects can be at the same time a curse and a blessing. If not appropriately recognized, they can introduce biases in the inferred binary parameters or, even worse, degradate the signal-to-noise ratio to the point where the event is missed altogether. Conversely, if accurately modelled, they can offer a precious insight into previously inaccessible astrophysical environments. Being able to correctly model, detect and distinguish environmental effects will be one of the new challenges of future GW astrophysics \cite{Cole:2022yzw}.

\vskip 0pt
Perhaps the primary reason for the interest in environmental effects is their relation to questions of fundamental physics. While a certain number of sources are expected to carry ``mundane'' environments, such as accretion disks, there are also proposals for more exotic and interesting scenarios. For example, intermediate or supermassive BHs at the centre of galactic haloes might be surrounded by a dense \emph{spike} of dark matter (DM), which forms as a consequence of the mass growth of the BH \cite{Gondolo:1999ef,Ullio:2001fb,Bertone:2024wbn}. The inspiral of a smaller object around the massive BH will then be affected by the interaction with the DM, through dynamical friction and other effects, which take energy away and thus speed up the inspiral. The observational impact of this process has been studied in a large number of papers \cite{Eda:2013gg,Eda:2014kra,Kavanagh:2020cfn,Coogan:2021uqv,Cole:2022ucw,Karydas:2024fcn,Kavanagh:2024lgq,Becker:2021ivq,Becker:2022wlo,Montalvo:2024iwq}, motivated by the hope that GWs can provide us with a window on new physics. The spirit of the present thesis is in many ways similar to these works, but focuses on a different environment, composed of a specific kind of hypothetical particles.

\section{Ultralight scalars}

\label{sec:ultralight-scalars}

We now take a small but necessary digression into some models of particle physics beyond the Standard Model. These form the core of the new physics that this thesis aims at discovering, and so it is important to be aware of them and their motivations, before putting a lot of work into phenomenological implications.

\vskip 0pt
The Standard Model (SM) of particle physics is a non-Abelian gauge theory with symmetry group $\text{U}(1)\times\text{SU}(2)\times\text{SU}(3)$. It has had extraordinary experimental success, to the point of being able to explain virtually any particle physics experiment ever performed. At the same time, while the SM does not appear to have any internal inconsistency, it comes with a small number of theoretical puzzles which are widely believed to be hints of a more complete theory. For example, the theory contains a large number of independent parameters, 19 to be precise, which require to be experimentally measured. The need for so many parameters to be put in by hand, without having any explanation for their values, is not intellectually satisfactory. But more concrete problems arise due to the seemingly unnatural values taken by some of these parameters. One famous example is the hierarchy problem. The only force not described by the SM, gravity, has a characteristic energy scale of $M\ped{Pl}\sim\SI{e+19}{GeV}$; this is much higher than the highest energy scale of the SM, the electroweak scale $\Lambda_\slab{ew}\sim\SI{e+2}{GeV}$. In particular, the mass of a scalar particle like the Higgs boson is sensitive to quantum corrections coming from all energy scales, and is thus expected to be of order $M\ped{Pl}$, rather than $\Lambda_\slab{ew}$ as seen in experiments. This is a problem of \emph{fine tuning}: the parameters of Nature seem to have values so unnatural from a theoretical perspective that scientists feel the need for a deeper theoretical explanation.

\vskip 0pt
A conceptually similar problem also arises in the strong sector of the SM, which corresponds to the $\text{SU}(3)$ group. One of the 19 parameters is the QCD vacuum angle $\theta_\slab{qcd}$, which appears as a coefficient of the CP-violating term
\beq
\mathcal L_\slab{sm}\supset\theta_\slab{qcd}\frac{g_s^2}{32\pi^2}G^a_{\mu\nu}\tilde G^{\mu\nu}_a\,,
\eeq
where $g_s$ is the $\text{SU}(3)$ gauge coupling, $G^a_{\mu\nu}$ is the gluon field strength and $\tilde G^a_{\mu\nu}$ is its dual. The value of $\theta_\slab{qcd}$ is constrained by measurements of the neutron electric dipole moment to be $\abs{\theta_\slab{qcd}}<10^{-10}$.\footnote{To be precise, the constraint holds for the sum of $\theta\ped{QCD}$ and the argument of the determinant of the quark mass matrix. We ingore this subtlety here.} Such an tiny value is puzzling because no SM symmetry is restored when $\theta_\slab{qcd}\to0$, and the issue is known as the strong CP problem. The most popular solution is the Peccei-Quinn (PQ) mechanism \cite{Peccei:1977hh,Weinberg:1977ma,Wilczek:1977pj}, where $\theta_\slab{qcd}$ is promoted to a dynamical field $a$, and the theory is assumed to be invariant under the PQ symmetry $a\to a+\text{const}$, which is spontaneously broken at low energies. The corresponding Nambu-Goldstone boson, called the \emph{axion}, obtains a nonzero mass because topological QCD configurations explicitly break the PQ symmetry. Furthermore, it acquires a nonzero vacuum expectation value that naturally cancels $\theta_\slab{qcd}$, thus solving the strong CP problem. From the low-energy effective action for the axion, its mass is found to be
\beq
\mu\sim\SI{6e-10}{eV}\biggl(\frac{\SI{e+16}{GeV}}{f_a}\biggr)\,,
\label{eqn:axion-mass}
\eeq
where $f_a$ is the axion decay constant. Because $f_a$ also dictates the strength of the axion's interactions with other particles, the mass of the axion must be very light (not above the $\si{keV}$ scale) to avoid experimental bounds. The interactions and phenomenology of the axion are slightly model-dependent \cite{Kim:1979if,Shifman:1979if,Dine:1981rt,Zhitnitsky:1980tq,DiLuzio:2020wdo}, but for our purposes all it matters is that it is a very light boson.

\vskip 0pt
Pseudoscalar fields qualitatively similar to the QCD axion generically arise in string theory \cite{Arvanitaki:2009fg}. The reason is that the compactification of extra dimensions leads to Kaluza-Klein (KK) zero modes, a large number of which (more than 100) is expected due to the topological complexity of six-dimensional Calabi-Yau manifolds. The potential of each KK zero mode then receives contributions from a variety of nonperturbative effects, such as gauge theory instantons as in the case of the QCD axion, endowing these particles with a nonzero mass. We thus see that not only the QCD axion fits well within string theory, but a more natural expectation would be to have many \emph{axion-like} particles (ALPs), as it would otherwise be very surprising that only one of them remains light. This scenario is known as the \emph{string axiverse}~\cite{Arvanitaki:2009fg}. These ALPs do not follow a relation like \eqref{eqn:axion-mass}, but their low-energy effective action is still described by the same two parameters, the mass $\mu$ and decay constant $f_a$. The starting point is an action of the kind
\beq
\mathcal L=-\frac{f_a^2}2\partial_\mu a\partial^\mu a-\Lambda^4V(a)\,,
\eeq
where $V(a)$ is a periodic potential, such as $V(a)=1-\cos a$. Because the potential is attained through nonperturbative (e.g.~instantonic) contributions, the energy scale $\Lambda$ is of the form $\Lambda^4\sim M\ped{Pl}^2\Lambda_\slab{s}^2e^{-S}$ \cite{Arvanitaki:2009fg,Hui:2016ltb}, where $\Lambda_\slab{s}$ measures a possible suppression of instanton effects due to supersymmetry and can vary over a wide range, from $\SI{e+4}{GeV}$ to $M\ped{Pl}$. After a field redefinition $\Phi=f_aa$, the Lagrangian can be expanded as
\beq
\mathcal L=-\frac12\partial_\mu\Phi\partial^\mu\Phi-\frac{\mu^2}2\Phi^2+\frac1{4!}\frac{\mu^2}{f_a^2}\Phi^4\,,
\eeq
where $\mu=\Lambda^2/f_a$ is the mass of the axion-like particle. An instanton action of about $S\gtrsim200$ can easily lead to an axion much lighter than any SM particle. The exponential dependence of $\mu$ on $S$ also makes it natural to expect the multitude of axions to have masses homogeneously distributed in logarithmic, rather than linear, scale.

\vskip 0pt
ALPs of the kind described above also enjoy a strong phenomenological motivation, as their feeble interactions make them excellent dark matter (DM) candidates. They fall in the class of \emph{wave dark matter}: when an ALP is assumed to have an abundance matching that of DM, the average interparticle separation must be much smaller than the de Broglie wavelength. The axion field appears then as a fluid, rather than a collection of distinct particles~\cite{Hui:2021tkt}. A particularly influential proposal belonging to this class has been that of \emph{fuzzy DM}~\cite{Hu:2000ke}, where a DM mass of $\mu\sim\SI{e-22}{eV}$ is assumed to resolve small-scale difficulties arising within the conventional particle DM scenario. While that mass window is now severely constrained by Lyman-$\alpha$ forest data~\cite{Irsic:2017yje,Rogers:2020ltq}, the ultralight DM paradigm remains alive and attractive. Among its appealing properties is the ability to relieve some of the known issues of particle DM models, such as the missing satellite~\cite{Klypin:1999uc}, cusp-core~\cite{Flores:1994gz} and ``too big to fail'' problems~\cite{Boylan-Kolchin:2011qkt}. Wave DM also exhibits interesting and testable phenomenology, such as the formation of solitons, vortices and interference patterns~\cite{Hui:2020hbq}.

\vskip 0pt
Even though all models of ultralight scalars involve a nonzero degree of self-interactions ($1/f_a\ne0$), in the rest of the thesis we will ignore this altogether. This is a simplification that can certainly be relaxed in future works. It is also worth mentioning that many searches for ultralight bosons rely on their direct, although feeble, couplings to SM particles, such as the axion-photon interaction $g_{a\gamma}aF_{\mu\nu}\tilde F^{\mu\nu}$, where $F_{\mu\nu}$ is the electromagnetic field strengh. These couplings are also ignored here, as only the gravitational phenomenology is studied.

\section{Black hole superradiance}

\label{sec:superradiance}

The fundamental mechanism that allows us to probe ultralight fields with gravitational waves is \emph{superradiance}. While the term can refer to several phenomena of wave amplification in various areas of physics, we focus here only on rotational black hole superradiance. In this section, we present a short overview of the mechanism, referring to \cite{Brito:2015oca} for a more comprehensive review.

\vskip 0pt
In Section~\ref{sec:black-holes}, we mentioned that the event horizon of a Kerr black hole is surrounded by the ergosphere, a region within which negative-energy objects are allowed to exist. The energy of a particle is, however, constant along the geodesic it traces during its motion, so a negative energy can only be achieved if some interactions are taking place inside the ergosphere, or if the particle is created therein. A gedanken experiment proposed by Penrose in 1971 \cite{Penrose:1971uk} thus involves a particle with energy $E>0$ which is sent in from infinity and decays inside the ergosphere into two particles, with energies $E_A$ and $E_B$, with $E=E_A+E_B$. Suppose now that $E_A<0$. While particle $B$ can leave the ergosphere, particle $A$ cannot, as no place in the outer spacetime allows it to have a negative energy. Hence, if particle $B$ escapes, while particle $A$ falls into the black hole, we have effectively extracted mass from the black hole, as the outcoming particle has more energy than the incoming one: $E_B=E-E_A>E$. At the same time, the black hole has lost a net amount of mass.

\vskip 0pt
The gedanken experiment described above has a classical field theory analog. Consider a real scalar field $\Phi$ with mass $\mu$, minimally coupled to gravity, in a Kerr background. The field obeys the Klein-Gordon equation of motion,
\beq
(\Box-\mu^2)\Phi=0\,.
\label{eqn:klein-gordon}
\eeq
Remarkably, the simple ansatz
\beq
\Phi=R(r)S(\theta)e^{im\phi}e^{-i\omega t}
\label{eqn:Phi-ansatz}
\eeq
separates equation \eqref{eqn:klein-gordon}, which reduces to two ordinary differential equations for the functions $R$ and $S$:
\beq
\Delta\frac\dd{\dd r}\biggl(\Delta\frac{\dd R}{\dd r}\biggr)+\bigl(\omega^2(r^2+a^2)^2-4maM\omega r-\mu^2r^2\Delta+m^2a^2-(\omega^2a^2+\lambda)\Delta\bigr)R=0\,,
\label{eqn:R}
\eeq
\beq
\frac1{\sin\theta}\frac\dd{\dd\theta}\biggl(\sin\theta\frac{\dd S}{\dd\theta}\biggr)+\biggl(\lambda+a^2(\omega^2-\mu^2)\cos^2\theta-\frac{m^2}{\sin^2\theta}\biggr)S=0\,,
\label{eqn:S}
\eeq
where $\lambda$ is a separation constant. Both \eqref{eqn:R} and \eqref{eqn:S} are classified as confluent Heun equations, meaning that they are obtained from an ordinary linear differential equation with four regular singular points, upon confluence of two of these singluarities, see~\cite{ronveaux1995heun,Fiziev:2009kh} and Appendix~\ref{app:heunc}.

\vskip 0pt
It is useful to introduce the \emph{tortoise coordinate}
\beq
r_*=r+\frac{2M}{r_+-r_-}\biggl(r_+\log\abs*{\frac{r-r_+}{r_+-r_-}}-r_-\log\abs*{\frac{r-r_-}{r_+-r_-}}\biggr)\,,
\eeq
which generalizes the one given in \eqref{eqn:v-r_*-finkelstein} for the Schwarzschild black hole and sends the event horizon $r=r_+$ to $r_*\to-\infty$. Equation \eqref{eqn:R} can then be written in the form
\beq
\frac{\dd^2R}{\dd r_*^2}+V\ped{eff}(r_*)R=0\,.
\eeq
Assuming that $V\ped{eff}$ tends asymptotically to $k\ped{H}^2$ for $r_*\to-\infty$ and to $k_\infty^2$ for $r_*\to+\infty$, the asymptotic solutions are\footnote{In reality, $V\ped{eff}$ does not fall off fast enough at $r\to\infty$ to write the solution in the form \eqref{eqn:asymptotic-scattering}. The actual solutions are not exponentials, but spherical Bessel functions, $R\sim\mathcal Ie^{-ik_\infty r_*}/r+\mathcal Re^{ik_\infty r_*}/r$. For the purpose of our schematic discussion, however, this technicality is not important.}
\beq
R(r_*)\sim\begin{cases}
\mathcal Te^{-ik\ped{H}r_*}+\mathcal Oe^{i\ped{H}r_*} & r_*\to-\infty\,,\\
\mathcal Ie^{-ik_\infty r_*}+\mathcal Re^{ik_\infty r_*} & r_*\to+\infty\,.
\end{cases}
\label{eqn:asymptotic-scattering}
\eeq
These represent an ``incoming'' wave ($\mathcal I$) from infinity, a ``reflected'' wave ($\mathcal R$) that moves towards infinity, a ``transmitted'' wave ($\mathcal T$) that goes into the horizon, and an ``outgoing'' wave ($\mathcal O$) from the horizon. Causal boundary conditions require $\mathcal O=0$. Because the Wronskian $(\dd R/\dd r_*)R^*-(\dd R^*/\dd r_*)R$ is independent of $r_*$, we get
\beq
\abs{\mathcal R}^2=\abs{\mathcal I}^2-\frac{k\ped{H}}{k_\infty}\abs{\mathcal T}^2\,.
\eeq
If the ratio $k\ped{H}/k_\infty$ is negative, the reflected wave can then have an amplitude that is larger than the incoming one, an effect known as \emph{superradiant amplification}.

\vskip 0pt
We now determine the condition for which $k\ped{H}/k_\infty<0$. At $r_*\to\infty$, we simply have $k_\infty=\sqrt{\omega^2-\mu^2}>0$. Close to $r=r_+$ instead, \eqref{eqn:R} can be written as
\beq
\frac{\dd^2R}{\dd r_*^2}+(\omega-m\Omega_+)^2R=0\,,
\eeq
and we thus conclude that superradiance occurs whenever the following condition is satisfied,
\beq
\omega<m\Omega_+\,.
\label{eqn:superradiance-condition}
\eeq
A similar derivation holds for all bosonic fields, but not for fermionic ones, as Pauli's exclusion principle forbids the creation of new particles in a state already occupied by existing ones. Kerr black holes thus have the ability to increase the amplitude of waves that scatter off them, if their frequency is smaller than the angular velocity of the horizon (multiplied by the azimuthal number $m$).

\vskip 0pt
While remarkable, this property would not be of much astrophysical interest per se. It was however noticed, by Press and Teukolsky \cite{Press:1972zz}, that if the field waves were somehow reflected back towards the black hole by a mirror, a positive-feedback mechanism could kick in and exponentially intensify the energy extraction. The mirror was envisioned to be a future technological wonder, but the authors also wrote that
\begin{quote}
``Others may care to speculate on the possibility that nature provides her own mirror.''
\end{quote}
This might, in fact, be the case. Fields with mass $\mu\ne0$ can exist in states that are gravitationally bound to the black hole. Making an analogy with the motion of particles, massive waves can be ``reflected'' at the outer inversion point of their orbit, allowing them to undergo continuous superradiant scattering and initiate a runaway process.

\vskip 0pt
The same conclusion is reached by solving equations \eqref{eqn:R} and \eqref{eqn:S} with ingoing boundary conditions at the horizon and a bound-state-like exponential decay of $R$ at infinity. Equivalently, we require $\mathcal O=\mathcal I=0$ in \eqref{eqn:asymptotic-scattering}. These two boundary conditions can only be satisfied on a discrete set of frequencies \cite{Dolan:2007mj},
\beq
\omega_{n\ell m}=E_{n\ell m}+i\Gamma_{n\ell m}\,.
\label{eqn:omega-E-Gamma}
\eeq
The imaginary part $\Gamma_{n\ell m}$ of the frequency determines the exponential growth or decay of the field, according to \eqref{eqn:Phi-ansatz}. Using a matched asymptotic expansion, where the near-horizon and far-field solutions of \eqref{eqn:R} are glued together in an intermediate region, one can find an approximation for the growth or decay rate \cite{Detweiler:1980uk,Baumann:2019eav}:
\beq
\Gamma_{n\ell m}=\frac{2r_+}MC_{n\ell}g_{\ell m}(m\Omega_+-\omega_{n\ell m})(\mu M)^{4\ell+5}\,,
\label{eqn:Gamma_nlm}
\eeq
where
\begin{align}
C_{n\ell}&=\frac{2^{4\ell+1}(n+\ell)!}{n^{2\ell+4}(n-\ell-1)!}\biggl(\frac{\ell!}{(2\ell)!(2\ell+1)!}\biggr)^2\,,\\
g_{\ell m}&=\prod_{k=1}^\ell\Bigl(k^2(1-\tilde a^2)+(\tilde am-2r_+\omega_{n\ell m})^2\Bigr)\,.
\end{align}
As expected, \eqref{eqn:Gamma_nlm} is positive when \eqref{eqn:superradiance-condition} is satisfied and negative otherwise.

\begin{figure}[t]
\centering
\includegraphics{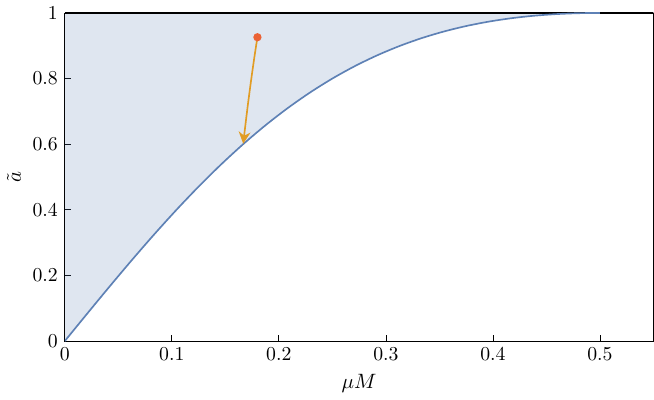}
\caption{Regge plane of black holes. The blue shaded area corresponds to the region where $m=1$ modes of a field with mass $\mu$ are superradiantly unstable, i.e, \eqref{eqn:superradiance-condition} is satisfied. The red circle denotes an example of possible initial parameters of a black hole. As the scalar field is amplified, the black hole loses mass and angular momentum, moving along the orange arrow. When the threshold of the instability region is reached, superradiance for $m=1$ modes shuts down.}
\label{fig:regge}
\end{figure}

\vskip 0pt
Superradiance can thus manifest itself as an instability of the Kerr solution against scalar field perturbations. A similar conclusion holds more generally for any boson field, but in this thesis we focus on scalars due to their simplicity and stronger theoretical motivation. To visualize the region of parameter space where the instability takes place, it is useful to introduce the so-called \emph{Regge plane} (by analogy with the Regge trajectories in QCD), where the $x$ and $y$ axes correspond to the mass and normalized spin of the black hole, as in Figure~\ref{fig:regge}. The condition \eqref{eqn:superradiance-condition} corresponds to the region above a certain \emph{threshold} spin. As superradiance extracts energy and angular momentum from the black hole to feed the scalar \emph{cloud}, the former moves in the Regge plane until it hits the threshold. At that point, the instability is saturated and the cloud stops growing.

\section{Gravitational atoms}

\label{sec:gravitational-atoms}

The endpoint of the superradiant instability is a meta-stable cloud of ultralight bosons around the central black hole. The system is also referred to as a \emph{gravitational atom}, for its resemblance with the familiar proton-electron system. The analogy is not merely aesthetic, as the rest of this section, as well as many parts of this thesis, will show. For the time being, inspired by this similarity, let us define the \emph{gravitational fine structure constant} as
\beq
\alpha\equiv\mu M\,.
\label{eqn:alpha}
\eeq
While $\alpha\ped{em}\approx1/137$ determines the strength of the proton-electron electrostatic attraction, $\alpha$ as in \eqref{eqn:alpha} determines the strength of the BH-boson gravitational attraction.

\subsubsection{Formation, mass and lifetime}

There are a few basic questions one might naturally ask about the gravitational atom, such as: How much time does the cloud take to form? How massive is it? What is its lifetime? Let us address these questions. Since the mass density carried by the scalar field is $T_{00}\sim2\mu^2\Phi^2,$ the $e$-folding time of the mass growth is $1/(2\Gamma_{n\ell m})$. Starting from a quantum fluctuation, the scalar field grows by a number of $e$-folds of about 175 before saturating the instability \cite{Hui:2022sri}. The largest possible value of $\Gamma_{n\ell m}$ is attained for $(n,\ell,m)=(2,1,1)$, $\alpha=0.42$ and a near-extremal BH, and equals about $\Gamma_{211}=1.5\times10^{-7}\mu$ \cite{Dolan:2007mj}. Together, these facts imply a growth time as short as a few hours for stellar mass black holes, and $10^5$--$10^6$ years for supermassive ones ($M\sim10^9M_\odot$). Different values of $\alpha$, or modes other than $(2,1,1)$, will lead to longer growth times, which can easily exceed the age of the Universe if the ``quantum numbers'' $(n,\ell,m)$ are increased. For small enough $n$, modes with $n-1=\ell=m$ are the fastest-growing ones for any given value of $m$.

\vskip 0pt
The total mass of the cloud can be computed by conservation of mass and angular momentum. By approximating $\omega\approx\mu$ in \eqref{eqn:superradiance-condition}, one can show that a black hole with initial mass $M$ and initial spin $\tilde a$ creates a cloud of mass \cite{East:2017ovw,Herdeiro:2021znw,Hui:2022sri}
\beq
M\ped{c}=M-\frac{m}{\mu}\frac{1-\sqrt{1-\tilde a'^2}}{\tilde a'}\,,\qquad\text{where}\qquad \tilde a'=4\biggl(\frac{\mu M}m\biggr)\biggl(1-\frac{\mu M\tilde a}m\biggr)\,.
\label{eqn:cloud-mass}
\eeq
From \eqref{eqn:cloud-mass}, the highest achievable cloud-to-BH mass ratio is found to be
\beq
\frac{M\ped{c}}{M-M\ped{c}}=10.8\%\,,
\eeq
for $\mu M\approx0.24$ and $\tilde a=1$.

\vskip 0pt
The lifetime of the cloud depends on the processes that can deplete it, or change its state. For instance, if the scalar field $\Phi$ is real, then the cloud has a time-varying quadrupole moment due to terms of the kind $T_{00}\sim\cos^2(\mu t)$ in its energy-momentum tensor.\footnote{Clouds of complex scalar fields instead have $T_{00}\sim\Phi^*\Phi\sim e^{i\mu t}e^{-i\mu t}$, which is time-independent. Hence, they do not emit gravitational waves.} As a consequence, the cloud releases approximately monochromatic gravitational waves at the frequency $2\mu$, a process that can be understood as annihilations of pairs of scalar particles into a graviton. The quadrupole formula \eqref{eqn:quadrupole-general} does not apply here, because the wavelength is much smaller than the size of the cloud. However, the GW luminosity can be computed analytically in a flat-space approximation, and leads to the following law for the cloud's mass decay~\cite{Yoshino:2013ofa,Brito:2017zvb},
\beq
M\ped{c}(t)=\frac{M\ped{c}(0)}{1+t/t\ped{c}}\,,
\eeq
where
\beq
t\ped{c}^{-1}=C_{n\ell}\frac{M\ped{c}(0)}{M^2}\alpha^{4\ell+10}\,,\qquad C_{n\ell}=\frac{16^{\ell+1}\ell(2\ell-1)(2\ell-2)!^2(\ell+n)!^2}{n^{4\ell+8}(\ell+1)\ell!^4(4\ell+2)!(n-\ell-1)!^2}\,.
\label{eqn:tgwc}
\eeq
Numerical results \cite{Yoshino:2013ofa} suggest that \eqref{eqn:tgwc} overestimates $t\ped{c}$ by a factor of $\sim10$ for $\alpha\lesssim0.2$, and underestimates $t\ped{c}$ for $\alpha\gtrsim 0.24$, up to a few orders of magnitude. By comparing the $\alpha$-dependence of \eqref{eqn:tgwc} and \eqref{eqn:Gamma_nlm}, we conclude that the cloud's growth is always much faster than its decay through GWs. Depending on the value of $\alpha$, however, $t\ped{c}$ can still be a relatively short timescale. For example, for $(n,\ell,m)=(2,1,1)$, equation \eqref{eqn:tgwc} gives
\beq
t\ped{c}\sim\SI{e+6}{yrs}\,\biggl(\frac{M}{10^4M_\odot}\biggr)\biggl(\frac{M\ped{c}/M}{0.01}\biggr)^{-1}\biggl(\frac\alpha{0.2}\biggr)^{-14}\,.
\eeq
We stress again that this formula loses accuracy for larger values of $\alpha$.

\vskip 0pt
It is important to realize that this steady depletion results in a polynomial, rather than exponential, decay. Furthermore, $t\ped{c}$ has a similar $\alpha$-scaling as the instability rate of the ``next'' superradiant state, with $\ell$ and $m$ increased by one unity. Hence, it is often the case that, if the cloud has had enough time to decay in GWs, then a cloud in a new state has also had the time to form. It is also worth mentioning that other effects can limit the cloud's mass, such as the scalar field's self-interactions \cite{Baryakhtar:2020gao}. For all these reasons, in this thesis we simply treat $M\ped{c}$ as an independent parameter, rather than linking it to the BH parameters or imposing a sharp bound on $\alpha$.

\subsubsection{Schrödinger equation}

Having established that gravitational atoms can form and remain stable over astrophysical timescales, we now describe their physical properties. The backreaction of the cloud on the geometry is of order $\mathcal O(M\ped{c}/M)$. We may thus ignore it and solve the unbackreacted Klein-Gordon equation \eqref{eqn:klein-gordon} on the Kerr background. For $\alpha\ll1$, corresponding to a cloud with most of its support in a region where relativistic effects are small, it is useful to adopt the following ansatz,
\beq
\Phi=\frac1{\sqrt{2\mu}}\biggl(\psi(t,\vec r)e^{-i\mu t}+\text{c.c.}\biggr)\,,
\eeq
which effectively trades $\Phi$ for a nonrelativistic \emph{wavefunction} $\psi$. Upon substitution in \eqref{eqn:klein-gordon}, we are left with a Schrödinger equation for $\psi$,
\beq
i\frac{\partial\psi}{\partial t}=\biggl(-\frac{\nabla^2}{2\mu}-\frac\alpha{r}+\mathcal \ldots\biggr)\psi\,.
\label{eqn:schrodinger}
\eeq
The leading term of the potential appearing in \eqref{eqn:schrodinger} is the familiar Coulomb (or Newton) attraction, because at large distances from the BH the $1/r$ tail dominates over all relativistic corrections. Subleading terms involve higher powers of $\alpha$ and $1/r$.

\begin{figure}[t]
\centering
\includegraphics[width=0.8\textwidth]{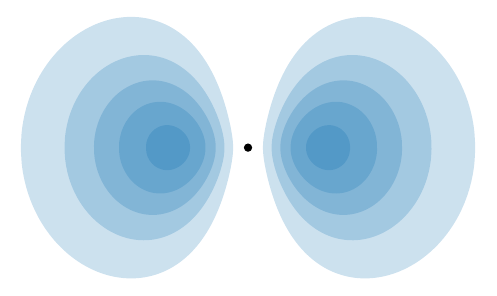}
\caption{Equatorial cross section of a gravitational atom in the $\ket{211}$ state. In this illustration, the color intensity is proportional to $\abs{\psi}^2$.}
\label{fig:atom}
\end{figure}

\subsubsection{Bound states}

To leading order in $\alpha$, the bound state solutions of \eqref{eqn:schrodinger} thus coincide with the familiar hydrogenic electron orbitals. These take the form
\beq
\psi_{n \ell m}(t, \vec{r})=R_{n \ell}(r) Y_{\ell m}(\theta, \phi) e^{-i\left(\omega_{n \ell m}-\mu\right) t}\,,
\label{eqn:eigenstates}
\eeq
where $Y_{\ell m}$ are the spherical harmonics and $R_{n\ell}$ are the hydrogenic radial functions. As is customary in atomic physics, we may interpret $n$, $\ell$, $m$ as the principal, angular momentum and azimuthal ``quantum numbers'', respectively, satisfying $n>\ell$, $\ell\ge0$ and $\ell\ge\abs{m}$. The explicit expression of the radial functions is
\begin{equation}
R_{n \ell}(r) = \sqrt{\left(\frac{2 \mu \alpha}{n}\right)^3 \frac{(n - \ell - 1)!}{2 n (n+ \ell)!}} \left(\frac{2 \alpha \mu r}{n}\right)^\ell \exp\!\left(\minus \frac{\mu \alpha r}{n}\right) L_{n - \ell - 1}^{2 \ell +1}\!\left(\frac{2 \mu \alpha r}{n}\right), \label{eq:boundWavefunctions}
\end{equation}
where $L_{n-\ell-1}^{2\ell+1}(x)$ is the associated Laguerre polynomial. For small values of $\alpha$, the radial profile peaks at a multiple of the ``Bohr radius''  $r\ped{c} \equiv (\mu \alpha)^{-1}$ and decays exponentially as $r \to \infty$. An equatorial cross section of the cloud in the $(2,1,1)$ state is shown in Figure~\ref{fig:atom}. For notational simplicity, it is convenient to lean on the quantum mechanical analogy and represent \eqref{eqn:eigenstates} using the bra-ket notation $\ket{n\ell m}$. The normalization of the bound states is chosen so that 
\beq
\braket{n\ell m|n'\ell'm'}=\int\dd^3r\,\psi^*_{n\ell m}(t,\vec r)\psi_{n'\ell'm'}(t,\vec r)=\delta_{nn'}\delta_{\ell\ell'}\delta_{mm'}\,.
\eeq
The cloud's mass density is determined from \eqref{eqn:eigenstates} as
\begin{equation}
\rho(t,\vec r)=\begin{cases}
M_\lab{c}\abs{\psi(t,\vec r)}^2 & \quad  \text{(complex field)\,,}\\[1pt]
2M_\lab{c}\abs{\Re\psi(t,\vec r)}^2 & \quad  \text{(real field)\,.}
\end{cases}
\label{eqn:cloud-density}
\end{equation}
The azimuthally-averaged density is instead identical in both cases.

\subsubsection{Unbound states}
    
The Schrödinger equation \eqref{eqn:schrodinger} also permits unbound (or continuum) state solutions. In addition to the orbital and azimuthal angular momentum $\ell$ and $m$, these solutions are labeled by a positive, real-valued wavenumber $k$,
\begin{equation}
\psi_{k; \ell m}(t,\vec r) = R_{k;\ell}(r) Y_{\ell m}(\theta, \phi) e^{- i \epsilon_{\ell m}(k) t}\,.
\end{equation}
We distinguish the continuous index by a trailing semicolon and use the bra-ket notation~$|k;\ell \es m \rangle$. In the hydrogen atom, these continuum states represent those states  in which the electron has been unbound from the proton, and can thus be thought of as scattering states. A similar interpretation applies to the gravitational atom: these states represent the situation in which the scalar field is not bound to the black hole. The continuum radial functions are given by
\begin{equation}
R_{k; \ell}(r) = \frac{2 k e^{\frac{\pi \mu \alpha}{2 k}} |\Gamma(\ell + 1 + \tfrac{i \mu \alpha}{k})|}{(2 \ell+1)!} (2 k r)^{\ell} e^{- i k r} {}_1 F_{1}(\ell + 1 + \tfrac{i \mu \alpha}{k}; 2 \ell + 2; 2 i k r)\,, \label{eq:contWavefunctions}
\end{equation}
where ${}_1 F_1(a; b; z)$ is the Kummer confluent hypergeometric function. In contrast to the bound states, these continuum states do not decay exponentially  as $r \to \infty$ and are not unit-normalizable. The normalization is instead chosen so that
\begin{equation}
\braket{ k; \ell m | k'; \ell' m'} = \int\dd^3 r\, \psi^*_{k; \ell m}(t, \mb{r})\, \psi_{k'; \ell' m'}(t, \mb{r}) = 2 \pi \delta(k - k') \delta_{\ell \ell'} \delta_{m m'}\,,
\label{eq:contNormalization}
\end{equation}
i.e., these continuum states are $\delta$-function normalized.

\subsubsection{Spectrum}

While the leading-order energy eigenvalues coincide with the hydrogenic spectrum, higher-order corrections are different. The bound state nonrelativistic energies $\epsilon_{n\ell m}=E_{n\ell m}-\mu$ up to $\mathcal O(\alpha^5)$ are \cite{Baumann:2019eav}
\beq
\epsilon_{n\ell m}=\mu\biggl(-\frac{\alpha^{2}}{2 n^{2}}-\frac{\alpha^{4}}{8 n^{4}}-\frac{(3 n-2 \ell-1) \alpha^{4}}{n^{4}(\ell+1 / 2)}+\frac{2 \tilde{a} m \alpha^{5}}{n^{3} \ell(\ell+1 / 2)(\ell+1)}+\mathcal{O}(\alpha^{6})\biggr)\,.
\label{eq:eigenenergy}
\eeq
It is useful to define different types of energy splittings. We say that two states have a \emph{Bohr} ($\Delta n\ne0$), \emph{fine} ($\Delta n=0$, $\Delta\ell\ne0$) or \emph{hyperfine} ($\Delta n=0$, $\Delta\ell=0$, $\Delta m\ne0$) splitting.

\begin{figure}[t]
\centering
\includegraphics[width=\textwidth]{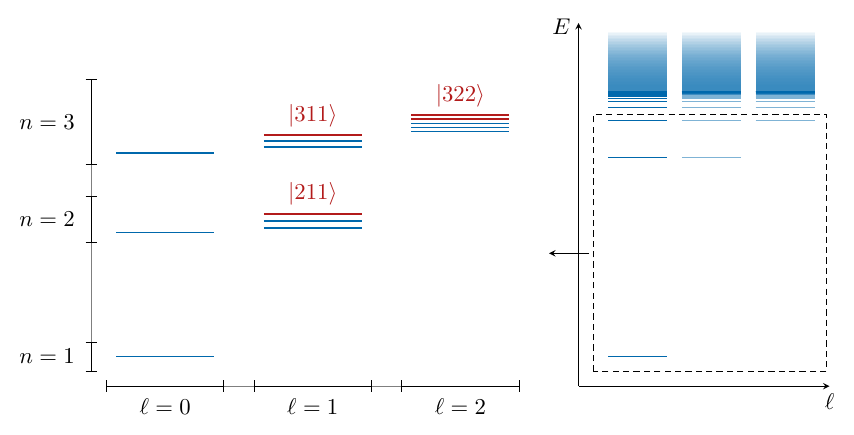}
\caption{Illustration of the spectrum of bound and unbound states of the gravitational atom. The states colored in red can be populated by superradiance.}
\label{fig:spectrum}
\end{figure}

\vskip 0pt
Since the boundary conditions for the continuum states are much less restrictive than those for the bound states, the exact eigenfrequencies are known and equal to $\omega(k) = \sqrt{\mu^2 + k^2}$. In the non-relativistic limit, $k \ll \mu$, this dispersion relation can be approximated as
\begin{equation}
\epsilon(k)\equiv\sqrt{\mu^2 + k^2}-\mu\approx\frac{k^2}{2 \mu}\,,
\label{eq:contEnergies}
\end{equation}
which is accurate enough for the purpose of this thesis.

\subsubsection{Searching for gravitational atoms}

Superradiance and gravitational atoms can be used to detect ultralight scalars, or impose constraints on their existence.

\vskip 0pt
Conceptually, the simplest such avenue consists in measuring the masses and spins of black holes. Given a value $\mu$ of the boson's mass, certain regions of the Regge plane (cf.~Figure~\ref{fig:regge}) correspond to black holes with fast superradiant instabilities, and are thus expected to be depopulated. An ``empty'' region in the Regge plane could thus signal the existence of a new light degree of freedom, while even a single measurement of a fast-spinning BH can put constraints on ultralight scalars in the corresponding window of masses. Measuring the spin of a BH is generally hard: the current best data comes from a handful of X-ray binaries and supermassive BHs.\footnote{GWs from binary inspirals also carry information about the spins. However, the capabilities of current detectors only allow to measure accurately the effective spin $\chi\ped{eff}$, which is a weighted average of the orthogonal components of the spins of the two bodies. The spin of the remnant is also well-measured, but it is not useful to place bounds on ultralight particles, because superradiance takes some time to extract the spin after the BH formation.} Several works \cite{Arvanitaki:2014wva,Stott:2018opm,Stott:2020gjj,Baryakhtar:2020gao,Mehta:2021pwf,Hoof:2024quk} have used these data to place bounds on the parameter space of ultralight fields, resulting in two disfavored regions, roughly corresponding to $\SI{2e-13}{eV}<\mu<\SI{6e-12}{eV}$ and $\SI{e-19}{eV}<\mu<\SI{e-18}{eV}$.\footnote{The precise values of the bounds differ slightly among authors. Furthermore, they also depend on the boson's decay constant, a small value of which could shut down superradiance early \cite{Baryakhtar:2020gao}.}

\vskip 0pt
Other proposed ways of placing observational constraints involve the rare spontaneous transitions between different bound states with associated GW burst emission~\cite{Arvanitaki:2014wva}, the tidal deformability of the cloud~\cite{DeLuca:2021ite}, and the direct detection of the GWs released by the boson cloud~\cite{LIGOScientific:2021rnv}. The subject of this thesis is another, more recently proposed, way of finding gravitational atoms, based on their signatures on the dynamics of binary systems and the ensuing gravitational waveform.

\chapter{Binary--cloud interaction}

\label{chap:binary-cloud-interaction}

While the final goal of this thesis is to discover new gravitational-wave signatures of boson clouds, its heart is a detailed treatment of the dynamics of a gravitational atom in the presence of a binary companion. We start the journey in this chapter.

\vskip 0pt
In Section~\ref{sec:perturbation} we study the gravitational perturbation induced by the binary on the cloud, establishing a framework that will be essential for most of the effects examined later. In Section~\ref{sec:capture} we find out how the presence of a boson cloud can affect the process of binary formation, by significantly enhancing the cross section for dynamical capture. In Sections~\ref{sec:resonances-review} and \ref{sec:ionization-review} we introduce the two most prominent and rich phenomena that occur in the cloud-binary system, namely resonances and ionization. Due to their importance and complexity, we will limit their analysis in this chapter to a simple introduction: the two effects will then become the protagonists of their own Chapter~\ref{chap:resonances} and Chapter~\ref{chap:ionization}, respectively. Finally, in Section~\ref{sec:accretion} we study the accretion of mass onto the companion object as it moves through the cloud.

\section{Gravitational perturbation}

\label{sec:perturbation}

Our main goal is to understand the dynamics of the cloud during a binary inspiral. To this end, we must describe the effect that the binary companion has on the cloud through its gravitational field.

\begin{figure}
\centering
\includegraphics[width=0.6\textwidth]{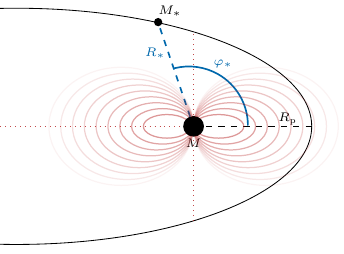}
\caption{Schematic illustration of the binary system. The primary object of the binary has mass $M$, while the companion has mass $M_*$. The motion of the companion on the equatorial plane is described by $\{R_{*}, \varphi_*\}$ and $R\ped{p}$ is the periapsis. The red lines schematically indicate the boson~cloud.}
\label{fig:IllustrationParabolic}
\end{figure}

\vskip 0pt
We consider a binary system where the primary object with mass $M$ hosts the cloud and is much heavier than its companion with mass $M_{*}$, such that the \textit{mass ratio} $q \equiv M_*/M \ll1$. We work in the reference frame of the central BH, where $\vec{r} = \{r, \theta, \phi \}$. The coordinates of the companion are $\vec{R}_{*} = \{R_{*},\theta_*, \varphi_{*} \}$, where $R_{*}$ is the binary's separation and $\theta_*$ is the polar angle with respect to the BH's spin. In this chapter, we assume that the orbit entirely lies in the equatorial plane; consequently, we have $\theta_*=\pi/2$, while $\varphi_*$ coincides with the $\emph{true anomaly}$. On a non-circular orbit we denote with $R\ped{p}$ the $\emph{periapsis}$, which is the distance of closest approach between the two components of the binary. In Figure~\ref{fig:IllustrationParabolic}, we show a schematic illustration of our setup, including the relevant parameters.

\vskip 0pt
The companion object interacts with the cloud gravitationally, introducing a perturbation $V_*$ to the right-hand side of the Schrödinger equation (\ref{eqn:schrodinger}). We can write it using the multipole expansion of the Newtonian potential as
\beq
\label{eqn:V_star}
V_*(t,\vec r)=-\frac{M_*\mu}{\abs{\vec R_*-\vec r}}=-\sum_{\ell_*=0}^\infty\sum_{m_*=-\ell_*}^{\ell_*}\frac{4\pi q\alpha}{2\ell_*+1}Y_{\ell_*m_*}(\theta_*,\varphi_*)Y_{\ell_*m_*}^*(\theta,\phi)\,F(r)\,,
\eeq
where
\beq
F(r)=
\begin{cases}
\dfrac{r^{\ell_*}}{R_*^{\ell_*+1}}\Theta(R_*-r)+\dfrac{R_*^{\ell_*}}{r^{\ell_*+1}}\Theta(r-R_*)&\text{for }\ell_*\ne1\,,\\[12pt]
\biggl(\dfrac{R_*}{r^2}-\dfrac{r}{R_*^2}\biggr)\Theta(r-R_*)&\text{for }\ell_*=1\,,
\end{cases}
\label{eqn:F(r)}
\eeq
and $\Theta$ is the Heaviside step function. The expressions for $\ell_*=1$ and $\ell_*\ne1$ differ because the frame of the larger BH is non-inertial, so a fictitious force is present. This requires the addition of $\delta V_*=q\vec{R_*}\cdot\vec{r}/R_*^3$, which only contributes to the $\ell_*=1$ term, which then takes form given in \eqref{eqn:F(r)}.

\vskip 0pt
The perturbation induces a mixing between the cloud's bound state $\ket{n_b \ell_b m_b}$ and another state $\ket{n \ell m}$, with matrix element
\beq
\braket{n\ell m|V_*(t,\vec r)|n_b\ell_bm_b}=-\sum_{\ell_*,m_*}\frac{4\pi\alpha q}{2\ell_*+1}Y_{\ell_*m_*}(\theta_*,\varphi_*)\, I_r \, I_{\Omega}\,,
\label{eqn:MatrixElement}
\eeq
where the \emph{radial} and \emph{angular} integrals are
\begin{align}
I_r&=\int_0^\infty F(r)R_{n\ell}(r) R_{n_b \ell_b}(r)\, r^2\dd r\,,\\
\label{eqn:I_Omega}
\begin{split}
I_{\Omega}&= \int Y_{\ell m}^*(\theta, \phi) Y_{\ell_* m_*}^*(\theta, \phi) Y_{\ell_bm_b}(\theta, \phi) \dd \Omega\\
&=\sqrt{\frac{(2 \ell+1)(2 \ell_*+1)(2 \ell_b+1)}{4 \pi}}\begin{pmatrix}
\ell& \ell_* & \ell_b \\
0 & 0 & 0
\end{pmatrix}\begin{pmatrix}
\ell& \ell_* & \ell_b \\
-m & -m_* & m_b
\end{pmatrix}\,.
\end{split}
\end{align}
The expression of $I_\Omega$ in terms of the Wigner-3j symbols implies the existence of \emph{selection rules} that need to be satisfied in order for (\ref{eqn:I_Omega}) to be non-zero:
\begin{align}
\label{eqn:S1}
(\text{S1})\qquad & {-m}-m_*+m_b=0\,,\\
\label{eqn:S2}
(\text{S2})\qquad & \ell+\ell_*+\ell_b=2 p,\qquad \text{for }p \in \mathbb{Z}\,,\\
\label{eqn:S3}
(\text{S3})\qquad & \abs{\ell_b-\ell} \leq \ell_* \leq \ell_b+\ell\,.
\end{align}
Due to the (quasi)-periodicity of $\varphi_{*}(t)$, the matrix element (\ref{eqn:MatrixElement}) can be decomposed into Fourier components as
\beq
\braket{n\ell m|V_{*}(t,\vec r)|n_b \ell_b m_b} = \sum_{g\in \mathbb{Z}} \eta^\floq{g} e^{-i g \Omega t}\,.
\label{eqn:def-eta}
\eeq
To make the notation clearest, we will often remove or add superscripts and subscripts to $\eta^\floq{g}$, depending on the context. Analogous formulae hold for the mixing of $\ket{n_b\ell_bm_b}$ with an unbound state $\ket{k;\ell m}$.

\section{Dynamical capture}
\label{sec:capture}

The first consequence of the binary-cloud interaction we investigate is its effect on the formation of the binary itself. The formation of compact binaries is an active area of research (see e.g.~\cite{Amaro-Seoane:2012lgq,Amaro-Seoane:2022rxf,Babak:2017tow} and references therein). One of the proposed mechanisms, \emph{dynamical capture}, allows the creation of a bound system through dissipation of energy in a burst of GWs during a close encounter between the two objects. The cross section for this process is \cite{1989ApJ...343..725Q,Mouri:2002mc}
\beq
\sigma_\slab{gw}=2\pi M^2\biggl(\frac{85\pi}{6\sqrt2}\biggr)^{2/7}q^{2/7}(1+q)^{10/7}v^{-18/7}\,,
\label{eqn:gw-cross-section}
\eeq
where the two compact objects have masses $M$ and $M_*=qM$, and $v$ is their relative asymptotic velocity before the close encounter.

\vskip 0pt
When one of the two objects is surrounded by a scalar cloud, then the energy during a dynamical capture is not only emitted via GWs, but also exchanged with the cloud. This phenomenon was first considered in \cite{Zhang:2019eid} and is akin to the ``tidal capture'' found in \cite{Cardoso:2022vpj}. In this section, we compute the energy exchanged with the bound states of the cloud\footnote{We thank the authors of \cite{Zhang:2019eid} for acknowledging in a private communication the discrepancy with their results.} and extend it to include unbound states as well. Then, we show how the formula (\ref{eqn:gw-cross-section}) for the cross section gets corrected, and discuss the impact for the merger rate in astrophysically realistic environments.

\subsection{Energy lost to the cloud}
\label{sec:E_lost}

In the same spirit as in the classical derivation of (\ref{eqn:gw-cross-section}), we consider a binary on a parabolic orbit.\footnote{For a non-zero $v$, the orbit is actually hyperbolic. However, approximating it with a parabola allows to correctly compute the leading order in $v$ of the cross section, while greatly simplifying the calculation.} The separation $R_*$ and azimuthal angle $\varphi_*$ can be parametrized as
\beq
R_*=R\ped{p}\biggl(\xi^2-1+\frac1{\xi^2}\biggr)\qquad\text{and}\qquad\varphi_*=2\arctan\biggl(\xi-\frac1\xi\biggr)\,,
\label{eqn:parabola}
\eeq
where $R\ped{p}$ is the periapsis of the orbit and
\beq
\xi\equiv\Bigl(\Omega\ped{p}t+\sqrt{1+\Omega\ped{p}^2t^2}\Bigr)^{1/3}\,,\qquad\text{with}\qquad\Omega\ped{p}\equiv\frac32\sqrt{\frac{(1+q)M}{2R\ped{p}}}\,.
\eeq
Under the gravitational perturbation of the binary, the cloud's wavefunction will evolve with time. It is useful to decompose it as
\beq
\ket{\psi(t)}=\sum_{n,\ell, m}c_{n\ell m}(t)\ket{n\ell m}+\int\frac{\dd k}{2\pi}\sum_{\ell, m}c_{k;\ell m}(t)\ket{k;\ell m}\,.
\eeq
As long as the perturbation is weak enough to keep $\abs{c_{n_b\ell_bm_b}}\approx1$ throughout the evolution, with all other coefficients remaining much smaller, the Schrödinger equation can be approximated as
\beq
i\frac{\dd c_{n\ell m}}{\dd t}\approx \braket{n\ell m|V_*(t,\vec r)|n_b\ell_bm_b}e^{i(\epsilon_{n\ell m}-\epsilon_b)t}\,,
\label{eqn:schrodinger-approx}
\eeq
where $\epsilon_b$ is the energy of $\ket{n_b\ell_bm_b}$. In the limit $t\to+\infty$, equation (\ref{eqn:schrodinger-approx}) can then be integrated to give
\beq
c_{n\ell m}=-i\int_{-\infty}^{+\infty}\dd t\braket{n\ell m|V_*(t,\vec r)|n_b\ell_bm_b}e^{i(\epsilon_{n\ell m}-\epsilon_b)t}\,.
\label{eqn:cnlm-b}
\eeq
An identical formula holds for unbound states, where the principal quantum number $n$ is replaced by the continuous wavenumber $k$.

\vskip 0pt
The coefficients $c_{n\ell m}$ and, especially, $c_{k;\ell m}$ are computationally expensive to determine, as they feature the radial integral $I_r$ nested inside the time integral appearing in (\ref{eqn:cnlm-b}). Restricting to an equatorial orbit, where $\theta_*=\pi/2$, they can be written as
\beq
c_{n\ell m}=-i\sum_{\ell_*}\frac{4\pi\alpha q}{2\ell_*+1}I_\Omega Y_{\ell_*g}\biggl(\frac\pi2,0\biggr)\times2\int_0^\infty\dd t\,I_r(t)\cos(g\varphi_*+(\epsilon_{n\ell m}-\epsilon_b)t)\,,
\eeq
and similarly for $c_{k;\ell m}$, where $g=\pm(m-m_b)$ for co/counter-rotating orbits, respectively. The radial integral $I_r$ depends on the time $t$ through $R_*$, as determined in (\ref{eqn:parabola}). Once $c_{n\ell m}$ and $c_{k;\ell m}$ are known, the total energy lost by the binary to the cloud is then given by
\beq
E\ped{lost}=\frac{M\ped{c}}\mu\sum_{n,\ell, m}(\epsilon_{n\ell m}-\epsilon_b)\abs{c_{n\ell m}}^2+\frac{M\ped{c}}\mu\int\frac{\dd k}{2\pi}\sum_{\ell, m}(\epsilon(k)-\epsilon_b)\abs{c_{k;\ell m}}^2\,.
\label{eqn:E_lost}
\eeq
Note that the contribution due to bound states can in principle be negative, due to the existence of states with lower energy, while the term associated to unbound states can only be positive.

\vskip 0pt
Equation (\ref{eqn:E_lost}) requires to sum over an infinite number of final states. For the first term, the one corresponding to transitions to other bound states, we truncate the sum when the addition of a term with higher $n$ would change the result by less than $0.1\%$. This typically requires including terms up to $n\sim10$ to $n\sim35$, depending on the chosen value of $R\ped{p}$. The second term is harder to handle, as there is yet another integral, over the wavenumber $k$. Moreover, for a fixed $k$, all values of $\ell$ are allowed. We evaluate the integrand at discrete steps in $k$, truncating the sum over $\ell$ when the addition of a new term would change the result by less than $0.01\%$. The size of the step depends on the value of $R\ped{p}$ and is chosen to be small enough to properly sample the integrand. The integral over $\dd k$ is then performed with a simple trapezoidal approximation.

\begin{figure}[th]
\centering
\includegraphics{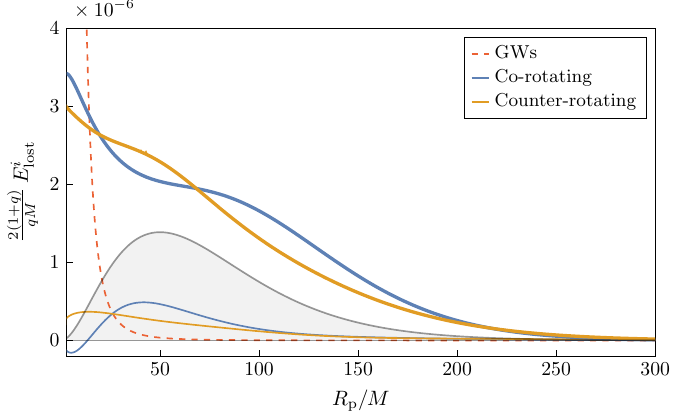}
\caption{Energy lost to the cloud (\ref{eqn:E_lost}) as function of the distance of closest approach $R\ped{p}$, for $\alpha=0.2$, $q=10^{-3}$, $M\ped{c}=0.01M$ and a cloud in the $\ket{211}$ state. The $i$ subscript denotes each of the two contributions to (\ref{eqn:E_lost}), so that $E\ped{lost} = \sum_iE\ped{lost}^{i}$. Thick (thin) lines refer to the energy lost to unbound (bound) states, while the colors differentiate between co-rotating and counter-rotating orbits. For comparison, we also show the density profile of the cloud, $\abs{\psi(R\ped{p})}^2$, in shaded gray, arbitrarily normalized.}
\label{fig:E_lost_211}
\end{figure}

\vskip 0pt
The results are shown in Figure~\ref{fig:E_lost_211}. Here, we plot $E\ped{lost}$, normalized by\footnote{The motivation behind this normalization will become clear in equation (\ref{eqn:E_lost-Kin}).} $qM/(2(1+q))$, for the state $\ket{211}$ and a fiducial set of parameters. Both the contribution due to bound states and to unbound states vanish exponentially for $R\ped{p}\to\infty$ and are largest when $R\ped{p}$ is roughly comparable to the size of the cloud. We also see that the dominant contribution to $E\ped{lost}$ is the one associated to unbound states. At very small radii, $E\ped{lost}$ has a finite limit, meaning that the cloud is only able to dissipate a certain maximum amount of energy. On the other hand, $E_\slab{gw}$ (i.e.\ the energy radiated in GWs) formally diverges for $R\ped{p}\to0$, implying that the high-$v$ limit is dominated by GWs, which become much more effective than the cloud at dissipating energy. Because $E_\slab{gw}$ decays polynomially for $R\ped{p}\to\infty$, GWs also dominate the low-$v$ limit.

\subsection{Scalings and cross section}
\label{sec:cross-section}

Although the values presented in Figure~\ref{fig:E_lost_211} are computed for a fiducial set of parameters, in the limit of small $q$ an approximate scaling relation allows us to predict the values for an arbitrary set of parameters. We can exploit the $\alpha$-scaling of the radial wavefunctions and of the overlap integrals to write
\beq
E\ped{lost}=M\ped{c}\alpha^2q^2\mathcal E(\alpha^2R\ped{p}/M)\,,
\label{eqn:E_lost-scaling}
\eeq
where $\mathcal E$ is a function that only depends on the initial state $\ket{n_b\ell_bm_b}$.

\vskip 0pt
Once $E\ped{lost}$ is known, we can use it to determine the total cross section $\sigma\ped{tot}$ for dynamical capture by requiring that it is larger than the total initial energy of the binary:
\beq
E\ped{lost}+E_\slab{gw}>\frac12\frac{qM}{1+q}v^2\,,
\label{eqn:E_lost-Kin}
\eeq
where we took into account the contribution due to GW emission,
\beq
E_\slab{gw}=\frac{85\sqrt2}{24}\frac{q^2M^{9/2}}{R\ped{p}^{7/2}}\,.
\eeq
If the left-hand side of (\ref{eqn:E_lost-Kin}) were a decreasing function of $R\ped{p}$, the inequality would hold for all $R\ped{p}<R\ped{p}(v)$ for some function $R\ped{p}(v)$. By relating this to the binary's impact parameter $b$, we would find the total cross section as
\beq
\sigma\ped{tot}=\pi b^2,\qquad\text{where}\qquad b^2=\frac{2MR\ped{p}}{v^2}\,.
\eeq
In reality, while $E_\slab{gw}$ is indeed a decreasing function, $E\ped{lost}$ in general is not. The inequality (\ref{eqn:E_lost-Kin}) will then hold in some finite intervals of $R\ped{p}$. Consequently, for some values of $v$, the cross section for dynamical capture should be geometrically interpreted as an annulus (or several concentrical annuli), rather than a circle.

\begin{figure}[t]
\centering
\includegraphics{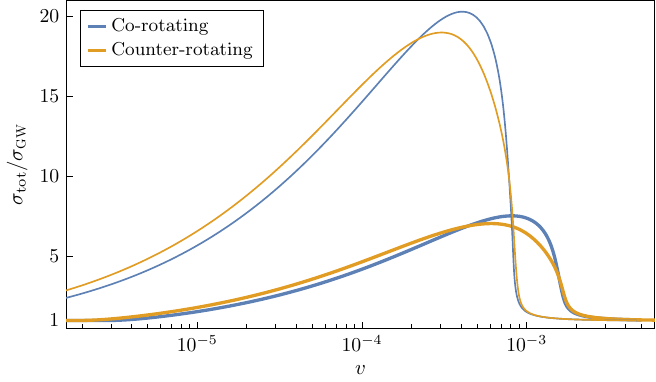}
\caption{Capture cross section $\sigma\ped{tot}$, including the energy lost to both the cloud and GWs, normalized by capture cross section (\ref{eqn:gw-cross-section}) due to GWs only. The cross section is shown as a function of the relative asymptotic velocity between the two objects, $v$. Thick lines are computed for the same set of parameters as in Figure \ref{fig:E_lost_211}, while thin lines show the result when $\alpha$ is decreased from $0.2$ to $0.1$.}
\label{fig:capture_cross_section}
\end{figure}

\vskip 0pt
The results are shown in Figure~\ref{fig:capture_cross_section}. As anticipated in Section~\ref{sec:E_lost}, the ratio $\sigma\ped{tot}/\sigma_\slab{gw}$ asymptotes to unity for very high and very low values of $v$. For intermediate velocities, instead, the cross section is significantly enhanced by the presence of the cloud, which dominates over GWs. The magnitude of the enhancement and the velocities at which it occurs depend on the chosen parameters. In general, the total cross section $\sigma\ped{tot}$ does not inherit any scaling relation akin to (\ref{eqn:E_lost-scaling}), because $E_\slab{gw}$ and $E\ped{lost}$ scale differently with the parameters. However, in the region of parameter space where $E\ped{lost}\gg E_\slab{gw}$, we can neglect the latter in (\ref{eqn:E_lost-Kin}) and derive an approximate scaling relation that reads
\beq
\frac{\sigma\ped{tot}}{\sigma_\slab{gw}}\sim\biggl(\frac{M\ped{c}}M\biggr)^{2/7}\alpha^{-10/7}\,\mathcal S\biggl(\frac{Mv^2}{M\ped{c}\alpha^2q}\biggr)\,,
\label{eqn:scaling-cross-section}
\eeq
where again $\mathcal S$ is a universal function to be found numerically. Equation (\ref{eqn:scaling-cross-section}) allows us to rescale the results of Figure~\ref{fig:capture_cross_section} for other values of the parameters. In particular, it shows that, for smaller values of $\alpha$, the relative enhancement of the cross section is greater and happens at lower values of $v$.

\subsection{Capture rate}
\label{sec:capture-rate}

An increased capture cross section like in Figure~\ref{fig:capture_cross_section} leads to a higher binary formation rate, and thus to an enhanced merger rate $\mathcal R$. In general, the latter can be computed as
\beq
\mathcal R=\int \dd V\dd M\dd M_*\:n_{\scalebox{0.65}{$M$}}n_{\scalebox{0.65}{$M_*$}}\braket{\sigma\ped{tot}v}\,,
\eeq
where $n_{\scalebox{0.65}{$M$}}$ and $n_{\scalebox{0.65}{$M_*$}}$ are the comoving average number densities of the primary and companion object, respectively, while the integral over $\dd V$ is performed over the volume one is interested in, e.g.\ the Milky Way or the LISA range. The term $\braket{\sigma\ped{tot}v}$ is the capture cross section weighted by some velocity distribution $P(v)$, that is,
\beq
\braket{\sigma\ped{tot}v}=\int \sigma\ped{tot}(v)P(v)v\dd v\,.
\eeq
Depending on the specific astrophysical environments under consideration (e.g.\ globular clusters or active galactic nuclei), a suitable velocity distribution must be chosen, from which the merger rate can be calculated. In practice, however, this approach hides many subtleties such as mass segregation \cite{OLeary:2008myb,Amaro-Seoane:2010dzj}, and the values for the merger rates are very uncertain \cite{Amaro-Seoane:2020zbo}.

\vskip 0pt
Giving a detailed account of these issues is beyond the scope of this thesis. We can, however, provide an estimate for the increase in the merger rate due to the presence of the cloud, based on the fact that $\mathcal R$ is directly proportional to $\braket{\sigma\ped{tot}v}$. The maximum increase happens when $P(v)$ has most of its support in correspondence of the peak of $\sigma\ped{tot}/\sigma_\slab{gw}$: in that case, one can expect the merger rate to be enhanced by a factor of $\mathcal O(10)-\mathcal O(100)$, depending on the parameters. Any other velocity distribution will give an increase by a factor from 1 up to that maximum value. For the parameters chosen in Figure~\ref{fig:capture_cross_section}, the peak is indeed located at values of $v$ close to the typical velocities found in the center of Milky Way-like galaxies: we can thus expect the rate of events with $q\sim10^{-3}$ to be significantly enhanced. On the other hand, from (\ref{eqn:scaling-cross-section}), we note that the peak shifts to lower values of $v$ when the mass ratio is reduced, hinting to a less significant increase for the rate of EMRIs with $q\ll10^{-3}$.

\section{Resonances}

\label{sec:resonances-review}

We now proceed to study the phenomenology subsequent to the binary formation. In~\cite{Baumann:2018vus,Baumann:2019ztm}, it was shown that the companion's gravitational perturbation can force the cloud to transition from one bound state to another. In this section, we briefly review what was known about resonant transitions from those work. A more extensive, general and self-consistent analysis will then be presented in Chapter~\ref{chap:resonances}.

\vskip 0pt
The matrix element \eqref{eqn:MatrixElement} of the gravitational perturbation $V_*$ between two states $\ket a=\ket{n_a\ell_am_a}$ and $\ket b=\ket{n_b\ell_bm_b}$ is an oscillatory function of the true anomaly of the orbit $\varphi_*$,
\beq
\braket{a|V_{*}(t)|b} = \sum_{g\in \mathbb{Z}} \eta^\floq{g} e^{ig \varphi_*}\,.
\label{eqn:eta-circular}
\eeq
On equatorial co-rotating quasi-circular orbits, the only nonzero term is $g=m_b-m_a\equiv\Delta m$ (on counter-rotating orbits, $g$ has opposite sign), and the coefficients $\eta^\floq{g}$ only depend on time through $\Omega(t)$. Restricting our attention to the two-state system, the Hamiltonian is thus
\beq
\mathcal H=\begin{pmatrix}-\Delta\epsilon/2 & \eta^\floq{g}e^{ig\varphi_*}\\ \eta^\floq{g}e^{-ig\varphi_*} & \Delta\epsilon/2\end{pmatrix}\,,
\label{eqn:schrodinger-hamiltonian}
\eeq
where $\Delta\epsilon=\epsilon_b-\epsilon_a$ is the energy difference between $\ket{b}$ and $\ket{a}$. As in \cite{Baumann:2019ztm}, it is useful to rewrite (\ref{eqn:schrodinger-hamiltonian}) in a \emph{dressed} frame, where the fast oscillatory terms $e^{ig\varphi_*}$ are traded for a slow evolution of the energies. This is done by means of a unitary transformation,
\beq
\begin{pmatrix}c_a\\ c_b\end{pmatrix}=\begin{pmatrix}e^{ig\varphi_*/2} & 0\\ 0 & e^{-ig\varphi_*/2}\end{pmatrix}\begin{pmatrix}\tilde c_a\\ \tilde c_b\end{pmatrix}\,,
\eeq
where $c_j=\braket{j|\psi}$ (with $j=a,b$) are the Schrödinger frame coefficients, while $\tilde c_a$ and $\tilde c_b$ are the dressed frame coefficients. Because $\abs{c_j}^2=\abs{\tilde c_j}^2$, we will drop the tildes in the following discussion. In the dressed frame, the Schrödinger equation reads
\beq
\frac\dd{\dd t}\!\begin{pmatrix}c_a\\ c_b\end{pmatrix}=-i\mathcal H\ped{D}\begin{pmatrix}c_a\\ c_b\end{pmatrix},\qquad\mathcal H\ped{D}=\begin{pmatrix}-(\Delta\epsilon-g\Omega)/2 & \eta^\floq{g}\\ \eta^\floq{g} & (\Delta\epsilon-g\Omega)/2\end{pmatrix}\,.
\label{eqn:dressed-hamiltonian}
\eeq
When $\Omega(t)\equiv\dot\varphi_*$ is specified, (\ref{eqn:dressed-hamiltonian}) determines the evolution of the population of the two states.

\begin{figure}
\centering
\includegraphics[width=\textwidth]{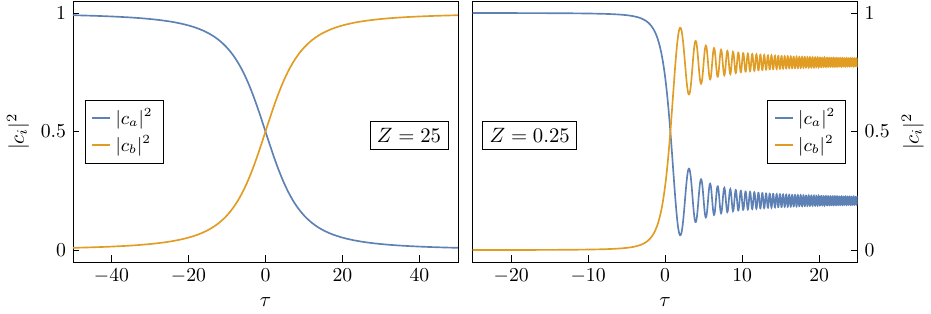}
\caption{Numerical solution of \eqref{eqn:dimensionless-dressed-schrodinger}. An adiabatic transition with full transfer from $\ket{a}$ to $\ket{b}$ is observed for $Z=25$ (\emph{left panel}), while a partial transfer is observed for for $Z=0.25$ (\emph{right panel}). In both cases, the final populations at $t\to+\infty$ asymptote those given in \eqref{eqn:lz}.}
\label{fig:unbackreacted}
\end{figure}

\vskip 0pt
Without including the backreaction of the resonance on the orbit, $\Omega(t)$ is exclusively determined by external factors, such as the energy losses due to GW emission or cloud ionization (to be discussed in Section~\ref{sec:ionization-review}), which induce a frequency chirp. These effects typically have a nontrivial dependence on $\Omega$ itself, widely varying in strength at different points of the inspiral. However, the resonances described by (\ref{eqn:dressed-hamiltonian}) are restricted to a bandwidth $\Delta\Omega\sim\eta^\floq{g}$. This is typically narrow enough to allow us to approximate the external energy losses, as well as any other $\Omega$-dependent function, with their value at the resonance frequency 
\beq
\Omega_0=\frac{\Delta\epsilon}{g}\,.
\label{eqn:resonant-frequency}
\eeq
Around $\Omega_0$, we can linearize the frequency chirp and write $\Omega=\gamma t$. For concreteness, in this section we assume that external energy losses are only due to GW emission, in which case
\beq
\gamma=\frac{96}5\frac{qM^{5/3}\Omega_0^{11/3}}{(1+q)^{1/3}}\,.
\label{eqn:gamma_gws}
\eeq
It is particularly convenient to rewrite the Schrödinger equation in terms of dimensionless variables and parameters:
\beq
\label{eqn:dimensionless-dressed-schrodinger}
\frac\dd{\dd\tau}\!\begin{pmatrix}c_a\\ c_b\end{pmatrix}=-i\begin{pmatrix}\omega/2 & \sqrt{Z}\\ \sqrt{Z} & -\omega/2\end{pmatrix}\begin{pmatrix}c_a\\ c_b\end{pmatrix}\,,
\eeq
where the frequency chirp now reads $\omega=\tau$, and we defined
\beq
\label{eqn:coefficients}
\tau=\sqrt{\abs{g}\gamma}\,t\,,\qquad\omega=\frac{\Omega-\Omega_0}{\sqrt{\gamma/\abs{g}}}\,,\qquad Z=\frac{(\eta^\floq{g})^2}{\abs{g}\gamma}\,.
\eeq
The initial conditions at $\tau\to-\infty$ we are interested in are those where only one state is populated, say $c_a=1$ and $c_b=0$. The only dimensionless parameter of \eqref{eqn:dimensionless-dressed-schrodinger} is the so-called ``Landau-Zener parameter'' $Z$, which determines uniquely the evolution of the system and its state at $\tau\to+\infty$. In fact, the populations at $\tau\to+\infty$ can be derived analytically and are given by the Landau-Zener formula \cite{zener1932non,landau1932theorie}:
\beq
\abs{c_a}^2=e^{-2\pi Z}\,,\qquad \abs{c_b}^2=1-e^{-2\pi Z}\,.
\label{eqn:lz}
\eeq
For $2\pi Z\gg1$ the transition can be classified as \emph{adiabatic}, meaning that the process is so slow that the cloud is entirely transferred from $\ket{a}$ to $\ket{b}$. Conversely, for $2\pi Z\ll1$, the transition is \emph{non-adiabatic}, with a partial or negligible transfer occurring.\footnote{The adiabaticity of a resonance is not related to the adiabaticity of the orbital evolution, which is always assumed to hold throughout our work.} A numerical solution of \eqref{eqn:dimensionless-dressed-schrodinger} is shown in Figure~\ref{fig:unbackreacted}, for two different values of $Z$.

\vskip 0pt
Dealing with two-state transitions is a good approximation as long as the frequency width of the resonance, $\Delta\Omega\sim\eta^\floq{g}$, is much narrower than the distance (in frequency) from the closest resonance. The latter becomes extremely small for hyperfine resonances, especially on generic orbits, where $g$ can take values different from $\Delta m$. In some cases formula \eqref{eq:eigenenergy} can indeed return an exact degeneracy of two resonances, up to $\mathcal O(\alpha^5)$. We have thoroughly checked, by numerical computation of the eigenfrequencies up to $\mathcal O(\alpha^6)$, that in all realistic cases the resonances are indeed narrow enough for the two-state approximation to hold.

\section{Ionization}

\label{sec:ionization-review}

In \cite{Baumann:2021fkf,Baumann:2022pkl} it was first shown that the interaction with the binary companion can cause the partial transfer of the cloud from its starting bound state $\ket{n_b\ell_bm_b}$ to any unbound state $\ket{k;\ell m}$. This process is referred to as \emph{ionization}, in analogy with atomic physics. In this section, we briefly introduce the phenomenon, following \cite{Baumann:2022pkl,Tomaselli:2023ysb}. A deeper analysis and several generalizations will then be the subject of Chapter~\ref{chap:ionization}.

\begin{figure}[t]
\centering
\includegraphics{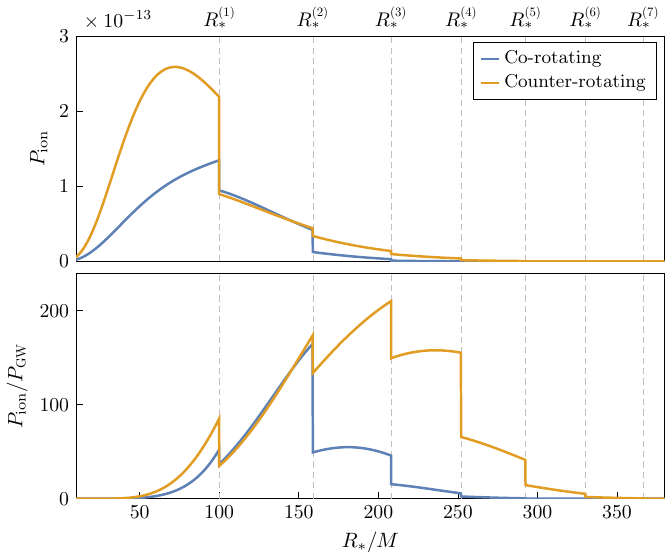}
\caption{Ionization power (\ref{eqn:pion}) as function of the orbital separation $R_*$, for $\alpha=0.2$, $q=10^{-3}$, $M\ped{c}=0.01M$ and a cloud in the $\ket{211}$ state. The top panel shows $P\ped{ion}$ in units where $M=1$. The bottom panel shows the ratio of $P\ped{ion}$ to $P_\slab{gw}$, the power lost due to GW emission (\ref{eq:p_gw}). We see that the energy lost due to ionization can dominate over GW emission.}
\label{fig:ionization_211}
\end{figure}

\vskip 0pt
As in Section~\ref{sec:E_lost}, it is useful to decompose the wavefunction as
\beq
\ket{\psi(t)}=c_{n_b\ell_bm_b}(t)\ket{n_b\ell_bm_b}+\int\frac{\dd k}{2\pi}\sum_{\ell, m}c_{k;\ell m}(t)\ket{k;\ell m}\,.
\eeq
Similar to (\ref{eqn:cnlm-b}), the coefficients $c_{k;\ell m}$ can be computed perturbatively as
\beq
c_{k;\ell m}(t)=-i\int_0^{t}\dd t'\braket{k;\ell m|V_*(t',\vec r)|n_b\ell_bm_b}e^{i(\epsilon(k)-\epsilon_b)t'}=i\,\eta\,\frac{1-e^{i(\epsilon(k)-\epsilon_b-g\Omega)t}}{i(\epsilon(k)-\epsilon_b-g\Omega)}\,,
\label{eqn:cnlm-u-ion}
\eeq
where $\eta$ is defined in (\ref{eqn:def-eta}). The last equality only holds on equatorial quasi-circular orbits. In order to obtain it, we exploited the selection rules of the angular integral $I_\Omega$, which hides inside the matrix element $\eta$: of all terms in the perturbation, only those that oscillate with frequency $g\Omega$ survive, where $g=m-m_b$ for co-rotating orbits and $g=m_b-m$ for counter-rotating orbits. When a long-time average of $\abs{c_{k;\ell m}}^2$ is taken, the time-dependent numerator of (\ref{eqn:cnlm-u-ion}) combines with the denominator to produce a delta function:
\beq
\abs{c_{k;\ell m}}^2=2\pi t\,\abs{\eta}^2\,\delta\bigl(\epsilon(k)-\epsilon_b-g\Omega\bigr)\,.
\label{eqn:fermis-golden-rule}
\eeq
Equation (\ref{eqn:fermis-golden-rule}) is nothing more than Fermi's Golden Rule. Summing over all unbound states yields the total ionization rate,
\beq
\frac{\dot M\ped{c}}{M\ped{c}}=-\sum_{\ell,g}\,\frac{\mu\abs{\eta^\floq{g}}^2}{k_\floq{g}}\Theta\bigl(k_\floq{g}^2\bigr)\,,
\label{eqn:rate}
\eeq
where we defined $k_\floq{g}=\sqrt{2\mu(\epsilon_b+g\Omega)}$, as well as the matrix element $\eta^\floq{g}$ of $V_*$ between the states $\ket{k_\floq{g};\ell,m_b\pm g}$ and $\ket{n_b\ell_bm_b}$. Similarly, one can define the rates of energy (``ionization power'') and angular momentum (``ionization torque'') transferred into the continuum as
\begin{align}
\label{eqn:pion}
P\ped{ion}&=\frac{M\ped{c}}\mu\sum_{\ell,g}\,g\Omega\,\frac{\mu\abs{\eta^\floq{g}}^2}{k_\floq{g}}\Theta\bigl(k_\floq{g}^2\bigr)\,,\\
\label{eqn:tion}
\tau\ped{ion}&=\frac{M\ped{c}}\mu\sum_{\ell,g}\,g\,\frac{\mu\abs{\eta^\floq{g}}^2}{k_\floq{g}}\Theta\bigl(k_\floq{g}^2\bigr)\,.
\end{align}

\vskip 0pt
When equations (\ref{eqn:rate}), (\ref{eqn:pion}) and (\ref{eqn:tion}) are evaluated numerically, two noteworthy features are found. First, we note that $P\ped{ion}$ is much larger than $P_\slab{gw}$ for a wide range of orbital separations (see bottom panel of Figure~\ref{fig:ionization_211}). This means that the backreaction of ionization dominates over the radiation reaction due to the emission of gravitational waves, which is the main driving force of the inspiral in vacuum.

\vskip 0pt
The second main feature of ionization are the sharp discontinuities exhibited by $P\ped{ion}$ (as well as by the rate and torque) at separations $R_*^\floq{g}$ corresponding to the orbital frequencies
\beq
\Omega^\floq{g}=\frac{\alpha^3}{2gMn_b^2}\,,\qquad g=1,2,3,\ldots
\label{eqn:omega-g}
\eeq
These can be interpreted as \emph{threshold frequencies}, in analogy to the ones found in the photoelectric effect. In our case, because the perturbation is not monochromatic, each different Fourier component produces a different jump. It is important to realize that, while $P\ped{ion}$ is indeed discontinuous in the limit where $\Omega$ is kept fixed, in reality the orbital frequency chirps as the binary inspirals. As a consequence, the discontinuities are replaced by smooth, although steep, transient oscillating phenomena, thoroughly described in Chapter~\ref{chap:ionization}.

\section{Accretion}

\label{sec:accretion}

\begin{figure}[h!]
      \centering
      \includegraphics[width=\textwidth]{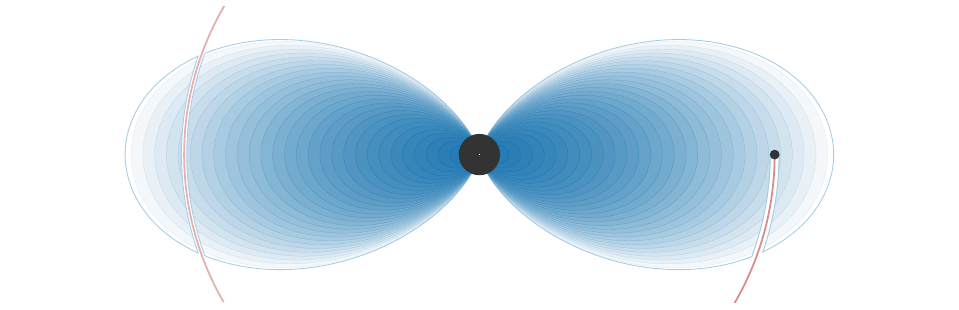}
      \caption{Cartoon illustrating the accretion of the boson cloud by the companion black hole. As explained in the text, the cloud responds rapidly and replenishes the local density behind the companion.}
      \label{fig:accretionSchematic}
    \end{figure}

So far, we have treated the perturbing object as pointlike and studied only its gravitational influence on the cloud.
In this section, we take the finite size of the companion into account and compute its absorption of the cloud (see Figure~\ref{fig:accretionSchematic}).\footnote{
The absorption cross section of a scalar field by a black hole has been studied extensively: in the massless case for rotating black holes in~\cite{Das:1996we,Higuchi:2001si,Macedo:2013afa,Cardoso:2019dte}, in the massive case for Schwarzschild black holes in \cite{Unruh:1976fm}, and more recently, in the massive case for charged and/or rotating black holes in \cite{Benone:2014qaa,Benone:2019all}.  Our analysis will be similar to that in \cite{Benone:2019all, Unruh:1976fm}.}  
 If the secondary object is a black hole of mass $M_*$ and spin $a_*$, then this absorption will play an important role in the binary's dynamics. 
 
\subsection{Motion in a uniform medium}
\label{sec:uniform-medium}

We start by solving the problem in the idealized case of a black hole moving with a constant velocity in a medium with a uniform density $\rho$. If the medium were made of small particles at rest at infinity, the problem would be relatively straightforward to solve via geodesic motion in the rest frame of the black hole. In the Schwarzschild case, the energy flux takes the form~\cite{Unruh:1976fm}
\begin{equation}
\frac{\dd M_*}{\dd t}= \frac{\pi \rho \es M_*^2}{2v^3}\,\frac{\bigl(4v^2+\sqrt{8v^2+1}-1\bigr)^3}{\bigl(\sqrt{8v^2+1}-1\bigr)^2} \sim \frac{ 4 \pi\rho\, (2M_*)^2}{v}\,,\qquad v \to 0\,,
\label{eqn:accretion-particles-gr}
\end{equation}
where $v$ is the asymptotic value of the relative velocity between the particles and the black hole.  The divergence at $v\to0$ signals the non-existence of a stationary configuration with $v=0$ where the density of the medium approaches a finite non-zero value at infinity.

\vskip 0pt
In the case of interest, the Compton wavelength of the medium is much larger than the gravitational radius, $r_{\lab{g},*}=M_*$, and therefore \eqref{eqn:accretion-particles-gr} does not hold. 
We expect the true answer to be smaller because the quantum pressure of the field suppresses small-scale overdensities.  
Because of the relative motion, the black hole will see the scalar field as having a  
wavenumber~$k\sim\mu v$. Besides the (reduced) Compton wavelength, $\lambda_\slab{c} = \mu^{- 1}$, the other
 relevant scale in the problem is then the (reduced) de Broglie wavelength, $\lambda_\lab{dB} = k^{- 1}$.  It will also be useful to define the  dimensionless ratios   $r_{\lab{g},*}/\lambda_\slab{c} = \mu M_*$ and $\lambda_\slab{c}/\lambda_\lab{dB} = k/\mu$. We are interested in the limit where both of these ratios are small,
\begin{equation}
\begin{split}
\mu M_* &\ll1\qquad\text{(``fuzzy'')\,,}\\
k/\mu&\ll 1\qquad\text{(``non-relativistic'')\,.}\\
\end{split}
\end{equation}
\label{equ:fuzzy}
We will see, in Section \ref{sec:accretion-realistic}, why these are the relevant limits in the realistic setting.

\vskip 0pt
Our goal is to compute the radial energy flux at the outer horizon $r=r_+$,
\beq
\frac{\dd M_*}{\dd t}=\int\! \dd\theta\dd\phi\,\sqrt{g_{\theta\theta}\es g_{\phi\phi}}\,\es T^r{}_0(r_+)\,,
\label{eqn:dM/dt-general}
\eeq
where the energy-momentum tensor $T_{\mu \nu}$ is that of 
the field profile $\Phi(t,\r)$.
Expanding this profile in modes with definite frequency $\omega^2 = \mu^2 + k^2$, we have (cf.~Appendix~\ref{app:heunc})
\begin{equation}
\Phi(t, \mb{r}) =\sum_{\ell_*,m_*}R_{k; \ell_* m_*}(r)S_{\ell_* m_*}(ka_*;\cos\theta)e^{-i\omega t+i m_* \phi}\, ,
\label{eqn:Phi-separation}
\end{equation}
where $S_{\ell_* m_*}(c;\cos\theta)$ are spheroidal harmonics with spheroidicity $c$, 
we can write the radial energy flux associated to this profile as
\begin{equation}
\label{eqn:Tr0-expanded}
  \begin{split}
    T^r{}_0 = \frac{2\omega(r-r_+)(r-r_-)}{r^2+a^2\cos^2\theta} \sum_{\ell_*,m_*} \Im(\partial_r R_{\ell_* m_*}^*R^{\phantom{*}}_{\ell_* m_*})\abs{S_{\ell_* m_*}}^2 + \cdots\,,
  \end{split}
\end{equation}
where the ellipses represent terms that mix different angular momenta and will vanish when integrated in (\ref{eqn:dM/dt-general}) to compute the radial energy flux. We denote the angular momentum quantum numbers measured with respect to the companion's position as $\ell_*$ and $m_*$, to distinguish them from those measured with respect to the parent black hole.

\vskip 0pt
The presence of the black hole deforms the field profile and determines its shape at the horizon, and thus the flux, as function of the boundary conditions at large distances. We work in the rest frame of the black hole and consider an incident monochromatic plane wave from infinity with wavevector $\mb{k}$. In Minkowski spacetime, the asymptotic field profile would be 
\begin{equation}
\begin{split}
\Phi(t, \mb{r})& \sim\sqrt{\frac{\rho}{2\omega^2}}\,e^{i\vec{k}\cdot\mb{r}}e^{- i\omega t}\\
&= \sqrt{\frac{\rho}{2\omega^2}}\,\sum_{\ell_*=0}^\infty (2\ell_*+1) i^{\ell_*} j_{\ell_*}\es\!(kr) P_{\ell_*}(\hat{\mb{k}} \cdot \hat{\mb{r}})\,e^{- i\omega t} \,,\quad r/M_*\to\infty\,,
\end{split}
\label{eqn:field-uniform-asymptotic-Mink}
\end{equation}
where $\omega=\sqrt{\mu^2+k^2}$, with $k = \mu v/\sqrt{1 - v^2}$. 
 In this expression, 
 $j_{\ell_*}\!\es (k r)$ is the spherical Bessel function, $P_{\ell_*}(\hat{\mb{k}} \cdot \hat{\mb{r}})$ is the Legendre polynomial and the normalization has been chosen so that $\rho\approx T_{00}=2\omega^2\Phi^*\Phi$. The long-range nature of the gravitational field, however, deforms the field; in a spherically symmetric spacetime, we have \cite{Benone:2014qaa}
\begin{equation}
\Phi(t, \mb{r}) \sim\sqrt{\frac{\rho}{2\omega^2}}\,\sum_{\ell_*=0}^\infty (2\ell_*+1)i^{\ell_*} j_{\ell_*}\!\es\big(kr+\delta(r)\big)P_{\ell_*}(\hat{\mb{k}} \cdot \hat{\mb{r}}) \,e^{- i\omega t}\,,\quad r/M_* \to\infty\,,
\label{eqn:field-uniform-asymptotic}
\end{equation}
where 
$\delta(r) = k M_* (1 + \omega^2/k^2) \log(2 k r) + \delta_{\ell_*}$, and $\delta_{\ell_*}$ is a constant phase shift. Although our case is not quite spherically symmetric, deviations from (\ref{eqn:field-uniform-asymptotic}) are controlled by the spheroidicity parameter, which is $ka_* \ll 1$ in the non-relativistic limit we are considering.

\begin{figure}
            \centering
            \includegraphics[scale=1, trim={0 2pt 0 0}]{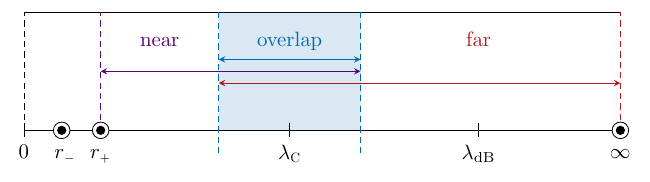}
            \caption{Schematic illustration of the near-field and far-field expansions, where $r_\pm$ are the inner and outer horizons of the black hole. The two asymptotic solutions are matched in the overlap region, $M_* \ll r\ll 1/k$.} 
            \label{fig:matching}
        \end{figure}

\vskip 0pt
To compute the energy flux at the horizon, we must understand the dependence of the near-field solution on the boundary condition (\ref{eqn:field-uniform-asymptotic}). This will be achieved by a \emph{matched asymptotic expansion}: the far-field and near-field solutions will be studied separately and matched in the overlap region, where both expansions hold. The boundary condition will then fix the overall amplitude of the solution. This procedure is schematically illustrated in Figure~\ref{fig:matching}.

\vskip 0pt
{\it Near-field solution}---With the ansatz (\ref{eqn:Phi-separation}), the Klein--Gordon equation is separable. The exact solution of the equation for $R_{k; \ell_* m_*}(r)$ can be expressed in terms of the confluent Heun function (see Appendix~\ref{app:heunc} and \cite{Bezerra:2013iha}). 
We expect the contributions from modes with $\ell_* \geq 1$  to be suppressed at radii smaller than about $\ell_*^2/(\mu^2M_*)$ (due to the angular momentum barrier), so that the $\ell_*=m_*=0$ mode dominates near the horizon. 
Expanding the confluent Heun function around $r=r_+$, one can show that
\begin{equation}
R_{k}(r)=C_{k}\es e^{-i\omega(\tilde r-r)-i m_* \tilde \phi}\bigl(1+\mathcal O(\mu M_*,kM_*)\bigr)\,,\quad\text{for}\quad r_+ \le r<r\ped{max}\,,
\label{eqn:R00-near-field}
\end{equation}
where we use $R_{k}(r) = R_{k; 00}(r)$ as a shorthand, the coefficient $C_{k}= C_{k; 00}$ defines the near-horizon amplitude of the $\ell_*=m_*=0$ mode, $\tilde r$ and $\tilde\phi$ are the radial and angular tortoise coordinates (defined in Appendix~\ref{app:heunc}),  
and the breakdown of the expansion is at
\begin{equation}
    \frac{r_\lab{max}}{M_*} \sim \min \left\{ \frac{1}{(\mu M_*)^2}, \frac{1}{k M_*}\right\} \gg 1\,.
\end{equation}
Using the explicit expressions of the tortoise coordinates, and plugging (\ref{eqn:R00-near-field}) into 
(\ref{eqn:dM/dt-general}), we get
\begin{equation}
  \frac{\dd M_*}{\dd t}= 4 M_*\hskip 1pt  r_+ \es\omega^2 \abs{C_{k}}^2\,.
  \label{eqn:flux-C00}
\end{equation}
We will now determine $C_{k}$ by matching (\ref{eqn:R00-near-field}) to the far-field solution.

\vskip 0pt
{\it Far-field solution}---Far from the companion, $r\gg M_*$, the equation for $R_{k}(r)$ becomes
\begin{equation}
\frac{\dd^2R_{k}}{\dd r^2}+\biggl(\frac2r+\cdots\biggr)\frac{\dd R_{k}}{\dd r}+\biggl(k^2+\frac{2M_*(\omega^2+k^2)}{r}+\cdots\biggr)R_{k}=0\,.
\end{equation}
This equation is solved by a linear combination of confluent hypergeometric functions, 
\begin{multline}
    e^{ikr}R_{k} =C_F\,{}_1F_1 \!\left(1 + i k M_* \big(1 + \omega^2/k^2\big); 2 \es ; 2ikr\right)\\
    +C_U \es U\!\left(1 + i k M_*\big(1 + \omega^2/k^2\big); 2; 2 i k r\right) . 
    \label{eqn:lin-comb-hypergeo}
\end{multline}
For $kr \ll 1$, this solution overlaps with the near-field solution (\ref{eqn:R00-near-field}). Expanding (\ref{eqn:lin-comb-hypergeo}) in this limit and matching to (\ref{eqn:R00-near-field}) then gives $C_F=C_{k}$ and $C_U\leq \mathcal O\bigl((\mu M_*)^2\bigr)$. To determine the overall amplitude of the solution, we then expand (\ref{eqn:lin-comb-hypergeo}) for $kr \gg 1$, where it reduces to a spherical Bessel function, $R_{k}(r)\propto j_0(kr+\delta(r))$, and compare it to the $\ell_*=0$ mode of the boundary condition (\ref{eqn:field-uniform-asymptotic}).  This gives
\begin{equation}
C_F = C_{k}= \frac{\sqrt{2 \pi \rho}}{\omega}  \left|\es\es \Gamma\!\left(1+i k M_* \big(1 + {\omega^2}/{k^2}\big) \vphantom{\tfrac{\omega^2}{k^2}}\right) e^{\frac{1}{2} \pi k M_* (1 + {\omega^2}/{k^2})}\right| .
\end{equation}
Plugging this back into \eqref{eqn:flux-C00}, we get
\begin{equation}
\begin{split}
\frac{\dd M_*}{\dd t}&=\mathcal{A}_* \es\es \rho\es \left|\es\es \Gamma\!\left(1+i k M_* \big(1 + {\omega^2}/{k^2}\big) \vphantom{\tfrac{\omega^2}{k^2}}\right)\right|^2 e^{\pi k M_* (1 + {\omega^2}/{k^2})}\\
&=\mathcal A_*\rho\frac{\pi k M_* (1 + {\omega^2}/{k^2}) e^{\pi k M_* (1 + {\omega^2}/{k^2})}}{\sinh(\pi k M_* (1 + {\omega^2}/{k^2}))}\, ,
\label{eqn:flux-analytical-gamma}
\end{split}
\end{equation}
where $\mathcal{A}_* \equiv 8\pi M_* r_{+,*}$ is the area of the outer horizon of the Kerr black hole. This is our final answer for the mass accretion rate.

\begin{figure}[t]
\centering
\includegraphics[width=0.8\textwidth]{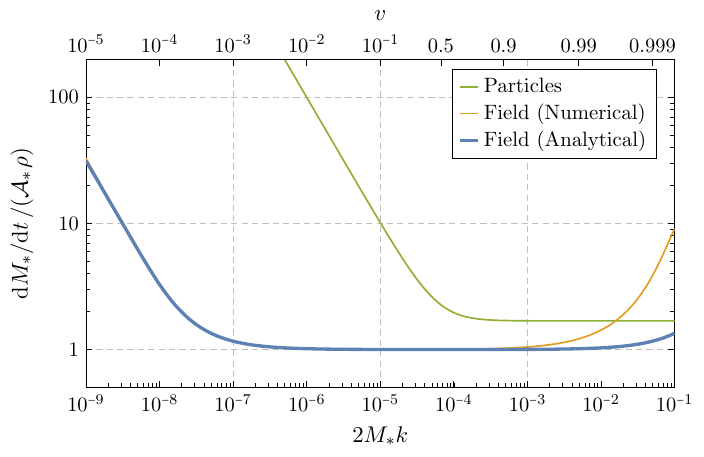}
\caption{Mass accretion rate of a Schwarzschild black hole computed analytically---from (\ref{eqn:flux-analytical-gamma})---and numerically for a scalar field with $2\mu M_* =10^{\protect - 4}$. Shown for comparison is also the accretion rate for particles given by (\ref{eqn:accretion-particles-gr}). }
\label{fig:accretion}
\end{figure}

\vskip 0pt
The result is shown in Figure~\ref{fig:accretion} for $2\mu M_* =10^{- 4}$. As anticipated, the flux is smaller than for particles, but still divergent for $v\to0$. For non-relativistic momenta, $k < \mu$, we can identify two different regimes 
\begin{equation}
\frac{\dd M_*}{\dd t}=\mathcal{A}_*\es\rho  \begin{cases}
\quad\,\, 1 & \text{for}\quad k \gg2\pi\mu^2M_* \,,\\[6pt]
\displaystyle \frac{2\pi\mu^2M_*}{k}\ & \text{for}\quad k\ll2\pi\mu^2M_*\,.
\end{cases}
\label{eqn:accretion-cases}
\end{equation}
It is worth noting that, at the cross-over point $k=2\pi\mu^2M_*$, the de Broglie wavelength of the scalar field equals the Bohr radius of the gravitational atom, $2\pi/k=r\ped{c}$.\footnote{To give an interpretation of this result, recall that a particle with impact parameter $b$ and velocity $v$ is scattered by an angle $\sim M/(v^2b)$ by the Coulomb interaction. Taking $b\sim\lambda_\lab{dB}$, we get an order-one deflection angle for $\lambda_\lab{dB} \sim r\ped{c}$. Scattering of waves with more (less) energy will be less (more) effective.} 
For $k\ll2\pi\mu^2M_*$, the energy flux diverges as $1/v$, just like in the particle case, but with a smaller normalization. For $k \gg2\pi\mu^2M_*$, instead, the energy flux is  independent of $v$ and takes the very natural form~$\mathcal{A}_*\es \rho c$, if we restore a factor of $c$. 
This indeed matches the result for the low-energy cross section for a massless field \cite{Das:1996we,Higuchi:2001si,Macedo:2013afa,Cardoso:2019dte}. The regime holds until relativistic corrections kick in at $k\sim\mu$, and our derivation breaks down.

\vskip 0pt 
{\it Numerical solution}---Figure~\ref{fig:accretion} also shows the result of a numerical approach to the problem. In the Schwarzschild case, we numerically integrated the confluent Heun function for different values of $k$ and $\ell_*$, with the main goal of confirming that  the $\ell_*= m_*=0$ mode indeed dominates in the fuzzy limit. This allowed us to determine the near-horizon amplitudes $C_{k; \ell_* m_*}$ of modes with $\ell_* \geq 1$ as a function of the asymptotic density $\rho$ by comparing the asymptotic limit of the confluent Heun function with the partial wave expansion of the boundary conditions (\ref{eqn:field-uniform-asymptotic}). 
The results are in remarkable agreement with the analytical estimate for all $\mu M_*\ll1$ and $k\ll\mu$, and explicitly show the suppression of $C_{k; \ell_* m_*}$ for $\ell_*\ge1$.

\subsection{Application to the realistic case}
\label{sec:accretion-realistic}

So far, we have studied an idealized model of a black hole moving through a uniform scalar field mass density. However, we would like to apply these results to the case we are actually interested in: a companion black hole of mass $M_* = q M$ moving through a non-uniform cloud that is bound to its parent black hole. This more realistic scenario has a few major complications over its idealized counterpart and in this section we confront them.

\vskip 0pt
First and foremost, the scalar field mass density can have nontrivial azimuthal structure and so the companion can experience different densities along a single orbit. For instance, if the cloud is composed of a \emph{real} scalar field occupying the $|2 \es 1 \es 1 \rangle$ state, its mass density \eqref{eqn:cloud-density} behaves as $\rho(\mb{r}) \propto \cos^2 \phi$. In contrast, if it is a  \emph{complex} scalar field occupying the same state (or any other pure eigenstate), its mass density does not vary along the orbit, $\rho(\mb{r}) = \rho(r, \theta)$. When the mass density has nontrivial $\phi$-dependence, we will assume that we can replace it with its azimuthal average, $\rho(r, \theta) = \frac{1}{2 \pi} \! \es \int_{0}^{2 \pi} \!\dd \phi\, \rho(r, \theta, \phi)$. In this case, both real and complex scalar fields are treated equally and give identical predictions. We do not expect this to be a bad approximation, as it is roughly akin to only tracking quantities that have been averaged over an orbit.

\vskip 0pt
Even assuming that we can azimuthally average the scalar field density, it is still non-uniform in the radial direction, and the relative asymptotic velocity between the companion and scalar field is ill-defined.  We will assume that accretion occurs dynamically in a region that is much smaller than the size of the cloud, so that we can define this velocity ``locally.'' We will later justify this assumption. This dynamical region is mesoscopic, in the sense that the dynamics is only sensitive to the local properties of the cloud (like its density and velocity), but the region is still much larger than the  size of the companion object.   In place of the asymptotic fluid density, we can then use the local density $\rho(\vec R_*)$ of the cloud at the position of the companion. Similarly, we define the local velocity to be the ratio of the probability current to the probability density, 
\begin{samepage}
\begin{equation}
\vec v_\lab{c}(\vec R_*)=\frac{i}{2\mu|\psi|^2}\bigl(\psi \nabla\psi^*-\psi^* \nabla\psi\bigr)=\frac{m}{\mu R_*^{\sperp}}\,\vec{\hat\phi}\,,
\label{eqn:v_c}
\end{equation}
where $m$ is the azimuthal angular momentum of the cloud and $R_*^\sperp$ is the length of the projection \end{samepage} of $\mb{R}_*$ on the equatorial plane, so that the difference between (\ref{eqn:v_c}) and the orbital velocity of the companion, $\mb{v}_* \sim \pm \sqrt{M/R_*} \es \hat{\bm{\phi}}$, is the relative fluid-black hole velocity. For equatorial circular orbits, with $R_*^\sperp=R_*$, this relative velocity is
\begin{equation}
  v  = \biggl|\sqrt{\frac{M}{R_*}}\mp\frac{m}{\mu R_*}\biggr| = \frac{\alpha}{\sqrt{R_*/r_\lab{c}}}\left|1 \mp \frac{m}{\sqrt{R_*/r_\lab{c}}}\right| ,
  \label{eqn:relative-v-c-BH}
\end{equation}
where the $-$ ($+$) sign  
refers to co-rotating (counter-rotating) orbits and $r_\lab{c} = (\mu \alpha)^{- 1}$ is the Bohr radius. We stress that the quantities $\rho(\mb{R}_*)$ and $\mb{v}_\lab{c}(\mb{R}_*)$ are computed without taking the backreaction of the companion into account. For small $q$, this is a good approximation.

\vskip 0pt
Under these assumptions, and for the systems we study, the mass accretion flux is approximately independent of velocity,
\begin{equation}
    \frac{\dd M_*}{\dd t} \approx \mathcal{A}_* \, \rho(\mb{R}_*)\,,\label{eqn:accretion-law}
\end{equation}
where $\mathcal{A}_* \approx 4 \pi(2 q M)^2$ is the area of the companion's horizon. From the discussion of the previous section (see the ``plateau'' in Figure~\ref{fig:accretion}), this approximation is valid as long as the relative fluid velocity is neither too slow nor too fast,
\begin{equation}
    2\pi q\alpha\ll v\ll1\,. \label{eqn:limits-on-v}
\end{equation}
From (\ref{eqn:relative-v-c-BH}), we see that this condition can be violated when either the orbital separation is very small, $R_* \sim \alpha^2 r_\lab{c}$, in which case the fluid is moving too quickly, $v \sim 1$, or when the orbital separation is very large, $R_* \sim r_\lab{c}/q^2$, in which case the fluid is moving too slowly, $v \ll 2 \pi \alpha q^2$. Both of these cases occur during a typical inspiral. However, for small $q$ and $\alpha$, the cloud is extremely dilute whenever (\ref{eqn:limits-on-v}) is violated, because the companion is either too close\footnote{We have assumed that the cloud has nontrivial angular momentum, which pushes the density of the cloud away from the parent black hole. This is a fair assumption, as these are the types of states prepared by superradiance. Moreover, we do not expect accretion to be significant for $\ell=0$ states anyway, since the time spent by the companion in the region $R_*\lesssim\alpha^2 r_\lab{c}$ is very short.} or too far away from the parent black hole to see an appreciable density, and so accretion is negligible whenever (\ref{eqn:accretion-law}) does not apply.\footnote{This reasoning can fail when the relative velocity (\ref{eqn:relative-v-c-BH}) vanishes and the companion orbits the parent black hole at the same local speed as the cloud, which occurs for co-rotating orbits at $R_* = m^2 r_\lab{c}$. 
In an orbital band of width $\Delta R_*\sim\pi qm^3r_\lab{c}$ around this special orbit, the constraint $2\pi q\alpha\ll v$ is violated and (\ref{eqn:accretion-law}) cannot be applied. Rather, the low-velocity limit of (\ref{eqn:accretion-cases}) must be used instead and accretion is enhanced.} 

\vskip 0pt
Finally, let us now check that the accretion process actually happens in a mesoscopic region where we can assume that the companion sees a uniform medium. 
The mass absorption formula (\ref{eqn:accretion-law}) can be written as
\begin{equation}
\frac{\dd M_*}{\dd t}=\big(\pi b\ped{max}^2\big)\es v\rho\,, 
\label{eqn:accretion-formula-M*}
\end{equation}
where $b\ped{max} \equiv 4qM/\sqrt{v}$ is the radius of the absorption cross section, or the maximum impact parameter for absorption in a particle analogy. To apply the idealized derivation, we need to satisfy two conditions: (1) the density and velocity of the cloud are approximately constant over a region of size $b\ped{max}$ and (2) the region of size $b\ped{max}$ is gravitationally dominated by the companion, i.e.~it is smaller than the radius of the Hill sphere $r\ped{Hill}=R_*(q/3)^{1/3}$.
These two conditions then require that
\begin{align}
&(1) \ \  
b\ped{max}\ll r_\lab{c} \hspace{9pt} \implies\ \frac{R_*}M\ll \big(4q\alpha^2\big)^{- 4}\, ,
\label{eqn:condition-a} \\
 &(2) \ \ b\ped{max}\ll r\ped{Hill} \implies\ \frac{R_*}M\gg \big(8 q/{\sqrt{3}}\big)^{8/9}\, .
\label{eqn:condition-b}
\end{align}
Both of these conditions are  easily satisfied for the typical values of $\alpha$, $q$ and $R_*$ that we are interested in.

\vskip 0pt
There are two ways the companion can fail to see such a uniform medium. The first is simply if the azimuthally-averaged density $\rho(\mb{R}_*)$ vanishes, or changes dramatically, at a particular orbital separation. This can occur when the cloud occupies a state $|n \es \ell \es m \rangle$, with $\ell \neq n -1$, for which the radial wavefunction has zeros away from the origin. In this case, we can think of the density that the companion sees as simply being the averaged density within a Hill sphere about the companion. Similarly, as illustrated in Figure~\ref{fig:accretionSchematic}, the companion itself changes the local density---it vacuums up the scalar field as it passes through the cloud and leaves an empty ``tube'' of diameter $\mathcal{O}(M_*)$. However, the cloud will respond and replenish this local density on a relatively short timescale. This perturbation excites modes with typical wavelength of $\mathcal{O}(M_*)$, whose frequencies $\omega^2 = \mu^2 + k^2$ scale as $\mathcal{O}\big(\mu/(\alpha q)\big)$. These modes respond extremely quickly, and we expect that this empty ``tube'' is rapidly filled in before companion can complete an orbit and encounter this locally depleted region again. So, the companion should see a relatively uniform medium throughout the inspiral, and we will thus use the approximation~(\ref{eqn:accretion-law}) throughout Chapter~\ref{chap:observational-signatures} to capture the effect accretion has on the binary's dynamics. 

\chapter{A deeper look at ionization}

In Section~\ref{sec:ionization-review} we have introduced and quantified ionization, or the excitation of unbound states due to the binary perturbation. Not only is this phenomenon striking already at first glance, due to its large magnitude compared to GW emission, but it also has dramatic and peculiar observational consequences. As such, it deserves its own chapter.

\vskip 0pt
In Section~\ref{sec:dynamical-friction} we shed new light on ionization by interpreting its backreaction on the orbit as dynamical friction, which is a well-known effect at play in many astrophysical scenarios. In Section~\ref{sec:scaling} we derive useful scaling formulae which allow to extend the results to arbitrary values parameters after they have been computed once for some fiducial values of the parameters, then in Section~\ref{sec:backreaction-ionization} give a quick look at the backreaction of ionization on the orbit. To better understand the discontinuous features observed in Figure~\ref{fig:ionization_211}, we present a different derivation of ionization, which consistently takes into account the chirp of the orbital frequency. We first perform this in a simplified toy model in Section~\ref{sec:warmup}, and then apply the same approach to the realistic case in Section~\ref{sec:realisticIonization}. Finally, in Sections~\ref{sec:eccentricity} and \ref{sec:inclination} we generalize the results to eccentric and inclined orbits, respectively.

\label{chap:ionization}

\section{Dynamical friction}

\label{sec:dynamical-friction}

As detailed in Section~\ref{sec:ionization-review}, ionization pumps energy into the scalar field. This must happen at the expense of the binary's total energy, meaning that ionization backreacts on the orbit by inducing an energy loss, or a ``drag force''. The effect peaks roughly when the orbital separation equals the distance at which the cloud is densest, as is clear from the top panel of Figure~\ref{fig:ionization_211}. This conclusion is hardly a surprise. The existence of a drag force acting on an object (in our case, the secondary body of mass $M_*$) that moves through a medium (the cloud) with which it interacts gravitationally is well-established, and goes under the name of \emph{dynamical friction}, as already shown in \eqref{eqn:Pdf-intro}. In this section, our goal is to give a detailed comparison between ionization and well-known results about dynamical friction, eventually showing that the two effects should be interpreted as one.

\vskip 0pt
Dynamical friction was first studied by Chandrasekhar in \cite{Chandrasekhar:1943ys} for a medium composed of collisionless particles. More recently, results have been found for the motion in an ultralight scalar field \cite{Hui:2016ltb,Traykova:2021dua,Vicente:2022ivh,Traykova:2023qyv,Buehler:2022tmr}, which is relevant for our case. For non-relativistic velocities, the dynamical friction power is found to be
\beq
P_\slab{df}=\frac{4\pi M_*^2\rho}v\bigl(\log(v\mu b\ped{max})+\gamma_\slab{e}\bigr)\,.
\label{eqn:df}
\eeq
We now define the parameters entering (\ref{eqn:df}), as well as highlight all the assumptions behind it.

\begin{enumerate}
\item At large distance from the object of mass $M_*$, the medium is assumed to be uniform with density $\rho$. The velocity $v$ of the object is measured with respect to the asymptotically uniform regions of the medium.
\item The motion of the object is assumed to be uniform and straight. In particular, this implies that its interaction with the medium started an infinitely long time in the past.
\item If the two previous assumptions are taken strictly, the result for $P_\slab{df}$ is logarithmically divergent. The reason is that, in the stationary configuration, the medium forms an infinitely extended wake of overdensity behind the moving body, whose gravitational pull on the object diverges. A regulator is thus introduced: the parameter $b\ped{max}$ sets an upper bound to the impact parameter of the elements of the medium whose interaction with the object is taken into account. The last factor of (\ref{eqn:df}) depends on $b\ped{max}$ (logarithmically), as well as on the mass of the scalar field $\mu$ and the Euler-Mascheroni constant $\gamma_\slab{e}\approx0.577$.
\end{enumerate}

\vskip 0pt
Before applying formula (\ref{eqn:df}) to the case of a gravitational atom in a binary, one must realize that these three points all fail or need modifications: (1) the medium is not uniform and has a finite size; as a consequence, the relative velocity $v$ must be redefined; (2) the object moves in a circle rather than in a straight line; (3) the finiteness of the medium acts as a natural regulator for the divergence of $P_\slab{df}$; as a consequence, the parameter $b\ped{max}$ (which would not be needed in a self-consistent calculation) must be fixed with a suitable choice. Nevertheless, formula (\ref{eqn:df}), as well as similar ones for other kinds of media, are routinely applied in similar astrophysical contexts \cite{Eda:2014kra,Macedo:2013qea,Barausse:2014tra,Zhang:2019eid,Kavanagh:2020cfn,Kim:2022mdj}, with the expectation that they capture the correct dependence on the parameters and provide a result which is correct up to factors of $\mathcal O(1)$.

\vskip 0pt
Let us now evaluate (\ref{eqn:df}) in our case, adopting choices for the various parameters that are common in the literature. We set $\rho$ equal to the local density of the cloud at the companion's position, $\rho=M\ped{c}\abs{\psi(\vec R_*)}^2$; we fix $v$ equal to the orbital velocity, $v=\sqrt{(1+q)M/R_*}$; finally, we choose $b\ped{max}=R_*$. Note that these choices are, strictly speaking, mutually inconsistent: for example, we are considering impact parameters as large as the size of the orbit, but ignoring that over such distance the cloud's density varies significantly compared to its local value.

\begin{figure}
\centering
\includegraphics{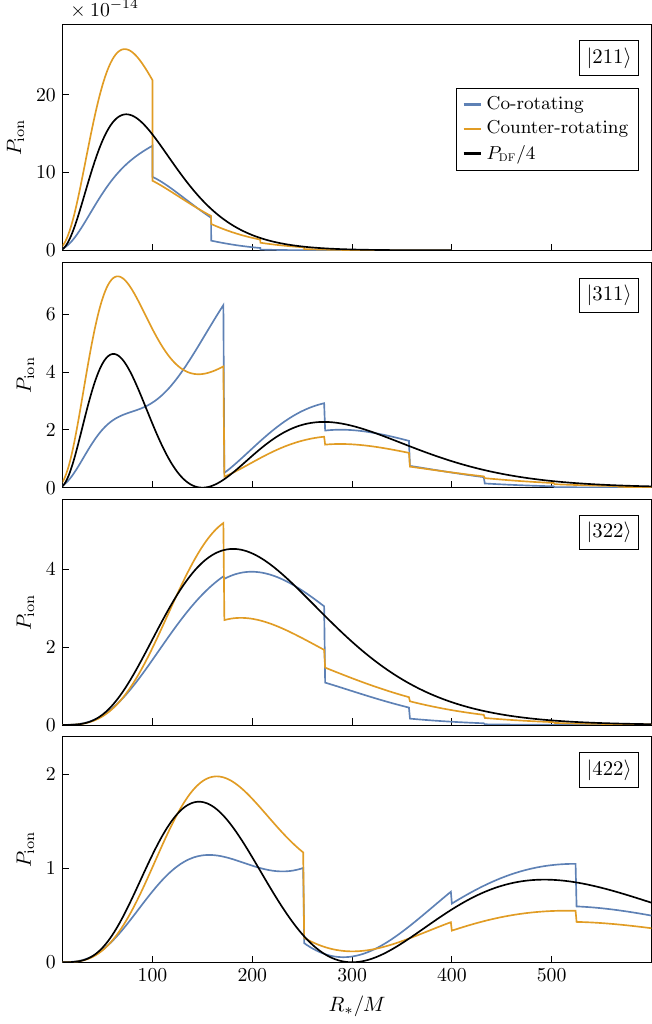}
\caption{Comparison of $P_\slab{df}$ (divided by 4 for clarity) with $P\ped{ion}$, for clouds in the states $\ket{311}$, $\ket{322}$ and $\ket{422}$. All the parameters and the units are the same as in Figure~\ref{fig:ionization_211}.}
\label{fig:dynamical-friction}
\end{figure}

\vskip 0pt
With these assumptions, we calculate $P_\slab{df}$ and compare it to $P\ped{ion}$ in Figure~\ref{fig:dynamical-friction}, for a selection of states $\ket{n_b\ell_bm_b}$. In all cases, $P_\slab{df}$ turns out to be a factor of a few larger than $P\ped{ion}$; for the sake of a better visual comparison, we plotted the fourth part of $P_\slab{df}$ instead. The various states have been selected not necessarily because they are expected to be populated by superradiance, but simply to exhibit the comparison between $P_\slab{df}$ and $P\ped{ion}$ on clouds with different profiles. Clearly, $P_\slab{df}$ possesses no discontinuities and, with the assumed values of its parameters, its value does not depend on the orientation of the orbit. In all cases, nevertheless, the two quantities have roughly the same overall shape, generally peaking in correspondence with the densest regions of the cloud and having minima elsewhere. This conclusion does not depend on the chosen values of the parameters: by plugging the assumed values of $\rho$, $v$ and $b\ped{max}$ in (\ref{eqn:df}), it is possible to show that the ratio $P_\slab{df}$ has exactly the same scaling as $P\ped{ion}$, which will be discussed in Section~\ref{sec:scaling}. This means that the ratio $P_\slab{df}/P\ped{ion}$ is universal, and roughly equal to a constant of $\mathcal O(1)$.

\vskip 0pt
Having demonstrated that $P_\slab{df}$ and $P\ped{ion}$ always give the same result, modulo the expected corrections of $\mathcal O(1)$ due to the ambiguities in fixing the parameters entering $P_\slab{df}$, we now briefly discuss, on theoretical grounds, in what sense dynamical friction must be interpreted as the backreaction of ionization. One way to derive $P_\slab{df}$ is to first solve the Schrödinger equation for the Coulomb scattering of the scalar field off the moving object, and then perform a surface integral of (some component of) the energy-momentum tensor of the medium \cite{Hui:2016ltb}. By Newton's third law, the drag force on the moving body is equal to the flux of momentum carried by the medium around it. On the other hand, the physical mechanism behind ionization, as well as the derivation of the result, is basically the same. Due to different boundary conditions, bound states carry no energy-momentum flux at infinity, while unbound states do. We solve perturbatively the Schrödinger equation and determine the rate at which the latter are populated: this defines~$P\ped{ion}$.

\vskip 0pt
The main physical difference between the two cases is the initial, unperturbed state of the medium: unbound for $P_\slab{df}$, bound around the larger object for $P\ped{ion}$. The finite energy jump that separates each bound state from the continuum is the cause of the discontinuities observed in $P\ped{ion}$ but not in $P_\slab{df}$. In this sense, we can say that ionization is sensitive to both local properties of the cloud (as it correlates with its density) and global ones (such as the bound states' spectrum) and is nothing but a self-consistent calculation of dynamical friction for the gravitational atom.

\section{Scaling formula}

\label{sec:scaling}

It is important to note that, for small $q$, the curves shown in Figures~\ref{fig:ionization_211} and \ref{fig:dynamical-friction} exhibit a universal scaling behavior. The radial wavefunctions $R_{n\ell}(r)$ and $R_{k;\ell}(r)$, given in (\ref{eq:boundWavefunctions}) and (\ref{eq:contWavefunctions}), only depend on the dimensionless variables $r/r_\lab{c}=\alpha^2r/M$ and $k r$, respectively. The wavelength $k_*$ appearing in (\ref{eq:realisticDeoccupation}) and (\ref{eq:realisticIonizationPower}) is also a function of $r/r_\lab{c}$ that scales as $\alpha^2$ and is independent of $q$, when $q \ll 1$. Because the matrix elements $\big|\eta_{K_* b}^\floq{g}(t)\big|^2$ are evaluated at $k_*$, every radial variable in the overlap integrals will therefore appear in the combination $\alpha^2r/M$. The overlaps themselves thus also inherit a homogeneous $\alpha$-scaling, which can be found by power counting. For the ionization power and the deoccupation rate, this leads to
    \begin{align}
    P_\lab{ion} &=\alpha^5 q^2 \frac{M_\lab{c}}M\, \mathcal{P}(\alpha^2 R_*/M)\, , \label{eqn:scaling-Pion} \\[4pt]
    \frac{P\ped{ion}}{P_\slab{gw}}&=\frac{M\ped{c}}M\alpha^{-5}\mathcal F(\alpha^2R_*/M)\,,
\label{eqn:pion-pgw-scaling}\\[4pt]
     \frac{\dot M\ped{c}}{M\ped{c}} &=\frac{\alpha^3 q^2}M\, \mathcal{R}(\alpha^2 R_*/M)\, , 
\end{align}
where $\mathcal{P}$, $\mathcal F$ and $\mathcal{R}$ are universal functions for each bound state $|n_b\es \ell_b\es  m_b\rangle$ that have to be found numerically.  These relations are particularly useful when results are needed for many points in parameter space, as we now only need to compute the relatively complicated functions $\mathcal{P}$ and $\mathcal{R}$ once for a fiducial set of parameters.

\section{Backreaction on the orbit}

\label{sec:backreaction-ionization}

We now briefly study the effect of ionization on a binary inspiral. This aspect is closely tied to the observational signatures, which will be discussed in Chapter~\ref{chap:observational-signatures}. For this reason, we limit our analysis here to the simplest possible exercise, i.e., solving the inspiral dynamics by including the cloud's ionization in addition to GW emission. We thus neglect resonances, accretion, and other subleading effects such as the gravitational field of the cloud.

\vskip 0pt
In a Newtonian approximation, on circular equatorial orbits, the evolution of the binary's separation is determined by energy conservation:
\begin{equation}
    \frac{qM^2}{2R_*^2}\frac{\dd R_*}{\dd t}={-P_\slab{gw}}-P_\lab{ion}\,,
    \label{eqn:evolution-R-ion}
\end{equation}
where $P_\slab{gw}$ is defined in \eqref{eqn:p_gw_circular} and $P\ped{ion}$ is given in \eqref{eqn:pion}. Ionization opens up a new channel for energy loss, which makes $R_*$ shrink faster than in vacuum, speeding up the inspiral. The magnitude of $P\ped{ion}$ depends on the cloud's mass $M\ped{c}$, which also decreases due to ionization,
\begin{equation}
\frac{\dd M_\lab{c}}{\dd t} = M_\lab{c}\biggl(\frac{\dot M\ped{c}}{M\ped{c}}\biggr)\ped{ion}\,, \label{equ:qc-evolve}
\end{equation}
where the ionization rate $(\dot M\ped{c}/M\ped{c})\ped{ion}$ is given in \eqref{eqn:rate}.

\begin{figure}
\centering
\includegraphics[trim={0 2pt 0 0}]{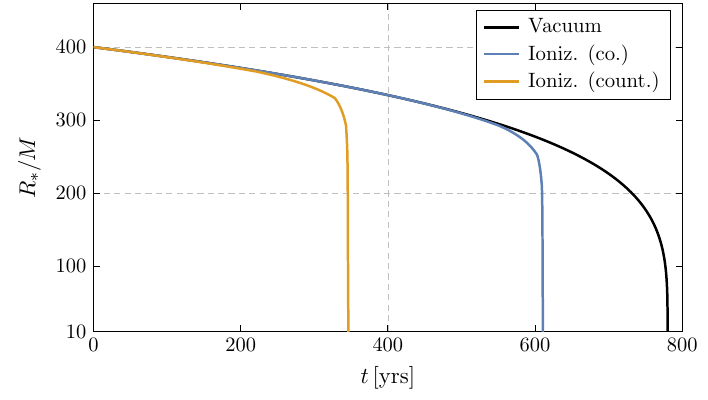}
\caption{Evolution of the separation $R_*$, for $M=10^4M_\odot$ and $\alpha=0.2$, with initial values of $R_*=400M$, $q=10^{-3}$ and $M_\lab{c}/M=0.01$ in a $|2 \es 1 \es 1\rangle$ state. Shown are the results for both co-rotating and counter-rotating orbits. The vacuum system, where no cloud is present, is shown for comparison. We see that ionization significantly reduces the merger time.}
\label{fig:separation1}
\end{figure}

\vskip 0pt
The coupled equations \eqref{eqn:evolution-R-ion} and \eqref{equ:qc-evolve} can be numerically solved for $R_*$ and $M\ped{c}$. We choose the same fiducial parameters we will use later in Chapters~\ref{chap:history} and \ref{chap:observational-signatures}, i.e., $M=10^4M_\odot$, $q=10^{-3}$ and $\alpha=0.2$. These make for a strong observational case, as the ionization shap features \eqref{eqn:omega-g} fall in the LISA band. We show in Figure~\ref{fig:separation1} the evolution of the parameters separately under the effect of ionization and GW emission, starting from a separation of $R_*=400M$. We observe a very significant shortening of the time to merger, with the orbits suddenly sinking as soon as the ionization energy losses overcome those in gravitational radiation. The dynamical evolution of the system is thus \emph{driven}, and not just perturbed, by the interaction of the binary with the cloud. The binary merges faster for counter-rotating orbits, since the ionization power is larger at large $R_*$.

\begin{figure}
\centering
\includegraphics[trim={0 2pt 0 0}]{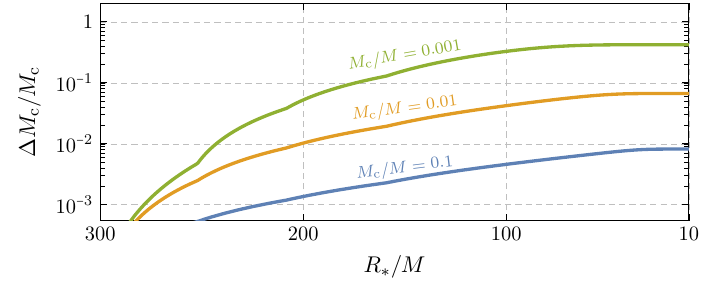}
\caption{Fractional changes of the mass of the cloud $M_\lab{c}$ due to ionization, for $M=10^4M_\odot$ and $\alpha=0.2$, with initial values of $R_*=400M$, $M_* =10^{-3} M$. Shown are the results for three different initial values of $M_\lab{c}$.  All curves refer to co-rotating orbits and a $|2\es1\es1\rangle$ bound state.}
\label{fig:q_and_M_c_ionization}
\end{figure}

\vskip 0pt
In Figure~\ref{fig:q_and_M_c_ionization} we show the fractional change of the mass of the cloud~$M_\lab{c}$, which is depleted during the inspiral due to ionization. We see that the total mass loss does not seem to depend sensitively on the initial value of $M_\lab{c}$, so that the fractional mass loss is larger for smaller clouds. In particular, numerical experiments suggest that the total ionized mass is roughly equal to the mass of the companion, $\Delta M\ped{c}\sim M_*$. In the example with $M_\lab{c}/M=0.1$, only about 1\% of the initial mass is lost at the end of the inspiral; instead, more than 50\% would be depleted for an initial $M_\lab{c}/M = 10^{- 3}$.

\section{A more thorough derivation}

\label{sec:thorough-derivation}

The derivation of ionization presented in Section~\ref{sec:ionization-review} assumes that the orbital frequency $\Omega$ stays constant, thus neglecting the frequency chirp due to the inspiral. This simplification is needed to apply Fermi's Golden Rule. It is natural, however, to ask whether the procedure leads to qualitatively and quantitatively accurate results. For instance, the nonzero chirp rate $\gamma$ plays a crucial role in the Landau-Zener transitions described Section~\ref{sec:resonances-review}, and one may wonder whether that is the case for ionization too. Answering this question requires us to take a step back. In this section, we re-derive the ionization rate and power by directly solving the Schrödinger equation with a nonzero chirp rate $\gamma$. By doing so, we are able not only to confirm the results of Section~\ref{sec:resonances-review}, but also to gather crucial insights on the nature of the discontinuities observed in Figures~\ref{fig:ionization_211} and~\ref{fig:dynamical-friction}.

\subsection{A toy model}

\label{sec:warmup}
          
Consider a single bound state $|b \rangle$, with energy $\epsilon_b < 0$, interacting with a semi-infinite continuum of states $|k \rangle$.  For simplicity, we will assume that the continuum states depend only on the wavenumber~$k$, with dispersion relation $\epsilon(k) = k^2/2 \mu$, and that they do not interact with one another. We will also assume that the interaction between the bound state and the continuum oscillates at a frequency $\dot{\varphi}_*(t)$ that grows slowly and linearly in time, $\ddot{\varphi}_*(t) = \gamma$. This is the simplest generalization of the familiar two-state Landau--Zener system to include the coupling to the continuum. Despite its simplicity, this toy model will illustrate many of the phenomena we will encounter in the more realistic scenario. 

\vskip 0pt
       The Hamiltonian of our toy model is\footnote{This is an extension of the Demkov--Osherov model~\cite{Demkov:1968san} to a single bound state interacting with a semi-infinite continuum. A similar model was studied in~\cite{Basko:2017lzs}, but with a different focus and using different techniques.}
        \begin{equation}
            \mathcal{H} = \epsilon_b \es |b \rangle \langle b| + \frac{1}{2 \pi} \int_0^{\infty}\!\dd k\, \Big[\eta(k) e^{- i \varphi_*(t)} |k \rangle \langle b | + \eta^*(k) e^{i \varphi_*(t)} |b \rangle \langle k | + \epsilon(k) |k \rangle \langle k | \Big]\, . \label{eq:toyHam}
        \end{equation}
        As in \eqref{eq:contNormalization}, the continuum states are normalized such that $\langle k | k' \rangle = 2 \pi \delta(k - k')$, while the phase is $\varphi_*(t) = \varphi_0 + \Omega_0 t + \gamma t^2/2$. A general state in the Hilbert space can be written as
        \begin{equation}
            |\psi \rangle = c_b(t) e^{- i\epsilon_b t} | b \rangle + \frac{1}{2 \pi} \int_{0}^{\infty}\!\dd k \, c_k(t) e^{- i \epsilon(k) t} |k \rangle\, , \label{eq:warmupState}
        \end{equation}
     where we have peeled off the standard oscillatory behavior caused by the non-zero energies of each state---this will help us isolate the effect of the interactions $\eta(k)$. The Schr\"{o}dinger equation associated to the Hamiltonian (\ref{eq:toyHam}) leads to the equations of motion
        \begin{align}
            i\hskip 1pt \frac{\dd{c}_b}{\dd t} &= \frac{1}{2 \pi} \int_{0}^{\infty}\!\dd k\, \eta^*(k) e^{i \varphi_*(t) + i (\epsilon_b - \epsilon(k)) t} c_k(t)\,, \label{eq:warmupBoundEom} \\
            i\hskip 1pt \frac{\dd {c}_k}{\dd t} &= \eta(k) e^{-i \varphi_*(t) + i(\epsilon(k) - \epsilon_b) t} c_b(t)\,. \label{eq:warmupContEom}
        \end{align}
        Our goal is to ``integrate out'' the continuum to find an approximate description of the system entirely in terms of the bound state's dynamics. We do so using the so-called Weisskopf--Wigner method; see e.g.~\cite{Weisskopf:1930bdn,scully1997quantum,Herring:2018afy}.

        \vskip 0pt
        Assuming that the system begins its life in the bound state, $c_k(t) \to 0$ as $t \to \minus \infty$, for all $k$, we can solve (\ref{eq:warmupContEom}) exactly, 
         \begin{equation}
            c_k(t) = -i \int_{- \infty}^{t}\!\dd t'\, \eta(k) e^{i (\epsilon(k) - \epsilon_b) t'-i \varphi_*(t')} c_b(t')\,. \label{eq:warmupContSol}
        \end{equation}
        Substituting this into (\ref{eq:warmupBoundEom}), we find an (integro-differential) equation for the dynamics of the entire system purely in terms of the bound state amplitude,
        \begin{equation}
          i \hskip 1pt \frac{ \dd{c}_b}{\dd t} = \int_{- \infty}^{t}\!\dd t'\, \Sigma_b(t, t') \hskip 1pt c_b(t')\,, \label{eq:toySelfEnergyEq}
        \end{equation} 
        where we have defined the \emph{self-energy}
        \begin{equation}
          \Sigma_b(t, t') \equiv \frac{1}{2 \pi i} \int_{0}^{\infty}\!\dd k\, |\eta(k)|^2\,  e^{i (\varphi_*(t) - \varphi_*(t'))-i (\epsilon(k) - \epsilon_b)(t - t')} \,. \label{eq:toySelfEnergyDef}
        \end{equation}
        This equation of motion is still quite complicated, but we can make significant progress via the \emph{Markov approximation}~\cite{Herring:2018afy}, wherein we integrate by parts and drop the remainder term. The bound state Schr\"{o}dinger equation then simplifies to 
        \begin{equation}
             i\hskip 1pt \frac{\dd{c}_b}{\dd t}  = \mathcal{E}_b(t) c_b(t)\,, \label{eq:toyEffSchro}
        \end{equation}
        where we have introduced the  
        \emph{induced energy}
        \begin{equation}
            \mathcal{E}_b(t) = \int_{- \infty}^{t}\!\dd t' \, \Sigma_b(t, t') = \frac{1}{2 \pi i} \int_{- \infty}^{t} \!\dd t' \int_0^{\infty}\!\dd k \, |\eta(k)|^2 \, e^{i(\varphi_*(t) - \varphi_*(t'))-i (\epsilon(k)- \epsilon_b)(t - t')}\,. \label{eq:effEnergy}
        \end{equation}
        As we discuss in Appendix~\ref{app:Markov}, this approximation consists of dropping terms that are higher order in $\mathcal{E}_b(t)$ and its time integrals. 
       The imaginary part of the induced energy completely determines the behavior of the bound state occupation probability, which may be approximated as
        \begin{equation}
            \frac{\dd \log |c_b(t)|^2}{\dd t} = 2 \Im \mathcal{E}_b(t) \approx - \frac{\mu |\eta(k_*(t))|^2}{k_*(t)} \, \Theta\big(k_*^2(t)\big)\,, \label{eq:toyDeoccupation}
        \end{equation}
        where $k_*(t) = \sqrt{2 \mu \left(\dot{\varphi}_*(t) + \epsilon_b\right)}$ and $\Theta(x)$ is the Heaviside function, with $\Theta\big(k^2_*(t)\big) = \Theta(\dot{\varphi}_*(t) + \epsilon_b)$. We will devote the rest of this section to understanding the time dependence of $\Im \mathcal{E}_b(t)$ and qualitatively justifying the approximation in (\ref{eq:toyDeoccupation}).

        \begin{figure}[t]
            \centering
            \includegraphics[trim={0 2pt 0 0}]{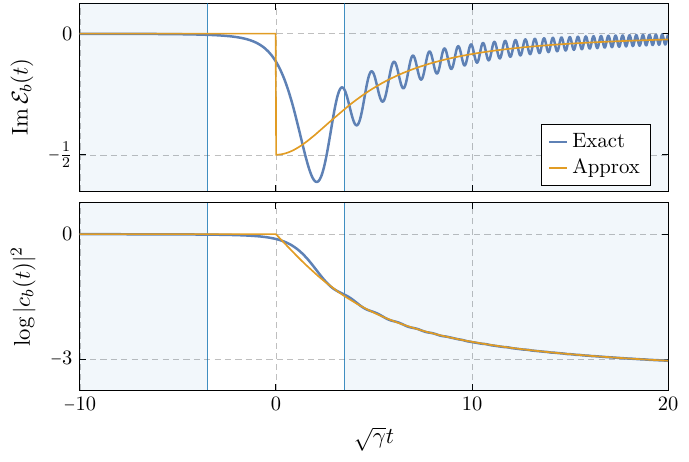}
            \caption{The imaginary part of the induced energy $\mathcal{E}_b(t)$ (\emph{top}) and the log occupation of the bound state~$\log |c_b(t)|^2$ (\emph{bottom}) as functions of dimensionless time $\sqrt{\gamma} t$, using both the exact expression (\ref{eq:effEnergy}) [{\color{Mathematica1}blue}] and our approximation (\ref{eq:toyDeoccupation}) [{\color{Mathematica2}orange}]. Here, we assume that the bound-to-continuum couplings take the form $|\eta(k)|^2 = (k/\mu)/[1 + k^4/(81 \mu^2 \gamma)]$.
            \label{fig:transientBehavior}}
        \end{figure}

\vskip 0pt
        As we might expect, the bound state only starts to significantly interact with the continuum when the frequency of the perturbation $\dot{\varphi}_*(t)$ is high enough to compensate for the bound state's binding energy, $\minus \epsilon_b$. 
        This is when the bound state starts to ``resonate" with the continuum and we can choose our time coordinate so that this resonance occurs at $t = 0$. This is not a resonance in the classic sense, but we find it useful to continue using this language. As illustrated in Figure~\ref{fig:transientBehavior}, the system (\ref{eq:toyEffSchro}) evolves on a time scale set by $\gamma^{- 1/2}$ and its behavior can be divided into three distinct stages. 

        \vskip 0pt
        Far before the resonance, in the left shaded region, where $\sqrt{\gamma} t \ll -1$, the perturbation cannot provide enough energy for the bound and continuum states to interact and so the population of the bound state is, to good approximation, completely unaffected by the presence of the continuum. This changes when $|\sqrt{\gamma} t | \lesssim 1$, in the unshaded transient region, where the system goes on resonance and develops a relatively complicated time dependence. We do not need to fully understand this complicated stage, other than to note that this region serves to smoothly interpolate between the $\sqrt{\gamma} t \ll -1$ stage and the final $\sqrt{\gamma} t \gg 1$ stage.

        \vskip 0pt
        In the right shaded region, where $\sqrt{\gamma} t \gg 1$, the system approaches a type of steady state where the imaginary part of the induced energy $\mathcal{E}_b(t)$ is well-approximated by two distinct behaviors. The first is a remaining transient oscillation whose amplitude decays in time and whose properties depend only on the behavior of the coupling $|\eta(k)|^2$ as $k \to 0$. As described in Appendix~\ref{app:approx}, when $|\eta(k)|^2$ goes to zero linearly in $k$ at the edge of the continuum, these oscillations decay as $(\sqrt{\gamma} t)^{- 1}$, and thus their effect on the solution $\log |c_b(t)|^2$ decays as $(\sqrt{\gamma} t)^{- 2}$. As illustrated in Figure~\ref{fig:transientBehavior}, these oscillations provide a subleading correction to the dominant behavior, which is a steady and smooth deoccupation of the cloud, whose instantaneous rate depends only on the properties of the continuum state that the system is currently ``resonating'' with, i.e. the continuum state whose energy is $\frac{1}{2 \mu} k_*^2(t) = \dot{\varphi}_*(t) + \epsilon_b$. This dominant contribution (\ref{eq:toyDeoccupation}) corresponds to the result found in Section~\ref{sec:ionization-review} and is the only one we will consider from this point onwards.

        \vskip 0pt 
            We are mostly interested in the total energy that has been ionized by the perturbation, as a function of time. Assuming that the system only occupies the bound state in the far past, this ionized energy can be defined as the total energy contained within the continuum,
        \begin{equation}
        E_\lab{ion}(t) \equiv \frac{1}{2\pi} \frac{M_\lab{c}}{\mu} \int_{0}^{\infty} \!\dd k\, (\epsilon(k) - \epsilon_b)|c_k(t)|^2\,.
        \end{equation}
        As we describe in Appendix~\ref{app:ionizedEnergy}, the rate at which energy is ionized $\dd E_\lab{ion}/\dd t$, corresponding to the \emph{ionization power} already defined in \eqref{eqn:pion}, can be expressed in a particularly simple form by again working in the Markov approximation and ignoring subleading transient contributions, 
        \begin{equation}
          P_\lab{ion}(t) 
           \approx  \frac{M_\lab{c}}{\mu} \!\left[\dot{\varphi}_*(t) \, \frac{  \mu  \big|\eta(k_*(t))\big|^2}{k_*(t)}\right] \!\Theta\big(k^2_*(t)\big)\, |c_b(t)|^2  \, .
                \label{equ:EionApprox}
        \end{equation}
        This is clearly evocative of (\ref{eq:toyDeoccupation}) and enjoys a simple interpretation: how quickly the ionized energy grows is equal to the rate at which the bound state ``resonates'' into the state at $k_*(t)$, namely $\mu |\eta(k_*(t))|^2/k_*(t)$, weighted both by the energy difference $\epsilon(k_*(t)) - \epsilon_b = \dot{\varphi}_*(t)$ between these states and by how much is still left in the bound state at that time,  $(M_\lab{c}/\mu) |c_b(t)|^2$.

        \vskip 0pt
        Perhaps the most striking phenomenon we encounter are the discontinuous jumps in the ionization power, which occur when the perturbation begins to resonate with the continuum---that is, when the perturbation's frequency is just enough to excite the bound state into the very edge of the continuum. These discontinuities are apparent in our approximation of the time evolution (\ref{eq:toyDeoccupation}), shown in Figure~\ref{fig:transientBehavior}, and are ultimately due to the behavior of the continuum wavefunctions as $k \to 0$. 
        As we explain in Appendix~\ref{app:zeroMode}, the long-range nature of the $r^{- 1}$ potential localizes this zero mode to a Bohr radius-sized region around $r = 0$ and, by a matching argument, this implies that the wavefunction's normalization scales like $\sqrt{k}$ as $k \to 0$, as do all matrix elements between the bound and continuum states. The combination $\mu|\eta(k_*(t))|^2/k_*(t)$ thus approaches a \emph{finite} limit for $k_*(t) \to 0$, when the bound state begins to resonate with the continuum, leading to an apparent discontinuity in our approximation (\ref{eq:toyDeoccupation}). Said differently, the coupling per unit energy $|\eta(\epsilon)|^2 = \dd k(\epsilon)/\dd \epsilon \hskip 1pt \big|\eta\big(k(\epsilon)\big)\big|^2$ is finite in the zero-energy limit because the zero-energy modes are still localized about the origin. Of course, this approximation does not capture the transient region shown in Figure~\ref{fig:transientBehavior}, which smooths out these apparent discontinuities on a timescale $\gamma^{- 1/2}$. 
        
        \vskip 0pt
        It is instructive to compare the timescale of the transition, $\gamma^{- 1/2}$, to the characteristic timescale of the inspiral, $\Omega_0/\gamma$, which measures how long it takes for the separation between the two black holes to change by a $\mathcal{O}(1)$ fraction. Using the definition of $\gamma$ in \eqref{eqn:gamma_gws},
        the ratio of the two timescales is\footnote{Here, we have ignored the backreaction of  ionization on the binary's dynamics, which can increase the effective chirp rate $\ddot{\varphi}_*(t) \approx \gamma$ by a factor of ${\cal O}(100)$.
        This changes the estimate (\ref{eqn:estimate-sqrtgamma}), which scales as $\gamma^{1/2}$, by an $\mathcal{O}(10)$ factor. However, for the values of $q$ and $\alpha$ we consider, this does not change the fact that these transitions are~fast.\label{ftnt:comment-backreaction}}
      \begin{equation}
   \frac{\gamma^{- 1/2}}{\Omega_0/\gamma} = \sqrt{\frac{96}{5}} \frac{q^{1/2}}{(1+q)^{1/6}} \left(\frac{\alpha \Omega_0}{\mu}\right)^{5/6} \propto \sqrt{q \alpha^3}\, ,
          \label{eqn:estimate-sqrtgamma}
       \end{equation}   
       where we used that the transitions occur for $\Omega_0 \sim \mu \alpha^2$ to get the scaling in the final equality. For small $q$ and $\alpha$, the transitions therefore are very fast on the timescale of the inspiral.

\subsection{Realistic case}

\label{sec:realisticIonization}
  
    Conceptually, extending our analysis to the realistic case of the gravitational atom requires very little extra work beyond what we have already done, the main complication being that there are simply many more states to keep track of. Our goal is again to integrate out the continuum states and encode their effects on the bound states in terms of a set of induced energies and couplings, analogous to (\ref{eq:effEnergy}). These effective interactions will be relatively complicated functions of time, but will contain a simple ``steady-state'' behavior similar to (\ref{eq:toyDeoccupation}).

\vskip 0pt
    We can write the Hamiltonian of the gravitational atom as
    \begin{equation}
      \mathcal{H} = \sum_{b} \epsilon_b(t) |b \rangle \langle b | + \sum_{a \neq b} \eta_{ab}(t) |a \rangle \langle b | + \sum_{K} \epsilon_K(t) |K \rangle \langle K| + \sum_{K, b} \big[\eta_{K b}(t) |K \rangle \langle b| + \lab{h.c.}\big]\,,
    \end{equation}
    where we use $a, b, \dots$ as a bound state multi-index,\footnote{In the previous subsection, we used the subscript $b$ to denote ``bound state'' whereas now we use it as a bound state \emph{index}, slightly abusing notation.}  $|a \rangle \equiv |n_a\es \ell_a \es m_a \rangle$ and $|b \rangle = |n_b \es \ell_b \es m_b \rangle$, while $K, L, \dots$ is a continuum state multi-index,  $|K \rangle \equiv |k; \ell \es m \rangle$. We take $\epsilon_b(t)$, $\epsilon_K(t)$, $\eta_{ab}(t)$ and $\eta_{K b}(t)$ as shorthands for the energies and couplings $\epsilon_{n_b \ell_b m_b}(t)$, $\epsilon_{\ell m}(k; t)$, $\eta_{n_a \ell_a m_a | n_b \ell_b m_b}(t)$ and $\eta_{k; \ell m | n_b \ell_b m_b}(t)$, respectively. Sums over multi-indices should be understood to include a sum over \emph{all} states of a given type. For instance, the analog of (\ref{eq:warmupState}) for a generic state is\footnote{Since the energies $\epsilon_b(t)$ and $\epsilon_K(t)$ depend on time, the appropriate ``integrating factor'' in this ansatz should be $\exp\big(\minus i \int \!\dd t' \epsilon_b(t')\big)$ instead of $\exp(\minus i \epsilon_b t)$, etc. However, the time dependence of these energies is extremely suppressed, $\dot{\epsilon}_b \sim \mathcal{O}\big(\gamma (q \alpha)^2\big)$, since it only arises from the radial dynamics of the companion. Such time-dependent terms are not critical to the resonant effects we discuss in this section, and only provide very small corrections to details like the time at which the resonance begins. We will thus ignore them.}
    \begin{equation}
      \begin{aligned}
        |\psi \rangle &= \sum_{b} c_b(t) e^{- i \epsilon_b t} |b \rangle + \sum_{K} c_{K}(t) e^{- i \epsilon_K t} | K \rangle \\
        &= \sum_{n, \ell, m} c_{n \ell m}(t) e^{- i \epsilon_{n \ell m} t} |n \es \ell \es m \rangle + \frac{1}{2 \pi} \sum_{\ell, m} \int_0^{\infty}\!\dd k\, c_{k; \ell m}(t) e^{- i \epsilon(k) t} |k; \ell \es m \rangle\,,
      \end{aligned}
    \end{equation}
    where $\epsilon_{n \ell m}$ and $\epsilon(k)$ are defined in \eqref{eq:eigenenergy} and~(\ref{eq:contEnergies}), respectively.

    \vskip 0pt
    In this abbreviated notation, the coefficients obey the following equations of motion
    \begin{align}
       i\hskip 1pt \frac{\dd{c}_b}{\dd t}  &= \sum_{a \neq b} \eta_{b a}(t) c_{a}(t) e^{i (\epsilon_b - \epsilon_{a}) t} + \sum_{K} \eta_{b K}(t) c_{K}(t) e^{i (\epsilon_b - \epsilon_K) t} \,,\label{eq:realisticBoundEom}\\
        i\hskip 1pt \frac{\dd{c}_K}{\dd t} &= \sum_{a} \eta_{K a}(t) c_{a}(t) e^{i (\epsilon_K - \epsilon_{a}) t}\,. \label{eq:realisticContEom}
    \end{align}
    Assuming that the continuum states are completely deoccupied in the far past, $t \to \minus \infty$, we can solve (\ref{eq:realisticContEom}) exactly,
    \begin{equation}
      c_{K}(t) = -i \int_{- \infty}^{t}\!\dd t' \left[\sum_{a} \eta_{K a}(t') c_{a}(t') e^{i(\epsilon_K - \epsilon_a) t'}\right] .
    \end{equation}
Substituting this into (\ref{eq:realisticBoundEom}) yields an integro-differential equation purely in terms of the bound states
    \begin{equation}
     i\hskip 1pt \frac{\dd{c}_b}{\dd t} =   \sum_{a \neq b} \eta_{b a}(t) c_a(t) e^{i(\epsilon_b - \epsilon_a) t}+ \sum_{a} \int_{- \infty}^{t}\!\dd t'\,\Sigma_{ba}(t, t') c_a(t')\,, \label{eq:realisticSelfEnergyEom}
    \end{equation}
    where we have defined the self-energies
    \begin{equation}
      \Sigma_{b a}(t, t') = -i \sum_{K} \eta_{b K}(t) \eta_{K a}(t')  e^{i(\epsilon_b - \epsilon_K) t + i (\epsilon_K - \epsilon_a) t'}\,,
    \end{equation}
    which generalize (\ref{eq:toySelfEnergyEq}) to include multiple bound states. The main complication, compared to the toy model presented in Section~\ref{sec:warmup}, is that the continuum can mediate transitions between different bound states, and will thus induce off-diagonal couplings.

    \vskip 0pt
    Again working in the Markov approximation, we can rewrite (\ref{eq:realisticSelfEnergyEom}) as an effective Schr\"{o}dinger equation for the bound states
    \begin{equation}
     i\hskip 1pt \frac{\dd{c}_b}{\dd t}  = \mathcal{E}_{b}(t) c_b(t) + \sum_{a \neq b} \left[\eta_{ba}(t)e^{i(\epsilon_b - \epsilon_a)t} + \mathcal{E}_{ba}(t)\right] c_{a}(t)\,, \label{eq:realisticEffSchro}
    \end{equation}
    where we have defined both the \emph{induced couplings}
    \begin{equation}
      \mathcal{E}_{ba}(t) = -i \int_{- \infty}^{t}\!\dd t'\, \sum_{K} \eta_{b K}(t) \eta_{K a}(t') e^{i( \epsilon_b - \epsilon_K) t + i (\epsilon_K - \epsilon_a) t'} \label{eq:inducedCouplings}
    \end{equation}
    and the \emph{induced energies} $\mathcal{E}_b(t) = \mathcal{E}_{bb}(t)$, the realistic analog of (\ref{eq:effEnergy}). As before, we have reduced the complicated problem of bound states interacting with a continuum to the analysis of a set of (complicated) time-dependent functions $\mathcal{E}_{ba}(t)$.

    \vskip 0pt
    These induced couplings take a much simpler form when we remember that both the bound and continuum states have definite azimuthal angular momentum, which we will denote as $m$ for the continuum state $K$ and $m_a$ or $m_b$ for the bound states $|a\rangle$ or $|b \rangle$, respectively. Since the couplings between the bound and continuum states $\eta_{K a}(t)$ reduce to a single Floquet component, $\eta_{Ka}(t)=e^{-i(m-m_a)\varphi_*(t)}\eta_{Ka}^\floq{m-m_a}(t)$, we can write the induced couplings appearing in (\ref{eq:realisticEffSchro}) as\footnote{ The Floquet components $\eta_{Ka}^{\floq{m \minus m_a}}$ inherit their time dependence purely from the radial motion of the companion. Though this slow radial motion is extremely important when it forces the frequency of the perturbation to slowly increase in time and cannot be ignored there, taking the adiabatic approximation $\eta_{K a}^\floq{m \minus m_a}(t') \approx \eta_{K a}^\floq{m \minus m_b}(t)$ only requires dropping subleading terms of~$\mathcal{O}(\gamma)$, and so we will use this approximation throughout.} 
    \begin{equation}
    \begin{aligned}
      \mathcal{E}_{ba}(t) = -i \int_{- \infty}^{t}\!\dd t'\, &\sum_{K} \eta^{*\floq{m \es\es \scalebox{0.9}{$-$} \es\es m_b}}_{K b}(t) \eta^{\floq{m \es\es \scalebox{0.9}{$-$} \es\es m_a}}_{K a}(t) \\
      &\times e^{i (m - m_b) \varphi_*(t) - i (m - m_a) \varphi_*(t') + i( \epsilon_b - \epsilon_K) t + i (\epsilon_K - \epsilon_a) t'}\,.
      \end{aligned}
    \end{equation}
    As we argue in Appendix~\ref{app:approx}, the off-diagonal terms oscillate as $\mathcal{E}_{ba}(t) \propto e^{i (\epsilon_b - \epsilon_a)t-i(m_b - m_a) \varphi_*(t)}$,
    just like the directly mediated transitions between the bound states
    \begin{equation}
      \eta_{ba}(t)\es e^{i( \epsilon_b - \epsilon_a)t} = \eta^\floq{m_b \es\es \scalebox{0.9}{$-$} \es\es m_a}_{ba}(t)\es e^{i (\epsilon_b - \epsilon_a)t-i(m_b - m_a) \varphi_*(t)}\,.
    \end{equation}
    The total coupling between the bound states $|a \rangle$ and $|b \rangle$, $\eta_{ba}(t) e^{i (\epsilon_b - \epsilon_a) t} + \mathcal{E}_{ba}(t)$, thus oscillates extremely rapidly unless the argument of this exponential becomes stationary, which occurs when
    \begin{equation}
      (m_b - m_a) \dot{\varphi}_*(t) = \epsilon_b - \epsilon_a\,.
    \end{equation}
    This is exactly the resonance condition \eqref{eqn:resonant-frequency} and so, even including the effects of the continuum, we can ignore transitions between bound states as long the system is not actively in resonance. That is, away from resonances the coupling between $|a \rangle$ and $|b \rangle$ oscillates rapidly enough so as to effectively average out to zero. In Section~\ref{sec:ion-at-resonance} we generalize this approach and show that the corrections from resonances have a very small impact anyway. At the end of the day, we can thus ignore the resonances, and dramatically simplify the effective Schr\"{o}dinger equation (\ref{eq:realisticEffSchro}) to
    \begin{equation}
     i\hskip 1pt \frac{\dd{c}_b}{\dd t} = \mathcal{E}_b(t) c_b(t)\,,
    \end{equation}
    where the induced energies,
    \begin{equation}
      \mathcal{E}_b(t) = -i \, \sum_{K}  \int_{- \infty}^{t}\!\dd t'\,  \big|\eta^{\floq{m \es\es \scalebox{0.9}{$-$} \es\es m_b}}_{K b}(t)\big|^2 \,e^{i (m - m_b) ( \varphi_*(t) - \varphi_*(t'))- i(\epsilon_K - \epsilon_b) (t- t')  }\,,
    \end{equation}
    are simply the generalization of (\ref{eq:effEnergy}) to include continuum states with different angular momenta.

    \vskip 0pt
    The dynamics of this effective Schr\"{o}dinger equation are very similar to those of the toy model. Assuming that the system occupies a single bound state and ignoring the transient oscillations as we discussed in Section~\ref{sec:warmup}, we may write the analog of (\ref{eq:toyDeoccupation}) as
    \begin{equation}
      \frac{\dot M\ped{c}}{M\ped{c}}=\frac{\dd \log |c_b(t)|^2}{\dd t} = 2 \, \lab{Im}\, \mathcal{E}_b(t) \approx -\sum_{\ell, g}\left[ \frac{\mu \big|\eta^\floq{g}_{K_* b}(t)\big|^2}{k_*^\floq{g}(t)} \Theta\big(k^\floq{g}_*(t)^2\big)\right] , \label{eq:realisticDeoccupation}
    \end{equation}
    with $K_* = \{k_*^{\floq{g}}(t), \ell, m = g+m_b\}$ and $k_*^\floq{g}(t) = \sqrt{2 \mu(g \dot{\varphi}_*(t) + \epsilon_b)}$, where the sum ranges from $\ell = 0,1, \dots, \infty$ and over all $g$ such that $|g + m_b| \leq \ell$. As before, the instantaneous rate of deoccupation only relies on the properties of the state that the system currently ``resonates'' with. However, in contrast to our toy model, there are two main complications. First, the perturbation oscillates at every overtone $g \in \mathbb{Z}$ of the base frequency $\dot{\varphi}_*(t)$. Second, the continuum state with energy $\frac{1}{2 \mu} k_*^2(t) = g \dot{\varphi}_* + \epsilon_b$ is infinitely degenerate. The sum over overtones is killed by the fact that (on equatorial orbits) the coupling oscillates with a definite frequency, but we still need to account for this infinite degeneracy, leading to the sum over total and azimuthal orbital angular momentum.

    \vskip 0pt
    The same simplifications apply to the ionization power, which we may write as
        \begin{equation}
      P_\lab{ion} \equiv  \frac{\dd E_\lab{ion}}{\dd t} \approx \sum_{\ell, g} \frac{M_\lab{c}}{\mu}\!\left[g \dot{\varphi}_*(t) \, \frac{\mu \big|\eta_{K_* b}^\floq{g}(t)\big|^2}{k_*^{\floq{g}}(t)}\right] \!\Theta\big(k_*^\floq{g}(t)^2\big)\, |c_b(t)|^2 \,, \label{eq:realisticIonizationPower}
    \end{equation}
    assuming that the system initially only occupies one bound state $|b \rangle$, where the sum is again over all states that can participate in the resonance. This expression precisely matches \eqref{eqn:pion}, and thus we recover the result from perturbation theory with stationary frequency.
  
      \vskip 0pt  
    The ionization power has been plotted as a function of the binary separation $R_*$ in Figure~\ref{fig:ionization_211}.
    As we explained in the previous subsection, the discontinuous jumps that appear in both panels are due to the bound state beginning to resonate with the continuum and the fact that all couplings $|\eta_{K b}|^2$ are $\propto k$ as $k \to 0$. The fact that the perturbation now has multiple overtones means that this resonance can occur at multiple points in the orbit.

\section{Eccentric orbits}

\label{sec:eccentricity}

A binary that forms via dynamical capture, as discussed in Section~\ref{sec:capture}, is initially characterized by very eccentric orbits. So far, our study of ionization has instead focused on quasi-circular orbits. Before this thesis, most previous works on resonances also made this same simplifying assumption \cite{Baumann:2018vus,Zhang:2018kib,Zhang:2019eid,Baumann:2019ztm,Takahashi:2021eso,Takahashi:2021yhy,Tong:2022bbl,Takahashi:2023flk}, with only \cite{Berti:2019wnn} considering non-zero eccentricity, at a time where, however, some physical aspects of the problem were not yet completely understood.

\vskip 0pt
In this section, we relax the assumption of circular orbits, generalizing the treatment of ionization to arbitrary eccentricity. We then discuss the evolution of eccentricity due to ionization and emission of GWs, explaining under what conditions the assumption of quasi-circular orbits is justified. However, we still assume for simplicity that the binary lies in the equatorial plane of the cloud: this assumption will be relaxed in Section~\ref{sec:inclination}.

\subsection{Ionization power and torque}

As reviewed in Section~\ref{sec:ionization-review}, neglecting the short transient phenomena that happen around the frequencies given in (\ref{eqn:omega-g}), the ionization rates can be found by applying Fermi's Golden Rule to a non-evolving orbit, which requires computing the matrix element 
\beq
\braket{k;\ell m|V_*(t,\vec r)|n_b\ell_bm_b}=-\sum_{\ell_*,m_*}\frac{4\pi\alpha q}{2\ell_*+1}Y_{\ell_*m_*}\biggl(\frac\pi2,\varphi_*\biggr)I_rI_\Omega\,.
\label{eqn:matrix-element-ionization}
\eeq
In the case of a circular orbit, the calculation is simplified by the fact that not only $I_\Omega$, but also $I_r$ is constant in time. The only time dependence of (\ref{eqn:matrix-element-ionization}) is then encoded in the spherical harmonics, each of which oscillates with a definite frequency, because $\varphi_*=\Omega t$ on circular orbits. This allows one to extract analytically the expression of the Fourier coefficient of the matrix element corresponding to a given oscillation frequency $g\Omega$.

\vskip 0pt
On an eccentric Keplerian orbit, the separation $R_*$ and the angular velocity $\dot\varphi_*$ vary with time. A useful parametrization is given in terms of the \emph{eccentric anomaly}~$E$:
\beq
R_*=a(1-\varepsilon\cos E)\,,\qquad(1-\varepsilon)\tan^2\frac{\varphi_*}2=(1+\varepsilon)\tan^2\frac{E}2\,,
\eeq
where $a$ is the semi-major axis and $\varepsilon$ is the eccentricity. The eccentric anomaly as function of time must then be found by solving numerically Kepler's equation,
\beq
\Omega t=E-\varepsilon\sin E\,.
\eeq
The matrix element is thus an oscillating function with period $\Omega$, which we can expand in a Fourier series as in (\ref{eqn:def-eta}),
\beq
\braket{k;\ell m|V_*(t,\vec r)|n_b\ell_bm_b}=\sum_{f\in\mathbb{Z}} \eta^\floq{f}e^{-if\Omega t}\,.
\label{eqn:fourier}
\eeq
If $k=\sqrt{2\mu(\epsilon_b+g\Omega)}\equiv k_\floq{g}$, Fermi's Golden Rule tells us that the only term of (\ref{eqn:fourier}) that gives a non-zero contribution to the ionization rate is the one that oscillates with a frequency equal to the energy difference between the two states, that is, the one with $f=g$. By comparison with equation \eqref{eqn:rate}, the ionization rate is
\beq
\frac{\dot M\ped{c}}{M\ped{c}}=-\sum_{\ell, m,g}\,\frac{\mu\abs{\eta^\floq{g}}^2}{k_\floq{g}}\Theta\bigl(k_\floq{g}^2\bigr)\,,
\label{eqn:rate-eccentric}
\eeq
where the sum runs over all continuum states of the form $\ket{k_\floq{g};\ell m}$. The Fourier coefficients $\eta^\floq{g}$ have an implicit dependence on $k_\floq{g}$ as well as on the orbital parameters. Similarly, the ionization power and torque (along the central BH's spin) are\footnote{Equations (\ref{eqn:rate-eccentric}), (\ref{eqn:pion-eccentric}) and (\ref{eqn:tion-eccentric}) are very similar to (\ref{eqn:rate}), (\ref{eqn:pion}) and (\ref{eqn:tion}). The difference is that now $\eta^\floq{g}$ no longer vanishes when $g\ne\pm(m-m_b)$, so we need to sum over $m$ and $g$ independently.}
\begin{align}
\label{eqn:pion-eccentric}
P\ped{ion}&=\frac{M\ped{c}}\mu\sum_{\ell, m,g}\,g\Omega\,\frac{\mu\abs{\eta^\floq{g}}^2}{k_\floq{g}}\Theta\bigl(k_\floq{g}^2\bigr)\,,\\
\label{eqn:tion-eccentric}
\tau\ped{ion}&=\frac{M\ped{c}}\mu\sum_{\ell, m,g}\,(m-m_b)\,\frac{\mu\abs{\eta^\floq{g}}^2}{k_\floq{g}}\Theta\bigl(k_\floq{g}^2\bigr)\,.
\end{align}

\vskip 0pt
An important difference with respect to the circular case is that it is no longer true that $P\ped{ion}=\Omega\tau\ped{ion}$. The equality held because, in that case, $Y_{\ell_*m_*}(\pi/2,\Omega t)$ was the only time-dependent term of (\ref{eqn:matrix-element-ionization}). This spherical harmonic oscillates with frequency $m_*\Omega$, which is fixed by the angular selection rules in $I_\Omega$ to be $\pm(m-m_b)\Omega$, depending on the orbit's orientation. For $\varepsilon>0$, instead, the factors entering $P\ped{ion}$ and $\tau\ped{ion}$ are independent. As we will see in Section~\ref{sec:evolution-e}, the evolution of the eccentricity will be determined by the ratio $\tau\ped{ion}/P\ped{ion}$.

\subsection{Numerical evaluation}
\label{sec:numerics-e}

The complexity of expressions (\ref{eqn:rate-eccentric}), (\ref{eqn:pion-eccentric}) and (\ref{eqn:tion-eccentric}) is hidden in the Fourier coefficients $\eta^\floq{g}$, which we evaluate numerically. Their expression,
\beq
\eta^\floq{g}=\int_0^{2\pi/\Omega}\braket{k;\ell m|V_*(t,\vec r)|n_b\ell_bm_b}e^{ig\Omega t}\dd t\supset\int_0^\pi\cos(m_*\phi_*+g\Omega t)I_r(t)\dd t\,,
\eeq
contains the overlap integral $I_r$ nested inside a time integral, as we made manifest in the last term, where we neglected all time-independent coefficients. In order to improve the convergence of the numerical routine, we write the time integrals as
\beq
\int_0^\pi\biggl[\cos[m_*(\phi_*-\Omega t)]I_r(t)\cos[(m_*+g)\Omega t]-\sin[m_*(\phi_*-\Omega t)]I_r(t)\sin[(m_*+g)\Omega t]\biggr]\dd t\,.
\eeq
The monochromatic oscillatory term $\cos[(m_*+g)\Omega t]$ multiplies a function, $\cos[m_*(\phi_*-\Omega t)]I_r(t)$, whose $\varepsilon\to0$ limit is time-independent (and similarly for the second term, replacing the cosine with a sine). This form makes it therefore particularly convenient to perform the integration using a routine optimized for definite Fourier integrals, as the $\varepsilon\to0$ limit is expected to be numerically smooth and should recover the result for circular orbits. The task is nevertheless computationally expensive: increasing the eccentricity requires to extend the sum to a larger number of final states to achieve a good numerical precision; moreover, the convergence of the integrals starts to degrade for $\varepsilon\gtrsim0.7$.

\begin{figure}[t]
\centering
\includegraphics{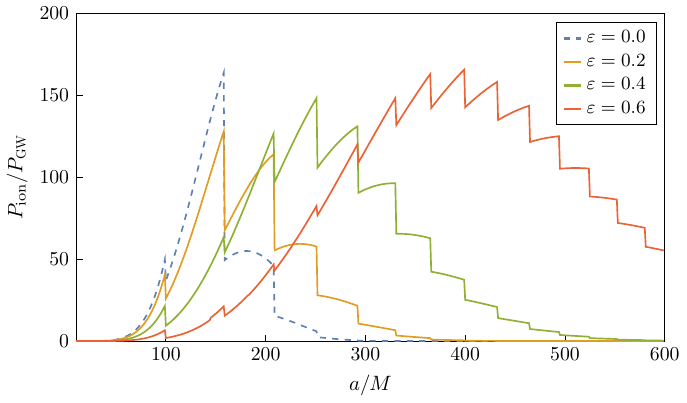}
\caption{Ionization power (\ref{eqn:pion-eccentric}) for different values of the eccentricity $\varepsilon$, as function of the semi-major axis $a$. The values are normalized by $P_\slab{gw}$, the average power emitted in gravitational waves on a correspondingly eccentric orbit, and are computed for $\alpha=0.2$, $q=10^{-3}$, $M\ped{c}=0.01M$, a cloud in the $\ket{211}$ state, and co-rotating equatorial orbits.}
\label{fig:P_ion_P_GW_eccentricity_211}
\end{figure}

\vskip 0pt
In Figure~\ref{fig:P_ion_P_GW_eccentricity_211}, we show $P\ped{ion}$ as function of the semi-major axis $a$, for different values of the eccentricity $\varepsilon$. We normalize the result by $P_\slab{gw}$, which itself depends on the eccentricity and is defined as an orbit-averaged value. The characteristic discontinuities of $P\ped{ion}$ remain at the same positions, as they are determined by the value of the orbital frequency (\ref{eqn:omega-g}), which is only a function of $a$. On the other hand, the peak of the curve shifts to larger values of $a$ for increasing $\varepsilon$. This implies that the effect of ionization is felt earlier on eccentric binaries. Similar calculations and considerations hold for the ionization rate (\ref{eqn:rate-eccentric}) and the torque (\ref{eqn:tion-eccentric}).

\subsection{Evolution of eccentricity}
\label{sec:evolution-e}

We now have all the ingredients to compute the backreaction of ionization on eccentric orbits. While a detailed solution of the evolution of the system should include the accretion of matter on the companion (if it is a BH) and the mass loss of the cloud, as well as its self gravity \cite{Ferreira:2017pth,Hannuksela:2018izj}, to first approximation we may neglect all of these effects. With respect to the case of circular orbits, the evolution of the semi-major axis does not present new insightful features: we can determine it with the energy balance equation alone,
\beq
\frac\dd{\dd t}\biggl(-\frac{qM^2}{2a}\biggr)=-P\ped{ion}-P_\slab{gw}\,,
\label{eqn:energy-balance-eccentric}
\eeq
where $P_\slab{gw}$ is defined in (\ref{eq:p_gw}) and $P\ped{ion}$ has the effect of making $a$ decrease faster than expected in vacuum. Much less trivial is the evolution of the eccentricity. In order to find it, we need the balance of angular momentum,
\beq
\frac\dd{\dd t}\sqrt{\frac{q^2M^3}{1+q}a(1-\varepsilon^2)}=-\tau\ped{ion}-\tau_\slab{gw}\,,
\label{eqn:L-balance-eccentric}
\eeq
where $\tau_\slab{gw}$ is defined in (\ref{eq:tau_gw}). This equation can then be used together with (\ref{eqn:energy-balance-eccentric}) to find $\dd\varepsilon/\dd t$.

\begin{figure}
\centering
\includegraphics{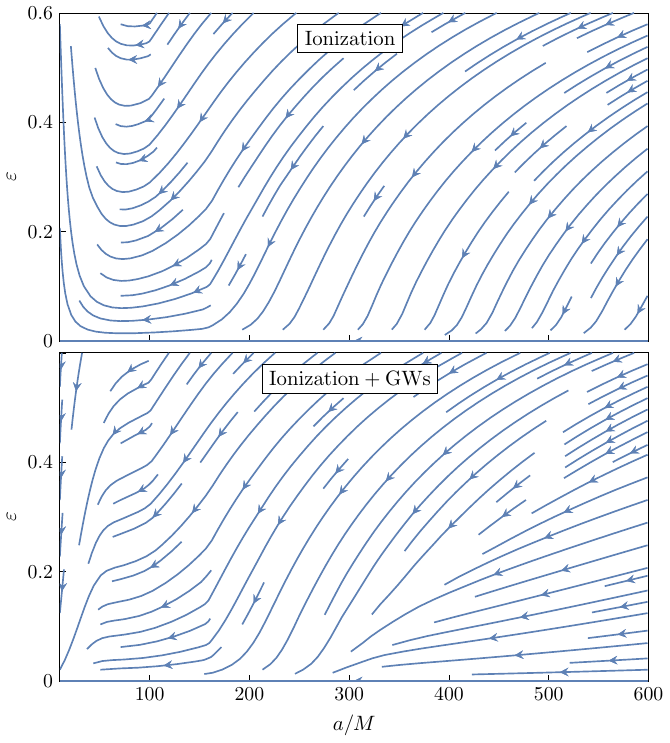}
\caption{Numerical solutions to (\ref{eqn:de/da}), for various different initial values of the semi-major axis and the eccentricity. The top panel neglects $P_\slab{gw}$ and $\tau_\slab{gw}$, while the bottom panel shows the solution to the complete equation. The values of the parameters and the orientation of the orbit are the same as in Figure~\ref{fig:P_ion_P_GW_eccentricity_211}.}
\label{fig:streamplot_eccentricity_unified}
\end{figure}

\vskip 0pt
The most pressing question is perhaps whether ionization acts to reduce or increase the binary's eccentricity. Besides being an interesting question per se, it is necessary to justify (or disprove) the assumption of quasi-circular orbits adopted in a number of works, such as \cite{Cole:2022yzw}. It is useful to combine (\ref{eqn:energy-balance-eccentric}) and (\ref{eqn:L-balance-eccentric}) as
\beq
\frac{\dd\varepsilon}{\dd(a/M)}=\frac{1-\varepsilon^2}{2\varepsilon(a/M)}-\frac{\sqrt{(1-\varepsilon^2)(1+q)}}{2\varepsilon(a/M)^{5/2}M}\frac{\tau\ped{ion}+\tau_\slab{gw}}{P\ped{ion}+P_\slab{gw}}\,,
\label{eqn:de/da}
\eeq
which allows to numerically integrate the eccentricity $\varepsilon$ as function of the semi-major axis $a$. We do this in Figure~\ref{fig:streamplot_eccentricity_unified}, where several curves corresponding to different initial values of $a$ and $\varepsilon$ are shown. In the top panel, we neglect $P_\slab{gw}$ and $\tau_\slab{gw}$ in (\ref{eqn:de/da}), while in the bottom panel we solve the full equation. Generally speaking, the binary undergoes circularization under the combined effect of ionization and gravitational wave emission. Nevertheless, when gravitational waves are neglected, for small enough $a$ the binary can experience eccentrification.

\vskip 0pt
This interesting behaviour has an insightful qualitative explanation. The density profile of the $\ket{211}$ state, shown in Figure~\ref{fig:E_lost_211}, has a maximum at a certain radius and goes to zero at the center and at infinity. Suppose that the companion is on a very eccentric orbit with semi-major axis larger than the size of the cloud, so that the density of the cloud at periapsis is much higher than at apoapsis. According to the interpretation as dynamical friction laid down in Section~\ref{sec:dynamical-friction}, the drag force experienced at periapsis will thus be much stronger than the one at apoapsis. To approximately model the fact that most of the energy loss is concentrated at the periapsis, we may imagine that the orbiting body receives a ``kick'' every time it passes through the periapsis, with the rest of the orbit being unperturbed. This way, the periapsis of successive orbits stays unchanged, while the apoapsis progressively reduces orbit by orbit: in other words, the binary is circularizing. Conversely, suppose that the semi-major axis is smaller than the size of the cloud. The situation is now reversed: the periapsis will be in a region with lower density, and successive kicks at the apoapsis will eccentrify the binary.

\vskip 0pt
The transition between circularization and eccentrification in the top panel of Figure~\ref{fig:streamplot_eccentricity_unified} happens indeed at a distance comparable with the size of the cloud, supporting the qualitative interpretation of the phenomenon. As is well-known, the emission of gravitational waves has a circularizing effect on binary systems. Indeed, when they are taken into account, the eccentrifying effect of ionization at small values of $a$ is reduced, especially for $a\to0$, where $P_\slab{gw}\gg P\ped{ion}$. It is worth noting, however, that while in Figure~\ref{fig:streamplot_eccentricity_unified} only circularization is allowed after the addition of GWs, it is in principle possible that part of the eccentrifiying effect survives, depending on the parameters (for example, a high enough mass of the cloud would guarantee a ``region'' of~eccentrification).

\section{Inclined orbits}

\label{sec:inclination}

Gravitational atoms are not spherically symmetric systems. Not only must the central BH be spinning around its axis to trigger superradiance, but the cloud itself is necessarily generated in a state with non-zero angular momentum, implying that it must have a non-trivial angular structure. Its impact on the evolution of a binary system will therefore depend on the inclination $\beta$ of the orbital plane with respect to equatorial plane defined by the spins of central BH and its cloud.

\vskip 0pt
Our discussion so far, as well as any other paper in the literature on gravitational atoms, has been restricted to equatorial orbits. In this section, we relax this assumption for the first time, by extending the treatment of ionization to the full range $0\leq\beta\leq\pi$. Precession of the orbital plane and evolution of the inclination angle will then be discussed. Motivated by the results of Section~\ref{sec:eccentricity}, in this section we will assume for simplicity that the orbits are quasi-circular.

\begin{figure}[t]
\centering
\includegraphics[scale=1]{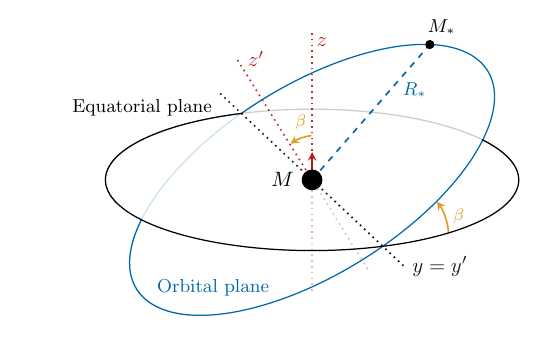}
\caption{Diagram of the coordinates used to describe inclined orbits. The orbital plane is obtained by rotating the equatorial plane by an angle $\beta$ around the $y$ axis.}
\label{fig:InclinedOrbit}
\end{figure}

\vskip 0pt
Before detailing the calculation, it is useful to state our conventions clearly. With reference to Figure~\ref{fig:InclinedOrbit}, we align the $z$ axis with the BH's spin and the $y$ axis with the intersection of the equatorial plane with the orbital plane. We use the $z$-$y$-$z$ convention for the Euler angles, so that the Euler angle $\beta$ is defined in the $x$-$z$ plane and is identified with the orbital inclination. The axes $x'$, $y'$ and $z'$, instead, will be aligned with the binary's orbit, with $y'\equiv y$.

\subsection{Ionization power and torque}

The most obvious way to compute the ionization power and torque on an inclined orbit is to simply evaluate the perturbation (\ref{eqn:V_star}) accordingly. As we assume a constant $R_*$, the only term that depends on the inclination angle $\beta$ is the spherical harmonic $Y_{\ell_*m_*}(\theta_*(t),\varphi_*(t))$. This can be written as \cite{wigner}
\beq
Y_{\ell_*m_*}(\theta_*(t),\varphi_*(t))=\sum_{g=-\ell_*}^{\ell_*}d_{m_*,-g}^\floq{\ell_*}(\beta)Y_{\ell_*,-g}\biggl(\frac\pi2,0\biggr)e^{-ig\Omega t}\,,
\label{eqn:rotated-Y}
\eeq
where $d_{m'm}^\floq{j}(\beta)$ is a Wigner small $d$-matrix, that in our conventions reads
\beq
d_{m'm}^\floq{j}(\beta)=\mathcal N\sum_{s=s\ped{min}}^{s\ped{max}}\frac{(-1)^{m'-m+s}\left(\cos\frac\beta2\right)^{2j+m-m'-2s}\left(\sin\frac\beta2\right)^{m'-m+2s}}{(j+m-s)!s!(m'-m+s)!(j-m'-s)!}\,,
\eeq
with $s\ped{min}=\max(0,m-m')$, $s\ped{max}=\min(j+m,j-m')$ and the normalization factor given by $\mathcal N=\sqrt{(j+m')!(j-m')!(j+m)!(j-m)!}$. As the expansion (\ref{eqn:rotated-Y}) separates the various monochromatic components, it is possible to proceed in a similar fashion to Fermi's Golden Rule, i.e.\ by only keeping the terms that survive a long-time average in first-order perturbation theory. In this way, we can find the total energy and angular momentum in the continuum. In order to find the ionization power and torque, however, one must subtract the energy and angular momentum remaining in the bound state, and this approach hides an important subtlety, as we will discuss now.

\vskip 0pt
On equatorial orbits, only one of the $2\ell_*+1$ terms in (\ref{eqn:rotated-Y}) is not zero and the binary's gravitational perturbation generates a transfer from the bound state $\ket{n_b\ell_bm_b}$ to continuum states. When a rotation is applied to the orbit, however, the new terms appearing in (\ref{eqn:rotated-Y}) can mediate transitions to the entire set of quasi-degenerate states $\ket{n_b\ell_bm'}$, with $m'\ne m_b$ (although not to states $\ket{n_b\ell'm'}$ with $\ell'\ne\ell_b$, as rotations do not mix different values of $\ell$). In other words, the quasi-degenerate states $\ket{n_b\ell_bm'}$ can be excited. The amount by which this happens is important in determining the ionization torque, as this is determined by the \emph{total} angular momentum carried by the scalar field, be it in continuum or bound states.

\vskip 0pt
In order to consistently describe the phenomenon, it is useful to take another approach and apply a rotation to the bound state, transforming it into a mixture of quasi-degenerate states,
\beq
\ket{n_b\ell_bm_b}\to\sum_{m'=-\ell_b}^{\ell_b}d_{m_bm'}^\floq{\ell_b}(\beta)\ket{n_b\ell_bm'}\,,
\label{eqn:rotated-state}
\eeq
which will then be perturbed by an equatorial orbit. It is important to realize that only in the limit where the Hamiltonian is invariant under rotations this approach is expected to be equivalent to the one where the orbit is rotated instead. Isotropy is only restored in the limit of vanishing BH spin, $\tilde a\to0$, while at finite spin a hyperfine splitting between the states, proportional to $\tilde a$, is present. Assuming that the ionization rate, power and torque for a given inclination angle $\beta$ are continuous in the limit $\tilde a\to0$, the two approaches will become approximately equivalent for sufficiently small BH spin. We can translate this observation into a requirement on the orbital separation by noting that there are only two relevant frequencies in the problem: the orbital frequency $\Omega=\sqrt{M(1+q)/R_*^3}$ and the hyperfine splitting $\Delta\epsilon$, which can be found from (\ref{eq:eigenenergy}). By requiring $\Delta\epsilon\ll\Omega$, we get
\beq
R_*\ll\biggl(\frac{\ell_b(\ell_b+1 / 2)(\ell_b+1)}{2\mu \tilde a\alpha^5\abs{m_b-m'} }\biggr)^{2/3}M^{1/3}(1+q)^{1/3}n_b^2\,.
\label{eqn:hyperfine_R_*}
\eeq
In other words, the rest of the discussion in this section, as well as all the results presented, will only be valid at orbital separations much smaller than the distance of the hyperfine resonance, defined by (\ref{eqn:hyperfine_R_*}). This is a well-justified assumption, as this region of space is parametrically larger than the ``Bohr'' region, where ionization peaks; for typical parameters, it is also larger than the region where $P\ped{ion}/P_\slab{gw}$ has most of its support.

\vskip 0pt
Let us therefore assume that the cloud is in the mixed state given in (\ref{eqn:rotated-state}) and consider its perturbation by an equatorial orbit. Because the matrix elements oscillate monochromatically, at fixed momentum $k$ and angular momentum $\ell$ of the final state, a state $\ket{n_b\ell_bm'}$ can only be ionized towards $\ket{k_\floq{g};\ell,m'+g}$, where $g\Omega=k_\floq{g}^2/(2\mu)-\epsilon_b$. Each of the $2\ell_b+1$ states appearing in (\ref{eqn:rotated-state}) is therefore ionized ``independently'', meaning that no interference terms are generated. We can thus find the total ionization rate, power and $z$ component of the torque by simply adding the contributions from all the $2\ell_b+1$ bound states:
\begin{align}
\label{eqn:rate-inclined}
\frac{\dot M\ped{c}}{M\ped{c}}&=-\sum_{\ell, g,m'}\,\bigl(d_{m_bm'}^\floq{\ell_b}(\beta)\bigr)^2\ \frac{\mu\big|\eta^\floq{g}_{m'}\big|^2}{k_\floq{g}}\Theta\bigl(k_\floq{g}^2\bigr)\,,\\
\label{eqn:pion-inclined}
P\ped{ion}&=\frac{M\ped{c}}\mu\sum_{\ell, g,m'}\,g\Omega\,\bigl(d_{m_bm'}^\floq{\ell_b}(\beta)\bigr)^2\ \frac{\mu\big|\eta^\floq{g}_{m'}\big|^2}{k_\floq{g}}\Theta\bigl(k_\floq{g}^2\bigr)\,,\\
\label{eqn:tauz'-inclined}
\tau\ped{ion}^{z'}&=\frac{M\ped{c}}\mu\sum_{\ell, g,m'}\,g\,\bigl(d_{m_bm'}^\floq{\ell_b}(\beta)\bigr)^2\ \frac{\mu\big|\eta^\floq{g}_{m'}\big|^2}{k_\floq{g}}\Theta\bigl(k_\floq{g}^2\bigr)\,.
\end{align}
In these expressions, we denoted by $\eta^\floq{g}_{m'}$ the matrix element of the perturbation $V_*$ between the states $\ket{k_\floq{g};\ell,m'+g}$ and $\ket{n_b\ell_bm'}$, with the same relation between $k_\floq{g}$ and $g$ as above. Note that it is very easy to go from the expression for $\dot M\ped{c}/M\ped{c}$ to the ones for $P\ped{ion}$ and $\tau\ped{ion}^{z'}$: because the states $\ket{n_b\ell_bm'}$ and $\ket{k;\ell m}$ are simultaneously eigenstates of the energy and of the $z'$ component of the angular momentum, we simply weight each term by the corresponding difference of the eigenvalues: $g\Omega$ for the energy, and $g$ for the angular momentum.

\vskip 0pt
The component $\tau\ped{ion}^{z'}$ of the torque given in (\ref{eqn:tauz'-inclined}) is relative to the axis $z'$, which is orthogonal to the orbital plane. In principle, however, there may also be components that lie in the orbital plane. Our basis does not include eigenstates of the $x'$ or $y'$ components of the angular momentum, meaning that finding the expressions for $\tau\ped{ion}^{x'}$ and $\tau\ped{ion}^{y'}$ requires a little more attention. First, remember that the matrix elements of the angular momentum operator are, in the Condon--Shortley convention, given by
\beq
L_\pm\ket{\ell,m}=\sqrt{\ell(\ell+1)-m(m\pm1)}\ket{\ell,m\pm1}\,,
\eeq
where
\beq
L_{x'}=\frac{L_++L_-}2\,,\qquad L_{y'}=\frac{L_+-L_-}{2i}\,.
\eeq
The time derivative of the angular momentum contained in the continuum states is thus
\beq
\tau\ped{out}^\pm=\frac{M\ped{c}}\mu\frac\dd{\dd t}\int\frac{\dd k}{2\pi}\sum_{\ell,m}\sqrt{\ell(\ell+1)-m(m\pm1)}\,c_{k;\ell,m\pm1}^*c_{k;\ell m}\,.
\eeq
Fermi's Golden Rule only gives the result for $\dd \abs{c_{k;\ell m}}^2/\dd t$. We thus have to go one step back and remember how the amplitudes evolve to first order in perturbation theory:
\beq
c_{k;\ell m}(t)=i\, d_{m_bm'}^\floq{\ell_b}(\beta)\,\eta_{m'}\,\frac{1-e^{i(\epsilon(k)-\epsilon_b-g\Omega)t}}{i(\epsilon(k)-\epsilon_b-g\Omega)}\,,
\label{eqn:c_kellm}
\eeq
where $\eta_{m'}$ is the matrix element of $V_*$ between $\ket{k;\ell m}$ and $\ket{n_b\ell_bm'}$. The time-dependent part of (\ref{eqn:c_kellm}) only depends on $g$, and is thus the same for $c_{k;\ell m}$ and $c_{k;\ell,m\pm1}^*$, while the prefactor differs. We can thus still apply Fermi's Golden Rule, the only difference with the previous cases being that the prefactor $\bigl(d_{m_bm'}^\floq{\ell_b}(\beta)\bigr)^2\big|\eta^\floq{g}_{m'}\big|^2$ will be replaced by its corresponding mixed product. This gives
\begin{align}
\label{eqn:tauoutx'-inclined}
\tau\ped{out}^{x'}&=\frac{M\ped{c}}\mu\sum_{\ell,g,m'}\,\sqrt{\ell(\ell+1)-m(m+1)}\,d_{m_b,m'+1}^\floq{\ell_b}(\beta)d_{m_bm'}^\floq{\ell_b}(\beta)\frac{\mu\,\eta^\floq{g}_{m'+1}\eta^\floq{g}_{m'}}{k_\floq{g}}\Theta\bigl(k_\floq{g}^2\bigr)\,,\\
\label{eqn:tauouty'-inclined}
\tau\ped{out}^{y'}&=0\,.
\end{align}
The vanishing of $\tau\ped{out}^{y'}$ is a consequence of the fact that, in our conventions, both the couplings $\eta_{m'}^\floq{g}$ and the Wigner matrices $d_{m'}^\floq{g}$ are real.

\vskip 0pt
Having computed $\tau\ped{out}^{x'}$ and $\tau\ped{out}^{y'}$, we still need to find the corresponding quantity for the angular momentum contained in the bound states,
\beq
\tau\ped{in}^\pm=\frac{M\ped{c}}\mu\frac\dd{\dd t}\sum_{m'}\sqrt{\ell_b(\ell_b+1)-m'(m'\pm1)}\,c^*_{n_b\ell_b,m'\pm1}c_{n_b\ell_bm'}\,.
\eeq
In this case, the evolution of the amplitude of each state is determined by its own ionization rate, via the requirement of unitarity (again, to first order in perturbation theory):
\beq
c_{n_b\ell_bm'}(t)=d_{m_bm'}^\floq{\ell_b}(\beta)\biggl(1-t\sum_{\ell, g}\,\frac{\mu\big|\eta^\floq{g}_{m'}\big|^2}{2k_\floq{g}}\Theta\bigl(k_\floq{g}^2\bigr)\biggr)\,.
\eeq
We thus find
\begin{multline}
\tau\ped{in}^\pm=-\sum_{\ell, g,m'}\sqrt{\ell_b(\ell_b+1)-m'(m'\pm1)}\,d_{m_b,m'\pm1}^\floq{\ell_b}(\beta)d_{m_bm'}^\floq{\ell_b}(\beta)\\
\times\biggl(\frac{\mu\big|\eta^\floq{g}_{m'\pm1}\big|^2}{2k_\floq{g}}+\frac{\mu\big|\eta^\floq{g}_{m'}\big|^2}{2k_\floq{g}}\biggr)\Theta\bigl(k_\floq{g}^2\bigr)\,,
\end{multline}
which can be expressed as
\begin{align}
\label{eqn:tauinx'-inclined}
\tau\ped{in}^{x'}&=-\frac{M\ped{c}}\mu\sum_{\ell, g,m'}J^\floq{\ell_b}_{m_bm'}(\beta)\,d_{m_bm'}^\floq{\ell_b}(\beta)\,\frac{\mu\big|\eta^\floq{g}_{m'}\big|^2}{k_\floq{g}}\Theta\bigl(k_\floq{g}^2\bigr)\,,\\
\label{eqn:tauiny'-inclined}
\tau\ped{in}^{y'}&=0\,,
\end{align}
where we defined the coefficient $J^\floq{\ell_b}_{m_bm'}(\beta)$ as
\begin{multline}
J^\floq{\ell_b}_{m_bm'}(\beta)\equiv\frac12\biggl(d_{m_b,m'+1}^\floq{\ell_b}(\beta)\sqrt{\ell_b(\ell_b+1)-m'(m'+1)}\\
+d_{m_b,m'-1}^\floq{\ell_b}(\beta)\sqrt{\ell_b(\ell_b+1)-m'(m'-1)}\biggr)\,.
\end{multline}
Finally, the contributions of the continuum and of the bound states can be added to get the total ionization torque: 
\beq
\tau\ped{ion}^{x'}=\tau\ped{out}^{x'}+\tau\ped{in}^{x'},\qquad\tau\ped{ion}^{y'}=0.
\label{eqn:tauion}
\eeq
To obtain the components of the torque in the $x$-$y$-$z$ frame, we simply need to apply a backwards rotation:
\beq
\label{eqn:rotation-tau}
\tau\ped{ion}^z=\tau\ped{ion}^{z'}\cos\beta-\tau\ped{ion}^{x'}\sin\beta\,,\qquad\tau\ped{ion}^x=\tau\ped{ion}^{x'}\cos\beta+\tau\ped{ion}^{z'}\sin\beta\,,\qquad\tau\ped{ion}^y=\tau\ped{ion}^{y'}=0\,.
\eeq
Note that, because $P\ped{ion}=\tau\ped{ion}^{z'}\Omega$, only one of the components of the torque is actually independent of $P\ped{ion}$. This is a direct consequence of having assumed a circular orbit for the binary.\footnote{Suppose that a force $\vec F$ acts on the companion, and let $\braket{\vec\tau}=\braket{\vec r\times\vec F}$ be the average torque over one orbit. Then, the average dissipated power is $\braket{P}=\braket{\vec v\cdot\vec F}=\braket{(\vec\Omega\times\vec r)\cdot\vec F}=\braket{(\vec r\times \vec F)\cdot\vec \Omega}=\braket{\vec\tau\cdot\vec\Omega}=\braket{\vec\tau}\cdot\vec\Omega$. This relation is identically satisfied by equations (\ref{eqn:pion-inclined}), (\ref{eqn:tauz'-inclined}) and (\ref{eqn:rotation-tau}). On the other hand, if we had continued with the approach outlined in (\ref{eqn:rotated-Y}), and neglected the change in the occupancy of the states with $m'\ne m_b$, we would have found a result that violates this identity, and that is therefore inconsistent.}

\subsection{Numerical evaluation}

\begin{figure}
\centering
\includegraphics{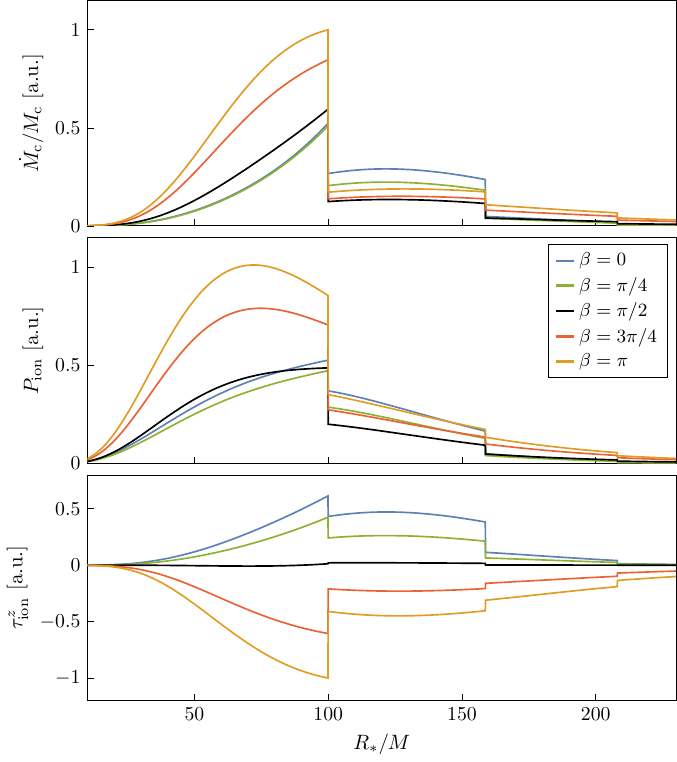}
\caption{Instantaneous ionization rate (\emph{top}), power (\emph{middle}) and torque along $z$ (\emph{bottom}) for a cloud in the $\ket{211}$ state, as function of the binary separation, for different values of the orbital inclination $\beta$. The $y$ axes are reported in arbitrary units (a.u.), while the $x$ axis has been normalized assuming $\alpha=0.2$.}
\label{fig:ionization_inclined_211}
\end{figure}

Expressions (\ref{eqn:rate-inclined}), (\ref{eqn:pion-inclined}), (\ref{eqn:tauz'-inclined}), (\ref{eqn:tauoutx'-inclined}) and (\ref{eqn:tauinx'-inclined}) can be evaluated numerically. In Figure~\ref{fig:ionization_inclined_211}, we show the ionization rate, power and $z$-component of the torque as function of the binary separation $R_*$, for selected values of the orbital inclination and a cloud in the $\ket{211}$ state.

\vskip 0pt
Varying $\beta$, each curves goes continuously from the equatorial co-rotating ($\beta=0$) to the equatorial counter-rotating ($\beta=\pi$) result derived earlier. Rather than interpolating monotonically between the two limits, however, the curve is generally seen to first \emph{decrease} in amplitude, reaching a minimum for some intermediate value of the inclination (which varies depending on $R_*$), then increase again. This behaviour has an easy qualitative interpretation: the angular structure of the $\ket{211}$ state is such that the cloud has its highest density on the equatorial plane. When the binary's orbit is inclined, the companion does not stay in this high density region all the time, instead it moves out of it during parts of its orbit. According to the interpretation from Section~\ref{sec:dynamical-friction}, ionization is thus expected to be less efficient, because the companion encounters, on average, a lower local scalar density.

\subsection{Evolution of inclination}
\label{sec:evolution-beta}

In the same spirit as Section~\ref{sec:evolution-e}, we can now study the backreaction of ionization on inclined orbits, in a simplified setup where self gravity and mass loss of the cloud, as well as accretion on the companion, are neglected.\footnote{If we wanted to track the mass loss of the cloud, an extra complication would be present. While (\ref{eqn:rate-inclined}) gives the \emph{total} ionization rate, the individual states $\ket{n_b\ell_bm'}$ are each ionized at a different rate. The cloud is then forced to go to a mixed state, and we would need to track the occupancies of all the $2\ell_b+1$ quasi-degenerate states. As some of them are not superradiant, it may be necessary to include their decay in the evolution too.} The energy balance equation reads, once again,
\beq
\frac{qM^2}{2R_*^2}\frac{\dd R_*}{\dd t}=-P\ped{ion}-P_\slab{gw}\,.
\label{eqn:energy-balance-inclined}
\eeq
Because we are considering circular orbits, equation (\ref{eqn:energy-balance-inclined}) is equivalent to the balance of angular momentum along the $z'$ axis. Instead, the other two components of the torque give new information. First of all, from (\ref{eqn:tauion}) and (\ref{eqn:rotation-tau}) we see that the torque lies in the $x$-$z$ plane, as its component along the $y$ axis vanishes identically. The orbital angular momentum also has a vanishing $y$ component, which will thus remain zero during the evolution of the system. In other words, we draw the conclusion that ionization induces \emph{no precession} of the orbital plane, and the orbit's axis will only rotate in the $x$-$z$ plane.

\vskip 0pt
This rotation, quantified by the evolution of the inclination angle $\beta$, is determined by the $x'$ component of the equation,
\beq
qM\sqrt{\frac{MR_*}{1+q}}\,\frac{\dd\beta}{\dd t}=-\tau\ped{ion}^{x'}\,.
\label{eqn:dbetadt}
\eeq
To understand the magnitude of the evolution of inclination, it is convenient to combine (\ref{eqn:energy-balance-inclined}) and (\ref{eqn:dbetadt}) as
\beq
\frac{\dd\beta}{\dd R_*}=\frac{\sqrt{(1+q)M}\,\tau\ped{ion}^{x'}}{2R^{5/2}(P\ped{ion}+P_\slab{gw})}\,,
\label{eqn:dbetadR}
\eeq
which allows us to compute $\beta$ as a function of $R_*$. This defines a ``trajectory'' in the $(R_*,\beta)$ plane that the binary follows through its evolution.

\begin{figure}
\centering
\includegraphics{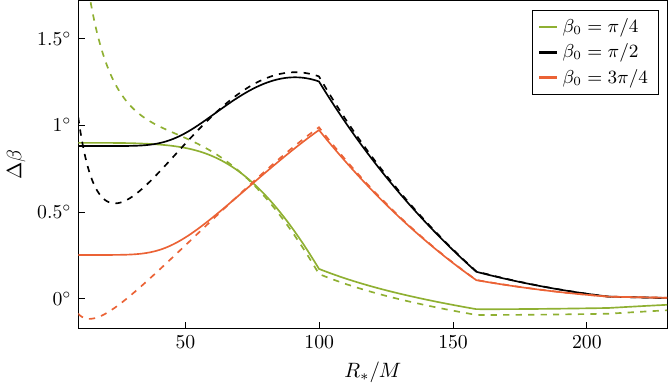}
\caption{Variation of the inclination angle $\Delta\beta\equiv\beta-\beta_0$, as function of the orbital separation $R_*$, for different values of the initial inclination $\beta_0$. The curves represent the evolution of $\Delta\beta$, from right to left, over the course of an inspiral. Solid lines are obtained by direct integration of (\ref{eqn:dbetadR}), with parameters $\alpha=0.2$, $q=10^{-3}$, $M\ped{c}=0.01M$ and a cloud in the $\ket{211}$ state. Dashed lines, instead, are computed by neglecting $P_\slab{gw}$ in (\ref{eqn:dbetadR}). In all cases, the inclination angle remains almost constant throughout the inspiral, with $\Delta\beta$ being at most of order \SI{1}{\degree}. We do not show trajectories with values of $\beta_0$ closer to 0 (co-rotating) of $\pi$ (counter-rotating), as the variation $\Delta\beta$ is even more limited in those cases.}
\label{fig:delta_beta_evolution}
\end{figure}

\vskip 0pt
As a general result, we find that the variation of the inclination angle $\beta$ is always very limited: over the course of a full inspiral, $\beta$ changes by at most a few degrees. It is useful to first consider the limit where ionization dominates the inspiral, thus neglecting $P_\slab{gw}$ in (\ref{eqn:dbetadR}). In this case, the  trajectory $\beta(R_*)$ only depends on the initial value $\beta_0\equiv\beta(R_*\to\infty)$, as well as on the state of the cloud. We show a few selected examples as dashed lines in Figure~\ref{fig:delta_beta_evolution}, where, for various choices of $\beta_0$, the total variation $\Delta\beta\equiv\beta-\beta_0$ is manifestly confined within a few degrees. When $P_\slab{gw}$ is included in (\ref{eqn:dbetadR}), the variation of $\beta$ is further limited: this case is shown with solid lines in Figure~\ref{fig:delta_beta_evolution}. This means that, as a simplifying approximation, the inclination angle $\beta$ can be treated as a fixed parameter in the evolution of the binary system.

\vskip 0pt
We conclude that, overall, ionization acts on inlined orbits in a simple way. The ionization rate (\ref{eqn:rate-inclined}) and power (\ref{eqn:pion-inclined}) need to be calculated for the specific value of the orbital inclination $\beta$ considered (see also Figure~\ref{fig:ionization_inclined_211}). The orbital plane, however, may be assumed to stay approximately fixed over time: the off-axis component of the torque induce no precession, and very little change in the value of $\beta$ over the course of an inspiral.

\chapter{A deeper look at resonances}

\label{chap:resonances}

In Section~\ref{sec:resonances-review} we have introduced the resonances between bound states, and associated Landau-Zener transitions, mediated by the binary perturbation. Resonances are crucial in shaping the history and evolution of the system, but the machinery of Section~\ref{sec:resonances-review} is not yet adequate to capture the whole phenomenology. This chapter serves to generalize the framework enough for that purpose.

\vskip 0pt
We start by extending the setup to eccentric and inclined orbits in Section~\ref{sec:eccentric-inclined-resonances}. Then, in Section~\ref{sec:backreaction}m we include the backreaction, thus coupling the resonating states to the evolution of the binary parameters. The phenomenology of the resulting nonlinear system is explored in Section~\ref{sec:floating} and Section~\ref{sec:sinking}, for the floating and sinking cases respectively. Finally, in Section~\ref{sec:types-of-resonances}, we determine the scaling of the relevant resonance parameters with $M$, $M\ped{c}$, $q$, $\alpha$ and $\tilde a$.

\section{Resonances on eccentric and inclined orbits}
\label{sec:eccentric-inclined-resonances}

Here we extend the treatment of Section~\ref{sec:resonances-review} to orbits with nonzero eccentricity or inclination, explaining the changes for the resonant frequencies and the overlap coefficients $\eta^\floq{g}$.

\vskip 0pt
Let us start with eccentric co-rotating orbits. In the quasi-circular case, equation (\ref{eqn:eta-circular}) manifestly separates a ``fast'' and a ``slow'' motion: the former originates from $\varphi_*$ varying over the course of an orbit, while the latter is due to the dependence of the coefficients $\eta^\floq{g}$ on $\Omega(t)$ (and can be safely neglected). It will be helpful to work with a variable that performs the same trick on eccentric orbits: the \emph{mean anomaly}
\beq
\tilde\varphi_*(t)=\int^t\Omega(t')\dd t'\,.
\eeq
Because $\varphi_*$ itself is an oscillating function of $\tilde\varphi_*$, we can write
\beq
\braket{a|V_{*}(t)|b} = \sum_{g\in \mathbb{Z}} \tilde\eta^\floq{g} e^{ig\tilde\varphi_*}\,,
\label{eqn:eta-tilde}
\eeq
where the coefficients $\tilde\eta^\floq{g}$ only depend on time through $\Omega(t)$. For simplicity, in the following discussion we will drop the tildes, with the different definition of $\eta^\floq{g}$ for nonzero eccentricity left understood.

\vskip 0pt
For a given eccentricity $\varepsilon\ne0$, multiple terms of (\ref{eqn:eta-tilde}), each corresponding to a different value of $g$, can be nonzero. As a consequence, a resonance between two given states can be triggered at different points of the inspiral, at the frequencies $\Omega_0^\floq{g}=\Delta\epsilon/g$, for any integer $g$ (provided that it has the same sign as $\Delta\epsilon$). The numerical evaluation of the coefficients $\eta^\floq{g}$ requires to Fourier expand $V_*$ in the time domain, at the orbital frequency $\Omega=\Omega_0^\floq{g}$. This can be done with techniques similar to Section~\ref{sec:eccentricity}, where the same matrix element was evaluated between a bound and an unbound state. The coefficient $\eta^\floq{\Delta m}$ is special because it is the only one with a finite, nonzero limit for $\varepsilon\to0$, where it reduces to its circular-orbit counterpart. For all other values of $g$, instead, $\eta^\floq{g}$ vanishes for $\varepsilon\to0$. Even at moderately large $\varepsilon$, the coefficient $\eta^\floq{\Delta m}$ remains significantly larger than all the others

\vskip 0pt
Let us now look at circular but inclined orbits. Here, the Fourier coefficients $\eta^\floq{g}$ acquire a dependence on the inclination angle $\beta$, where $\beta=0$ and $\beta=\pi$ correspond to the co-rotating and counter-rotating scenarios. The functional dependence can be readily extracted by evaluating the perturbation \eqref{eqn:V_star} using the the identity\footnote{Equation \eqref{eqn:Y-decomposed} is the same as \eqref{eqn:rotated-Y}, but the convention on the sign of $g$ is opposite. This reflects the fact that we have denoted a transition between bound states $\ket{a}\to\ket{b}$ with the matrix element $\braket{a|V_*|b}$, and a transition from a bound to an unbound state $\ket{b}\to\ket{K}$ with $\braket{K|V_*|b}$.} \cite{wigner}
\beq
Y_{\ell_*m_*}(\theta_*,\varphi_*)=\sum_{g=-\ell_*}^{\ell_*}d^\floq{\ell_*}_{m_*,g}(\beta)Y_{\ell_*g}\biggl(\frac\pi2,0\biggr)e^{ig\Omega t}\,.
\label{eqn:Y-decomposed}
\eeq
Here, $d^\floq{\ell_*}_{m_*,g}(\beta)$ is a Wigner small $d$-matrix and is responsible for the angular dependence of the coupling, $\eta^\floq{g}\propto d^\floq{\ell_*}_{m_*,g}(\beta)$. Its functional form takes on a simple expression in many of the physically interesing cases, as we will discuss explicitly in Section~\ref{sec:types-of-resonances}. We thus see that inclined orbits also trigger resonances at $\Omega=\Omega_0^\floq{g}=\Delta\epsilon/g$, but this time $g$ can only assume a finite number of different values. Similar to the eccentric case, $g=\Delta m$ is special, because it is the only case where $d^\floq{\ell_*}_{m_*,g}(\beta)$ does not vanish for $\beta\to0$, as the resonance survives in the equatorial co-rotating limit. Similarly, in the counter-rotating case $\beta\to\pi$, the only surviving value is $g=-\Delta m$.

\vskip 0pt
Similar techniques can be applied in the eccentric \emph{and} inclined case, where the overlap can be expanded in two sums, each with its own index, say $g_\varepsilon$ and $g_\beta$. We do not explicitly compute $\eta^\floq{g}$ in the general case, as the understanding developed so far is sufficient to move forward and characterize the phenomenology in realistic cases.

\section{Backreaction on the orbit}
\label{sec:backreaction}

We now include the backreaction on the orbit, allowing for generic nonzero eccentricity \emph{and} inclination. During a resonance, the energy and angular momentum contained in the cloud change over time: this variation must be compensated by an evolution of the binary parameters, the (dimensionless) frequency $\omega$, eccentricity $\varepsilon$ and inclination $\beta$. In turn, this backreaction impacts the Schrödinger equation \eqref{eqn:dimensionless-dressed-schrodinger}, which directly depends on $\omega$. The result is a coupled nonlinear system of ordinary differential equations, describing the co-evolution of the cloud and the binary, which we derive in this section.

\vskip 0pt
To describe the evolution of $\omega$, $\varepsilon$ and $\beta$ we need three equations. These are the conservation of energy and of two components of the angular momentum: the projection along the BH spin and the projection on the equatorial plane. The conservation of energy reads
\beq
\frac\dd{\dd t}\bigl(E+E\ped{c}\bigr)=-\gamma f(\varepsilon)\,\frac{qM^{5/3}}{3(1+q)^{1/3}\Omega_0^{1/3}}\,,
\label{eqn:E-balance}
\eeq
where $\gamma$ was defined in \eqref{eqn:gamma_gws} and the binary's and cloud's energies are
\beq
E=-\frac{qM^{5/3}\Omega^{2/3}}{2(1+q)^{1/3}}\,,\qquad E\ped{c}=\frac{M\ped{c}}\mu(\epsilon_a\abs{c_a}^2+\epsilon_b\abs{c_b}^2)\,,
\eeq
while the function
\beq
f(\varepsilon)=\frac{1+\frac{73}{24}\varepsilon^2+\frac{37}{96}\varepsilon^4}{(1-\varepsilon^2)^{7/2}}
\eeq
quantifies the dependence of GW energy losses on the eccentricity, cf.~\eqref{eq:p_gw}. Similarly, the conservation of the angular momentum components requires
\begin{align}
\label{eqn:Lz-balance}
\frac\dd{\dd t}\bigl(L\cos\beta+S\ped{c}\bigr)=-h(\varepsilon)\gamma\,\frac{qM^{5/3}}{3(1+q)^{1/3}\Omega_0^{4/3}}\cos\beta\,,\\
\label{eqn:Lx-balance}
\frac\dd{\dd t}\bigl(L\sin\beta\bigr)=-h(\varepsilon)\gamma\,\frac{qM^{5/3}}{3(1+q)^{1/3}\Omega_0^{4/3}}\sin\beta\,.
\end{align}
where
\beq
L=\frac{qM^{5/3}}{(1+q)^{1/3}}\frac{\sqrt{1-\varepsilon^2}}{\Omega^{1/3}},\qquad S\ped{c}=\frac{M\ped{c}}\mu(m_a\abs{c_a}^2+m_b\abs{c_b}^2)\,,
\eeq
and
\beq
h(\varepsilon)=\frac{1+\frac78\varepsilon^2}{(1-\varepsilon^2)^2}\,.
\eeq
Before proceeding, there are two issues the reader might worry about. First, depending on the resonance, the spin of the cloud during the transition might also have equatorial components, and should thus appear in \eqref{eqn:Lx-balance}. Second, the BH spin breaks spherical symmetry, therefore the equatorial projection of the angular momentum should not be conserved. Clearly, in the Newtonian limit this is not a problem, but one might still question whether it is consistent to treat within this framework hyperfine resonances, whose very existence is due to a nonzero BH spin in the first place. We address both these issues in Appendix~\ref{sec:hyperfine-angular-momenutm}, where we justify our assumptions, and proceed here to study the dynamics of the previous equations.

\vskip 0pt
Equations \eqref{eqn:E-balance}, \eqref{eqn:Lz-balance} and \eqref{eqn:Lx-balance} can be put in a dimensionless form as follows,
\begin{align}
\label{eqn:dimensionless-omega-evolution}
\frac{\dd\omega}{\dd\tau}&=f(\varepsilon)-B\frac{\dd\abs{c_b}^2}{\dd\tau}\,,\\
\label{eqn:dimensionless-eccentricity-evolution}
C\frac\dd{\dd\tau}\sqrt{1-\varepsilon^2}&=\sqrt{1-\varepsilon^2}\biggl(f(\varepsilon)-B\frac{\dd\abs{c_b}^2}{\dd\tau}\biggr)+B\frac{\Delta m}g\frac{\dd\abs{c_b}^2}{\dd\tau}\cos\beta-h(\varepsilon)\,,\\
\label{eqn:dimensionless-inclination-evolution}
C\sqrt{1-\varepsilon^2}\frac{\dd\beta}{\dd\tau}&=-B\frac{\Delta m}g\frac{\dd\abs{c_b}^2}{\dd\tau}\sin\beta\,,
\end{align}
where we defined the dimensionless parameters
\beq
B=\frac{3M\ped{c}}M\frac{\Omega_0^{4/3}((1+q)M)^{1/3}}{q\alpha\sqrt{\gamma/\abs{g}}}(-g)\,,\qquad C=\frac{3\Omega_0}{\sqrt{\gamma/\abs{g}}}\,.
\label{eqn:BC}
\eeq
The Schrödinger equation \eqref{eqn:dimensionless-dressed-schrodinger} remains unchanged, but it should be kept in mind that $Z$ now depends on $\varepsilon$ and $\beta$ through $\eta^\floq{g}$ (instead, the dependence on $\omega$ can still be neglected if the resonance is narrow enough).

\vskip 0pt
The parameter $B$ controls the strength of the backreaction. As can be seen from \eqref{eqn:dimensionless-omega-evolution}, a positive $B>0$ (i.e., $g<0$ and $\Delta\epsilon<0$) will slow down the frequency chirp, giving rise to a \emph{floating} orbit and generally making the resonance more adiabatic. Conversely, $B<0$ (i.e., $g>0$ and $\Delta\epsilon>0$) induces \emph{sinking} orbits and makes resonances less adiabatic. By extension, we will refer to floating resonances and sinking resonances to denote the type of backreaction they induce. A summary of the main variables used to describe the resonances and their backreaction is given in Table~\ref{tab:parameters} for the reader's convenience.

\begin{table}
\centering
\begin{tabular}{clc}
\toprule
\textbf{Symbol} & \textbf{Meaning} & \textbf{Reference} \\
\midrule
$\varepsilon$ & Binary eccentricity & \\
$\beta$ & Binary inclination & \\
$g$ & Overtone number & \eqref{eqn:eta-circular} \\
$\gamma$ & Frequency chirp rate induced by GWs/ionization & \eqref{eqn:gamma_gws} \\
$\tau$ & Dimensionless time & \eqref{eqn:coefficients} \\
$\omega$ & Dimensionless frequency & \eqref{eqn:coefficients} \\
$Z$ & Landau-Zener parameter & \eqref{eqn:coefficients} \\
$B$ & Backreaction of a resonance & \eqref{eqn:BC} \\
$C$ & Inertia of $\varepsilon$ and $\beta$ w.r.t.~resonance backreaction & \eqref{eqn:BC} \\
$D$ & Distance parameter, $D=B/C$ & \eqref{eqn:D} \\
$\Gamma$ & Dimensionless decay width of the final state & \eqref{eqn:schrodinger-Gamma}\\
\bottomrule
\end{tabular}
\caption{Summary of the key resonance variables used throughout Chapters~\ref{chap:resonances} and Chapter~\ref{chap:history}.}
\label{tab:parameters}
\end{table}

\section{Floating orbits}
\label{sec:floating}

Backreaction of the floating type ($B>0$) turns out to be the most relevant case for realistic applications, so we make a detailed study of its phenomenology here. When the backreaction is strong, the evolution of the system exhibits a very well-defined phase of floating orbit. We are then concerned with three aspects.
\begin{enumerate}
\item Under what conditions is a floating resonance initiated? We answer this question with a simple analytical prescription, which is found and discussed in Section~\ref{sec:floating-adiabatic}.
\item How does the system evolve during the float? This is addressed in Section~\ref{sec:floating-evolution-e-beta}, where we study the evolution of the eccentricity and inclination.
\item When does a floating resonance end? In Section~\ref{sec:resonance-breaking} we show that several phenomena can \emph{break} (and end) the resonance before the transition from $\ket{a}$ to $\ket{b}$ is complete, and compute accurately the conditions under which this phenomenon happens.
\end{enumerate}

\subsection{Adiabatic or non-adiabatic}
\label{sec:floating-adiabatic}

From Section~\ref{sec:resonances-review}, we know that if $B=0$, then a fraction $1-e^{-2\pi Z}$ of the cloud is transferred during the resonances. For $2\pi Z\gg1$, this value is already very close to 1. Adding the backreaction does not change this conclusion: the resonance stays adiabatic and a complete transfer from $\ket{a}$ to $\ket{b}$ is observed. Assuming for simplicity quasi-circular orbits ($\varepsilon=0$), the duration of the floating orbit can be easily read off~(\ref{eqn:dimensionless-omega-evolution}):
\beq
\Delta t\ped{float}=\frac{B}{\sqrt{\abs{g}\gamma}}=\frac{3M\ped{c}}M\frac{\Omega_0^{4/3}((1+q)M)^{1/3}}{q\alpha\gamma}(-g)\,.
\label{eqn:t-float}
\eeq
This is independent of the strength of the perturbation $\eta^\floq{g}$, and corresponds to the time it takes for the external energy losses to dissipate the energy of the two-state system. For nonzero eccentricity instead, one must integrate $f(\varepsilon)$ over time to determine the duration of the float.

\begin{figure}[t]
\centering
\includegraphics[width=\textwidth]{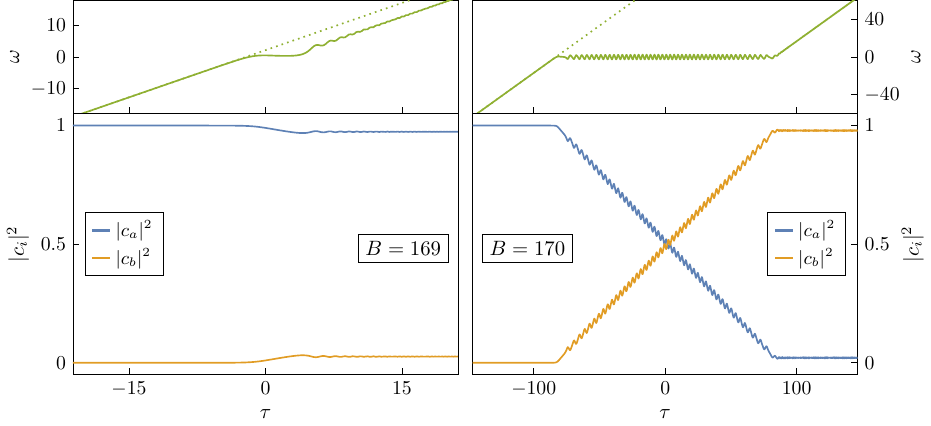}
\caption{Numerical solution of the nonlinear system (\ref{eqn:dimensionless-dressed-schrodinger})-(\ref{eqn:dimensionless-omega-evolution}). In both panels we set $Z=0.001$, whereas we choose the values of $B$ to be $169$ (\emph{left panel}) and $170$ (\emph{right panel}), slightly below or above the adiabaticity threshold (the limit value of $ZB$ differs slightly from the one given in (\ref{eqn:1/2piadiabatic}), due to finite-$Z$ corrections). In the left panel, a non-adiabatic transition is observed. Conversely, in the right panel, we find an adiabatic transition and the consequent formation of a floating orbit, whose duration matches the predicted $\Delta t\ped{float}=B/\sqrt{\abs{g}\gamma}$. The dotted lines represent the evolution of $\omega$ in absence of backreaction.}
\label{fig:floating-backreacted-resonance}
\end{figure}

\vskip 0pt
The situation for $2\pi Z\ll1$ is in principle much less clear: with $B=0$ the resonance would be non-adiabatic, but backreaction tends to make it more adiabatic. Let us once again restrict to quasi-circular orbits for simplicity. By careful numerical study of equations (\ref{eqn:dimensionless-dressed-schrodinger}) and (\ref{eqn:dimensionless-omega-evolution}), we find that the long-time behavior of the system is predicted by the parameter $ZB$ alone. Depending on its value, two qualitatively different outcomes are possible:
\beq
\text{if}\quad2\pi Z\ll1\quad \mathrm{and} \quad
\begin{cases}
ZB<0.1686\ldots\quad\longrightarrow\quad\text{very non-adiabatic,}\\[12pt]
ZB>0.1686\ldots\quad\longrightarrow\quad\text{very adiabatic.}
\end{cases}
\label{eqn:1/2piadiabatic}
\eeq
In the upper case, a negligible fraction of the cloud is transferred and the time evolution of $\omega$ is almost exactly linear. Conversely, in the bottom case, the cloud is entirely transferred from $\ket{a}$ to $\ket{b}$ and $\omega$ is stalled for an amount of time $\Delta t\ped{float}$ given by (\ref{eqn:t-float}), during which it oscillates around zero. Intermediate behaviors are not possible, unless the value of $ZB$ is extremely fine tuned. The two cases are illustrated in Figure~\ref{fig:floating-backreacted-resonance}.

\vskip 0pt
We can give an approximate derivation of the previous result as follows. As long as $\abs{c_b}^2$ is small enough, the backreaction term in (\ref{eqn:dimensionless-omega-evolution}) is negligible, hence $\omega$ evolves linearly and the final populations approximate the Landau-Zener result (\ref{eqn:lz}), giving $\abs{c_b}^2\approx2\pi Z$. As the unbackreacted transition happens in the time window $\abs{\tau}\lesssim1$, we see from (\ref{eqn:dimensionless-omega-evolution}) that the backreaction becomes significant when $1\lesssim B\cdot2\pi Z\implies ZB\gtrsim1/(2\pi)\approx0.159\ldots$ Given the minimal numerical difference between this coefficient and the one given in (\ref{eqn:1/2piadiabatic}), for simplicity we will often write the relevant condition for an adiabatic resonance simply as $2\pi ZB\gtrless1$. The slow-down effect on the evolution of $\omega(\tau)$ enjoys a positive-feedback mechanism: the slower $\omega$ evolves, the more the transition is adiabatic, meaning that $\abs{c_b}^2$ is larger, which further slows down $\omega(\tau)$, and so on. This explains why no intermediate behaviors are observed: once the backreaction goes over a certain critical threshold, the process becomes self-sustaining.

\vskip 0pt
The picture outlined so far changes slightly when the eccentricity is nonzero. First, if the binary had a constant eccentricity $\varepsilon_0$, we could simply replace $\gamma\to\gamma f(\varepsilon_0)$ to conclude that the critical threshold for adiabaticness becomes
\beq
2\pi ZB\gtrless f(\epsilon_0)^{3/2}\,.
\label{eqn:2piZB-epsilon0}
\eeq
When the eccentricity is allowed to vary starting from the initial value $\varepsilon_0$, equation \eqref{eqn:2piZB-epsilon0} still correctly predicts whether the system enters a floating orbit phase. However, the transfer might no longer be complete, as the resonance might \emph{break}. This aspect will be discussed in Section~\ref{sec:resonance-breaking}.

\subsection{Evolution of eccentricity and inclination}
\label{sec:floating-evolution-e-beta}

\begin{figure}[t]
\centering
\includegraphics[width=\textwidth]{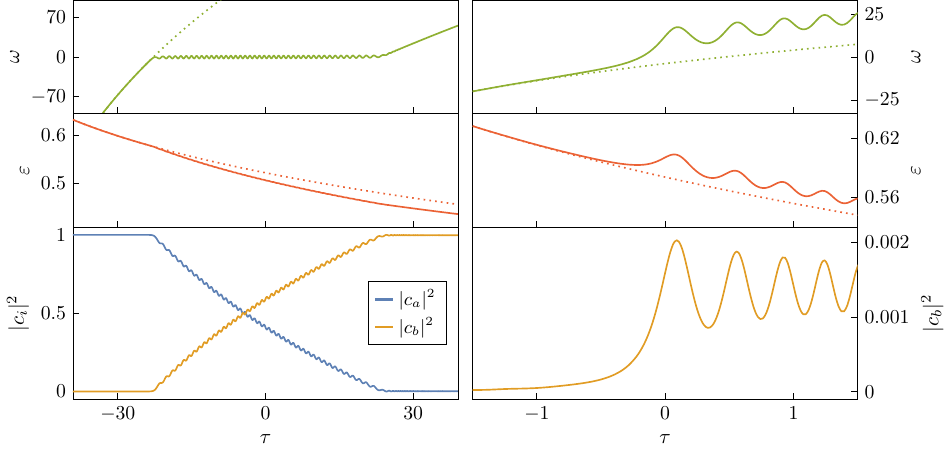}
\caption{Floating (\emph{left panel}) and sinking (\emph{right panel}) resonances on eccentric orbits, with $\Delta m/g=1$. We display the value of the frequency $\omega$, the eccentricity $\varepsilon$ and the populations $\abs{c_a}^2$ and $\abs{c_b}^2$ as function of $\tau$, obtained by solving equations (\ref{eqn:dimensionless-dressed-schrodinger}), (\ref{eqn:dimensionless-omega-evolution}) and (\ref{eqn:dimensionless-eccentricity-evolution}) with $\beta=0$ numerically. The parameters used for the floating case are $Z=0.03$, $B=250$, $C=1000$, while for the sinking case we used $Z=0.01$, $B=-10000$, $C=100$. The dotted lines represent the evolution of $\omega$ and $\varepsilon$ in absence of backreaction. Even though the impact of the resonance on the eccentricity might look mild, the effect is actually dramatic when seen as function of $\omega$, as shown in Figure~\ref{fig:omega-eccentricity}.}
\label{fig:eccentric-backreacted-resonance}
\end{figure}

\begin{figure}[t]
\centering
\includegraphics[width=\textwidth]{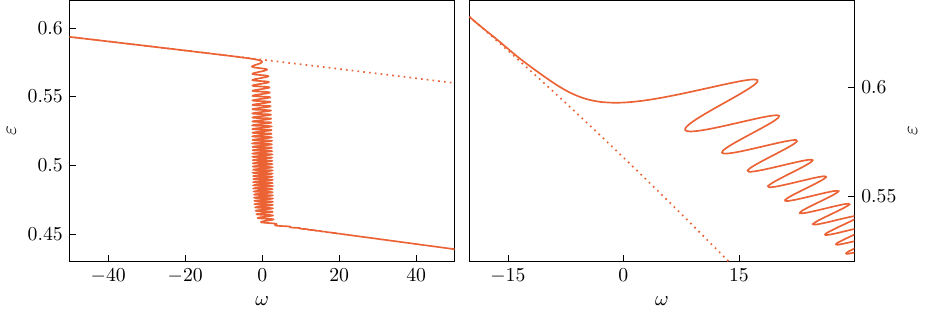}
\caption{Same resonances as in Figure~\ref{fig:eccentric-backreacted-resonance}, but now the evolution of eccentricity is shown as function of the frequency, for floating (\emph{left panel}) and sinking (\emph{right panel}) orbits. The dashed lines represent the vacuum evolution.}
\label{fig:omega-eccentricity}
\end{figure}

The left panel of Figure~\ref{fig:eccentric-backreacted-resonance} shows a numerical solution of the coupled nonlinear equations \eqref{eqn:dimensionless-dressed-schrodinger}, \eqref{eqn:dimensionless-omega-evolution}, \eqref{eqn:dimensionless-eccentricity-evolution} and \eqref{eqn:dimensionless-inclination-evolution}, for an equatorial co-rotating ($\beta=0$) but eccentric ($\varepsilon\ne0$) system, undergoing a floating orbit with $g=\Delta m$. The state dynamics is largely similar to what we described in Section~\ref{sec:floating-adiabatic}. The most interesting new effect concerns the evolution of the eccentricity, which can be seen to decrease during the float, at a rate faster than the circularization provided by GW emission. The same numerical solution is shown as function of frequency in Figure~\ref{fig:omega-eccentricity}. 

\vskip 0pt
The evolution of the eccentricity during a float can be studied analytically by plugging $\dd\omega/\dd\tau\approx0$ in equations \eqref{eqn:dimensionless-eccentricity-evolution} and \eqref{eqn:dimensionless-inclination-evolution}, which become
\begin{align}
\label{eqn:eccentricity-evolution-floating}
C\frac\dd{\dd\tau}\sqrt{1-\varepsilon^2}&=\frac{\Delta m}gf(\varepsilon)\cos\beta-h(\varepsilon)\,,\\
\label{eqn:inclination-evolution-floating}
C\sqrt{1-\varepsilon^2}\frac{\dd\beta}{\dd\tau}&=-\frac{\Delta m}gf(\varepsilon)\sin\beta\,.
\end{align}
For resonances with $\beta=0$ and $g=\Delta m$, such as the one shown in Figures~\ref{fig:eccentric-backreacted-resonance} and~\ref{fig:omega-eccentricity}, a small-$\varepsilon$ expansion leads to the following solution:
\beq
\varepsilon(t)\approx\varepsilon_0\,e^{-\frac{22}{18}\gamma t/\Omega_0}\,.
\label{eqn:circulatization-floating}
\eeq
This result should be compared to the GW-induced circularization in absence of backreaction,
\beq
\varepsilon(t)\approx\varepsilon_0\,e^{-\frac{19}{18}\gamma t/\Omega_0}\,.
\label{eqn:circulatization-GW}
\eeq
Therefore, not only is the orbit stalled at $\Omega(t)\approx\Omega_0$ for a potentially long time, given in (\ref{eqn:t-float}), during which the eccentricity keeps reducing; but it also goes down at a faster rate than in the vacuum, as can be seen comparing (\ref{eqn:circulatization-floating}) with (\ref{eqn:circulatization-GW}). The longer the resonance, the more the binary is circularized.

\vskip 0pt
This result holds for co-rotating resonances with $g=\Delta m$, which are the only ones surviving in the small-$\varepsilon$ limit and usually have the largest coupling $\eta^\floq{g}$ even at moderately large eccentricities. The dynamics is different in other cases. Remaining in the equatorial co-rotating case ($\beta=0$), eccentric binaries can also undergo (usually weaker) resonances where $g\ne\Delta m$. In this case, \eqref{eqn:eccentricity-evolution-floating} has a different behavior: if $\abs{\Delta m/g}<1$, there is a fixed point $\bar\varepsilon>0$ such that if $\varepsilon<\bar\varepsilon$ then $\varepsilon$ increases, while if $\varepsilon>\bar\varepsilon$ then $\varepsilon$ decreases. For example, for $\Delta m/g=1/2$, we have $\bar\varepsilon\approx0.46$ and the eccentricity approaches the fixed point according to
\beq
\varepsilon(t)\approx0.46+(\varepsilon_0-0.46)e^{-3.49\gamma t/\Omega_0}\,.
\label{eqn:eccentrification-floating-DeltamOverg=0.5}
\eeq
Floating resonances with $\abs{\Delta m/g}>1$ will instead circularize the binary even quicker than \eqref{eqn:circulatization-floating}. As $\varepsilon$ decreases, however, so does $Z$: eventually, the perturbation becomes too weak and the resonance stops, generically leaving the cloud in a mixed state as the inspiral resumes. This aspect will be discussed in Section~\ref{sec:resonance-breaking}.

\begin{figure}[t]
\centering
\includegraphics[width=\textwidth]{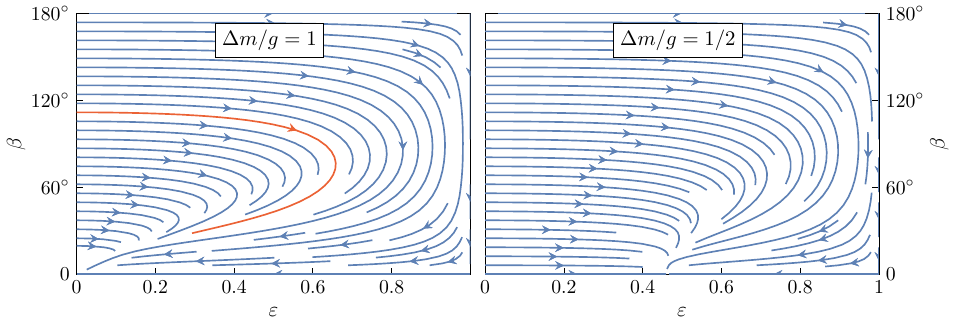}
\caption{Flow in the eccentricity-inclination plane $(\varepsilon,\beta)$ determined by equations \eqref{eqn:dimensionless-eccentricity-evolution} and \eqref{eqn:dimensionless-inclination-evolution} under the assumption that the system is on a floating orbit, i.e., $\dd\abs{c_b}^2/\dd\tau=f(\varepsilon)/B$, for two different values of $\Delta m/g$. The highlighted arrow [{\color{Mathematica4}red}] roughly depicts the trajectory followed by the system in Figure~\ref{fig:eccentric-inclined-backreacted-resonance}.}
\label{fig:streamplot_inclination-eccentricity}
\end{figure}

\begin{figure}[t]
\centering
\includegraphics[width=\textwidth]{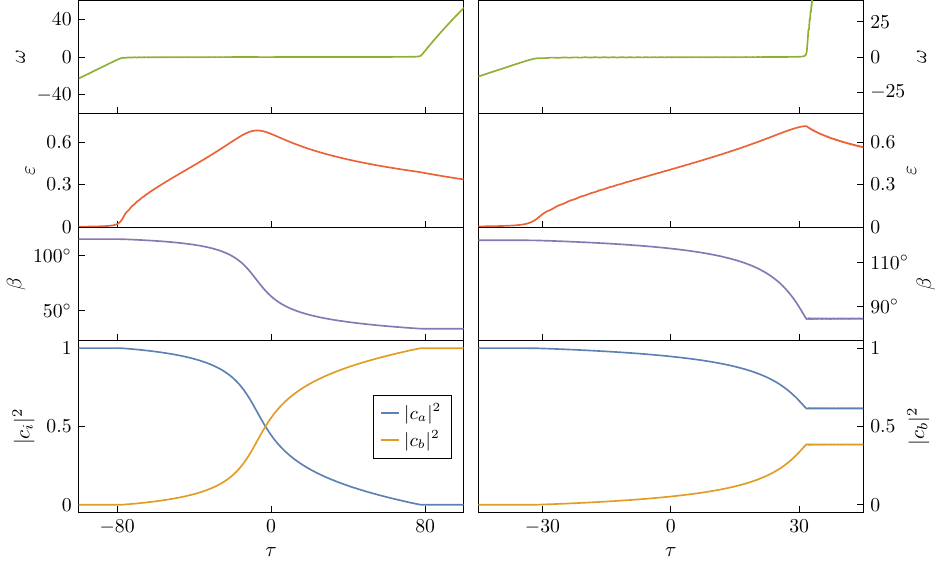}
\caption{Numerical solution of equations \eqref{eqn:dimensionless-dressed-schrodinger}, \eqref{eqn:dimensionless-omega-evolution}, \eqref{eqn:dimensionless-eccentricity-evolution} and \eqref{eqn:dimensionless-inclination-evolution} with parameters $Z=0.001$, $B=1000$, $D=4/3$ and $g=\Delta m$. For simplicity we ignore that in realistic cases $Z$ depends on the eccentricity, and we keep it constant instead. The system is initialized with eccentricity $\varepsilon_0=0$. A complete transition is achieved when the initial inclination is $\beta_0=\SI{115}{\degree}$ (\emph{left panel}), while a ``broken resonance'' is observed when $\beta_0=\SI{120}{\degree}$ (\emph{right panel}), with the float abruptly ending when \eqref{eqn:epsilon-breaking} is satisfied. In both cases, the system follows the trajectories indicated in Figure~\ref{fig:streamplot_inclination-eccentricity} until the resonance ends or breaks.}
\label{fig:eccentric-inclined-backreacted-resonance}
\end{figure}

\vskip 0pt
The possibilities described so far are a particular case of the general dynamics, which includes the evolution of the inclination $\beta$. The flow induced by equations \eqref{eqn:eccentricity-evolution-floating} and \eqref{eqn:inclination-evolution-floating} in the $(\varepsilon,\beta)$ plane is shown in Figure~\ref{fig:streamplot_inclination-eccentricity}, where the dynamics on the $x$ axis is described by equations \eqref{eqn:circulatization-floating} (\emph{left panel}) and \eqref{eqn:eccentrification-floating-DeltamOverg=0.5} (\emph{right panel}). Perhaps the most striking feature of Figure~\ref{fig:streamplot_inclination-eccentricity} is the fact that the system is violently pulled away from inclined circular orbits ($y$ axis). In fact, $\dd\varepsilon/\dd\tau$ diverges for $\varepsilon\to0$ and finite $\beta$, meaning that the validity of equations \eqref{eqn:eccentricity-evolution-floating} and \eqref{eqn:inclination-evolution-floating} must somehow break down in that limit. The explanation for this behavior is that it is inconsistent to assume that the system undergoes an adiabatic floating resonance on inclined circular orbits: eccentricity \emph{must} increase before the onset of the resonance. This is precisely the behavior observed in Figure~\ref{fig:eccentric-inclined-backreacted-resonance}, where equations \eqref{eqn:dimensionless-dressed-schrodinger}, \eqref{eqn:dimensionless-omega-evolution}, \eqref{eqn:dimensionless-eccentricity-evolution} and \eqref{eqn:dimensionless-inclination-evolution} are solved numerically starting from $\varepsilon_0=0$ and $\beta_0\ne0$. If $2\pi ZB>f(\varepsilon_0)^{3/2}$, then the system enters the floating orbit and starts to follow the trajectories shown in Figure~\ref{fig:streamplot_inclination-eccentricity}. The total ``distance'' in the $(\beta,\varepsilon)$ plane travelled by the system by the time the transition completes depends on a single dimensionless ``distance parameter'',
\beq
D\equiv\frac{B}C=\frac{\gamma\Delta t\ped{float}}{\Omega_0}\,.
\label{eqn:D}
\eeq
However, it is also possible that the transition stops before fully completing, as shown in Figure~\ref{fig:eccentric-inclined-backreacted-resonance} (\emph{right panel}). This is the subject of Section~\ref{sec:resonance-breaking}.

\subsection{Resonance breaking}
\label{sec:resonance-breaking}

When some parameters are allowed to vary with time, the floating orbit dynamics described in Section~\ref{sec:floating-adiabatic} and \ref{sec:floating-evolution-e-beta} features a new phenomenon, which we call \emph{resonance breaking}, and has been shown already in Figure~\ref{fig:eccentric-inclined-backreacted-resonance} (\emph{right panel}). The goal of this section is to determine analytically under which conditions a floating resonance breaks. Three different cases of parameter variation are encountered in realistic scenarios.
\begin{enumerate}
\item The binary eccentricity $\varepsilon$ changes with time, as seen in Figure~\ref{fig:eccentric-inclined-backreacted-resonance}. The eccentricity is the only binary parameter that appears explicitly in \eqref{eqn:dimensionless-dressed-schrodinger} and \eqref{eqn:dimensionless-omega-evolution}, while a change in $\beta$ only acts through a variation of $Z$.
\item As a consequence of changing $\varepsilon$ and $\beta$, the strength of the perturbation $\eta^\floq{g}$, and thus the Landau-Zener parameter $Z$, changes as well.
\item The total mass of the cloud changes with time if state $\ket{b}$ has $\Im(\omega_{n\ell m})\ne0$: as a consequence, the Schrödinger equation \eqref{eqn:dimensionless-dressed-schrodinger} is modified to
\beq
\frac\dd{\dd\tau}\!\begin{pmatrix}c_a\\ c_b\end{pmatrix}=-i\begin{pmatrix}\omega/2 & \sqrt{Z}\\ \sqrt{Z} & -\omega/2-i\Gamma\end{pmatrix}\begin{pmatrix}c_a\\ c_b\end{pmatrix}\,,
\label{eqn:schrodinger-Gamma}
\eeq
where $\Gamma \equiv \Im(\omega_{n\ell m})/\sqrt{\gamma/\abs{g}}$, and care must be paid in the definition of $B$.
\end{enumerate}
All three effects come with two possible signs, one of which ``weakens'' the resonance and potentially breaks it, while the other ``reinforces'' it: in the first category we have the increase of eccentricity, the decrease of $Z$ and the cloud decay ($\Gamma<0$).\footnote{The superradiant amplification of state $\ket{b}$, that is, $\Gamma>0$, is never encountered for floating resonances anyway.}

\vskip 0pt
To understand under what conditions a resonance breaks, it is insightful to study the evolution of $\omega$ during the float. To zeroth order, $\omega$ is identically zero, but Figures~\ref{fig:floating-backreacted-resonance}, \ref{fig:eccentric-backreacted-resonance} and \ref{fig:eccentric-inclined-backreacted-resonance} hint towards a nontrivial dynamics to higher order, with small oscillatory features of varying frequency. Let us try to find an equation of motion for the sole $\omega$, in the vanilla case with $\dd\varepsilon/\dd\tau=\dd Z/\dd\tau=\Gamma=0$, where no resonance break is expected. By taking the derivative of \eqref{eqn:dimensionless-omega-evolution} and repeatedly using Schrödinger's equation, we find
\beq
\frac{\dd^2\omega}{\dd\tau^2}=-B\frac{\dd\abs{c_b}^2}{\dd\tau^2}=-2ZB(1-2\abs{c_b}^2)+\sqrt ZB(c_a^*c_b+c_ac_b^*)\omega\,.
\label{eqn:d2omegadt2}
\eeq
Remarkably, the equation of motion obeyed by $\omega$ closely resembles a harmonic oscillator whose (squared) frequency is $-\sqrt ZB(c_a^*c_b+c_ac_b^*)$. It is thus natural to study this quantity: by directly applying Schrödinger's equation, we find
\beq
\sqrt Z\frac\dd{\dd\tau}(c_a^*c_b+c_ac_b^*)=\omega\frac{\dd\abs{c_b}^2}{\dd\tau}\,.
\label{eqn:dcacbcbcadt}
\eeq
We notice that equations \eqref{eqn:d2omegadt2} and \eqref{eqn:dcacbcbcadt} form a closed system of ordinary differential equations (because in the vanilla case $\abs{c_b}^2=(\tau-\omega)/B$), through which it is possible to prove mathematically a number of interesting properties of the system, such as the fact that at small $Z$ the evolution is entirely determined by $ZB$, as thoroughly described in Section~\ref{sec:floating-adiabatic}.

\vskip 0pt
For the scope of this section it is however sufficient to assume that the quantity $c_a^*c_b+c_ac_b^*$ evolves slowly during a float, with a timescale of $\Delta t\ped{float}$, similar to $\abs{c_b}^2$. Equation \eqref{eqn:d2omegadt2} can then be solved in a WKB approximation as
\beq
\omega\approx\frac{2\sqrt Z(1-2\abs{c_b}^2)}{c_a^*c_b+c_ac_b^*}+\frac{A\,Z^{-1/8}B^{-1/4}}{(-c_a^*c_b-c_ac_b^*)^{1/4}}\cos\biggl(Z^{1/4}B^{1/2}\int_0^\tau\sqrt{-c_a^*c_b-c_ac_b^*}\dd\tau'+\delta\biggr)\,,
\label{eqn:omega-solution-oscillator}
\eeq
where $A$ is a constant and $\delta$ is a phase. As the fast oscillations average out, we can plug the first, non-oscillatory, term of \eqref{eqn:omega-solution-oscillator} into \eqref{eqn:dcacbcbcadt} and integrate to find $c_a^*c_b+c_ac_b^*\approx-\sqrt{1-(1-2\abs{c_b}^2)^2}$. The resulting solution for $\omega$,
\beq
\omega\approx-\frac{2\sqrt Z(1-2\abs{c_b}^2)}{\sqrt{1-(1-2\abs{c_b}^2)^2}}+\text{oscillatory terms}\,,
\eeq
is well-behaved for the entire duration of the float, only diverging before ($\abs{c_b}^2=0$) or after ($\abs{c_b}^2=1$) the resonance.

\vskip 0pt
The same analytical approach can be applied to the cases mentioned above, with varying $\epsilon$ or $Z$, or $\Gamma\ne0$. A ``master equation'', where all three effects are turned on at the same time, is derived and shown in Appendix~\ref{sec:breaKING}. Here, we find it more illuminating to study them one at a time. The outcome in realistic cases may then be approximated by only retaining the strongest of the three effects.

\vskip 0pt
When the eccentricity is not a constant, the time derivative of \eqref{eqn:dimensionless-omega-evolution} contains the additional term $\dd f(\varepsilon)/\dd\tau$. As a result, the equation of motion for $\omega$ and the expression of $c_a^*c_b+c_ac_b^*$ are both modified. The final result, which has been thoroughly checked against numerical solutions of the full system \eqref{eqn:dimensionless-dressed-schrodinger}-\eqref{eqn:dimensionless-omega-evolution}-\eqref{eqn:dimensionless-eccentricity-evolution}-\eqref{eqn:dimensionless-inclination-evolution}, is
\beq
\omega\approx\frac{\frac{\dd f(\varepsilon)}{\dd\tau}-2ZB(1-2\abs{c_b}^2)}{\sqrt{ZB^2(1-(1-2\abs{c_b}^2)^2)-(f(\varepsilon)^2-f(\varepsilon_0)^2)}}+\text{oscillatory terms}\,.
\eeq
If $\varepsilon$ increases from its initial value $\varepsilon_0$, the denominator can hit zero before the transition is complete, and the resonance breaks. The population remaining in state $\ket{a}$ and the binary eccentricity at resonance breaking satisfy
\beq
4ZB^2(\abs{c_a}^2-\abs{c_a}^4)=f(\varepsilon)^2-f(\varepsilon_0)^2\,,
\label{eqn:epsilon-breaking}
\eeq
which can be compared with the numerical solution in Figure~\ref{fig:eccentric-inclined-backreacted-resonance} (\emph{right panel}). Despite the simplicity of \eqref{eqn:epsilon-breaking}, a numerical integration is still needed, in principle, to determine $\varepsilon$ as function of $\abs{c_a}^2$, and so whether a resonance will break. We can however make a simple conservative estimate by noting that the left-hand-side can be at maximum $ZB^2$. If the system follows a trajectory in the $(\varepsilon,\beta)$ plane (cf.~Figure~\ref{fig:streamplot_inclination-eccentricity}) that significantly increases its eccentricity, such that
\beq
f(\varepsilon)>\sqrt ZB
\label{eqn:epsilon-breaking-simple}
\eeq
at some point, then the resonance must necessarily break.

\begin{figure}[t]
\centering
\includegraphics[width=\textwidth]{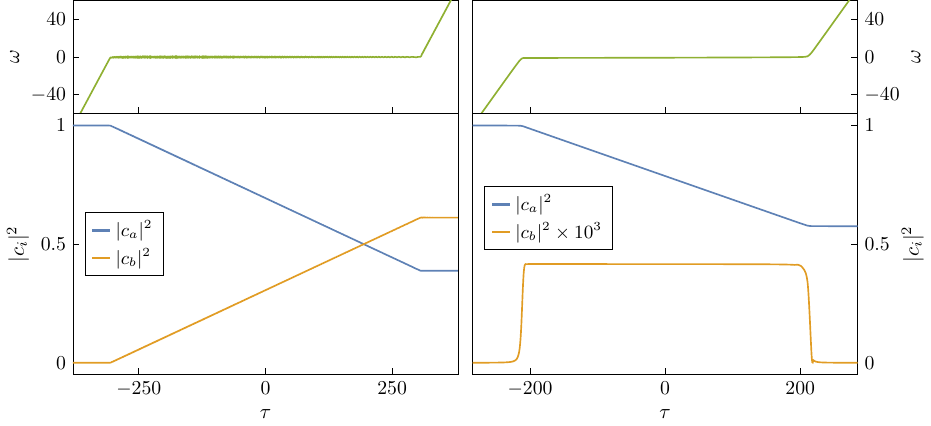}
\caption{Numerical solution of equations \eqref{eqn:dimensionless-dressed-schrodinger} and \eqref{eqn:dimensionless-omega-evolution} with (initial) parameters $Z=0.001$ and $B=1000$. A $Z$-breaking occurs when $Z$ is slowly reduced over time, with the resonance ending when \eqref{eqn:Z-breaking} is satisfied (\emph{left panel}). A $\Gamma$-breaking is observed when $Z$ is kept fixed but state $\ket{b}$ is given a nonzero decay width $\Gamma=1.2$, with the resonance ending when \eqref{eqn:gamma-breaking} is satisfied (\emph{right panel}).}
\label{fig:broken-resonance}
\end{figure}

\vskip 0pt
If instead $Z$ is allowed to vary while $\varepsilon$ is kept constant, new terms appear when taking the time derivative of the Schrödinger equation (used in the second equality of \eqref{eqn:d2omegadt2}), and \eqref{eqn:omega-solution-oscillator} becomes a damped harmonic oscillator. Similar to the previous case, the resonance breaks when $c_a^*c_b+c_ac_b^*=0$, which is equivalent to
\beq
4ZB^2(\abs{c_a}^2-\abs{c_a}^4)=f(\varepsilon)^2\biggl(1-\frac{Z}{Z_0}\biggr)\,,
\label{eqn:Z-breaking}
\eeq
see Figure~\ref{fig:broken-resonance} (\emph{left panel}). Analogous considerations as before can be applied to extract from \eqref{eqn:Z-breaking} the approximate point of resonance breaking without performing a numerical integration.

\vskip 0pt
Taking into account a nonzero decay width $\Gamma$, while keeping $\epsilon$ and $Z$ constant, requires more care. Because $\abs{c_a}^2+\abs{c_b}^2$ is no longer a constant, equation \eqref{eqn:dimensionless-omega-evolution} is now written as
\beq
\frac{\dd\omega}{\dd\tau}=f(\varepsilon)-\frac{B}{\Delta\epsilon}\biggl(\epsilon_a\frac{\dd\abs{c_a}^2}{\dd\tau}+\epsilon_b\frac{\dd\abs{c_b}^2}{\dd\tau}+2\Gamma\epsilon_b\abs{c_b}^2\biggr)=f(\varepsilon)+B\frac{\dd\abs{c_a}^2}{\dd\tau}\,,
\label{eqn:dimensionless-omega-evolution-Gamma}
\eeq
where the constant parameter $B$ is computed according to \eqref{eqn:BC}, using the value of the mass of the cloud before the start of the resonance. Furthermore, due to the modified Schrödinger equation, formula \eqref{eqn:dcacbcbcadt} becomes
\beq
\sqrt Z\frac\dd{\dd\tau}(c_a^*c_b+c_ac_b^*)=-\omega\frac{\dd\abs{c_a}^2}{\dd\tau}-\Gamma\sqrt Z(c_a^*c_b+c_ac_b^*)\,.
\label{eqn:dcacbcbcadt_Gamma}
\eeq
As we will show later (cf.~Figure~\ref{fig:tfloat_tdecay}), in almost all realistic cases state $\ket{b}$ decays much faster than the duration of the resonance, i.e., $\tau\ped{decay}\equiv(2\Gamma)^{-1}\ll B$. As a consequence, its population $\abs{c_b}^2$ during a floating orbit stays approximately constant, at a value $\abs{c_b}^2=f(\varepsilon)/(2\Gamma B)$, where the state decay is balanced by the transitions from $\ket{a}$ to $\ket{b}$. As this saturation value is typically very small, we will neglect it. Under this assumption, we can solve \eqref{eqn:dcacbcbcadt_Gamma} as
\beq
c_a^*c_b+c_ac_b^*\approx\sqrt{\frac{2ZB\abs{c_a}^2-\Gamma}{\Gamma ZB^2}}\,,
\eeq
and conclude that the resonance breaks when the remaining population in the initial state is
\beq
\abs{c_a}^2\approx\frac\Gamma{2ZB}\,,
\label{eqn:gamma-breaking}
\eeq
see Figure~\ref{fig:broken-resonance} (\emph{right panel}). Resonances where this quantity is larger than 1 do not exhibit a floating orbit at all, showing an ``immediate'' breaking.

\vskip 0pt
We refer to the three types of resonance breaking as $\varepsilon$-breaking, $Z$-breaking and $\Gamma$-breaking. A summary of the respective conditions is given below.

\begin{center}
\begin{tabular}{c|c|c} 
$\varepsilon$-breaking & $Z$-breaking & $\Gamma$-breaking \\ 
\hline
$f(\varepsilon)\gtrsim\sqrt ZB$ & $Z/Z_0\lesssim1-ZB^2/f(\varepsilon)^2$ & $\abs{c_a}^2\lesssim\Gamma/(2ZB)$ \\ 
\end{tabular}
\end{center}

\section{Sinking orbits}
\label{sec:sinking}

\vskip 0pt
Let us now turn our attention to sinking orbits, corresponding to $B<0$, where backreaction tends to make the resonance less adiabatic. This case turns out to not be as dramatically relevant as floating orbits for the resonant history of the system. However, it is important for direct GW signatures. For this reason, we will only study the aspects of it with observational consequences.

\vskip 0pt
All the observable sinking resonances have $2\pi Z\ll1$. In this case, the final population in state $\ket{b}$, as predicted by \eqref{eqn:lz}, is very small, and this quantity is further reduced by the backreaction. In the regime where this correction is dominant, we can find a rough approximation for the total population transferred by only keeping the backreaction term in (\ref{eqn:dimensionless-omega-evolution}). Further assuming $\abs{c_a}^2\approx1$ and $\dot c_b\approx0$, we can substitute in the second component of (\ref{eqn:dimensionless-dressed-schrodinger}) and obtain $\abs{c_b}^2\approx(Z/B^2)^{1/3}$, where we assumed for simplicity quasi-circular orbits.\footnote{The validity of the assumption will become clear in Section \ref{chap:history}.} This result is confirmed by numerical tests, modulo a multiplicative factor: we find
\beq
B\ll -\frac1Z,\qquad\abs{c_b}^2\approx3.7\,\biggl(\frac{Z}{B^2}\biggr)^{1/3}\,.
\label{eqn:sinkingpopulation}
\eeq
This formula is accurate for $2\pi Z\ll1$ and provides a slight under-estimate of the final population for moderately large $Z$.

\vskip 0pt
Sinking orbits backreact on the orbit by increasing both the orbital frequency and the binary eccentricity, as shown in Figure~\ref{fig:eccentric-backreacted-resonance} and \ref{fig:omega-eccentricity} (\emph{right panels}). At the same, time both $\Omega$ and $\varepsilon$ feature long-lived oscillations after the resonance. These oscillations slowly die out, so that a ``jump'' in the $\Omega$ and $\varepsilon$ is the only mark left after a long time. The non-monothonic behavior of $\Omega$ was already observed in \cite{Baumann:2019ztm}, where it was also speculated that sinking orbits could yield large eccentricities (becoming ``kicked orbits''). Our results confirm that the oscillations are not an artifact of having considered quasi-circular orbits and further show that the increase of the eccentricity is also not monothonic. However, for the realistic cases analyzed in Section~\ref{chap:history}, the increase in eccentricity due to sinking orbits turns out to be negligible.

\section{Three types of resonances}
\label{sec:types-of-resonances}

Resonances can be divided in three distinct categories, depending on the energy splitting between the two states, as computed from (\ref{eq:eigenenergy}). \emph{Hyperfine} resonances occur between states with same $n$ and $\ell$ but different $m$; they have the smallest energy splitting and thus occur the earliest in the inspiral, as the corresponding resonant orbital frequency is smallest. Then, \emph{fine} (same $n$, different $\ell$) and \emph{Bohr} resonances (different $n$) follow, the latter having the largest splittings. The tools developed so far in this chapter apply to all of them: the character of a resonance is only determined by the parameters $2\pi Z$ and $B$, its impact on eccentricity and inclination is quantified by $D$, and its duration (in case it is a floating resonance) is $\Delta t\ped{float}$. In principle, the recipe to determine the co-evolution of the binary and the cloud is clear: (1) pick the earliest resonance, (2) determine its character and backreaction by computing $Z$, $B$, $D$ and $\Delta t\ped{float}$, (3) update the state of binary and cloud accordingly, (4) move to the next resonance and repeat. We will indeed execute this algorithm in Chapter~\ref{chap:history}. To be as generic as possible and explore a wide parameter space, it will prove useful to find the scalings of the relevant quantities with $M$, $M\ped{c}$, $q$, $\alpha$ and $\tilde a$. Different types of resonances have different scalings, so we analyze them here systematically.

\subsection{Hyperfine resonances}
\label{sec:hyperfine}

Let us start with hyperfine resonances. From (\ref{eq:eigenenergy}), we see that the energy splitting (and thus the resonant frequency) scales as $\Omega_0\propto M^{-1}\alpha^6\tilde a$. The corresponding orbital separation is $R_0\propto M\alpha^{-4}\tilde a^{-2/3}$. This strong $\alpha$-dependence places hyperfine resonances at distances parametrically much larger than the cloud's size. At such large orbital separations, the cloud's ionization is very inefficient, and thus the only significant mechanism for energy loss is the GW emission. As this too is a very weak effect, other phenomena might potentially be relevant, including astrophysical interactions connected to the binary formation mechanism. We will postpone the discussion of these complications to Chapter~\ref{chap:history}, and assume for now that formula (\ref{eqn:gamma_gws}) applies, giving a chirp rate of $\gamma\propto qM^{-2}\alpha^{22}\tilde a^{11/3}$. This information is already enough to determine the scaling of three key quantities:
\beq
B\propto\frac{M\ped{c}}Mq^{-3/2}\alpha^{-4}\tilde a^{-1/2}\,,\qquad \Delta t\ped{float}\propto M\ped{c}q^{-2}\alpha^{-15}\tilde a^{-7/3}\,,\qquad D\propto \frac{M\ped{c}}Mq^{-1}\alpha\tilde a^{1/3}\,.
\label{eqn:hyperfine-B-tfloat-exponent}
\eeq

\vskip 0pt
The scaling of the Landau-Zener parameter $Z$ depends instead on the overlap coefficient $\eta^\floq{g}$. Given the hierarchy of length scales, $R_0\gg r\ped{c}$, the ``inner'' term in (\ref{eqn:F(r)}) dominates the radial integral $I_r$. At fixed $\ell_*\ne1$, we thus have
\beq
\begin{split}
\eta^\floq{g}&\propto q\alpha\,d^\floq{\ell_*}_{\Delta m,g}(\beta)\, I_r=q\alpha\,d^\floq{\ell_*}_{\Delta m,g}(\beta)\int_0^\infty \frac{r^{\ell_*}}{R_0^{\ell_*+1}}R_{n\ell}(r)^2r^2\dd r\\
&\propto M^{-1}q\alpha^{2\ell_*+5}\tilde a^{2(\ell_*+1)/3}d^\floq{\ell_*}_{\Delta m,g}(\beta)\,,
\end{split}
\eeq
and so
\beq
Z\propto q\alpha^{4\ell_*-12}\tilde a^{(4\ell_*-7)/3}\bigl(d^\floq{\ell_*}_{\Delta m,g}(\beta)\bigr)^2\,.
\eeq
The dipole $\ell_*=1$ is an exception for two reasons: (a) its ``inner'' term in (\ref{eqn:F(r)}) vanishes, (b) its ``outer'' term is not simply $r^{\ell_*}/R_*^{\ell_*+1}$. However, hyperfine resonances connect states with same $\ell$: from the selection rule \eqref{eqn:S2}, only even values of $\ell_*$ contribute. We can thus safely ignore the dipole. The rest of the multipole expansion can be seen as a power series in the small parameter $r\ped{c}/R_0$, the smallest $\ell_*$ giving the strongest contribution. Because selection rules require $\ell_*\ge \abs{g}=-g$,\footnote{Strictly speaking, this constraint only applies on circular orbits. In general, the same inequality applies to $g_\beta$ instead.} a resonance with a given value of $g$ will be be dominated by $\ell_*=-g$. The only two cases we will encounter in Chapter~\ref{chap:history} are
\begin{align}
\label{eqn:hyperfine-g=-2-Z}
g=-2&\qquad Z\propto q\alpha^{-4}\tilde a^{1/3}\bigl(d^\floq{2}_{\Delta m,g}(\beta)\bigr)^2\,,\\
\label{eqn:hyperfine-g=-4-Z}
g=-4&\qquad Z\propto q\alpha^4\tilde a^{3}\bigl(d^\floq{4}_{\Delta m,g}(\beta)\bigr)^2\,.
\end{align}
Furthermore, the assumption $\ell_*=-g$ allows us to write the explicit expression for the angular dependence of $Z$ as
\beq
d^\floq{-g}_{\Delta m,g}(\beta)\propto\sin^{\Delta m-g}(\beta/2)\cos^{-\Delta m-g}(\beta/2)\,.
\label{eqn:dgg}
\eeq

\subsection{Fine resonances}

Most of the assumptions made for hyperfine resonances work in the fine case too. The resonant frequency now scales as $\Omega_0\propto M^{-1}\alpha^5$ and, similar to before, we arrive to
\beq
B\propto\frac{M\ped{c}}Mq^{-3/2}\alpha^{-7/2}\,,\qquad \Delta t\ped{float}\propto M\ped{c}q^{-2}\alpha^{-38/3}\,,\qquad D\propto \frac{M\ped{c}}Mq^{-1}\alpha^{2/3}\,.
\eeq
The scaling of the overlap coefficient reads $\eta^\floq{g}\propto qM^{-1}\alpha^{(4\ell_*+13)/3}$ and we get
\beq
Z\propto q\alpha^{(8\ell_*-29)/3}\bigl(d^\floq{\ell_*}_{\Delta m,g}(\beta)\bigr)^2\,.
\eeq
The main difference with the previous case resides in the possible values of $\ell_*$. Fine resonances connect states with different values of $\ell$, and most of the cases we will study in Chapter~\ref{chap:history} will have odd values of $\ell_*$. For $g=-3$, all the previous arguments apply and the octupole $\ell_*=3$ is the dominant contribution. For $g=-1$, the extreme weakness of the dipole at large distances again leaves the octupole as the most important term; because now $\ell_*\ne-g$, however, the angular dependence will have a form different from (\ref{eqn:dgg}), which we will describe on a case-by-case basis in Chapter~\ref{chap:history}. There is one further exception to this: if $\ell_a+\ell_b=1$, the selection rule \eqref{eqn:S3} forbids all $\ell_*\ge2$. Only in this case (corresponding to the $\ket{211}\to\ket{200}$ resonance) the dipole is entirely responsible for the coupling between the two states. Its anomalous expression (\ref{eqn:F(r)}) endows $\eta^\floq{g}$ (and thus $Z$) with a non-power-law dependence on $\alpha$: given the peculiarity of this case, we will treat it separately from the others.

\subsection{Bohr resonances}\label{sec:Bohr_res}

\begin{figure}[t]
\centering
\includegraphics[width=\textwidth]{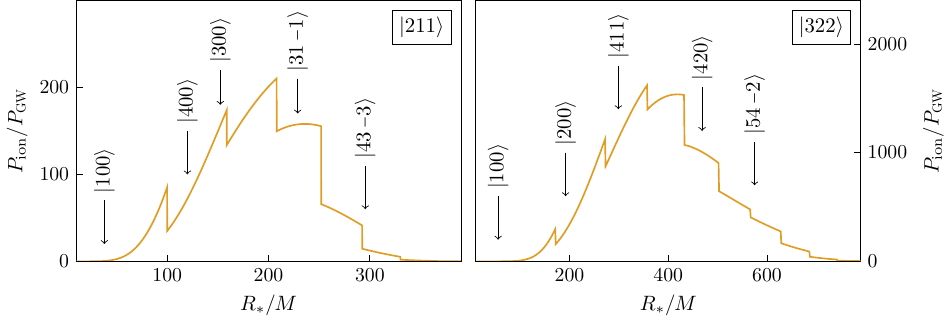}
\caption{Position of a few selected Bohr resonances, compared to $P\ped{ion}/P_\slab{gw}$, i.e., the ratio of the ionization power to the power emitted in GWs, shown here for a counter-rotating orbit and for a cloud in the $\ket{211}$ (\emph{left panel}) or $\ket{322}$ (\emph{right panel}) state. We assumed $M\ped{c}/M=0.01$ and $\alpha=0.2$, but the relative position of the resonances and the shape of the curve do not depend on the parameters.}
\label{fig:ionization-bohr-resonances}
\end{figure}

Bohr resonances are a different story. States with different principal quantum number $n$ have different energies to leading order, meaning that the resonant orbits are placed at distances comparable to the cloud's size. There is no parametric separation between the two, as now $R_0\propto M\alpha^{-2}\propto r\ped{c}$. At these orbital distances, the cloud's ionization is generally a more effective mechanism for energy loss than GWs. We prove this point in Figure~\ref{fig:ionization-bohr-resonances}, where the position of several Bohr resonances is shown on top of the ionization-to-GWs power ratio. This latter quantity scales as
\beq\label{eqn:P_ion}
\frac{P\ped{ion}}{P_\slab{gw}}\bigg|_{R_*=R_0}\propto\frac{M\ped{c}}M\alpha^{-5}\,.
\eeq
With the possible exception of transitions to $\ket{100}$, as they happen extremely late in the inspiral, Bohr resonances and ionization thus happen \emph{at the same time}. This observation raises two points.
\begin{enumerate}
\item Formula (\ref{eqn:gamma_gws}) for the chirp rate $\gamma$ is no longer accurate, as ionization must now be included.
\item The derivation of the expression for $P\ped{ion}$ laid down in Section~\ref{sec:thorough-derivation} assumes that the system is away from bound-to-bound state resonances.
\end{enumerate}
In Appendix~\ref{sec:ion-at-resonance} we extend the framework of Section~\ref{sec:thorough-derivation} to describe the ionization of a system actively in resonance. Although this requires the addition of new terms, their effect is generally negligible for realistic parameters. It is thus a good approximation to simply adjust the value of $\gamma$ by a factor $1+P\ped{ion}/P_\slab{gw}\approx P\ped{ion}/P_\slab{gw}$, where $P\ped{ion}$ is computed as in \eqref{eqn:pion}. The last approximation holds whenever $P\ped{ion}\gg P_\slab{gw}$ and is always satisfied, unless the resonance involves $\ket{100}$ or the value of $\alpha$ is exceptionally large.

\vskip 0pt
Under these assumptions, we arrive to
\beq\label{eqn:B_bohr}
B\propto\sqrt{\frac{M\ped{c}}M}q^{-3/2}\,,\qquad \Delta t\ped{float}\propto Mq^{-2}\alpha^{-3}\,,\qquad D\propto \frac{M\ped{c}}Mq^{-1}\,.
\eeq
These quantities now also have a $\beta$-dependence, due to $P\ped{ion}$ having different values for different inclinations. However, we will see in Chapter~\ref{chap:history} that this detail is not relevant, so we neglect it here. As for the overlap $\eta^\floq{g}$, there is now no clear hierarchy of multipoles. Luckily, $R_0$ has the same $\alpha$-scaling as the argument of the hydrogenic wavefunctions $R_{n\ell}$: with an appropriate change of variable, we can show that
\beq
\eta^\floq{g}\propto M^{-1}q\alpha^3d^\floq{\ell_*}_{\Delta m,g}(\beta)\,.
\label{eqn:eta-bohr}
\eeq
The $\beta$-dependence in (\ref{eqn:eta-bohr}) can be written in terms of a Wigner small $d$-matrix only when there is a single value of $\ell_*$ that contributes. As this is the case for many of the Bohr resonances we will encounter in Chapter~\ref{chap:history}, we keep that factor explicit here. Finally, the Landau-Zener parameter scales as
\beq\label{eqn:Z_bohr}
Z\propto \frac{M}{M\ped{c}}q\bigl(d^\floq{\ell_*}_{\Delta m,g}(\beta)\bigr)^2\,.
\eeq
One particularly interesting aspect of Bohr resonances is the disappearance of any $\alpha$-dependence from the Landau-Zener parameter $Z$ and from the backreaction $B$. This is in contrast with the steep power-laws found for hyperfine and fine resonances, and it means that the character of Bohr resonances is much more \emph{universal}.

\chapter{The resonant history}

\label{chap:history}

In this chapter we draw a consistent picture of the co-evolution of the cloud and the binary, using the tools developed in Chapter~\ref{chap:resonances}. The results we derive here are needed to understand the state of the system by the time it becomes observable: for example, when it enters the LISA band. First, we discuss the generic behavior of the different types of resonances in Section \ref{sec:generalB}; then, in Section \ref{sec:evolution-211} and \ref{sec:evolution-322}, we study explicitly the history for a cloud initialized in the state $\ket{211}$ or $\ket{322}$.

\vskip 0pt
Throughout Chapters~\ref{chap:history} and \ref{chap:observational-signatures}, we write many of our results in an explicit scaling form, with respect to the following set of benchmark parameters: $M=10^4M_\odot$, $M\ped{c}=0.01M$, $q=10^{-3}$ and $\alpha=0.2$. These make for a strong observational case for future millihertz gravitational-wave detectors.

\section{General behavior}
\label{sec:generalB}

The initial state $\ket{a}=\ket{n_a\ell_am_a}$ of the cloud, populated by superradiance, generally has $m_a=\ell_a=n_a-1$. Within the multiplet of states $\ket{n_a\ell_am}$ with $m\le m_a$, this is the one with highest energy, as can be readily seen from (\ref{eq:eigenenergy}). Hyperfine resonances, which occur the earliest in the inspiral, thus necessarily have $\Delta\epsilon<0$ and are of the floating type. To understand their behavior, it is important to keep in mind a few key points.
\begin{description}
\item[Adiabaticity.] The first question to answer is whether a given hyperfine resonance is adiabatic or not. We can apply the results of Section~\ref{sec:floating-adiabatic}. If $2\pi ZB>f(\varepsilon)^{3/2}$ the resonance is adiabatic: the binary starts evolving as described in Section~\ref{sec:floating-evolution-e-beta} until the transition completes after a time $\Delta t\ped{float}$, or the resonance breaks due to any of the conditions derived in Section~\ref{sec:resonance-breaking}. Almost all hyperfine resonances turn out to be adiabatic in the entire parameter space, except in a narrow interval of almost counter-rotating inclinations, say $\pi-\delta_1<\beta\le\pi$, where $\delta_1$ is the size of the interval. This is because, on floating orbits, the resonance condition $\Omega_0^\floq{g}=\Delta\epsilon/g$ forces $g$ to be negative; on the other hand, $\Delta m=m-m_a<0$, and from (\ref{eqn:dgg}) we see that for $\beta\to\pi$ the parameter $Z$ goes to zero as a (high) power of $\cos(\beta/2)$. The explicit determination of the angle $\delta_1$ as function of the parameters will be performed in Sections~\ref{sec:evolution-211} and~\ref{sec:evolution-322}.

\item[Cloud's decay and $\Gamma$-breaking.] After saturation of the dominant superradiant mode $\ket{n_a\ell_am_a}$, all states of the multiplet $\ket{n_a\ell_am}$ with $m\ne m_a$ have $\Im(\omega)<0$, meaning that they decay back in the BH with an $e$-folding time $t\ped{decay}\equiv\abs{2\Im(\omega_{n_a\ell_am})}^{-1}$. It is thus necessary to compare $\Delta t\ped{float}$ and $t\ped{decay}$. One of the most important results of this chapter is the following: for intermediate or extreme-mass ratios, and typical values of $M\ped{c}$ and $\alpha$, the decay timescale $t\ped{decay}$ is many orders of magnitude smaller than floating timescale, $\Delta t\ped{float}$. It is not easy to prove this statement in full generality, due to the complicated dependence of $t\ped{decay}$ on the parameters. Nevertheless, for small $\alpha$ and $\tilde a$, using the Detweiler approximation \cite{Detweiler:1980uk} and the results from Section~\ref{sec:hyperfine}, we have
\beq
\frac{t\ped{decay}}{\Delta t\ped{float}}\propto\frac{M}{M\ped{c}}q^2\alpha^{10-4\ell_a}\tilde a^{4/3}\,,
\label{eqn:t-decay-t-float}
\eeq
where $\tilde a\propto\alpha$ at the superradiant threshold. For $\alpha\to0$ and small enough values of $\ell_a$, this ratio becomes very small. In fact, for small $q$, any possible value of $\alpha$ results in $t\ped{decay}\ll\Delta t\ped{float}$. A more detailed comparison is given in Figure~\ref{fig:tfloat_tdecay}, where the numerical coefficients are properly taken into account and the spin $\tilde a$ is set to correspond to the boundary of the BH superradiant region.

\begin{figure}[t]
\centering
\includegraphics[width=\textwidth]{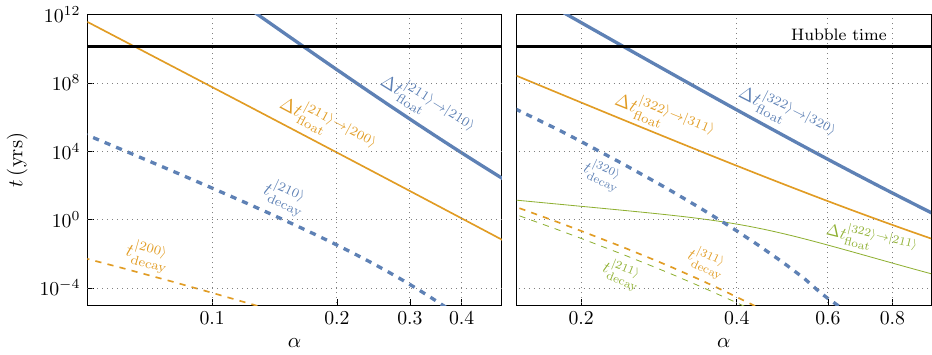}
\caption{Floating timescale $\Delta t\ped{float}$ (solid lines), compared to the decay timescale $t\ped{decay}$ (dashed lines) of the final state, for some selected resonances. We use benchmark parameters and determine the decay rate independently through Leaver's continued fraction method \cite{Leaver:1985ax,Cardoso:2005vk,Dolan:2007mj,Berti:2009kk} and the Chebyshev method in \cite{Baumann:2019eav}. Two resonances for a $\ket{211}$ initial state are shown (\emph{left panel}), namely $\ket{211}\to\ket{210}$ [{\color{Mathematica1}blue}] and $\ket{211}\to\ket{200}$ [{\color{Mathematica2}orange}]. Similarly, three resonances for a $\ket{322}$ initial state are shown (\emph{right panel}), namely $\ket{322}\to\ket{320}$ [{\color{Mathematica1}blue}], $\ket{322}\to\ket{311}$ [{\color{Mathematica2}orange}] and $\ket{322}\to\ket{211}$ [{\color{Mathematica3}green}]. The thick, normal and thin lines indicate hyperfine, fine or Bohr resonances respectively. Note the Bohr resonance falling outside of the ionization regime for large $\alpha$, changing the scaling of $\Delta t\ped{float}$ from $\alpha^{-3}$ to $\alpha^{-8}$, as predicted by \eqref{eqn:B_bohr} and~\eqref{eqn:P_ion}.}
\label{fig:tfloat_tdecay}
\end{figure}

This result has a dramatic consequence: hyperfine transitions are never able to change the state of the cloud. Instead, the portion that is transferred to state $\ket{b}$ decays immediately back into the BH.\footnote{\label{fn:bh-parameters}As a consequence, the mass and spin of the BH change. Our framework is not able to capture this effect, which we accordingly ignore.} The analysis of Section~\ref{sec:resonance-breaking} then applies, and the resonance $\Gamma$-breaks when the fraction of the cloud remaining in state $\ket{a}$ falls below the threshold determined in \eqref{eqn:gamma-breaking}. In a relatively large portion of parameter space, generally around counter-rotating orbits, that formula returns $\abs{c_a}^2>1$, meaning that the resonance $\Gamma$-breaks immediately. The outcome is effectively similar to a non-adiabatic resonance, that never even starts the floating phase. Similar to before, we will define an angular interval $\pi-\chi_1<\beta\le\pi$, within which the resonance is not effective. The $\varepsilon$-breaking and $Z$-breaking are instead less relevant for realistic parameters.

\item[The strongest resonance.] As shown in Section~\ref{sec:eccentric-inclined-resonances}, on eccentric and inclined orbits a resonance between two given states is excited at many different orbital frequencies, depending on the value\footnote{As briefly mentioned in Section~\ref{sec:eccentric-inclined-resonances}, two separate indices, say $g_\varepsilon$ and $g_\beta$ are necessary when both eccentricity and inclination are not zero. However, this technicality is not crucial in understanding the history of the system.} of $\abs{g}=1,2,3,\ldots$ The strength of the coupling also depends on $\varepsilon$ and $\beta$. Keeping track of so many different resonances would be very complicated. However, the hierarchy $t\ped{decay}\ll\Delta t\ped{float}$ implies that as soon as an adiabatic floating resonance is encountered (and does not break early), the cloud is destroyed. This means that studying the ``strongest'' resonance (the one that destroys the cloud in the largest portion of parameter space) actually suffices to determine the fate of the cloud.

Up to moderate values of the eccentricity, the coupling $\eta^\floq{g}$ that remains nonzero in the limit of circular orbit is much larger than all the others. We can then approximate the ``strongest resonance'' by ignoring eccentricity altogether. Regarding inclined orbits instead, we observe that higher values of $g$ require contributions from higher values of the multipole index $\ell_*$: at the separations of hyperfine resonances, the lowest value of $\ell_*$ (typically the quadrupole $\ell_*=2$) produces the strongest coupling. Given two states, we will then study the resonance with the smallest value of $\abs{g}$.
\end{description}

Applying the previous considerations to each possible hyperfine resonance, we are able to determine whether the cloud is destroyed in the process or survives to later stages of the inspiral. However, the binary might be able to ``skip'' hyperfine resonances for other reasons. This is because some of them are placed at extremely large binary separations: typically $R_*/M\gtrsim\mathcal O(10^3)$ for a $\ket{211}$ initial state, and $R_*/M\gtrsim\mathcal O(10^4-10^5)$ for $\ket{322}$. These distances are large enough that not only other kinds of astrophysical interactions may play a role, but their presence is in some cases necessary, in order to bring the binary close enough for the merger to happen within a Hubble time. Quantitatively, for a quasi-circular inspiral, the initial separation as function of the time-to-merger $t_0$ is given by
\beq
\frac{R_*}M=\SI{2.3e+4}{}\,\biggl(\frac{t_0}{\SI{e+10}{yrs}}\biggr)^{1/4}\biggl(\frac{M}{10^4M_\odot}\biggr)^{-1/4}\Bigl(\frac{q}{10^{-3}}\Bigr)^{1/4}\,.
\eeq
In other words: if we want the binary to merge within a Hubble time, we might be forced to assume that it ``starts'' its evolution too close for hyperfine resonances to be encountered, especially for a cloud initialized in the $\ket{322}$ state. This can be achieved by a variety of formation mechanisms, including dynamical capture \cite{Amaro-Seoane:2012lgq,LISA:2022yao} and \emph{in-situ} formation \cite{LISA:2022yao,Stone:2016wzz,Bartos:2016dgn,McKernan_2018,Levin:2006uc}.

\vskip 0pt
If the system is able to skip through hyperfine resonances, because they are either all non-adiabatic, or they $\Gamma$-break early, or the binary is formed at small enough separations, the cloud can be present when fine resonances are encountered. Their phenomenology is largely similar to hyperfine ones, as they too are all of the floating type. We defer the discussion of some state-dependent aspects to Sections~\ref{sec:evolution-211} and~\ref{sec:evolution-322}. For the purpose of the present general discussion, it suffices to say that, once again, the cloud can survive this stage if $\pi-\delta_2<\beta\le\pi$ (for some angle $\delta_2$ to be determined), if the resonance $\Gamma$-breaks early in an interval $\pi-\chi_2<\beta\le\pi$, or if the binary is formed $\emph{in situ}$ at very small radii.

\begin{figure}[t]
\centering
\includegraphics[width=\textwidth]{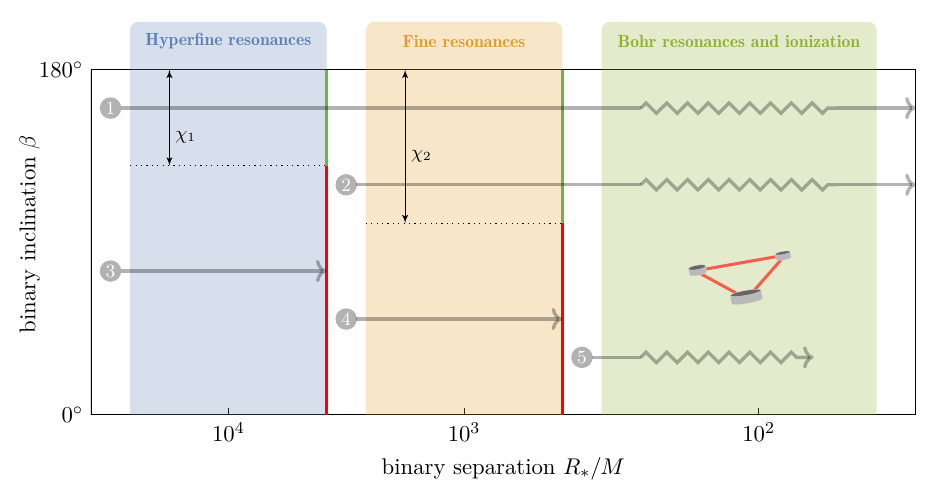}
\caption{Illustration of the possible outcomes of the resonant history of the cloud-binary system. The inspiral starts with the cloud in its initial state, either $\ket{211}$ or $\ket{322}$. Only systems {\normalsize \textcircled{\footnotesize 1}}--{\normalsize \textcircled{\footnotesize 2}} whose inclination angle is within intervals $\chi_1$ and $\chi_2$ from the counter-rotating ($\beta=\SI{180}{\degree}$) configuration are able to move past the hyperfine and fine resonances with the cloud still intact (green vertical lines). These later give rise to observational signatures in the form of ionization and Bohr resonances. Others {\normalsize \textcircled{\footnotesize 3}}--{\normalsize \textcircled{\footnotesize 4}} are destroyed by the hyperfine or fine resonances (red vertical lines). Binary systems that form at small enough separations may be able to skip early resonances~{\normalsize \textcircled{\footnotesize 5}}.}
\label{fig:history_211}
\end{figure}

\vskip 0pt
Finally, if the cloud makes it to this point, it becomes potentially observable: the ``Bohr region'' can be in the LISA band and is rich of signatures of the cloud. These come in the form of ionization and Bohr resonances, the vast majority of which are sinking and non-adiabatic. State-dependent details will be discussed in Sections~\ref{sec:evolution-211} and~\ref{sec:evolution-322} and a summary of the observational signatures will be given in Chapter~\ref{chap:observational-signatures}. A diagrammatic representation of the three stages of the resonant history is shown in Figure~\ref{fig:history_211}.

\vskip 0pt
As a concluding remark, we note that the results derived here and in Chapter~\ref{chap:resonances} are specific to resonances involving 2 states only. We have explicitly checked that this is the case for the resonances discussed in the next sections, so we apply the results of Chapter~\ref{chap:resonances} without further modification.

\section{Evolution from a $\ket{211}$ initial state}
\label{sec:evolution-211}

The $\ket{211}$ state is the fastest-growing superradiant mode, and represents therefore a natural assumption for the initial state of the cloud. The requirements that the superradiant amplification takes place, and does so on timescales no longer than a Gyr, set a constraint on $\alpha$:
\beq
0.02\biggl(\frac{M}{10^4M_\odot}\biggr)^{1/9}\lesssim\alpha<0.5\,.
\eeq
Once grown, the cloud will decay in GWs with a rate roughly proportional to $M\ped{c}^2\alpha^{14}$, assuming the scalar field is real. The resulting decay of $M\ped{c}$ is polynomial, rather than exponential in time; as such, we will not impose a further sharp bound on $\alpha$, and treat $M\ped{c}/M$ as an additional free parameter.

\vskip 0pt
There are two possible hyperfine resonances, with the states $\ket{210}$ and $\ket{21\,\minus1}$. Following the line of reasoning laid down in Section~\ref{sec:generalB}, we ignore the fact that the same resonances can be triggered at multiple points if the orbit is eccentric. Both resonances are then mediated by $g=-2$ and they are positioned at
\begin{align}
\ket{211}\overset{g=-2}{\longrightarrow}\ket{210} & \qquad \frac{R_0}M=\SI{8.3e+3}{}\,\Bigl(\frac\alpha{0.2}\Bigr)^{-4}\Bigl(\frac{\tilde a}{0.5}\Bigr)^{-2/3}\,,\\[8pt]
\ket{211}\overset{g=-2}{\longrightarrow}\ket{21\,\minus1} & \qquad \frac{R_0}M=\SI{5.2e+3}{}\,\Bigl(\frac\alpha{0.2}\Bigr)^{-4}\Bigl(\frac{\tilde a}{0.5}\Bigr)^{-2/3}\,,
\end{align}
where the value of the spin should be set equal to the threshold of superradiant instability of $\ket{211}$, that is, $\tilde a\approx4\alpha/(1+4\alpha^2)$. Both resonances become non-adiabatic in an interval $\pi-\delta_1<\beta\le\pi$, with the strongest constraint on $\delta_1$ given by $\ket{211}\to\ket{210}$. The value of $\delta_1$ is determined from \eqref{eqn:2piZB-epsilon0}: this means setting $2\pi ZB=f(\varepsilon_0)^{3/2}$, where $\varepsilon_0$ is the eccentricity at the onset of the resonance, and solving for $\beta$ as function of the parameters. Making use of the relations (\ref{eqn:hyperfine-B-tfloat-exponent}), (\ref{eqn:hyperfine-g=-2-Z}) and (\ref{eqn:dgg}), and evaluating numerically the overlap $\eta^\floq{2}$ between the two states, we find
\beq
\delta_1=\SI{7.5}{\degree}\,\biggl(\frac{M\ped{c}/M}{10^{-2}}\biggr)^{-1/6}\biggl(\frac{q}{10^{-3}}\biggr)^{1/12} \biggl(\frac{\alpha}{0.2}\biggr)^{4/3} \biggl(\frac{\tilde{a}}{0.5}\biggr)^{1/36}f(\varepsilon_0)^{1/4}\,.
\label{eqn:211-delta1}
\eeq
Although $\ket{211}\to\ket{210}$ is also non-adiabatic in a neighbourhood of $\beta=0$, such a co-rotating binary would still encounter the adiabatic floating resonance $\ket{211}\to\ket{21\,\minus1}$ later, so that the only ``safe'' inclinations are in the neighbourhood of counter-rotating determined in (\ref{eqn:211-delta1}).

\begin{figure}[t]
\centering
\includegraphics[width=\textwidth]{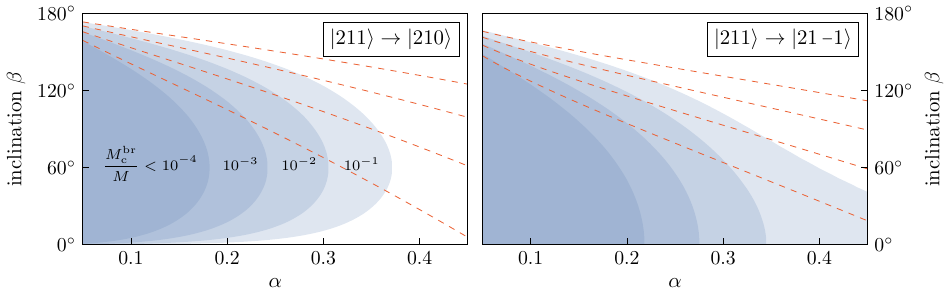}
\caption{Mass of the cloud $M\ped{c}\ap{br}$ at resonance $\Gamma$-breaking, as function of $\alpha$ and $\beta$, for the two hyperfine resonances from the initial state $\ket{211}$. The mass of the cloud decreases during the resonance from its initial value $M\ped{c}$, and the resonance breaks when the value $M\ped{c}\ap{br}$ is reached. Values $M\ped{c}\ap{br}>M\ped{c}$ indicate that the resonance breaks immediately as it starts. The contours [{\color{Mathematica1}blue}] are calculated on circular orbits, as this gives a good approximation for the strongest constraint on $M\ped{c}\ap{br}$ even when overtones due to orbital eccentricity (i.e., higher values of $\abs{g}$ for the resonance between two given states) are taken into account. Due to the inaccuracy of the analytical approximations for the decay width $(\omega_{211})\ped{I}$, especially at large $\alpha$, we have determined the contours with Leaver's \cite{Leaver:1985ax,Cardoso:2005vk,Dolan:2007mj,Berti:2009kk} and Chebyshev \cite{Baumann:2019eav} methods. The dashed lines [{\color{Mathematica4}red}] are analytical approximations to the blue contours in the proximity of $\beta=\pi$, based on~\eqref{eqn:211_chi_1}.}
\label{fig:deltas211}
\end{figure}

\vskip 0pt
Having determined when hyperfine resonances can be adiabatic, we now calculate where they break, using the results of Section~\ref{sec:resonance-breaking}. As anticipated in Section~\ref{sec:generalB}, the $\Gamma$-breaking is the most relevant mechanism of resonance breaking. To assess its impact, we observe that, because $B\propto M\ped{c}$, equation \eqref{eqn:gamma-breaking} can be written as a relation for the final mass of the cloud at resonance breaking, $M\ped{c}\ap{br}=M\ped{c}\abs{c_a}^2$, which can be computed as function of $\alpha$ and $\beta$. If $M\ped{c}\ap{br}>M\ped{c}$ is found, then the resonance breaks immediately as it starts, as if it was non-adiabatic. We show the result in Figure~\ref{fig:deltas211}. Note that in principle the resonance always $\Gamma$-breaks before the cloud is completely destroyed, but its observational impact becomes negligible when $M\ped{c}\ap{br}$ is too small.

\vskip 0pt
The combined constraints due to the $\Gamma$-breaking of $\ket{211}\to\ket{210}$ and $\ket{211}\to\ket{21\,\minus1}$ imply that the cloud survives in a neighborhood of $\beta=\pi$, say $\pi-\chi_1<\beta<\pi$, similar to what we found for the adiabaticity of the resonances. An analytical approximation of $\chi_1$ for $\ket{211}\to\ket{210}$ based on Detweiler's formula \cite{Detweiler:1980uk} is
\beq
\chi_1\approx\SI{38}{\degree}\biggl(\frac{M\ped{c}\ap{br}/M}{10^{-2}}\biggr)^{-1/6}\biggl(\frac\alpha{0.2}\biggr)^{7/6}\biggl(\frac{\tilde a}{0.5}\biggr)^{-5/18}\,,
\label{eqn:211_chi_1}
\eeq
which significantly underestimates the result for large $\alpha$, as shown in Figure~\ref{fig:deltas211}. Because $\chi_1>\delta_1$, this angular interval overwrites \eqref{eqn:211-delta1} as the portion of parameter space where the cloud survives hyperfine resonances.

\vskip 0pt
Finally, we check whether hyperfine resonances can $\varepsilon$-break or $Z$-break. Both $\varepsilon$ and $Z$ can vary significantly during the float, so we use the relation \eqref{eqn:epsilon-breaking} as an order-of-magnitude estimate. For generic values of the inclination, both hyperfine resonances have
\beq
\sqrt ZB\sim10^6\biggl(\frac{M\ped{c}/M}{10^{-2}}\biggr)\biggl(\frac{q}{10^{-3}}\biggr)^{-1}\biggl(\frac\alpha{0.2}\biggr)^{-6}\biggl(\frac{\tilde a}{0.5}\biggr)^{-1/3}\,.
\eeq
The resonances $\varepsilon$-breaks if $f(\varepsilon)=\sqrt ZB$, which is only satisfied at very high eccentricities, not smaller than $0.95$ for typical parameters. Such extreme eccentricities are only reachable if the initial inclination is very close to $\beta=\pi$, as can be seen from Figure~\ref{fig:streamplot_inclination-eccentricity}. But, as proved in \eqref{eqn:211-delta1} and \eqref{eqn:211_chi_1}, near-counter-rotating binaries do not undergo floating orbits at all, due to the resonances being either non-adiabatic or $\Gamma$-breaking immediately. As for the $Z$-breaking, one can conservatively ignore the term $Z/Z_0$ in \eqref{eqn:Z-breaking}, falling back to the same relation as \eqref{eqn:epsilon-breaking}.

\vskip 0pt
We conclude that the survival of the cloud to later stages of the inspiral is exclusively determined by the $\Gamma$-breaking. If the binary is outside the regions colored in Figure~\ref{fig:deltas211}, and computed in \eqref{eqn:211_chi_1}, it encounters the only possible fine resonance:
\beq
\ket{211}\overset{g=-1}{\longrightarrow}\ket{200}\qquad\frac{R_0}M=\SI{3.4e+2}{}\,\Bigl(\frac\alpha{0.2}\Bigr)^{-10/3}\,,
\label{eqn:R_0-211-200}
\eeq
whose angular dependence is determined through (\ref{eqn:dgg}) as usual. This resonance, however, has anomalous behavior for two reasons:
\begin{enumerate}
\item it is entirely mediated by the dipole $\ell_*=1$;
\item depending on the value of $\alpha$, it may fall inside the ionization regime ($P\ped{ion}\gtrsim P_\slab{gw}$) despite not being a Bohr resonance.
\end{enumerate}
As a consequence, its Landau-Zener parameter $Z$ does not scale as a pure power-law in $\alpha$ (nor $M\ped{c}$), and must be computed numerically. We therefore determine the angle $\delta_2$, such that for $\pi-\delta_2<\beta\le\pi$ the resonance is non-adiabatic, as
\beq
\delta_2=\SI{6.7}{\degree}\,\biggl(\frac{M\ped{c}/M}{10^{-2}}\biggr)^{-1/4}\biggl(\frac{q}{10^{-3}}\biggr)^{1/8} \biggl(\frac{\alpha}{0.2}\biggr)^{7/8}F(\alpha,M\ped{c})\,,
\label{eqn:211-delta2}
\eeq
where the formula holds for small $\delta_2$ and the function $F(\alpha,M\ped{c})$ is calculated numerically and shown in Figure~\ref{fig:211_200_F(alpha,Mc)}. Similar to hyperfine resonances, we can compute angular intervals $\delta_2$ and $\chi_2$ where the resonance is non-adiabatic and $\Gamma$-breaks, respectively. The extremely large decay width of $\ket{200}$ (as all states with $\ell=0$), however, makes $\chi_2$ as large as to correspond with the whole possible range of inclinations, from $\SI{0}{\degree}$ to $\SI{180}{\degree}$. Fine resonances are thus effectively never excited for a cloud in a $\ket{211}$ state.

\begin{figure}[t]
\centering
\includegraphics{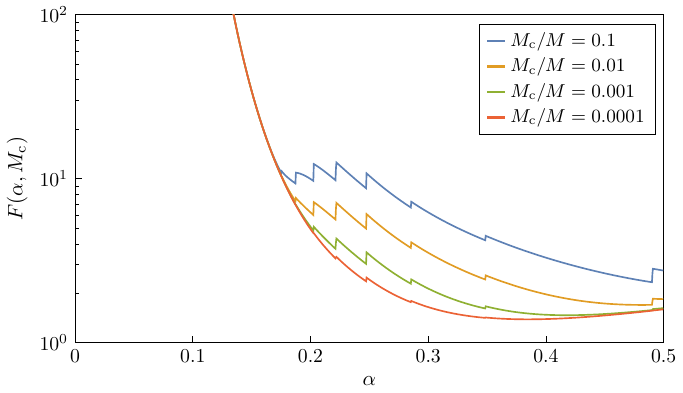}
\caption{Function $F(\alpha,M\ped{c})$ appearing in equation (\ref{eqn:211-delta2}), which defines the angular interval $\delta_2$ around a counter-rotating orbit where the resonance $\ket{211}\to\ket{200}$ is not adiabatic.}
\label{fig:211_200_F(alpha,Mc)}
\end{figure}

\begin{figure}[t]
\centering
\includegraphics[width=\textwidth]{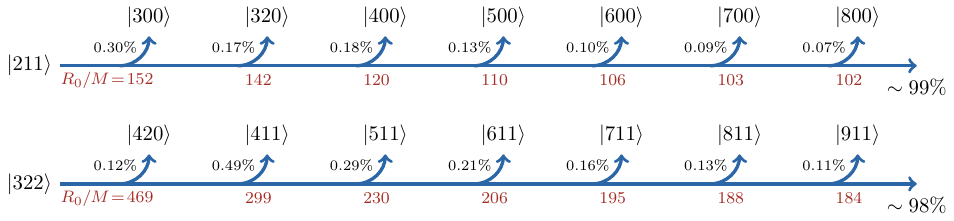}
\caption{Strongest sinking Bohr resonances on a counter-rotating orbit for a clound in the $\ket{211}$ or $\ket{322}$ state. The percentages next to each resonance are the values of $\abs{c_b}^2$ for benchmark parameters, and they scale with $M\ped{c}$ and $q$ according to \eqref{eqn:scaling-cb2}, while the red numbers below are the resonant orbital separations $R_0$, in units of $M$.}
\label{fig:sinking-from-211-and-322}
\end{figure}

\vskip 0pt
Finally, if the binary arrives to the Bohr region with the cloud still intact, it encounters the Bohr resonances, all of which are of the sinking type and fall inside the ionization regime (with the exception of $\ket{211}\to\ket{100}$). No extra circularization is provided by the hyperfine resonances, if they do not significantly destroy the cloud. Nevertheless, by the time the binary arrives to the Bohr regime, not only has it presumably evolved for a long time under the circularizing effect of GW radiation, but it also starts ionize the cloud, further suppressing the eccentricity (cf.~Section~\ref{sec:eccentricity}). We will therefore assume that quasi-circular orbits are a good approximation by this point. The final population after each sinking resonance can be found using the approximation (\ref{eqn:sinkingpopulation}), which, together with the scaling relations \eqref{eqn:B_bohr} and \eqref{eqn:Z_bohr}, implies
\beq
\abs{c_b}^2\approx3.7\,\biggl(\frac{Z}{B^2}\biggr)^{1/3}\propto\frac{M}{M\ped{c}}q^{4/3}\,.
\label{eqn:scaling-cb2}
\eeq
For the benchmark parameters, the values of $\abs{c_b}^2$ for the strongest sinking resonances (which are typically with states of the form $\ket{n00}$) are summarized in Figure~\ref{fig:sinking-from-211-and-322}, where we have assumed for simplicity a perfectly counter-rotating configuration ($\beta=\pi$). This is generally a good approximation, due to the relative smallness of the angle $\chi_1$. We see that all resonances are very non-adiabatic, in total transferring less than $1\%$ of the cloud to other states. Hence, ionization of $\ket{211}$ happens with minimal disturbance from Bohr resonances.

\vskip 0pt
The only floating Bohr resonance is $\ket{211}\to\ket{100}$. It is worth noting that this is also the only Bohr resonance falling outside the ionization regime (see Figure~\ref{fig:ionization-bohr-resonances}), and that recent numerical studies \cite{Brito:2023pyl} have shown that it has a resonance width much larger than all other resonances. This last observation means that the resonance might partially evade the present analysis, due to the nonlinear dependence of $P_\slab{gw}$ on $R_*$ playing an important role. In any case, we expect the extremely large decay width of $\ket{100}$ to $\Gamma$-break the resonance in most or all realistic cases, preventing the float from happening. 

\section{Evolution from a $\ket{322}$ initial state}
\label{sec:evolution-322}

The second-fastest growing mode is $\ket{322}$. In this case, the constraint on $\alpha$---imposing that the superradiance timescale is shorter than a Gyr---is
\beq
0.09\biggl(\frac{M}{10^4M_\odot}\biggr)^{1/13}\lesssim\alpha<1\,,
\eeq
while the rate of cloud decay in GWs is proportional to $M\ped{c}\alpha^{18}$.

\begin{figure}[htb]
\centering
\includegraphics[width=\textwidth]{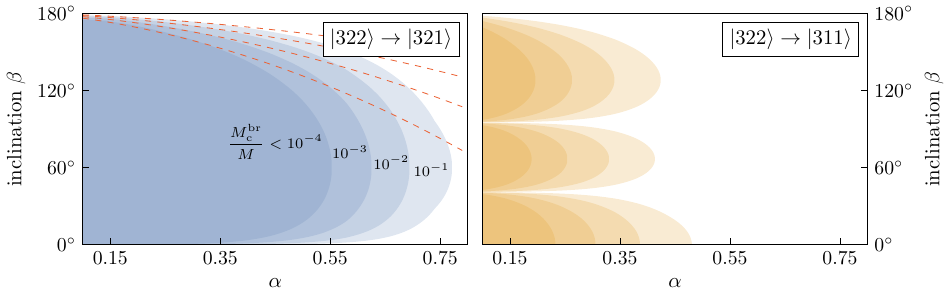}
\caption{Same as Figure~\ref{fig:deltas211}, for the strongest hyperfine (\emph{left panel}) and fine (\emph{right panel}) resonances from a $\ket{322}$ state. The analytical approximations of the contours are not shown in the latter case, as they quickly become inaccurate for moderate values of $\alpha$.}
\label{fig:deltas322}
\end{figure}

\vskip 0pt
Compared to Section~\ref{sec:evolution-211}, a larger number of hyperfine resonances are possible, with any state of the form $\ket{32m_b}$, with $-2\le m_b\le1$. All of these can happen with $g=-4$, in which case the hexadecapole $\ell_*=4$ is entirely responsible for the mixing of the states. However, the cases $m_b=0$ and $m_b=1$ can also resonate, at different separations, with $g=-2$: these are dominated by the quadrupole $\ell_*=2$ instead, which makes these resonances much stronger than the others. Their positions are
\begin{align}
\ket{322}\overset{g=-2}{\longrightarrow}\ket{321} & \qquad \frac{R_0}M=\SI{5.4e+4}{}\,\Bigl(\frac\alpha{0.2}\Bigr)^{-4}\Bigl(\frac{\tilde a}{0.5}\Bigr)^{-2/3}\,,\\[8pt]
\ket{322}\overset{g=-2}{\longrightarrow}\ket{320} & \qquad \frac{R_0}M=\SI{3.4e+4}{}\,\Bigl(\frac\alpha{0.2}\Bigr)^{-4}\Bigl(\frac{\tilde a}{0.5}\Bigr)^{-2/3}\,,
\end{align}
which should be evaluated at $\tilde a\approx2\alpha/(1+\alpha^2)$. The most stringent constraint on $\delta_1$ is given by $\ket{322}\to\ket{321}$ and equals
\beq
\delta_1=\SI{5.4}{\degree}\,\biggl(\frac{M\ped{c}/M}{10^{-2}}\biggr)^{-1/6}\biggl(\frac{q}{10^{-3}}\biggr)^{1/12} \biggl(\frac{\alpha}{0.2}\biggr)^{4/3} \biggl(\frac{\tilde{a}}{0.5}\biggr)^{1/36}f(\varepsilon_0)^{1/4}\,.
\label{eqn:322-delta1}
\eeq
The angle $\chi_1$ within which the same resonance $\Gamma$-breaks is instead
\beq
\chi_1\approx\SI{4.8}{\degree}\biggl(\frac{M\ped{c}\ap{br}/M}{10^{-2}}\biggr)^{-1/6}\biggl(\frac\alpha{0.2}\biggr)^{11/6}\biggl(\frac{\tilde a}{0.5}\biggr)^{-5/18}\,,
\label{eqn:322_chi_1}
\eeq
also more accurately numerically computed and shown in Figure~\ref{fig:deltas322} (\emph{left panel}). Similar to the resonant history of $\ket{211}$, some resonances (such as $\ket{322}\to\ket{321}$) become weak around $\beta=0$, yet other resonances (such as $\ket{322}\to\ket{320}$) do not, thereby eliminating any possible ``safe interval'' around a co-rotating configuration.

\vskip 0pt
Differently from Section~\ref{sec:evolution-211}, there is no clear hierarchy between $\delta_1$ and $\chi_1$. Which one is largest depends not only on $\alpha$, but also on the chosen value of $M\ped{c}\ap{br}$. The angular interval that leads to the survival of the cloud in appreciable amounts is however generally dominated by the $\Gamma$-breaking, as even very light clouds, say $M\ped{c}\ap{br}/M<10^{-4}$, are able to give clear signatures in the Bohr region (cf.~Section~\ref{sec:direct-signatures}).

\vskip 0pt
As in the $\ket{211}$ case, the $\varepsilon$-breaking and $Z$-breaking prove to not be relevant for the resonant history: the value
\beq
\sqrt ZB\sim10^7\biggl(\frac{M\ped{c}/M}{10^{-2}}\biggr)\biggl(\frac{q}{10^{-3}}\biggr)^{-1}\biggl(\frac\alpha{0.2}\biggr)^{-6}\biggl(\frac{\tilde a}{0.5}\biggr)^{-1/3}\,.
\eeq
requires extremely high eccentricities ($\varepsilon\gtrsim0.98$), to give rise to a resonance breaking. The corresponding initial inclinations are extremely close to $\beta=\pi$ and would fall in the interval \eqref{eqn:322-delta1}, where the resonance is not adiabatic.

\vskip 0pt
A cloud in the $\ket{322}$ state can experience fine resonances with states with $\ell\ne0$. Their decay width is smaller than those of the states with $\ell=0$: as a consequence, fine resonances can destroy a significant portion of the cloud before they $\Gamma$-break. The fine resonance that gives the most stringent constraints on $\delta_2$ and $\chi_2$ is
\beq
\ket{322}\overset{g=-1}\longrightarrow\ket{311}\qquad\frac{R_0}M=\SI{2.3e+3}{}\,\Bigl(\frac\alpha{0.2}\Bigr)^{-10/3}\,.
\eeq
Analytical approximations for $\beta\approx\pi$ give\footnote{For $\alpha\gtrsim0.5$, this resonance may marginally fall inside the ionization regime. However, the value of $P\ped{ion}$ never becomes much larger than $P_\slab{gw}$. We therefore ignore this detail, which only slightly increases the value of $\delta_2$ compared to the one presented in (\ref{eqn:322-delta2}).}
\beq
\delta_2=\SI{3.2}{\degree}\,\biggl(\frac{M\ped{c}/M}{10^{-2}}\biggr)^{-1/4}\biggl(\frac{q}{10^{-3}}\biggr)^{1/8} \biggl(\frac{\alpha}{0.2}\biggr)^{31/24}
\label{eqn:322-delta2}
\eeq
and
\beq
\chi_2\approx\SI{9}{\degree}\biggl(\frac{M\ped{c}\ap{br}/M}{10^{-2}}\biggr)^{-1/4}\biggl(\frac\alpha{0.2}\biggr)^{3/2}\,,
\label{eqn:322_chi_2}
\eeq
while a more accurate numerical determination of the mass of the cloud at resonance breaking is given in Figure~\ref{fig:deltas322} (\emph{right panel}). It is worth noting that the strength of the $\ket{322}\to\ket{311}$ resonance has a complicated $\beta$ dependence, due to the octupole $\ell_*=3\ne-g$ being the dominant term. Consequently, this resonance becomes weak not only around $\beta=\SI{180}{\degree}$, but also around $\beta=\SI{41}{\degree}$ and $\SI{95}{\degree}$ (as visible from Figure~\ref{fig:deltas322}). However, other fine resonances remain strong at these intermediate inclinations and so, once again, the cloud can only reach the Bohr region if the inclination is in a narrow interval around the counter-rotating configuration.

\vskip 0pt
In the Bohr region, the system encounters several sinking resonances, the strongest of which are with states of the form $\ket{n11}$. The final populations $\abs{c_b}^2$ are displayed in Figure~\ref{fig:sinking-from-211-and-322}. For benchmark parameters, about $2\%$ of the cloud is lost in the process. None of the floating resonances, with $n=1$ or $n=2$ states, becomes adiabatic within the interval of inclinations discussed above.

\vskip 0pt
Finally, in case the binary is formed at radii small enough to avoid constraints on the inclination coming from fine resonances, an interesting scenario opens up. The strongest floating Bohr resonance is $\ket{322}\overset{g=-1}\longrightarrow\ket{211}$, which becomes adiabatic, for benchmark parameters, for $\beta<\SI{155}{\degree}$.\footnote{Due to the weakness of the resonance compared to most the (hyper)fine ones, it is not possible to expand around $\beta=\pi$ and get a simple formula for the upper limit on the angle as function of the parameters. Nevertheless, a good approximation is given by the following cubic equation: $(\pi-\beta)^4+2.8(\pi-\beta)^6>0.056\times(10^5M\ped{c}q/M)^{1/2}$.} Among all possible scenarios we considered in Sections~\ref{sec:evolution-211} and \ref{sec:evolution-322}, this is the only case where the binary's evolution in the Bohr region features a new phenomenon, beyond ionization and non-adiabatic sinking resonances: namely, an adiabatic floating resonance. The companion's motion continues to ionize the cloud while this resonance takes place, potentially changing $M\ped{c}$ significantly before its end. This is also the only floating resonance with the actual potential to partially move the cloud to a different state, rather than merely destroying it: as can be seen in Figure~\ref{fig:tfloat_tdecay} (\emph{right panel}), the hierarchy $\Delta t\ped{float}\gg t\ped{decay}$ is not valid in the entire parameter space. Hence, depending on the parameters, when the resonance ends, the inspiral can either continue without the cloud, or with a cloud in a (decaying) $\ket{211}$ state and a reduced value of $M\ped{c}$. In the latter case, the discussion in Section~\ref{sec:evolution-211} applies from this point onwards.

\chapter{Observational signatures}

\label{chap:observational-signatures}

The previous chapters have explored in great detail the phenomenology and dynamics of the cloud-binary system, which turned out to be surprisingly intricate. In particular, in Chapter~\ref{chap:history} we used the results to determine when the cloud is entirely destroyed in the early inspiral, when it loses some of its mass upon resonance breaking, and when it remains intact until the binary enters the Bohr region.

\vskip 0pt
The present chapter is devoted to the two main ways the cloud can leave an imprint on the GW waveform: (1) modifications of the waveform due to interaction with the cloud, in case it is still present in the late stages of the inspiral (Section~\ref{sec:direct-signatures}); (2) permanent consequences on the binary parameters left by a cloud destroyed early in the inspiral (Section~\ref{sec:indirect-evidence}). A partially destroyed cloud, left by a broken resonance, may be able to combine both kinds of signatures.

\section{Direct signatures of the cloud}
\label{sec:direct-signatures}

As discussed extensively in Chapter~\ref{chap:history}, the requirement that the cloud survives the hyperfine and fine resonances forces either the inclination angle to be within $\mathcal{O}(10^\circ)$ of a counter-rotating configuration, or the binary to form at radii too small to ever excite those resonances. Then, most phenomena producing direct observational evidence of the cloud happen when the binary reaches the Bohr region. Due to their different nature (continuous vs discrete), we distinguish the effects of accretion and ionization from the ones of resonances.

\subsection{Ionization and accretion}

In Section~\ref{sec:backreaction-ionization} we have given a quick look at the backreaction of ionization on the orbit. We now include the effects of accretion into the analysis and focus on the ensuing gravitational wave signal. We neglect here the effect of Bohr resonances, which will be addressed in Section~\ref{sec:bohr-resonances-obs}. According to the results of Chapter~\ref{chap:history}, the expected binary inclination is near-counter-rotating. Nevertheless, we will still display some of the results for co-rotating orbits, for comparison.

\begin{figure}[p]
\centering
\includegraphics[width=0.81\textwidth, trim = 0pt 8pt 0pt 0pt]{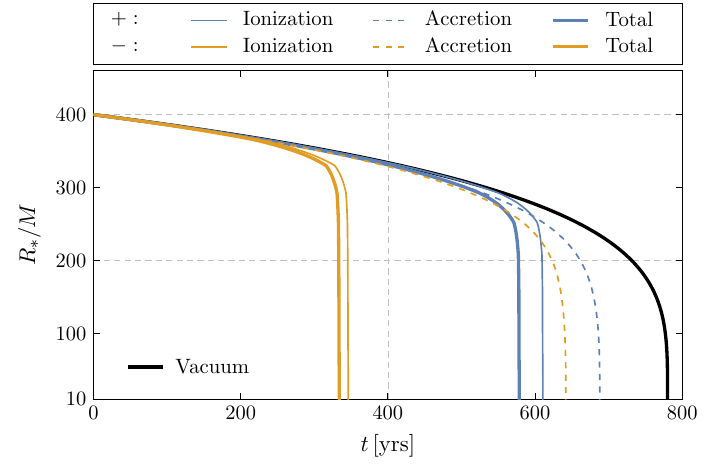}
\caption{Same as in Figure~\ref{fig:separation1}, with the addition of full solutions of \eqref{eqn:evolution-M*}, \eqref{equ:qc-evolve-acc} and \eqref{eqn:evolution-R-ion-acc} (thick lines). For comparison, solutions where ionization is neglected and only accretion is kept are also shown (dashed lines). The $+$ ($-$) sign refers to co-rotating (counter-rotating) orbits.}
\label{fig:separation_tot}
\end{figure}

\begin{figure}[p]
\centering
\includegraphics[width=0.81\textwidth, trim = 0pt 8pt 0pt 0pt]{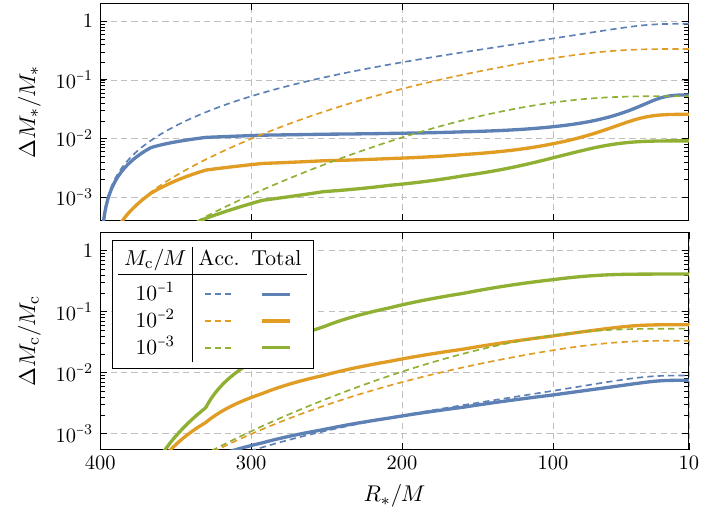}
\caption{Fractional change of the mass $M_*$ of the secondary and the mass $M\ped{c}$ of the cloud. The parameters are the same as Figure~\ref{fig:q_and_M_c_ionization}, but the orbit is assumed to be counter-rotating. Dashed lines only include the effect of accretion, while solid likes include ionization as well.}
\label{fig:deltaM_tot}
\end{figure}

\vskip 0pt
Assuming that the secondary object is a nonspinning black hole, its mass changes according to \eqref{eqn:accretion-law},
\beq
\frac{\dd M_*}{\dd t}=16\pi M_*^2\rho(\vec R_*)\,.
\label{eqn:evolution-M*}
\eeq
By mass conservation, the same term must be subtracted from the cloud's mass evolution equation (cf.~\eqref{equ:qc-evolve}),
\begin{equation}
\frac{\dd M_\lab{c}}{\dd t} = -\frac{\dd M_*}{\dd t}+M_\lab{c}\biggl(\frac{\dot M\ped{c}}{M\ped{c}}\biggr)\ped{ion}\,.\label{equ:qc-evolve-acc}
\end{equation}
While conservation of energy can no longer be applied (due to the dissipative nature of accretion), the conservation of angular momentum gives an equation similar to \eqref{eqn:evolution-R-ion-acc},
\begin{equation}
    \frac{qM^2}{2R_*^2}\frac{\dd R_*}{\dd t}={-P_\slab{gw}}-P_\lab{ion}-P\ped{acc}\,,
    \label{eqn:evolution-R-ion-acc}
\end{equation}
where, neglecting subleading terms for $q\ll1$, we defined
\beq
P\ped{acc}=
    \biggl(\sqrt{MR_*}+\frac{mM}{\alpha}\biggr) \biggl(\frac{M}{R_*} \bigg)^{3/2}\frac{\dd q}{\dd t}\,.
\eeq
Equations \eqref{eqn:evolution-M*}, \eqref{equ:qc-evolve-acc} and \eqref{eqn:evolution-R-ion-acc} can be solved numerically for $M_*$, $M\ped{c}$ and $R_*$. We show in Figures~\ref{fig:separation_tot} and \ref{fig:deltaM_tot} the separation $R_*$ as function of time, as well as the fractional changes in $M_*$ and $M\ped{c}$, for our set of benchmark parameters. While accretion is a large effect on its own, capable of significantly increasing the mass of the secondary object, ionization turns out to be the dominant effect.

\begin{figure}[t]
            \centering
            \includegraphics{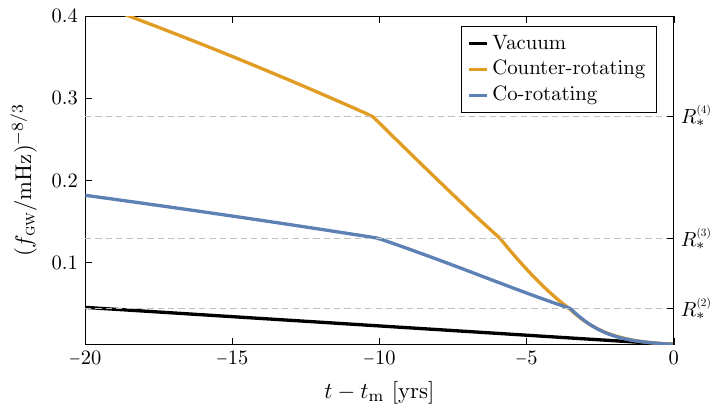}
            \caption{Evolution of the GW frequency as a function of the remaining time to merger, $t-t_\lab{m}$, for $M=10^4M_\odot$ and $\alpha=0.2$, with initial values of $R_*=400M$, $q=10^{-3}$ and $M_\lab{c}/M = 10^{-3}$
             in a $\ket{211}$ state. The central region of the range shown on the $y$ axis corresponds to a few millihertz, falling inside the LISA sensitivity band. The ``kinks'' at separations $R_*^{(g)}$ correspond to the discontinuities in the ionization power, see Figure~\ref{fig:ionization_211}.}
    \label{fig:evolution-frequency}
\end{figure}

\vskip 0pt
To assess the observability of the cloud, it is useful to compare the gravitational-wave frequency, $f_\slab{gw}=\Omega/\pi$, with that of a vacuum inspiral. This is done in Figure~\ref{fig:evolution-frequency} for the very conservative case of initial $M_\lab{c}/M = 10^{-3}$, demonstrating that even a tiny cloud can have a strong impact on the inspiral. In the plot, the scale of the frequency axis has been chosen such that the non-relativistic vacuum evolution, $f_\slab{gw}\propto(t_\lab{m}-t)^{\sminus 3/8}$, where $t_\lab{m}$ is the merger time, becomes a straight line. It is apparent that the shape of $f_\slab{gw}(t)$ deviates significantly from a straight line: a decisive role is played by the ``kinks'' appearing at the frequencies where the ionization power $P_\lab{ion}$ is discontinuous, cf.~Figure~\ref{fig:ionization_211}. From \eqref{eqn:omega-g}, kinks appear at the frequencies
\begin{equation}
  \begin{aligned}
    f_\slab{gw}^{(g)} &= \frac{6.45\,\text{mHz}}{g}\left(\frac{10^4M_\odot}{ M}\right)\!\left(\frac{\vphantom{10^{4} M_\odot}\alpha}{0.2 \vphantom{M}}\right)^{\!3}\!\left(\frac{2}{n}\right)^{\!2} \\
    &=\frac{33.5 \,\text{mHz}}{g} \left(\frac{M}{10^4M_\odot}\right)^{\!2}\!\left(\frac{\vphantom{M}\mu}{10^{\sminus 14}\,{\rm eV}}\right)^{\!3}\!\left(\frac{2}{n}\right)^{\!2}\! ,
  \end{aligned}
  \label{eqn:f-discontinuities}
\end{equation}
where the overtone number $g$ ranges over positive integers and $n$ is the principal number of the cloud's state. These kinks thus  constitute a sharp observational signature of ionization caught in the act. If only a region between two kinks is observed, then the evolution is likely to be more degenerate with a signal from a vacuum system, whose parameters would however differ from the true parameters of the binary.

\begin{figure}[t]
            \centering
            \includegraphics{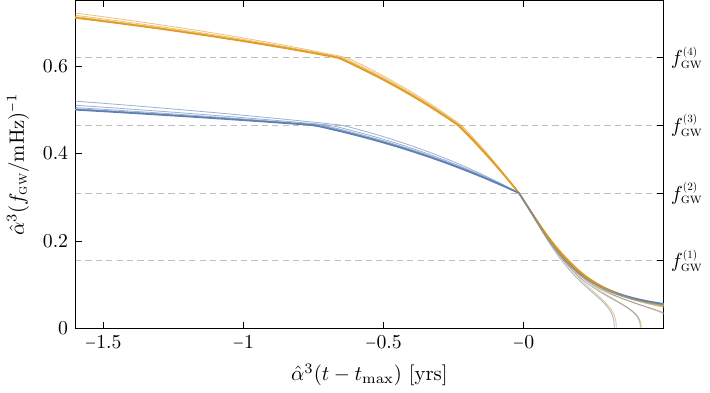}
            \caption{Evolution of the (inverse) frequency $f_\slab{gw}$ for $M=10^4M_\odot$ and $\alpha=0.04,0.08,\ldots,0.28$, with initial $q=10^{\sminus 3}$ and $M_{\lab{c}}/M=0.01$ in a $|211\rangle$ state. The axes are rescaled according to Eq.~(\ref{eqn:solution-scaled-f(t)}),  with $\hat \alpha \equiv \alpha/0.2$. The curves have been horizontally shifted to match at $t=t_\lab{max}$, which has been chosen close to the peak of $P_\lab{ion}/P_\slab{gw}$. Shown are the results for co-rotating [{\color{Mathematica1} blue}] and counter-rotating [{\color{Mathematica2} orange}] orbits.}
    \label{fig:frequency-scaling}
\end{figure}

Figure~\ref{fig:evolution-frequency} presented the evolution of the system for a specific choice of parameters. In the regime of interest, $P_\lab{ion} \gg P_\slab{gw}$, the dependence on these parameters can be determined analytically using a scaling symmetry of the evolution equations.
Neglecting 
other forces and changes in $q$ and $M_\lab{c}$ throughout the inspiral, we can obtain an approximate equation for the evolution of $f_\slab{gw}$ under the effect of ionization only:
\beq
\frac{\dd f_\slab{gw}^{2/3}}{\dd t} \approx \frac{2}{\pi^{2/3}} \frac{P_\lab{ion}}{q M^{5/3}}\, .
\eeq
Using the ionization power's scaling behavior (\ref{eqn:scaling-Pion}), 
we can write this as 
\beq
\frac{\dd z^{2/3}}{\dd \tau} \approx 2 \hskip 1pt \mathcal P(z^{\sminus2/3})\,, 
\eeq
where we have defined the dimensionless variables $z \equiv (M/\alpha^3) \hskip 1pt \pi f_\slab{gw}$ and $\tau \equiv \alpha^3qM_\lab{c} t/M^2$.  The solution can thus be written as
\beq
f_\slab{gw}(t) = \frac{\alpha^3}{M} f(\tau(t))\, ,
\label{eqn:solution-scaled-f(t)}
\eeq
where $f(\tau)$ is a universal function that depends on the shape of $P_\lab{ion}$ for a given state $\ket{n\ell m}$ of the cloud. The region of validity of this formula increases with larger $M_\lab{c}$. In Figure~\ref{fig:frequency-scaling}, we confirm that the solutions of the full system of equations \eqref{eqn:evolution-M*}, \eqref{equ:qc-evolve-acc} and \eqref{eqn:evolution-R-ion-acc} indeed are described by a universal shape, when both $f_\slab{gw}$ and $t$ are appropriately rescaled. The curves depart from each other only when the approximation $P_\lab{ion}\gg P_\slab{gw}$ fails (that is, very close and very far from the merger) or when corrections due to the varying $M_*$ and $M\ped{c}$ become important.

\subsection{Bohr resonances}

\label{sec:bohr-resonances-obs}

\vskip 0pt
On top of the characteristic frequency evolution induced by ionization (and, secondarily, by accretion), the large number of Bohr sinking resonances can cause non-negligible upward ``jumps'' of $f_\slab{gw}$ due to their backreaction\footnote{Here we only study the backreaction on the orbital parameters. When including the  backreaction on the geometry as well, the cloud's transitions could also cause ``resonant'' features in the emitted GWs, see e.g.~Figure~1 of \cite{Duque:2023cac}.}, even if they are strongly non-adiabatic. For a Bohr resonance $\ket{n_a\ell_am_a}\to\ket{n_b\ell_bm_b}$, they are located at
\beq
f_\slab{gw}\ap{res}=\frac{\SI{26}{mHz}}g\biggl(\frac{10^4M_\odot}{M}\biggr)\biggl(\frac\alpha{0.2}\biggr)^3\biggl(\frac1{n_a^2}-\frac1{n_b^2}\biggr)\,,
\label{eqn:position-resonances}
\eeq
where $g=m_b-m_a$, and thus fall inside the LISA band for benchmark parameters.\footnote{Formula~\eqref{eqn:position-resonances}, with $n_b\to\infty$, also describes the position of the $g$-th ``kink'' of the function $f$ appearing in \eqref{eqn:f-discontinuities}, corresponding to the $g$-th discontinuity of $P\ped{ion}$.}

\vskip 0pt
The amplitude of the jump can be computed explicitly from \eqref{eqn:dimensionless-omega-evolution} (assuming quasi-circular orbits):
\beq
\Delta f_\slab{gw}=\frac{\SI{0.61}{mHz}}{\Delta m^{1/3}}\,\biggl(\frac{10^4M_\odot}{M}\biggr)\biggl(\frac{M\ped{c}/M}{0.01}\biggr)\biggl(\frac{q}{10^{-3}}\biggr)^{-1}\biggl(\frac\alpha{0.2}\biggr)^3\biggl(\frac1{n_a^2}-\frac1{n_b^2}\biggr)^{4/3}\biggl(\frac{\abs{c_b}^2}{10^{-3}}\biggr)\,,
\label{eqn:jump-resonances}
\eeq
where the values of $\abs{c_b}^2$ and their dependence on the parameters are given in Figure~\ref{fig:sinking-from-211-and-322} and equation \eqref{eqn:scaling-cb2}. The increase in frequency comes with smaller, long-lived oscillations of the frequency, and with a slight increase of the eccentricity; both these effects have been shown in the right panels of Figure~\ref{fig:eccentric-backreacted-resonance} for an arbitrary choice of parameters. The dephasing introduced by a single sinking resonance on top of the one coming from ionization is $\Delta\Phi_\slab{gw}\approx\pi f_\slab{gw}\ap{res}\Delta f_\slab{gw}/\gamma$. This is of the order of thousands of radiants, although the exact number can vary by a few orders of magnitude in different regions of the parameter space. Not only is this well above the expected LISA precision of $\Delta\Phi_\slab{gw}\sim2\pi$, but such a dephasing would happen in a very narrow frequency range, in contrast to most other environmental effects, including ionization. This unique behavior would aid parameter estimation by directly linking the cloud's parameters with $\Delta\Phi_\slab{gw}$ via \eqref{eqn:position-resonances} and \eqref{eqn:jump-resonances}, especially if multiple jumps are observed within one signal.

\vskip 0pt
As discussed in Sections~\ref{sec:evolution-211} and \ref{sec:evolution-322}, the only cases where a floating resonance can be observed in the Bohr region require a binary formation at very small radii, so that all early resonances are skipped without a strict requirement on the inclination angle. Resonances of the type $\ket{n_a\ell_am_a}\to\ket{100}$ happen very late in the inspiral (see Figure~\ref{fig:ionization-bohr-resonances}), where relativistic corrections are expected to be more important \cite{Brito:2023pyl}. The only other floating Bohr resonance encountered in Section~\ref{chap:history} is $\ket{322}\to\ket{211}$. This is an interesting case, because it may not entirely destroy the cloud. The expected GW signal is a constant frequency $f_\slab{gw}$ given by equation \eqref{eqn:position-resonances}, for a total floating time of\footnote{This value assumes a quasi-circular co-rotating orbit. Moderate nonzero values of eccentricity or inclination introduce $\mathcal O(1)$ variations in $\Delta t\ped{float}$.}
\beq
\Delta t\ped{float}=\SI{5.8}{yrs}\,\biggl(\frac{M}{10^4M_\odot}\biggr)\biggl(\frac{q}{10^{-3}}\biggr)^{-2}\biggl(\frac\alpha{0.2}\biggr)^{-3}\,.
\label{eqn:deltatfloat-322-211}
\eeq
Although the cloud's mass is continuously reduced by ionization while the resonance takes place, the value given in \eqref{eqn:deltatfloat-322-211} remains independent of $M\ped{c}$ as long as it is large enough to guarantee $P\ped{ion}\gg P_\slab{gw}$. 

\section{Indirect signatures: impact on binary parameters}
\label{sec:indirect-evidence}

For sufficiently small orbital inclinations, as seen in Figures~\ref{fig:history_211}, \ref{fig:deltas211} and \ref{fig:deltas322}, the cloud can be destroyed during one of the floating resonances in the early inspiral, to a level where it no longer affects the binary dynamics in an observable way. Then, by the time the system enters in band, its evolution is expected to follow the rules of vacuum General Relativity. Nevertheless, the binary still carries the marks of the previously existing boson cloud, and of the resonance that destroyed it. These are due to the backreaction on the orbit from that floating resonance, and come in the form of a change in the eccentricity and tilt of the inclination angle.

\begin{figure}[t]
\centering
\includegraphics[width=\textwidth]{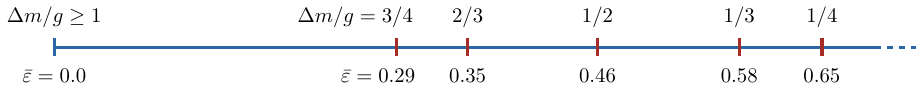}
\caption{Example values of the fixed point $\bar{\varepsilon}$ depending on $\Delta m/g$. Numbers can be found by solving equations \eqref{eqn:dimensionless-eccentricity-evolution} and \eqref{eqn:dimensionless-inclination-evolution} on a floating orbit.}
\label{fig:fixedpointsresonances}
\end{figure}

\vskip 0pt
While in Chapter~\ref{chap:history} we could simplify the analysis by studying only the ``strongest resonance'', the impact on the orbital parameters strongly depends on which overtone (that is, which value of $g$) mediated the last adiabatic resonance encountered by the system.\footnote{If the system undergoes multiple floats, for example because broken resonances leave a cloud massive enough to excite other adiabatic resonances, the evolution of the eccentricity follows several nontrivial steps. Here, however, we focus on the last of those as it has the most direct observational consequences.} As shown in Figure \ref{fig:streamplot_inclination-eccentricity}, the orbital parameters follow specific sets of trajectories on the $(\varepsilon,\beta)$ plane, until the resonance breaks or completes. While floating orbits \emph{always} tilt the inclination angle towards a co-rotating configuration, the eccentricity is forced towards a fixed point, whose value depends on $\Delta m/g$. Some examples of the value of this fixed point are shown in Figure \ref{fig:fixedpointsresonances} for different values of $\Delta m / g$.

\begin{figure}[t]
\centering
\includegraphics[width=\textwidth]{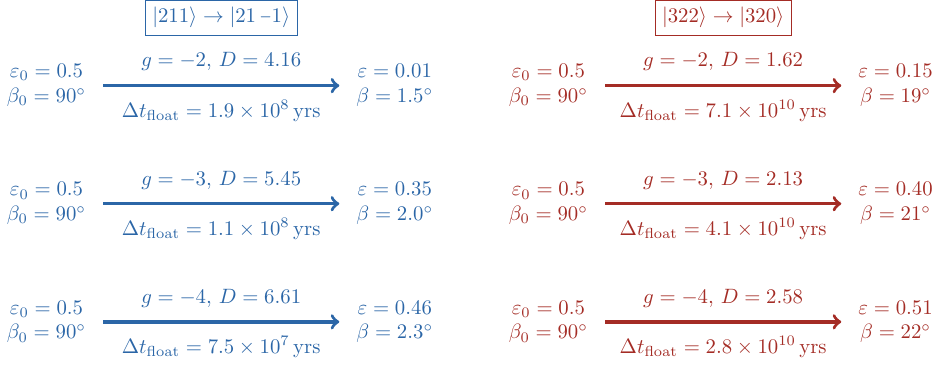}
\caption{Examples of backreaction on the eccentricity $\varepsilon$ and inclination $\beta$ during floating orbits that destroy the cloud entirely. We show the ``strongest resonance'' ($\Delta m/g=1$) and two overtones in each scenario, using the benchmark parameters $\alpha=0.2$, $q=10^{-3}$, $M\ped{c}=0.01M$ and $M=10^4M_{\odot}$. Each case is initialized with $\varepsilon_0=0.5$ and $\beta_0=\SI{90}{\degree}$ for illustrative purposes, but we observed that the final values of $\varepsilon$ and $\beta$ are very robust against the choice of different initial conditions. The final values of $\varepsilon$ and $\beta$, as well as $\Delta t\ped{float}$, are computed integrating numerically equations \eqref{eqn:dimensionless-dressed-schrodinger}, \eqref{eqn:dimensionless-omega-evolution}, \eqref{eqn:dimensionless-eccentricity-evolution} and \eqref{eqn:dimensionless-inclination-evolution}. For benchmark parameters, the floating time of resonances from $\ket{322}$ exceeds the Hubble time; however, it strongly depends on the parameters, as derived in \eqref{eqn:hyperfine-B-tfloat-exponent}.}
\label{fig:examples-backreaction-orbit}
\end{figure}

\vskip 0pt
Assuming that the resonance does not break prematurely, the distance travelled by the binary in the $(\varepsilon,\beta)$ plane depends on the parameter $D$ alone, introduced in \eqref{eqn:D}, which takes the value
\beq
D=D_0\biggl(\frac{-g}2\biggr)^{\!2/3}\biggl(\frac{M_{\rm c}/M}{10^{-2}}\biggr)\biggl(\frac{q}{10^{-3}}\biggr)^{\!-1}\biggl(\frac\alpha{0.2}\biggr)\biggl(\frac{\tilde a}{0.5}\biggr)^{\!1/3}\,.
\label{eqn:D-scaling}
\eeq
For the two strongest hyperfine resonances from $\ket{211}$ ($\ket{322}$), the prefactor $D_0$ assumes the values $3.30$ and $4.16$ ($1.28$ and $1.62$) respectively. Very roughly, the system gets $e^D$ times closer to the eccentricity fixed point than it was before the resonance started. The inverse proportionality of $D$ on $q$ implies that only intermediate or extreme mass ratio binaries change significantly their orbital parameters during a floating resonance. Example of variations of the parameters during a floating orbit are reported in Figure~\ref{fig:examples-backreaction-orbit} for the resonances $\ket{211}\to\ket{21\,\minus1}$ and $\ket{322}\to\ket{320}$.

\begin{figure}[htb]
\centering
\includegraphics[width=\textwidth]{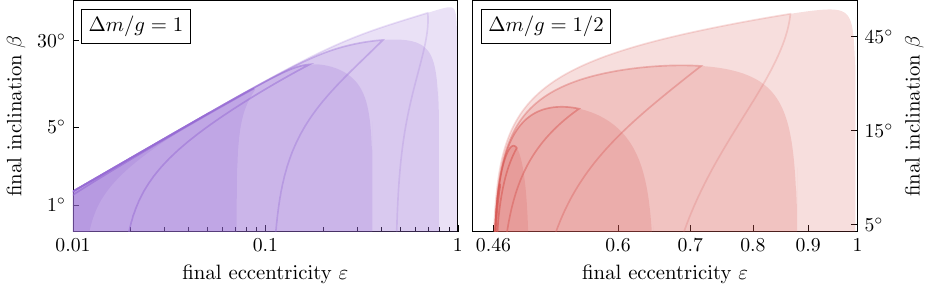}
\caption{The shaded regions show the possible values of eccentricity $\varepsilon$ and inclination $\beta$ at the completion of a floating resonance, starting from any initial values $\varepsilon_0$ and $\beta_0$, for different values of $D$. The values used, from the outermost to the innermost region, are $D=1,1.5,2,2.5,3$ (\emph{left panel}) and $D=1.8,2.6,3.4,4.2,5$ (\emph{right panel}), cf.~\eqref{eqn:D}. When the initial inclination is required to satisfy the conditions necessary to sustain the float, a smaller portion of each region is reachable. We enclosed in solid lines the reachable portions for $\beta_0\le\SI{128}{\degree}$ (\emph{left panel}) and $\beta_0\le\SI{142}{\degree}$ (\emph{right panel}), which are the thresholds for $\ket{211}\to\ket{21\,\minus1}$ and $\ket{211}\to\ket{210}$, for the reference parameters used in \eqref{eqn:211_chi_1}.}
\label{fig:parameter-range}
\end{figure}

\vskip 0pt
We show in Fig.~\ref{fig:parameter-range} the possible values of $\varepsilon$ and $\beta$ at the end of a floating resonance, as function of $D$ and for two values of $\Delta m/g$. The float brings the orbit significantly close to the equatorial plane, even for large initial inclinations. An abundance of quasi-planar inspiral events can thus be indirect evidence for boson clouds. Whether the formation mechanisms of the binary, or other astrophysical processes, also lead to a natural preference for small inclinations is still subject to large uncertainties \cite{Amaro-Seoane:2012lgq,Pan:2021ksp,Pan:2021oob}. Additionally, the eccentricity is suppressed by the main tones ($g=\Delta m$) and brought close to, or above, a nonzero fixed point by overtones ($g>\Delta m$). The latter scenario is especially interesting for binaries that are not dynamically captured, such as in the case of comparable mass ratios, because they are generally expected to be on quasi-circular orbits. The past interaction with a cloud can overturn this prediction. The float-induced high eccentricities are mitigated by the subsequent GW emission, but the binary will remain more eccentric than it would have been otherwise, even in late stages of the inspiral. For a related analysis exploring binary eccentricity as signature of boson clouds, see \cite{Boskovic:2024fga}.

\vskip 0pt
Lastly, we note that the extremely long floating time associated with some hyperfine or fine resonances can stop many binaries from getting in band at all, consequently reducing the merger rate. For example, for our choice of benchmark parameters, the hyperfine resonances from the $\ket{322}$ state, shown in Figure~\ref{fig:examples-backreaction-orbit}, float for longer than the Hubble time.

\chapter{Conclusions}

\label{chap:conclusions}

The groundbreaking detection of GWs by the LIGO-Virgo Collaboration has garnered significant interest from high-energy physicists due to the potential of GW astronomy to become a new tool for fundamental physics. In this context, BH environments have long been proposed as ideal laboratories to test properties of both visible and the dark matter in the Universe. This thesis focuses on a specific environment that has recently attracted much attention: ultralight bosons. These hypothetical particles are both theoretically well-motivated and observationally interesting. The ability to probe them with GWs is particularly exciting because it means that GWs can complement traditional methods, such as particle colliders, which are blind to the weakly-coupled frontier of new physics.

\vskip 0pt
Arguably, the most natural way of producing ultralight bosons around BHs is through superradiance, a spontaneous process that endows BHs with a ``cloud'' resembling a giant electron orbital. This quantum mechanical analogy is not merely cosmetic. Instead, it plays a crucial role throughout most of the thesis, as many of the problems we solve are equivalent to the study of single-particle quantum mechanical systems. This very particular aspect of gravitational atoms sets their attributes apart from other types of BH environments, introducing a rich array of distinctive phenomena. The investigation of this topic began before this thesis in \cite{Baumann:2018vus,Baumann:2019ztm}.

\vskip 0pt
The present work discovers many new effects and delves deep into the intricacies of the dynamics of the system, drawing for the first time a satisfactory and self-consistent picture of its evolution. The most important types of cloud-binary interactions can be distinguished in two separate categories: orbital resonances and dynamical friction (or ionization). Roughly speaking, the former determine the ``history'' of the system, while the latter is most relevant for the observational signatures of the cloud.

\vskip 0pt
Several recent studies \cite{Takahashi:2021eso,Zhang:2018kib,Ding:2020bnl,Tong:2021whq,Du:2022trq,Tong:2022bbl,Fan:2023jjj,Takahashi:2023flk} realized the importance of resonances in shaping the history of the system. Despite this, the combination of cloud states and binary configurations compatible with this kind of evolution had not yet been determined before this thesis. Extending the framework to orbits with any eccentricity and inclination, as well as taking into account the backreaction of the resonances on the orbit, turns out to be crucial to finalize the program. In principle, one might have expected the evolution of the system to be extremely complicated. The S-matrix approach developed in \cite{Baumann:2019ztm} suggests a tree of populated states branching more and more, every time a new resonance is encountered. In practice, however, the hierarchy of the timescales at play clarifies the picture dramatically. The conclusion is then remarkably simple: only binaries close enough to a counter-rotating configuration, where the early resonances are very weak, are able to carry the cloud up to the point where it becomes observable; if so, the cloud remains in its original state.

\vskip 0pt
The behavior of dynamical friction is also unexpected and very interesting. Although simple formulae derived for uniform media capture the order of magnitude of the effect \cite{Zhang:2019eid}, they miss its most characteristic feature, namely the quasi-discontinuous dependence on the binary's frequency. This kind of behavior is uncommon in astrophysics. We thus find it instructive to stick with a name, ionization, that is reminiscent of atomic physics and better highlights the underlying physical process. Most interestingly, these sharp features remain directly imprinted on the binary's gravitational waveform, signaling in a clear way the existence of particles with very light mass.

\vskip 0pt
Finally, even those clouds which do not become directly observable deliver a pleasant surprise. Their destruction is in fact not inconsequential, as it brings the eccentricity and inclination of the binary's orbit close to some fixed points. One of the reasons this signature is particularly interesting is that it is less reliant on the small mass-ratio approximation, $q\ll1$, compared to ionization. Given the absence of such a requirement, indirect signatures can already be looked for using the equal-mass binaries detected by LIGO-Virgo. Conversely, for $q\sim1$ ionization is so efficient that the cloud significantly loses mass before the companion reaches its densest regions. Furthermore, a low-frequency detector such as LISA is better suited to observe ionization's sharp features.

\vskip 0pt
For simplicity, the study of observational signatures in this thesis has neglected a number of subleading effects. These include the cloud's self gravity \cite{Ferreira:2017pth,Hannuksela:2018izj}, the backreaction of the cloud's decay in GWs \cite{Cao:2023fyv} and the non-resonant overlap between growing and decaying states \cite{Tong:2022bbl,Fan:2023jjj}. Most of these can be straightforwardly included in the evolution of the system, allowing for a more accurate determination of the waveform. Most importantly, however, we derived the entirety of our results within a nonrelativistic framework.\footnote{The determination of the accretion cross section presented in Section~\ref{sec:accretion} is fully relativistic, but the way it is included in equation \eqref{eqn:evolution-R-ion-acc} is not.} This approximation is particularly desirable to perform analytical work, because it allows the use of techniques borrowed from quantum mechanics. The nonrelativistic limit is expected to be a good approximation, because most of the interesting phenomena take place at radii where gravity is weak. For example, the peak of the cloud's density, as well as the positions of the ionization jumps and of the Bohr resonances receive relativistic corrections of order $\mathcal O(\alpha^2)$. Nevertheless, relativistic extensions of our results are still a natural direction for future work, and the first steps have already been taken in \cite{Brito:2023pyl,Cannizzaro:2023jle,Duque:2023cac}. The algorithms are currently too computationally expensive to allow for a detailed comparison with the results presented here, but some of the critical features, such as the ionization jumps, have been retrieved and confirmed.

\vskip 0pt
Even without the inclusion of any of the corrections mentioned above, the results contained in this thesis already allow for concrete studies of the detectability of gravitational atoms. For example, the imprints of ionization and accretion on the waveform were considered in \cite{Cole:2022yzw} for a simulated year-long LISA signal. The study clearly showed that the parameters of the system, such as the mass of the boson and the total mass of the cloud, can be measured with great accuracy if the correct waveform templates are used. Furthermore, gravitational atoms can be markedly distinguished from both vacuum systems and different kinds of environments, such as dark matter spikes or accretion disks. Possible next steps include the implementation in dedicated packages for the computation of inspiral waveforms, such as \cite{Katz:2021yft}. The road forward is now paved, and we hope to eventually reach its destination: using GW observations to advance fundamental physics.

\appendix

\chapter{Scalars around Kerr}
\label{app:heunc}

The aim of this appendix is to present a self-contained overview of the exact solutions for the definite frequency modes of a massive scalar field around a Kerr black hole.

\section{Definite frequency solutions}

  The Kerr geometry has two relevant isometries: time translations and azimuthal rotations. This suggests that we choose an ansatz for the scalar field profile, with a definite frequency, $\omega$, and azimuthal angular momentum, $m \in \mathbb{Z}$: 
  \begin{equation}
    \Phi(t, \mb{r}) = e^{-i \omega t + i m \phi} R(r) S(\theta)\, .\label{eqn:Phi-separation-no-sum}
  \end{equation}
  It is a special property of the Kerr background that this ansatz separates the Klein--Gordon equation \eqref{eqn:klein-gordon} into the angular spheroidal equation \eqref{eqn:S} and the radial equation \eqref{eqn:R}. The latter can be written as
  \begin{equation}
    \begin{aligned}
      0 =\ & \frac{1}{\Delta R} \frac{\dd}{\dd r}\!\left(\!\Delta \frac{\dd R}{\dd r}\right) + k^2  + \frac{P_+^2}{(r - r_+)^2} + \frac{P_-^2}{(r - r_-)^2}  \\
     & - \frac{A_+}{(r_+ - r_-) (r - r_+)} + \frac{A_-}{(r_+ - r_-)(r - r_-)}\,,
   \end{aligned}\label{eqn:radial}
\end{equation}
where we have introduced $k^2 = \omega^2 - \mu^2$ and the parameter combinations
\begin{equation}
  \begin{aligned}
    P_\pm  &= \frac{m a - 2 M \omega r_\pm }{r_+ - r_-} \,, \\
    A_\pm  &= P_+^2 + P_-^2 + \gamma_\pm ^2 + \lambda \,,
  \end{aligned} \label{eq:radialParameters}
\end{equation}
with $\gamma^2_\pm  = \mu^2 r_\pm ^2  - \omega^2 (4 M^2 + 2 M r_\pm  + r_\pm ^2)$. 

\vskip 0pt
Requiring the solution to be regular at $\theta = 0$ and $\pi$, forces the spheroidal eigenvalue $\lambda = \lambda_{\ell m}(c)$ to take a set of discrete values, depending on the spheroidicity parameter $c = k a$ and labeled by $\ell =0, 1, \dots$ and $|m| \leq \ell$. The corresponding angular functions $S(\theta) = S_{\ell m}(c; \cos \theta)$ are the ``spheroidal harmonics," which reduce to the ordinary spherical harmonics for $c=0$.

  \vskip 0pt
The radial equation (\ref{eqn:radial}) has three singularities: one at the outer horizon $r = r_+$ controlled by the parameter $P_+^2$, one at the inner horizon $r = r_-$ controlled by $P_-^2$, and an irregular singularity at $r = \infty$ controlled by $k^2$, which can be understood as the confluence of two regular singularities. This uniquely identifies the radial equation as a form of the ``confluent Heun equation," and we expect the radial solutions $R(r)$ to be proportional to the confluent Heun function, which we will define now.\footnote{It is useful to compare this to the Schr\"{o}dinger equation of the hydrogen atom, which has both a regular singularity at $r = 0$ and an irregular singularity at $r = \infty$ that can also be understood as the confluence of two regular singularities. Any linear differential equation with three regular singular points can be mapped to the hypergeometric equation with singularities at $z = 0$, $1$ and $\infty$. The solution to this equation that is regular about $z = 0$ is the familiar hypergeometric function ${}_2 F_1(a, b; c;z)$. Upon the confluence of the singularities at $z = 1$ and $z = \infty$, this turns into the confluent hypergeometric equation, and the regular solution ${}_2 F_{1}(a, b; c; z)$ turns into the confluent hypergeometric function ${}_1 F_{1}(a; c; z)$. An analogous story applies to the radial equation in the Kerr background, except it has an additional regular singularity at the inner horizon $r= r_-$. Any linear differential equation with four regular singular points can be mapped to the Heun equation, and upon a confluence of two singularities this reduced to the confluent Heun equation.}

\vskip 0pt
  Our goal is to find solutions on $r \in [r_+, \infty)$ that are purely ingoing at the outer horizon $r = r_+$, since no physical mode can escape from the black hole. Near the outer horizon, the singularity forces solutions to behave as $R(r) \sim (r - r_+)^{\pm i P_{\vphantom{|}\scalebox{0.6}{$+$}}}$, where the plus sign in the exponent corresponds to purely ingoing modes. Similarly, the singularity at $r = \infty$ forces the modes to behave as $R(r) \sim e^{\pm i k r}$. It will be convenient to define $z \equiv -(r - r_+)/(r_+ - r_-)$ and peel these asymptotic behaviors from the solution,
  \begin{equation}
    R(r) = e^{- i k(r - r_+)} z^{i P_{\vphantom{|}\scalebox{0.6}{$+$}}} (z - 1)^{- i P_{\vphantom{|}\scalebox{0.55}[0.6]{$-$}}} H(z)\,.
  \end{equation} 
  The function $H(z)$ then satisfies the confluent Heun equation \cite{ronveaux1995heun,Fiziev:2009kh}:
  \begin{equation}
    \frac{\dd^2 H}{\dd z^2} + \left( \alpha + \frac{1 + \beta}{z} + \frac{1 + \gamma}{z - 1}\right) \frac{\dd H}{\dd z} + \left(\frac{\mu}{z} + \frac{\nu}{z - 1}\right) H = 0 \,, \label{eqn:heunc-usual}
  \end{equation}
  where
  \beq
  \begin{aligned}
  \mu &= \frac{1}{2}(\alpha - \beta - \gamma + \alpha \beta - \beta \gamma) - \eta \,,\\
  \nu &= \frac{1}{2}( \alpha + \beta + \gamma + \alpha \gamma + \beta \gamma) + \delta + \eta\,,
  \end{aligned}
 \eeq
 with $ \alpha = 2 i k (r_+ - r_-)\,,$ $\beta = 2 i P_+\,,$ $ \gamma = -2 i P_-\,,$ $\delta = A_+ - A_-\,,$ and  $\eta = \minus A_+\,.$
Equation~(\ref{eqn:heunc-usual}) has a solution that is regular at the origin, $H(0) = 1$, called the confluent Heun function, $H(z) = \HeunC(\alpha, \beta, \gamma, \delta, \eta; z)$, and one which behaves as $z^{- 2 i P_{\vphantom{|}\scalebox{0.55}[0.6]{$+$}}}$ as $z \to 0$. Since we impose purely ingoing boundary conditions, we discard the latter and find that
  \begin{equation}
    \begin{aligned}
      \Phi(t, \mb{r}) &= R_{k; \ell m}(r) S_{\ell m}(k a; \cos \theta) e^{-i \omega t+i m \phi} \\
      &= \mathcal{C} \es e^{-i \omega t - i k(r -r_+) + i m \phi} z^{i P_{\vphantom{|}\scalebox{0.6}{$+$}}} (z - 1)^{- i P_{\vphantom{|}\scalebox{0.55}[0.6]{$-$}}} \HeunC(\alpha, \beta, \gamma, \delta, \eta; z) S_{\ell m}(k a; \cos \theta)\,, \label{eq:phiExactSol}
    \end{aligned}
  \end{equation}
  where $\mathcal{C}$ is a normalization constant.

  \vskip 0pt
Using the tortoise coordinates, 
  \begin{equation}
    \begin{aligned}
      \tilde{r} &=  \frac{2 M}{r_+ - r_-} \left[ r_+ \log \!\left(\frac{r - r_+}{r_+ - r_-}\right) - r_- \log\!\left(\frac{r - r_-}{r_+ - r_-}\right)\right] + r \,, \\
      \tilde{\phi} &= \frac{a}{r_+ - r_-}\left[ \log\!\left(\frac{r - r_+}{r_+ - r_-}\right) - \log\!\left(\frac{r - r_+}{r_+ - r_-}\right) \right] ,
    \end{aligned}
  \end{equation}
the solution can be written as
  \begin{equation}
    \Phi(t, \bm{r}) = \mathcal{C}\es e^{-i k(r -r_+) - i \omega(t + \tilde{r} - r) + i m (\phi + \tilde{\phi})} \HeunC(\alpha, \beta, \gamma, \delta, \eta; z) S_{\ell m} (k a; \cos \theta)\,.
  \end{equation}
  Since the combination $\tilde{r} - r$ increases as we move away from the outer horizon, this mode indeed represents a purely ingoing wave.

  \vskip 0pt
  There are two classes of solutions that we use throughout the main text. The first are the quasi-bound states, which are purely ingoing at the outer horizon and exponentially decaying as $r \to \infty$. These two boundary conditions can only be satisfied for a discrete set of frequencies $\omega_{n \ell m} = E_{n \ell m} + i \es \Gamma_{n \ell m}$, cf.\ \eqref{eqn:omega-E-Gamma}, and so these mode only come in a discrete set. The second are the unbound continuum states, which are purely ingoing at the outer horizon, but oscillate as $r \to \infty$. Since we impose only one boundary condition, these unbound modes comprise a continuous set with frequencies $\omega^2 = \mu^2 + k^2$. 

\section{Nonrelativistic limit}

The first four parameters of the confluent Heun function 
are either first order ($\alpha$, $\beta$, $\gamma$) or second order ($\delta$) in the dimensionless combinations $\mu M$ and $k M$. The fifth parameter, on the other hand, is generally $\eta = \mathcal{O}(1)$, because
\begin{equation}
    \lambda_{\ell m}(c) = \ell(\ell + 1) - \frac{1}{2}\left[1 - \frac{(2 m - 1)(2m+1)}{(2 \ell - 1) (2 \ell + 3)}\right] c^2 + \mathcal{O}\big(c^4\big)\,.
\end{equation}
The only exception is when $\ell=0$, where $\eta$ is second order in both $\mu M$ and $kM$. Modes with non-zero angular momentum see a centrifugal barrier which forces the field away from the black hole, suppressing its amplitude at radii below $\sim \ell^2/\big(\mu^2 M\big)$. This is not the case for the $\ell = 0$ mode, whose amplitude is not suppressed near the horizon. 

\vskip 0pt
In the main text, we need the profile of the $\ell = 0$ mode in the non-relativistic  ($k M \ll 1$) and fuzzy~($\mu M \ll 1$) limits. In this case, the confluent Heun function can be expanded to second order in $\mu M$ and $k M$, but at fixed $z$, as\footnote{Here, $\dilog(1-z)=\Li_2(z)=\sum_{n=1}^\infty z^n/n^2$.}${}^,$\footnote{The procedure consists in finding a recurrence relation among the coefficients of the power series $\HeunC(\alpha, \beta, \gamma, \delta, \eta; z) =\sum_{n=0}^\infty a_nz^n$, of the form $P_na_n=Q_na_{n-1}+R_na_{n-2}$, see e.g.~\cite{ronveaux1995heun,Hui:2019aqm}. After solving it to second order in $\alpha$, $\beta$, $\gamma$ and first order in $\delta$, $\eta$, the series can be resummed to give (\ref{eqn:heunc-expansion}).}
\begin{equation}
  \begin{split}
      \HeunC(\alpha&,\beta,\gamma,\delta,\eta;z)=\\
      & 1-\frac{1}{2} \alpha z +\frac{1}{6} \alpha^2 z^2- \frac{1}{24} \big(\alpha^2 +12 \delta\big)z +\frac{1}{4}\big(\alpha\beta+\alpha\gamma\big)z\log(1-z) \\
      & - \frac{1}{2}(\beta + \gamma) \log(1 - z)  +\frac{1}{4}\big(\gamma^2-\beta^2\big)\dilog(1-z) + \frac{1}{4}\big(\beta\gamma+ \gamma^2\big)\log^2(1-z)\\
      & - \frac{1}{24} \big(\alpha^2 - 6 \beta^2 - 6 \gamma^2 + 24 \eta + 12 \delta\big) \log(1 - z) +\cdots  \, .\label{eqn:heunc-expansion}
    \end{split}
\end{equation}
In the first line, we have grouped terms that are dominant as $z \to - \infty$, while the next two lines contain terms that are subdominant and can be ignored. Given that $\alpha\sim \mathcal{O}(kM)$ and $\delta\sim\beta^2\sim\gamma^2\sim \mathcal{O}(\mu^2M^2)$, we see that the confluent Heun function is approximately constant
\begin{equation}
\HeunC(\alpha, \beta, \gamma, \delta, \eta; z)  \sim 1+\mathcal O\bigl(\mu M,kM\bigr)\,, \qquad r_+ \le r<r\ped{max}\,,
\end{equation}
until the linear or quadratic terms in the first line of (\ref{eqn:heunc-expansion}) become $\mathcal{O}(1)$. This occurs at the radius
\begin{equation}
\frac{r\ped{max}}{M} \sim \min\biggl\{\frac1{(\mu M)^2},\frac1{kM}\biggr\}\gg 1\,.
\end{equation}
We use this approximation to derive the accretion rate in Section~\ref{sec:accretion}.

\chapter{More on ionization}

\label{app:more-on-ionization}

\section{Integrating out the continuum}
\label{app:approx} 

As explained in Section~\ref{sec:thorough-derivation}, the dynamics of the gravitational atom in a binary, including both bound and continuum states,  can be captured by integrating out the continuum and incorporating its effects in terms of a set of induced couplings and energies for the bound states alone. This process yields an effective Schr\"{o}dinger equation for the bound states that describes the behavior of the entire system. In this appendix, we justify the approximations we used to derive these continuum-induced couplings. First, we explain how our approximation for the fractional deoccupation rate (\ref{eq:toyDeoccupation}) in the toy model arises from the large time asymptotics of the induced energy. This derivation relies on ignoring the transitions between continuum states, so we then justify this assumption. Next, we discuss the complications that arise in the more realistic case, which includes many more bound and continuum states. We then discuss the effects of a nonlinearly ramping frequency $\dot{\varphi}_*(t)$ on our approximations. Finally, we extend the results to include the effects of resonances on ionization.

\subsection{Saddle point approximation}

\label{sec:saddle-point}
        
 We are interested in the asymptotic behavior of the induced energy 
          \begin{equation}
            \mathcal{E}_b(t) = \int_{- \infty}^{t}\!\dd t' \, \Sigma_b(t, t') = \frac{1}{2 \pi i} \int_{- \infty}^{t} \!\dd t' \int_0^{\infty}\!\dd k \, |\eta(k)|^2 \, e^{-i (\epsilon(k)- \epsilon_b)(t - t') + i(\varphi_*(t) - \varphi_*(t'))}\,, \label{eq:inducedEnergyApp}
        \end{equation}
where $\epsilon(k) = k^2/(2 \mu)$.
        Without loss of generality, we can absorb the bound state energy into our reference frequency, $\varphi_*(t) = - \epsilon_b t + \gamma t^2/2$, and assume that $\gamma > 0$. The bound state then begins to ``resonate'' with the continuum for $t \gtrsim 0$, and we would like to determine the asymptotic behavior of this function before and after this time,~$|\sqrt{\gamma} t| \gg 1$, as a way of approximating its behavior away from the complicated transient region around $t = 0$. 

        \vskip 0pt 
        There are two representations of this function that will be useful. We can either first perform the integral over $t'$ to find
        \begin{equation}
                \mathcal{E}_b(t) = \frac{1}{\sqrt{8 \pi \gamma}} \int_0^{\infty}\!\dd \epsilon\, |\eta(\epsilon)|^2 \es \exp\!\left[\frac{i(\epsilon-\gamma t )^2}{2 \gamma} - \frac{3 \pi i }{4}\right] \erfc\!\left[\frac{e^{\frac{i \pi}{4}} (\epsilon-\gamma t)}{\sqrt{2 \gamma}}\right], \label{eq:selfEnergyErfc}
        \end{equation}
        or we can define $z = \sqrt{\gamma}(t - t')$ and write 
        \begin{equation}
          \mathcal{E}_b(t) = \frac{1}{2 \pi i \sqrt{\gamma}} \int_{0}^{\infty}\!\dd z\, e^{i \tau z}\, \mathcal{K}(z)\,, \label{eq:selfEnergyLaplace}
        \end{equation}
        where we introduced the dimensionless time $\tau \equiv \sqrt{\gamma} t$ and the kernel
        \begin{equation}
          \mathcal{K}(z) \equiv e^{- \frac{1}{2} i z^2} \int_0^{\infty} \!\dd \epsilon \, e^{- i \epsilon z/\sqrt{\gamma}} \,|\eta(\epsilon)|^2\,. \label{eq:laplaceKernel}
        \end{equation}
        The former has the benefit of making the ``resonance'' behavior much clearer, while the latter is useful for understanding the large time $|\tau| \gg 1$ asymptotics since it has the form of a standard Laplace-like integral. In both representations, we have transformed the integral over momenta $k$ into an integral over the energy $\epsilon$ and defined $|\eta(\epsilon)|^2 = \dd k(\epsilon)/\dd \epsilon\, \big|\eta(k(\epsilon))\big|^2 = \mu |\eta(k)|^2/k$. In the cases of interest, $|\eta(\epsilon)|^2$ approaches a constant as $\epsilon \to 0$ and decays algebraically as $\epsilon \to \infty$, so that the ``total coupling'' of the bound state to the continuum $\int_{0}^{\infty}\!\dd\epsilon \, |\eta(\epsilon)|^2$ is finite.

        \begin{figure}[t]
          \centering 
          \includegraphics[trim={0 6pt 0 0}]{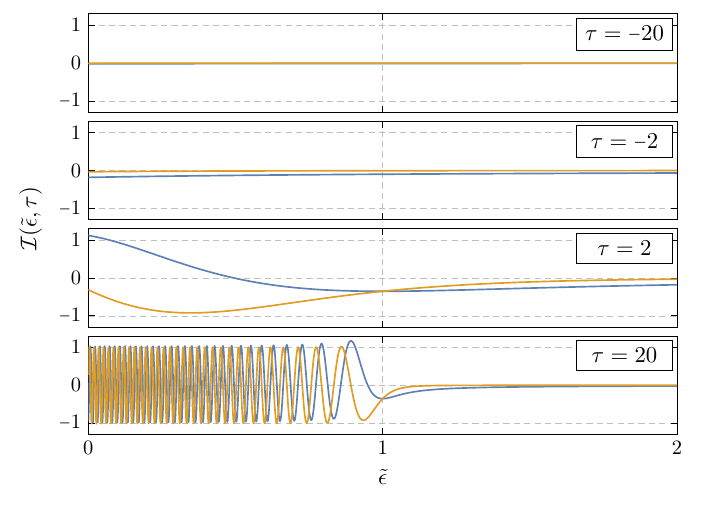}
          \caption{The real {\color{Mathematica1} [blue]} and imaginary {\color{Mathematica2} [orange]} parts of the modulating function $\mathcal{I}(\tilde{\epsilon}, \tau)$, for several values of the dimensionless time $\tau$. For large negative values of $\tau$, the integrand of (\ref{eq:inducedEnergyMapped}) is highly suppressed for $\tilde{\epsilon} \in [0, \infty)$. For large positive times $\tau \gg 1$, the integrand oscillates rapidly when $\tilde{\epsilon} \in [0, 1]$, slowing down when $\tilde{\epsilon} \sim 1$, and is then again highly suppressed for $\tilde{\epsilon} \gg 1$. \label{fig:intPlot}} 
        \end{figure}

        \vskip 0pt
        To get a sense for the behavior of this function, it is useful to first rescale the integral in (\ref{eq:selfEnergyErfc}) by taking $\epsilon \to \sqrt{\gamma} |\tau| \tilde{\epsilon}$\,,
        \begin{equation} 
            \mathcal{E}_b(\tau) = \frac{|\tau|}{\sqrt{2 \pi}} \int_{0}^{\infty}\!\dd \tilde{\epsilon} \, \big|\eta(\sqrt{\gamma} |\tau| \tilde{\epsilon})\big|^2 \,\es \mathcal{I}(\tilde{\epsilon}, \tau)\,, \label{eq:inducedEnergyMapped}
        \end{equation}
        where we defined the kernel
        \begin{equation}
            \mathcal{I}(\tilde{\epsilon}, \tau) \equiv \frac{1}{2}\es e^{\frac{i \tau^2}{2} (\tilde{\epsilon}- \sgn \tau )^2 - \frac{3 \pi i}{4}} \erfc\!\left[\tfrac{|\tau|}{\sqrt{2}}{e^{\frac{i \pi}{4}}(\tilde{\epsilon}-\sgn \tau)}\right] .
        \end{equation}
        We plot this kernel for several values of $\tau$ in Figure~\ref{fig:intPlot}. We see that, for $\tau \to \minus \infty$, the integrand of (\ref{eq:inducedEnergyMapped}) is strongly suppressed throughout the entire integration region, and so both the real and imaginary parts of the induced energy will be small. In the opposite limit, $\tau \to + \infty$, the integrand oscillates rapidly in the interval $\tilde{\epsilon} \in (0, 1)$, so we expect only the end point~$\tilde{\epsilon}=0$ and the region around $\tilde{\epsilon} = 1$
to contribute to the integral. For $\tilde{\epsilon} \in (1, \infty)$, the integrand no longer oscillates, but instead decays algebraically. The integrand---and especially the real part in {\color{Mathematica1} [blue]}---has a very heavy tail which the saddle point approximation is not able to fully capture. Instead, we will need to use the Laplace-like form (\ref{eq:selfEnergyLaplace}) to compute these additional contributions.

        \vskip 0pt
        Keeping in mind that the saddle point approximation does not capture the full behavior of the induced energy as $\tau \to \infty$, we will apply it anyway. As stated before, there are two relevant contributions---from the endpoint at $\tilde{\epsilon} = 0$ and from the ``saddle point'' at $\tilde{\epsilon} = 1$. From Figure~\ref{fig:intPlot}, we expect that the contribution at $\tilde{\epsilon} = 0$ produces an oscillatory \emph{ringing} that is left over from when the bound state first hits the edge of the continuum, and how quickly these oscillations decay depends on how the bound state couples to the lowest energy continuum modes, i.e.~how $|\eta(\epsilon)|^2$ scales as $\epsilon \to 0$. In contrast, the saddle point at $\tilde{\epsilon} = 1$ gives a non-oscillatory decay which only depends on the coupling between the bound state and the particular continuum state it is ``resonating with,''~$|\eta(\epsilon = \gamma t)|^2$. Assuming that $|\eta(\epsilon)|^2$ approaches a constant $|\eta|^2$ as $\epsilon \to 0$, we find that
        \begin{equation}
            \mathcal{E}_b(t) \sim - \frac{i \mu \big|\eta(k_*(t))\big|^2}{2 k_*(t)} -\frac{|\eta|^2 \lab{e}^{\frac{1}{2} i \gamma t^2 - \frac{i\pi}{4}}}{2\sqrt{2 \pi \gamma} t} \left[1 + \erf\!\left(\frac{\lab{e}^{\frac{i \pi}{4}} \!\sqrt{\gamma} t}{\sqrt{2}} \right)\right]  ,  \quad \sqrt{\gamma} t \to +\infty\,,
        \end{equation}
        where we have switched back to parameterizing the system in terms of the momentum and introduced $k_*(t) = \sqrt{2 \mu \gamma t}$, the momentum of the state at the saddle point. 

        \vskip 0pt
        To find the dominant behavior of $\Re\mathcal{E}_b(t)$ as $\tau \to \pm \infty$, we can use (\ref{eq:selfEnergyLaplace}) and repeatedly integrate by parts in $z$ to generate an expansion in powers of $\tau^{- 1}$. However, the aforementioned heavy tail can hinder this iterative process. Each integration by parts generates higher derivatives of the kernel evaluated at $z = 0$, 
        but these derivatives are not necessarily finite. From (\ref{eq:laplaceKernel}), we see that $\partial^k_z \mathcal{K}(z)|_{z = 0}$ contains a term proportional to $\int_{0}^{\infty}\!\dd \epsilon \, \epsilon^k |\eta(\epsilon)|^2$, and since $|\eta(\epsilon)|^2$ decays only algebraically, sufficiently high derivatives will diverge. This signals that $\mathcal{K}(z)$ has terms of the form $z^k \log^n z$, which produce asymptotic behavior of the form $\log^n \tau/\tau^{k+1}$, i.e. logarithmic behavior that is not captured in the standard saddle point approximation. 

        \vskip 0pt
        For our purposes, we will only concentrate on the leading order $|\tau| \to \infty$ behavior. This is governed by the total coupling $\mathcal{K}(0) = \int_0^{\infty}\!\dd \epsilon\, |\eta(\epsilon)|^2 = \int_{0}^{\infty}\!\dd k\, |\eta(k)|^2$, and direct integration yields
        \begin{equation}
          \mathcal{E}_b(t) \sim \frac{1}{2 \pi \gamma t} \left[\int_{0}^{\infty}\!\dd k \, |\eta(k)|^2\right] + \cdots\,.
        \end{equation}
        As $\tau \to - \infty$, this is the dominant contribution and gives an accurate approximation---as the effective energy gap between the bound and continuum states shrinks, the coupling to the continuum induces a correction to the bound state's energy. There is, however, no appreciable deoccupation of the bound state until after the transition at $\tau = 0$. As $\tau \to +\infty$, the integral picks up an additional saddle point and the induced energy is well approximated by
        \begin{equation}
        \begin{split}
          \mathcal{E}_b(t) \sim & - \frac{i \mu \big|\eta(k_*(t))\big|^2}{2 k_*(t)} -\frac{|\eta|^2 \lab{e}^{\frac{1}{2} i \gamma t^2 - \frac{i \pi}{4}}}{2\sqrt{2 \pi \gamma} t} \left[1 + \erf\!\left(\frac{\lab{e}^{\frac{i \pi}{4}} \!\sqrt{\gamma} t}{\sqrt{2}} \right)\right] \\
          & + \frac{1}{2 \pi \gamma t} \left[\int_{0}^{\infty}\!\dd k \, |\eta(k)|^2\right] + \cdots\, .\label{eq:inducedApproximation}
          \end{split}
        \end{equation} 
        Since we are mainly concerned with the imaginary part of this expression, we use the first term in (\ref{eq:inducedApproximation})  throughout the main text.

    \subsection{Unbound-unbound transitions}
  \label{sec:Unbound}
    
      It will be helpful to address our assumption that we can ignore the transitions between the continuum states in our analysis of the ionization process. We will do so in the toy model studied above and in Section~\ref{sec:warmup}. Numerical experiments show that the bound state's dynamics are relatively unaffected if we include these transitions and is still well-described by the first term in (\ref{eq:inducedApproximation}).  We can understand better why they may be ignored, and justify our assumption, by including  these couplings in the toy Hamiltonian (\ref{eq:toyHam}) and arguing that they should, at least at weak coupling, provide a subleading correction to the effective Schr\"{o}dinger equation (\ref{eq:toyEffSchro}). 

      \vskip 0pt
      A nontrivial coupling between continuum states $\eta(k, k') = \langle k | \mathcal{H} | k'\rangle$, for $k \neq k'$, changes the solution (\ref{eq:warmupContSol}) for the continuum amplitudes to
         \begin{equation}
         \begin{aligned}
            c_{k}(t) = &-i \int_{- \infty}^{t}\!\dd t'\, \eta(k)\, e^{- i \varphi_*(t') + i(\epsilon(k) - \epsilon_b) t'} c_b(t') \\
            &+ \frac{1}{2 \pi i}\int_{- \infty}^{t}\! \dd t' \int_0^\infty\!\dd k'\, \eta(k, k') \,e^{i (\epsilon(k) - \epsilon(k'))t'} c_{k'}(t')\,. 
            \end{aligned}\label{eq:warmupContSolMod}
        \end{equation}
        Importantly, both the bound-to-unbound couplings $\eta(k)$ and unbound-to-unbound couplings $\eta(k, k'; t)$ are $\mathcal{O}(q \alpha)$ and we work exclusively in the $q \alpha \ll 1$ regime. By plugging this solution back into itself, we can generate a solution purely in terms of the bound state amplitude, with the first correction to the $\eta(k, k') \to  0$ limit of (\ref{eq:warmupContSolMod}) being
        \begin{multline}
          c_k(t) \supset - \frac{1}{2 \pi }\int_{- \infty}^{t}\! \dd t_1 \int_{- \infty}^{t_1}\!\dd t_2 \int_0^\infty\!\dd k'\, \eta(k, k')\eta(k')\\
          \times \,e^{i (\epsilon(k) - \epsilon(k'))t_1 - i \varphi_*(t_2) + i(\epsilon(k') - \epsilon_b) t_2}c_b(t_2)\,,
        \end{multline}
        which is $\mathcal{O}(q^2 \alpha^2)$, while other corrections are higher order. 

        \vskip 0pt
        In the bound state Schr\"{o}dinger equation (\ref{eq:toyEffSchro}), this correction contributes a term involving the chain of matrix elements $\langle b | \mathcal{H} | k \rangle \langle k | \mathcal{H} | k' \rangle \langle k' | \mathcal{H} |b\rangle$, while the leading-order solution only involves the chain of elements~$\langle b | \mathcal{H} | k \rangle \langle k | \mathcal{H} |b \rangle$. Clearly, the leading-order contribution only accounts for the system transitioning into the continuum and then back to the bound state, while higher-order corrections involve the system going into the continuum and then bouncing around between different continuum states before returning to the bound state. Each of these transitions is thus penalized by an additional factor of $q \alpha$ and so we expect that they provide a subleading effect, especially at weak coupling $q \alpha \ll 1$.

        \vskip 0pt
        We might worry that, over long times, a substantial enough continuum population can be built up so that the second term in (\ref{eq:warmupContSolMod}) can overcome its $\mathcal{O}\big(q^2 \alpha^2\big)$-suppression and compete with the first. However, this sort of coherent effect is extremely unlikely in light of the oscillatory factors in (\ref{eq:warmupContSolMod}), which serve to randomize the ``direction'' of this perturbation and suppress its effects on long time scales. These arguments can be trivially extended to the more realistic case discussed in the next section, so we will ignore continuum-to-continuum transitions throughout our analysis and focus only on how the bound states interact with the continuum.

    \subsection{Extension to the realistic case}
    
    \label{sec:extension-realistic}

      The main complication in going to the more realistic case is that there are many more bound and continuum states, and the continuum now mediates transitions between different bound~states. These effects appear in the form of off-diagonal induced couplings:
      \begin{equation}
        \mathcal{E}_{ba}(t) = -i \sum_{K} \eta^{*\floq{\Delta m_b}}_{K b}(t) \eta^{\floq{\Delta m_a}}_{K a}(t) \int_{- \infty}^{t}\!\dd t'\,  e^{i \Delta m_b \varphi_*(t) - i \Delta m_a \varphi_*(t') + i( \epsilon_b - \epsilon_K) t + i (\epsilon_K - \epsilon_a) t'}\, , \label{eq:appInducedCouplings}
      \end{equation}
       where we have introduced the shorthand $\Delta m_a \equiv m - m_a$ and $\Delta m_b \equiv m - m_b$. We would like to understand the general behavior of these off-diagonal terms and argue that they can be ignored whenever the resonance condition between the states $|a \rangle$ and $|b \rangle$ is not satisfied. On resonance, they provide a small correction compared to the direct coupling between these states, on which we further elaborate in Appendix~\ref{sec:ion-at-resonance}, and so they can be neglected.

      \vskip 0pt
      Assuming that the frequency $\dot{\varphi}_*(t)$ is linear, we can again define the variable $z \equiv t - t'$ and write (\ref{eq:appInducedCouplings}) as
      \begin{equation}
        \begin{aligned}
        \mathcal{E}_{ba}(t) = &\, e^{i (\epsilon_b - \epsilon_a) t - i (m_b - m_a) \varphi_*(t)} \\
          & \times \left[-i \sum_{K}  \int_{0}^{\infty}\!\dd z\,  e^{-\frac{1}{2} i \Delta m_a \gamma z^2 + i (\Delta m_a \dot{\varphi}_*(t) - (\epsilon_K - \epsilon_a)) z } \eta^{*\floq{\Delta m_b}}_{K b}(t) \eta^{\floq{\Delta m_a}}_{K a}(t)\right] . \label{eq:appInducedCouplingsLaplace}
        \end{aligned}
      \end{equation}
      The term in braces is of a similar form to the induced energy (\ref{eq:selfEnergyLaplace}), whose behavior we have already analyzed in (\ref{eq:inducedApproximation}). It contains both oscillating and smoothly decaying  terms. Ignoring these oscillating terms for now, we see that the induced couplings oscillate rapidly with phase $\exp\!\big[i (\epsilon_b - \epsilon_a) t - i (m_b - m_a) \varphi_*(t)\big]$. As we argue in Section~\ref{sec:realisticIonization}, the direct couplings between $|a\rangle$ and $|b \rangle$ also oscillate with this phase, and if these oscillations are too rapid the contribution to the bound state solution will quickly average out. Of course, this oscillation slows down when the resonance condition $(m_b - m_a) \dot{\varphi}_*(t) = (\epsilon_b - \epsilon_a)$ is satisfied, but again these induced couplings, which are $\mathcal{O}(q^2 \alpha^2)$, must compete with the $\mathcal{O}(q \alpha)$ direct couplings $\eta_{ba}$, and so even then they have a small effect on the behavior of the resonance for $q \alpha \ll 1$.

      \vskip 0pt
      We might worry about the oscillations that arise in (\ref{eq:inducedApproximation}) as transients when the state $|a \rangle$  begins to resonate with the continuum might spoil this story, and that these induced couplings might become relevant. Fortunately, this is not the case. These transient oscillations ``start'' when the companion can excite $|a \rangle$ into the continuum, $\Delta m_a \dot{\varphi}_*(t) = -\epsilon_a$, and if they are present they modify the overall exponential in (\ref{eq:appInducedCouplingsLaplace}) to
      \begin{equation}
        \exp\!\left[-i (m_b - m_a) \varphi_*(t) + i (\epsilon_b - \epsilon_a) t +i (\Delta m_a\dot{\varphi}_*(t) + \epsilon_a)^2/(2 \Delta m_a \gamma)\right] .
      \end{equation}
      This term can  contribute appreciably when the argument of the exponential slows down, that is when $\Delta m_b \dot{\varphi}_*(t) = -\epsilon_b$. 
      The two conditions $\Delta m_i \dot{\varphi}_*(t) = - \epsilon_i$, for $i=a,b$,
      can only simultaneously satisfied when $(m_b - m_a) \dot{\varphi}_*(t) = \epsilon_b - \epsilon_a$, i.e.~exactly on resonance. So, the transient oscillatory terms in (\ref{eq:inducedApproximation}) may ``smear out'' the resonance slightly, but again since they are $\mathcal{O}(q^2 \alpha^2)$ and must compete with the $\mathcal{O}(q \alpha)$ direct couplings $\eta_{ba}(t)$, we do not expect that they provide a qualitative change in behavior in the dynamics, and away from resonance we can ignore the induced couplings entirely.

      \vskip 0pt
      With this out of the way, we can focus entirely on the diagonal terms, $\mathcal{E}_{b}(t) \equiv \mathcal{E}_{bb}(t)$, which are much simpler:
      \begin{align}
         \mathcal{E}_{b}(t) &= -i  \int_{- \infty}^{t}\!\dd t'\, \sum_{K}  \big|\eta^{\floq{\Delta m_b}}_{K b}(t)\big|^2 e^{i \Delta m_b (\varphi_*(t) -  \varphi_*(t')) - i( \epsilon_K - \epsilon_b) (t-t')} \nonumber \\
         &= \frac{1}{2 \pi i} \sum_{\ell, m} \int_{- \infty}^{t}\!\dd t'\int_{0}^{\infty}\!\dd k \, \big|\eta^{\floq{\Delta m_b}}_{K b}(t)\big|^2 e^{i \Delta m_b (\varphi_*(t) -  \varphi_*(t')) - i( \epsilon(k) - \epsilon_b) (t-t')}\,.
      \end{align}
      This is nothing more than a sum over integrals of the form we have already analyzed, and we can use the same techniques as before to attack this. In particular, the integral over $t'$ yields
      \begin{align}
        \mathcal{E}_{b}(t)&=   \frac{1}{2 \pi}\sum_{\ell, m}\sqrt{\frac{\pi }{2 \Delta m_b \gamma}}  \int_{0}^{\infty}\!\dd k\,    \big|\eta_{K b}^{\floq{\Delta m_b}}(t)\big|^2  \exp\!\left(\frac{i(\Delta m_b \dot{\varphi}_*(t) - (\epsilon(k) - \epsilon_b))^2}{2 \Delta m_b \gamma} - \frac{3 \pi i}{4} \right) \nonumber \\
         &\qquad\qquad \times  \left[ \lab{sgn}\, \Delta m_b \gamma + \erf\!\left(\frac{e^{\frac{i \pi}{4}} (\Delta m_b \dot{\varphi}_*(t) - (\epsilon(k)- \epsilon_b))}{\sqrt{2 \Delta m_b \gamma}}\right)\right] .
      \end{align}
      As discussed previously, we can think of the imaginary part as getting a saddle point contribution at $k_*^\floq{g}(t) = \sqrt{2 \mu(g \dot{\varphi}_*(t) + \epsilon_b)}$, which again only contributes if $k_*^\floq{g}(t)^2 > 0$. For this to ever happen (since $\epsilon_b < 0$), we must have that $\Delta m_b \gamma = (m - m_b) \gamma > 0$. Thus, ignoring the oscillatory terms and other transients, we have
      \begin{equation}
        \mathcal{E}_b(t) \approx -\sum_{\ell, g}\left[ \frac{i\mu \big|\eta^\floq{g}_{K_* b}(t)\big|^2}{2k_*^\floq{g}(t)} \Theta\big(k^\floq{g}_*(t)^2\big)\right] , \label{eq:realisticDeoccupationApp}
      \end{equation}
      with $K_* = \{k_*^{\floq{g}}(t), \ell, m = g+m_b\}$ and $k_*^\floq{g}(t) = \sqrt{2 \mu(g \dot{\varphi}_*(t) + \epsilon_b)}$, where the sum ranges from $\ell = 0,1, \dots, \infty$ and over all $g$ such that $|g + m_b| \leq \ell$. This is the extension of the first term in (\ref{eq:inducedApproximation}) to include other sectors of continuum states, with different angular momenta, connected to the bound state by perturbations that oscillate at different frequencies.
      
    \subsection{Nonlinear chirp frequency} \label{app:nonlinearChirp}

      Throughout Section~\ref{sec:thorough-derivation}, we have assumed that we can linearize the frequency and write the phase as  $\varphi_*(t) = -\epsilon_b t + \gamma t^2/2$. It will be useful to justify this approximation.

      \vskip 0pt
      Let us return to (\ref{eq:inducedEnergyApp}) and try to understand the behavior of the $t'$ integral,
      \begin{equation}
        \int_{- \infty}^{t}\!\dd t'\, e^{i( \epsilon - \epsilon_b) t' - i \varphi_*(t')}\,,
      \end{equation}
      for a phase $\varphi_*(t)$ with general time dependence. This integral has essentially two contributions. One comes from the end point, which we can isolate through integration by parts,
      \begin{equation}
        \int_{- \infty}^{t}\!\dd t'\, e^{i( \epsilon - \epsilon_b) t' - i \varphi_*(t')} \supset \frac{ie^{i( \epsilon - \epsilon_b) t - i \varphi_*(t)}}{\dot{\varphi}_*(t) - (\epsilon - \epsilon_b)}  + \cdots\,,
      \end{equation}
      while another can arise if $\dot{\varphi}(t_*) = \epsilon - \epsilon_b$
      for some $t' = t_*$ in the integration interval.  When such a time exists, the integral receives an additional contribution
      \begin{equation}
        \int_{- \infty}^{t}\!\dd t'\, e^{i( \epsilon - \epsilon_b) t' - i \varphi_*(t')} \supset \sqrt{\frac{2 \pi}{\ddot{\varphi}_*(t_*)}} e^{i(\epsilon - \epsilon_b) t_* - i \varphi_*(t_*) - \frac{i \pi}{4}}\,,
      \end{equation}
      which we should divide in half when $t = t_*$. We obtain a rough approximation for the $t'$ integral,
      \begin{equation}
         \int_{- \infty}^{t}\!\dd t'\, e^{i( \epsilon - \epsilon_b) t' - i \varphi_*(t')} \approx
         \begin{dcases} 
         \frac{ie^{i( \epsilon - \epsilon_b) t - i \varphi_*(t)}}{\dot{\varphi}_*(t) - (\epsilon - \epsilon_b)}\,, & t < t_* \\[4pt] \sqrt{\frac{\pi}{2 \ddot{\varphi}_*(t_*)}} e^{i(\epsilon - \epsilon_b) t_* - i \varphi_*(t_*) - \frac{i \pi}{4}}\,, & t = t_* \\[4pt]
          \frac{ie^{i( \epsilon - \epsilon_b) t - i \varphi_*(t)}}{\dot{\varphi}_*(t) - (\epsilon - \epsilon_b)} + \sqrt{\frac{2\pi}{\ddot{\varphi}_*(t_*)}} e^{i(\epsilon - \epsilon_b) t_* - i \varphi_*(t_*) - \frac{i \pi}{4}}\,, & t > t_* 
         \end{dcases} \label{eq:approxGenTime}
      \end{equation}
      by adding these different contributions.

      \vskip 0pt
      If we use $\varphi_*(t) = -\epsilon_b t + \gamma t^2/2$ and consider the exact answer, we find that
      \begin{equation}
        \sqrt{\frac{\pi}{2 \gamma}} e^{\frac{i \epsilon^2}{2 \gamma} - \frac{i \pi}{4}} \lab{erfc}\!\left[\frac{e^{\frac{i \pi}{4}} (\epsilon - \gamma t)}{\sqrt{2 \gamma}}\right]\approx
         \begin{dcases} 
         \frac{ie^{-\frac{1}{2} i \gamma t^2 + i \epsilon t}}{\gamma t}\,, & t \ll t_* \\[4pt] \sqrt{\frac{\pi}{2 \gamma}} e^{\frac{i\epsilon^2}{2 \gamma}- \frac{i \pi}{4}}\,, & t = t_* \\[4pt] \frac{ie^{-\frac{1}{2} i \gamma t^2 + i \epsilon t}}{\gamma t} + \sqrt{\frac{2\pi}{ \gamma}} e^{\frac{i\epsilon^2}{2 \gamma}- \frac{i \pi}{4}}\,, & t \gg t_* 
         \end{dcases} \,,
      \end{equation}
      where $t_* = \epsilon(k)/\gamma$. We see that (\ref{eq:approxGenTime}) accurately captures the large $|t|$ asymptotics of the integral, and that the complicated error function is merely present to interpolate between these three regimes. Furthermore, the relevant chirp rate for the induced energy (\ref{eq:inducedEnergyApp}) is just the instantaneous chirp rate $\ddot{\varphi}_*(t)$ which we can, to excellent approximation, replace with the chirp rate defined in \eqref{eqn:gamma_gws} associated to the frequency $\Omega_0 = -\epsilon_b$ of the energy gap between the bound state and the continuum.

      \vskip 0pt
      We see then that the linearization of $\dot \varphi_*(t)$ is not such a dramatic approximation. The integrand in (\ref{eq:selfEnergyErfc}) will still have a similar form as to the one considered there, and we would still be able to do a saddle point computation isolating the large $|\tau|$ asymptotics and get effectively the same results we have found in the main text, up to corrections in the (small) nonlinearities we have ignored.

    \section{Markov approximation} \label{app:Markov} 
    
    In Section~\ref{sec:thorough-derivation}, we studied how the cloud is ionized by first constructing an effective Schr\"{o}dinger equation (\ref{eq:realisticEffSchro}) for the bound states, fully integrating out the dynamics of the continuum states and incorporating their effects in the induced couplings (\ref{eq:inducedCouplings}). This was valid in the so-called ``Markov approximation," which we justify in this appendix.

    \vskip 0pt
    Let us review how the Markov approximation comes about for a single bound state interacting with the continuum. We argued in Section~\ref{sec:realisticIonization} that we can ignore the continuum-induced interactions between the bound states off-resonance, and so this truncation to a single bound state still accurately captures the true dynamics of the system, especially when the orbital frequency is too high for any resonance to occur. By solving (\ref{eq:realisticContEom}) for the continuum state amplitudes and plugging the result into  (\ref{eq:realisticBoundEom}), we arrive at a single equation for the bound state amplitude
    \begin{equation}
      i \frac{\dd c_b}{\dd t} = \int_{- \infty}^{t}\!\dd t'\, \Sigma_{b}(t, t') c_b(t')\,, \label{eq:appIntegDiff}
    \end{equation}
    in terms of the self-energy
    \begin{equation}
      \Sigma_{b}(t, t') \equiv -i \sum_{K} \eta_{b K}(t) \eta_{K b}(t') e^{-i(\epsilon_{K} - \epsilon_b)(t - t')}\,.
    \end{equation}
   Assuming that the couplings between the continuum states vanish and ignoring the transitions into other bound states, this equation of motion is exact. We then implement the Markov approximation by first integrating by parts,
    \begin{equation}
      i \frac{\dd c_b}{\dd t} = \mathcal{E}_{b}(t) c_b(t) - \int_{- \infty}^{t}\!\dd t_1\, \mathcal{E}_b(t, t_1) \, \frac{\dd c_b(t_1)}{\dd t_1}\,,\label{eq:eomMarkov}
    \end{equation}
    and dropping the second term, which we will argue can be neglected. Here, we have defined
    \begin{equation}
      \mathcal{E}_b(t, t') = \int_{- \infty}^{t'}\!\dd t_1\, \Sigma_{b}(t, t_1)\,,
    \end{equation}
    and the induced energy $\mathcal{E}_b(t) \equiv \mathcal{E}_b(t, t)$.
    
    \begin{figure}[t]
      \centering
      \includegraphics[trim={0 6pt 0 0}]{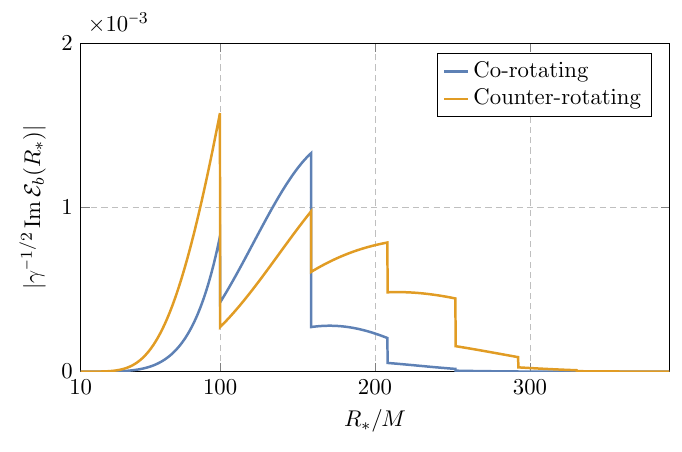}
      \caption{The dimensionless ratio $\big|\gamma^{\protect - 1/2} \Im \mathcal{E}_b(R_*)\big|$ as a function of the orbital separation $R_*$, using our approximation (\ref{eq:realisticDeoccupation}) as an estimate, for an inspiral with $q = 10^{\protect - 3}$ and $\alpha = 0.2$, where $\gamma$ is the instantaneous chirp rate $\gamma = \ddot{\varphi}_*(t)$, defined in \eqref{eqn:gamma_gws} with $\Omega_0^2 R_*^3 = (1+q)M$.  \label{fig:approxPlot}}
    \end{figure}

    \vskip 0pt
    Our goal now is to estimate the effect of the second term in (\ref{eq:eomMarkov}). To do this, we first strip off the first-order behavior by defining $\tilde{c}_b(t) = e^{i \varphi_b(t)} c_b(t)\,,$
    where $\varphi_b(t) = \int_{- \infty}^{t}\!\dd t_1 \, \mathcal{E}_b(t)$ is the time-dependent phase induced at first order by the continuum. Plugging this into (\ref{eq:eomMarkov}) yields
    \begin{equation}  
      i \frac{\dd \tilde{c}_b(t)}{\dd t} = i \int_{- \infty}^{t}\!\dd t_1 \, e^{i \varphi_b(t) - i \varphi_b(t_1)} \left[\mathcal{E}_b(t, t_1) \mathcal{E}_b(t_1, t_1) \tilde{c}_b(t_1) + i \mathcal{E}_b(t, t_1) \dot{\tilde{c}}_b(t_1) \right] . \label{eq:eom2Markov}
    \end{equation}
    Defining the second-order induced energy
    \begin{equation}
      \mathcal{E}_b^\floq{2}(t, t') = i \int_{- \infty}^{t'}\!\dd t_1\, e^{i \varphi_b(t) - i \varphi_b(t_1)} \mathcal{E}_b(t, t_1) \es \mathcal{E}_b(t_1, t_1)\,,
    \end{equation}
    with $\mathcal{E}^\floq{2}_b(t) \equiv \mathcal{E}^\floq{2}_b(t, t)$, integrating the first term in (\ref{eq:eom2Markov}) by parts, and dropping terms that contain factors of $\dd \tilde{c}_b/\dd t$, (\ref{eq:eom2Markov}) reduces to
    \begin{equation}
        i\frac{\dd \tilde{c}_b}{\dd t} = \mathcal{E}_b^{\floq{2}}(t)\es \tilde{c}_b(t)\,,
    \end{equation}
    As long as we can argue that this contribution is small compared to the first-order motion, this step of dropping terms containing $\dd \tilde{c}_b/\dd t$ is consistent. In principle, we could also iterate this process to find ever more accurate approximations to the true dynamics.

    \vskip 0pt
    It will be helpful to write the second-order induced energy as
    \begin{equation}
      \mathcal{E}_b^\floq{2}(t) = i \int_{- \infty}^{t}\!\dd t_1\, e^{-\Im\left[\varphi_b(t) - \varphi_b(t_1)\right] + i \Re\left[\varphi_b(t) - \varphi_b(t_1)\right]} \, \mathcal{E}_b(t, t_1) \es\mathcal{E}_b(t_1, t_1)\,.
    \end{equation}
    Of particular importance is the oscillating phase factor, which depends on the real part of the induced phase difference $\Re\left[\varphi_b(t) - \varphi_b(t_1)\right]$. Contributions to this integral will cancel unless $t_1$ is close to $t$. Since the relevant time scale of the transition is 
    \begin{equation}
        \gamma^{- 1/2} =  \sqrt{\frac{5}{96}} \frac{\alpha }{\mu}\frac{q^{- \frac{1}{2}}}{(1 + q)^{\frac{3}{4}}}\! \left(\!\es\frac{ \mu R_*}{\alpha}\!\es\right)^{\!\frac{11}{4}}\,,
    \end{equation}
    we can think of $\mathcal{E}_b^\floq{2}(t)$ as being on the same order as $\gamma^{- 1/2} \mathcal{E}_b(t, t)^2$. These second-order corrections are thus small as long as $\big|\gamma^{- 1/2} \es \mathcal{E}_b(t)^2\big| \ll |\es \mathcal{E}_b(t)|$. Since there is typically not a hierarchy between the real and imaginary parts of $\mathcal{E}_b(t)$, we can instead write this condition as $\big|\gamma^{- 1/2} \Im \mathcal{E}_b(t)\big| \ll 1$. We plot this quantity in Figure~\ref{fig:approxPlot} for the parameter values we consider in the main text and we see that it is comfortably small, so the Markov approximation is justified.

\section{Ionization power} \label{app:ionizedEnergy}
  
    In this appendix, we justify our approximation of the ionization power $P_\lab{ion} \equiv \dd E_\lab{ion}/\dd t$ in the toy model of Section~\ref{sec:warmup}. The extension to the realistic case is conceptually trivial.

    \vskip 0pt
    The total ionized energy is defined as
    \begin{equation}
        E_\lab{ion}(t) = \frac{1}{2 \pi}\frac{M_\lab{c}}{\mu}\int_{0}^{\infty}\!\dd k\, (\epsilon(k) - \epsilon_b) |c_k(t)|^2\,,
    \end{equation}
    where $M_\lab{c}/\mu$ represents the total occupation number of the cloud. We will set this to one and restore it at the end of the calculation.
    By taking a single time derivative we can express the ionization power as,
    \begin{equation}
  P_\lab{ion}  = \frac{1}{2 \pi}\int_{0}^{\infty}\!\dd k\, (\epsilon(k) - \epsilon_b) \left[\dot{c}_k^*(t) c_k(t) + c_k^*(t) \dot{c}_k(t)\right] .
    \end{equation} 
    and inserting both the Schr\"{o}dinger equation (\ref{eq:warmupContEom}) and the solution (\ref{eq:warmupContSol}), we can find an equation of motion for the ionized energy purely in terms of the bound state
    \begin{multline}
        P_\lab{ion}= \frac{1}{2 \pi}\int_{0}^{\infty}\!\dd k \int_{- \infty}^{t}\!\dd t'  \Bigl[   (\epsilon(k) - \epsilon_b) |\eta(k)|^2 \\
        \times e^{i(\varphi_*(t) - \varphi_*(t')) - i (\epsilon(k) - \epsilon_b)(t- t')} c^*_b(t) c_b(t')+ \lab{c.c.}\Bigr] .
    \end{multline}
    This has a very similar flavor to the effective bound state equation of motion (\ref{eq:toySelfEnergyEq}), and we can implement the Markov approximation by integrating by parts and dropping the remainder,
    \begin{equation}
           P_\lab{ion} = 2\Re\!\left[\frac{1}{2 \pi}\int_{0}^{\infty}\!\dd k \int_{- \infty}^{t}\!\dd t'\, (\epsilon(k) - \epsilon_b)\,  |\eta(k)|^2  e^{i(\varphi_*(t) - \varphi_*(t')) - i (\epsilon(k) - \epsilon_b)(t- t')}\right] \!|c_b(t)|^2 \,\,. \label{eq:appEIonApprox}
    \end{equation}
    This equation of motion is very similar to (\ref{eq:toyDeoccupation}), though now the term analogous to the induced energy $\mathcal{E}_b(t)$ is weighted with the energy difference $\epsilon(k) - \epsilon_b$. 

    \vskip 0pt
    This expression for the ionization power can be analyzed with the same techniques as used in Appendix~\ref{app:approx}---ignoring the transient region around $\dot{\varphi}_*(t) + \epsilon_b = 0$ and the subleading oscillatory terms, we can approximate (\ref{eq:appEIonApprox}) with its steady-state growth
    \begin{equation}
         P_\lab{ion} \approx \frac{M_\lab{c}}{\mu}\left[\frac{\mu \dot{\varphi}_*(t) |\eta(k_*(t))|^2}{k_*(t)}\right] |c_b(t)|^2 \,\Theta(k_*(t))\,,
    \end{equation}
    where we have replaced $\epsilon(k_*(t)) - \epsilon_b = \dot{\varphi}_*(t)$.

\section{Zero mode} \label{app:zeroMode}

\begin{figure}[t]
          \centering 
          \includegraphics[trim={0 6pt 0 0}]{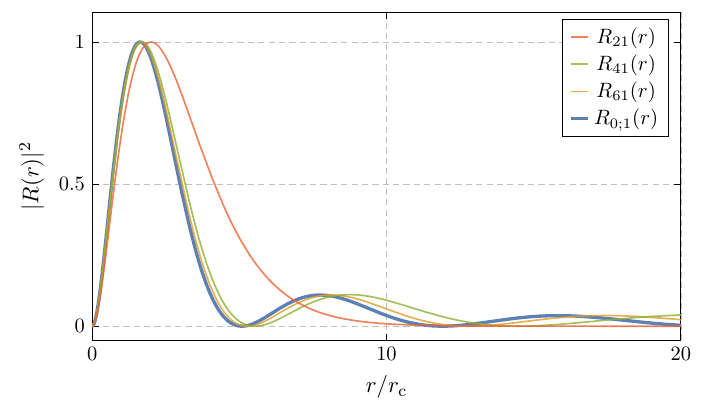}
          \caption{The radial zero mode density $\lim_{k \to 0} | k^{\protect - 1/2} R_{k; 1}(r)|^2$ compared to several bound state densities, all with orbital angular momentum $\ell = 1$. Here, $r_\lab{c} = (\mu \alpha)^{\protect - 1}$ is the typical radius of the cloud, and we have normalized each density so that it has unit maximum. Ignoring the overall normalization, the zero mode wavefunction can also be thought of as the limit of the bound state wavefunctions as $n \to \infty$. \label{fig:zeroMode}}  
        \end{figure}

        As we explained in the main text, the dramatic ``discontinuous'' behavior of the ionization power~$P_\lab{ion}$ is due to the fact that the coupling function $|\eta(k)|^2$ goes to zero linearly in $k$ as $k \to 0$. We mentioned there that this is because the long-range Coulombic potential keeps the zero mode relatively well-localized about the origin, as illustrated in Figure~\ref{fig:zeroMode}, such that the couplings in energy $|\eta(\epsilon)|^2 \equiv \dd k(\epsilon)/\dd \epsilon\, |\eta(k(\epsilon))|^2$ are finite as $\epsilon \to 0$. In this appendix, we discuss the zero mode of the hydrogen atom, its normalization, and the role the long-ranged $1/r$ potential plays in its radial behavior.

        \vskip 0pt
        In order to determine the overall normalization of the zero mode, we begin by writing the normalized continuum radial wavefunctions (\ref{eq:contWavefunctions}) as
        \begin{equation}  
          R_{k; \ell}(r) = \frac{2 k \es i^{\ell} e^{\frac{\pi \mu \alpha}{2 k}} \big|\Gamma\big(\ell + 1 + \tfrac{i \mu \alpha}{k}\big)\big|}{(\minus 2 i k r)^{\frac{1}{2}}\es\Gamma\big(\ell+1 + \frac{i \mu \alpha}{k}\big)} \,  e^{- i k r}\! \int_{0}^{\infty}\!\dd \zeta\, e^{- \zeta + \frac{i \mu \alpha}{k} \log \zeta} \zeta^{- \frac{1}{2}} J_{2 \ell+1}\!\left(2 \sqrt{\minus 2 i k r \zeta}\right) ,
        \end{equation}
        where we have used a standard integral representation of the confluent hypergeometric function in terms of the Bessel function of the first kind $J_{\nu}(z)$. 
        As $k \to 0$, the integral is localized around its saddle point $\zeta = i \mu \alpha/k$ and asymptotes to
        \begin{equation}
          R_{k; \ell}(r) \sim \sqrt{\frac{4 \pi k}{r}} J_{2 \ell+1}\big(2 \sqrt{2 \mu \alpha r} \hskip 1pt\big)\,,\mathrlap{\qquad k \to 0\,.} \label{eq:zeroAsympt}
        \end{equation}
      It is then clear that any matrix element between a continuum state and a bound state will also scale as $\sqrt{k}$ for $k \to 0$, so that $|\eta(k)|^2/k$ approaches a finite, non-zero limit as $k \to 0$.

        \vskip 0pt
        We can understand this scaling in a less opaque way by considering the Schr\"{o}dinger equation with a potential that asymptotes to a generic power law, $V(r) \sim 1/r^{\Delta}$ as $r \to \infty$, with $\Delta > 0$. Defining $\rho = 1/r$, the radial Schr\"{o}dinger equation for a state with energy $\epsilon(k) = {k^2}/{2 \mu}$ can then be written as
        \begin{equation}
          \left(-\frac{\dd^2}{\dd \rho^2} + \frac{\ell(\ell+1)}{\rho^2} - \frac{2 \alpha \mu^{2}}{\rho^{4}} \frac{\rho^\Delta}{\mu^\Delta} - \frac{k^2}{\rho^4}\right)R_{k;\ell}(\rho) = 0 \,, \label{eq:schroInv}
        \end{equation}
where we have introduced additional factors of $\mu$ to keep $\alpha$ dimensionless.
        We will only be concerned with the behavior of the solutions as $\rho \to 0$ or, analogously, as $r \to \infty$, so we have replaced the potential with its dominant long-distance behavior. If $\Delta > 2$, then the potential term is subleading to the centrifugal $\ell(\ell+1)/r^2$ term and the asymptotics of $R_{k; \ell}(\rho)$ are identical to that of a free particle.

        \vskip 0pt 
        For long-ranged potentials, $0 < \Delta < 2$, we can determine the overall normalization of the continuum wavefunctions as $k \to 0$ via a matching procedure. The basic idea is that the potential singularity $2 \alpha \mu^{2 -\Delta}/\rho^{4 - \Delta}$ in (\ref{eq:schroInv}) dominates over the energy singularity $k^2/\rho^4$ in the region $\rho \gtrsim \mu \big[ (k/\mu)^2/\alpha\big]^{1/\Delta}$. When $\rho$ is smaller than this, the energy singularity dominates, so we can construct asymptotic approximations to $R_{k;\ell}(\rho)$ that are valid in these two different regions. When $k$ is very small, the region $\rho \gtrsim \mu \big[ (k/\mu)^2/\alpha\big]^{1/\Delta}$ comprises most of space, and so this is the relevant solution in the $k \to 0$ limit. However, the overall normalization of the continuum wavefunctions is set for $\rho \lesssim \mu \big[ (k/\mu)^2/\alpha\big]^{1/\Delta}$, and so we must deduce the overall normalization in the $k \to 0$ limit by matching. Our goal then is to first determine the asymptotic behavior of $R_{k;\ell}(\rho)$ around each of these singularities and then match them.

        \vskip 0pt
        Depending on the value of $\Delta$, the asymptotic behavior of $R_{k;\ell}(\rho)$ in the region near the energy singularity can be relatively complicated,
        \begin{equation}
          R_{k; \ell}(\rho) \sim A \es \rho \es \sin \!\left(\frac{k}{\rho} + \sum_{n = 1}^{n \Delta \leq 1} \frac{(\minus \frac{1}{2}\alpha)^n (2n)!}{(2n-1)(n!)^2} \frac{(k/\mu)^{1- 2n}}{n \Delta -1} \frac{\rho^{n \Delta -1}}{\mu^{n \Delta-1}} + \delta\right),\label{eq:zeroModeLess}
        \end{equation}
        which holds for
        \beq
        \rho \lesssim \mu \left[ \frac{(k/\mu)^2}{\alpha}\right]^{1/\Delta}
        \eeq
        and the sum is over all $n$ such that $n \Delta \leq 1$, and a $n \Delta = 1$ term should be understood to give a logarithmic correction. Here, $A$ and $\delta$ are the overall normalization and phase, respectively. For example, the asymptotic behavior of wavefunctions for the Coulombic potential, with $\Delta = 1$, is  
        \begin{equation}
          R_{k; \ell}(\rho) \sim A \es \rho \es \sin \!\left(\frac{k}{\rho} + \frac{\mu \alpha}{k} \log \frac{k}{\rho} + \delta\right) ,
        \end{equation}
        and demanding these wavefunctions are appropriately normalized, $\langle k; \ell\es m| k'; \ell \es m \rangle = 2\pi\delta(k - k')$, sets the overall amplitude in this region to $A = 2$. In contrast, the asymptotic behavior of $R_{k;\ell}(\rho)$ in the region where the potential singularity dominates is relatively simple,
        \begin{equation}
          R_{k; \ell}(r) \sim A' \es \rho^{1 - {\Delta/4}} \sin \!\left(\frac{2 \sqrt{2 \alpha (\rho/\mu)^{\Delta - 2}}}{2 - \Delta} + \delta'\right) , \ \quad \rho \gtrsim \mu \left[ \frac{(k/\mu)^2}{\alpha}\right]^{1/\Delta}\,, \label{eq:zeroModeMore}
        \end{equation}
        where again $A'$ and $\delta'$ are an undetermined amplitude and phase.

        \vskip 0pt
        In the limit $k \to 0$, the region of (\ref{eq:zeroModeLess})'s validity, $\rho \lesssim \mu \big[ (k/\mu)^2/\alpha\big]^{1/\Delta}$, shrinks to a point, and the continuum wavefunctions are well approximated by (\ref{eq:zeroModeMore}) as $\rho \to 0$. However, we do not yet know its amplitude $A'$ or, specifically, the $k$-scaling of its amplitude. We can determine this scaling by matching the amplitudes of (\ref{eq:zeroModeLess}) and (\ref{eq:zeroModeMore}) in the  region where both expansions apply,~$\rho \sim \mu \big[ (k/\mu)^2/\alpha\big]^{1/\Delta}$. We find that the continuum wavefunctions then behave as
        \begin{equation}
          R_{k; \ell}(r) \propto \frac{\sqrt{k}}{r^{\frac{1}{4}(4-\Delta)}} \sin\!\left(\frac{2 \sqrt{2\alpha (\mu r)^{2 - \Delta}}}{2 - \Delta} + \tilde{\delta}\right), \qquad \begin{aligned} k &\to 0 \\[-6pt] r &\to \infty \end{aligned}\ ,
        \end{equation}
        for arbitrary $0 < \Delta < 2$, with $\tilde{\delta}$ an undetermined phase. As long as the potential is sufficiently long-ranged, $\Delta < 2$, the continuum wavefunctions therefore asymptote to a fixed radial function multiplied by an overall factor of $\sqrt{k}$ as $k \to 0$. This implies that, for $\Delta < 2$, the potential is sufficiently long-ranged enough to localize the zero mode.
        We can compare this general result with the asymptotic expansion of (\ref{eq:zeroAsympt}), in which case $\Delta = 1$ and 
        \begin{equation}
          R_{k; \ell}(r) \sim \frac{2 \sqrt{k}}{(2 \mu \alpha)^{1/4} r^{3/4}} \sin \!\left(2 \sqrt{2 \mu \alpha r} - \pi \ell - \frac{\pi }{4}\right), \qquad \begin{aligned} k &\to 0 \\[-6pt] r &\to \infty \end{aligned}\,\,,
        \end{equation}
        in agreement with our predicted scaling.
        
        \vskip 0pt
        This scaling can be contrasted with that of a free particle. In this case, the effective potential due to angular momentum $\ell(\ell+1)/\rho^2$ dominates the $\rho \to 0$ limit, and, for $k \to 0$, the radial wavefunction behaves as
        \begin{equation}
          R_{k; \ell}(\rho) \sim  C_1 \rho^{\ell + 1} + C_2 \rho^{- \ell}\,, \qquad \begin{aligned} k &\to 0 \\[-6pt] \rho&\to 0 \end{aligned}\,\,.    
        \end{equation}
        The appropriate $k \neq 0$ continuum wavefunctions are, instead, just the spherical Bessel functions,
        \begin{equation}
          R_{k; \ell}(\rho) = 2 k j_{\ell} (k/\rho)\,,
        \end{equation}
        which obey the asymptotic scaling
        \begin{equation}
          R_{k; \ell}(\rho) \sim \frac{2^\ell \ell! \, k}{\big(\ell + \frac{1}{2}\big) (2 \ell)!} \left(\frac{k}{\rho}\right)^{\ell}\,,\qquad k \to 0\,.
        \end{equation}
        Unlike for potentials with $0 < \Delta < 2$, these continuum wavefunctions do not have a normalization that scales as $\sqrt{k}$ as $k \to 0$, and indeed are \emph{not} localized near the origin. We see that $\Delta = 2$ represents a qualitative dividing line in the behavior of the continuum modes in the $k \to 0$ limit. The matrix elements between a bound state and the zero mode of a potential with $\Delta \geq 2$ obeys $|\eta(k)|^2/k \to 0$, while this approaches a finite limit for potentials with $0 < \Delta < 2$.
        
        \section{Ionization at resonance}
\label{sec:ion-at-resonance}

The expressions for the ionization rate and power derived above are valid under the assumption that the frequency $\Omega$ of the perturbation is away from any bound-to-bound state resonance, as further justified in Appendix~\ref{sec:extension-realistic}. We now relax this assumption by computing the new term contributing at resonance and confirming that its effect is ultimately negligible.

\vskip 0pt
Let us go back to \eqref{eq:realisticEffSchro}, which we report here for convenience,
\beq
\label{eq:IonResoccupation}
i \frac{\dd c_{b}}{\dd t}=\mathcal{E}_{b}c_{b}(t)+\sum_{a \neq b}\left[\eta_{ba}(t) e^{i(\epsilon_b-\epsilon_a) t}+\mathcal{E}_{ba}(t)\right] c_{a}(t)\,,
\eeq
where
\beq
\label{eq:inducedcouplingdef}
\mathcal{E}_{ba}(t)\equiv-i \int_{-\infty}^{t} \dd t^{\prime} \sum_{K} \eta^{*}_{K b}(t) \eta_{K a}(t^{\prime}) e^{-i(\epsilon_K-\epsilon_b)t+i(\epsilon_K-\epsilon_a)t'}\,.
\eeq
and $\mathcal E_b\equiv\mathcal E_{bb}$. The first term in \eqref{eq:IonResoccupation} controls the ionization of state $\ket{b}$, while the first term in the parenthesis is responsible for the $\ket{b}\to\ket{a}$ resonance. The last term, which is the focus of this appendix, is a coupling between $\ket{b}$ and $\ket{a}$ induced via the interaction with the continuum. Because $\mathcal{E}_{ba}(t)$ oscillates very rapidly unless $(m_b-m_a)\dot\varphi_*=\epsilon_b-\epsilon_a$, the parenthesis in \eqref{eq:IonResoccupation} can be neglected altogether whenever the system is not actively on resonance.

\vskip 0pt
Let us study what happens when this is the case instead. The same saddle-point approximation done in Appendix~\ref{sec:saddle-point} can be applied to the case $a\ne b$, arriving to
\beq
\mathcal E_{ba}(t)=e^{i(\epsilon_b-\epsilon_a)t-i(m_b-m_a)\varphi_*(t)}\sum_{\ell,m}\biggl[-\frac{i\mu\,\eta_{Kb}^{*\floq{g_b}}\eta_{Ka}^\floq{g_a}}{2k_*^\floq{g_a}}\,\Theta\bigl((k_*^\floq{g_a})^2\bigr)\biggr]\,.
\eeq
Here, we defined $g_a=m-m_a$, evaluated $\ket{K}$ at $k_*^\floq{g_a}=\sqrt{2\mu((m-m_a)\dot\varphi_*+\varepsilon_a)}$, and expanded the bound-continuum coupling in its Floquet components, $\eta_{Ka}=\eta_{Ka}^\floq{g_a}e^{i(m-m_a)t}$ (and similarly for $a\leftrightarrow b$). To understand the effect of the induced coupling $\mathcal E_{ba}$, we can temporarily set $\eta_{ba}=0$ and write \eqref{eq:IonResoccupation} as
\beq
\frac{\dd\abs{c_b}^2}{\dd t}=\sum_{a}\sum_{\ell,m}\frac{\mu}{k_*^\floq{g_a}}\Theta\bigl((k_*^\floq{g_a})^2\bigr)\Re\Bigl[e^{i(\epsilon_b-\epsilon_a)t-i(m_b-m_a)\varphi_*(t)}\eta_{Kb}^{*\floq{g_b}}\eta_{Ka}^\floq{g_a}c_b^*(t)c_a(t)\Bigr]\,.
\label{eqn:ionization+mixedcoupling}
\eeq
Here, the term with $a=b$ reproduces the ionization term $\mathcal{E}_{b}c_{b}(t)$ in \eqref{eq:IonResoccupation}. Moreover, the evolution of state $\ket{a}$ is determined by the same formula, swapping $b\leftrightarrow a$. For $a\ne b$, however, this operation transforms the term in brackets into its complex conjugate, so its real part stays unchanged. We thus see that the induced coupling $\mathcal E_{ba}$ does \emph{not} contribute to a $\ket{b}\to\ket{a}$ transition alongside $\eta_{ba}$, as one might have expected from \eqref{eq:IonResoccupation}. Instead, both $\abs{c_b}^2$ and $\abs{c_a}^2$ experience an \emph{identical} depletion (in addition to ionization) or recombination, depending on the sign of the real part appearing in \eqref{eqn:ionization+mixedcoupling}; both cases are possible.

\vskip 0pt
We have validated the previous results by comparing them to an explicit numerical integration of the Schrödinger equation, with the continuum states modelled as a large set of discrete states, quadratically spaced in energy. By tuning the parameters to make the impact of the induced coupling clearly visible, we found that \eqref{eqn:ionization+mixedcoupling} gives indeed a very accurate description of the evolution of the populations around the resonance. In Chapter~\ref{chap:resonances}, in particular for Bohr resonances, we are mainly concerned with the correction from the induced coupling to a naive approach where the contributions of ionization and the resonance are simply summed up. To determine its importance, we assume for simplicity that $\eta_{ab}=0$, $\abs{c_b}^2=1$ and $\abs{c_a}^2=0$ at $t=-\infty$, and employ a (further) saddle-point approximation in \eqref{eq:IonResoccupation} around the time $t_0$ such that $\dot\varphi_*=\Omega_0=(\epsilon_b-\epsilon_a)/(m_b-m_a)$. The population at $t=+\infty$ is then
\beq
\abs{c_a}^2=\frac{2\pi}{\abs{m_b-m_a}\gamma}\,\abs*{\sum_{\ell, m}\frac{\mu\,\eta_{Kb}^{*\floq{g_b}}\eta_{Ka}^\floq{g_a}}{2k_*^\floq{g_b}}\Theta\bigl((k_*^\floq{g_b})^2\bigr)}^2\,,
\label{eqn:final-saddle-point}
\eeq
where the couplings and $k_*^\floq{g_b}$ have to be evaluated at $\Omega=\Omega_0$. Similar to the argument in Section~\ref{sec:extension-realistic}, this quantity $\abs{c_a}^2$ is $\mathcal O(q^3\alpha^4)$ and it has to compete with the $\eta^2/\gamma\sim\mathcal O(q\alpha^2)$ contributions due to direct coupling $\abs{\eta_{ba}}^2/\gamma$. Once again, we have validated \eqref{eqn:final-saddle-point} by comparing it to a direct numerical integration of the Schrödinger equation, and evaluated it for a typical Bohr resonance, finding a final population of $\mathcal O(10^{-11})$. We conclude that simply adding the steady deoccupation introduced by ionization on top of the resonant transition studied in the Chapter~\ref{chap:resonances} is a good approximation for our purposes.

\chapter{Resonance phenomenology}

\label{app:resonance-phenomenology}

This appendix contains additional results regarding the study of resonances performed in Chapter~\ref{chap:resonances}. We first justify the assumption, used in Section~\ref{sec:backreaction}, of conservation of the total angular momentum, and then present a more general treatment of the resonance breaking, compared to the disucssion in Section~\ref{sec:resonance-breaking}.

\section{Hyperfine resonances and angular momentum}
\label{sec:hyperfine-angular-momenutm}

A nonzero black hole spin is responsible for the existence of the hyperfine energy splitting, as it breaks the spherical symmetry of the background spacetime. At the same time, we study the backreaction of resonances (hyperfine or not) on the orbit in the Newtonian approximation, assuming the conservation of the total angular momentum, which leads to equations \eqref{eqn:dimensionless-omega-evolution}, \eqref{eqn:dimensionless-eccentricity-evolution} and \eqref{eqn:dimensionless-inclination-evolution}. This methodology might appear as fundamentally inconsistent, so let us inspect it more closely.

\vskip 0pt
The weak-field approximation of the Kerr metric, which is valid at large distances, reads
\beq
ds^2=-\biggl(1-\frac{2M}r\biggr)\dd t^2+\biggl(1+\frac{2M}r\biggr)\dd r^2+r^2(\dd\theta^2+\sin^2\theta\dd\phi^2)-\tilde aM\frac{4M}r\sin^2\theta\dd t\dd\phi\,.
\eeq
The last term is known to give rise to the Lense-Thirring precession, as the equation of motion of a scalar particle can be put in the form
\beq
\frac{\dd^2\vec r}{\dd t^2}=-\frac{M}{r^3}\vec r+4\frac{\dd\vec r}{\dd t}\times\vec B\,,
\eeq
where the \emph{gravitomagnetic field} $\vec B$ is related to the black hole spin as
\beq
\vec B=\vec\nabla\times\vec A\,,\qquad\vec A=-\frac{\vec J\times\vec r}{2r^3},\qquad\vec J=\tilde aM^2\hat z\,.
\eeq
The corresponding Hamiltonian is
\beq
H=\frac{(\vec p-4\mu\vec A)^2}{2\mu}-\frac{\mu M}r\approx\frac{\vec p^2}{2\mu}-\frac\alpha{r}+\frac{2\tilde aM^2}{r^3}L_z\,,
\label{eqn:hamiltonian-gravitomagnetism}
\eeq
where $\mu=\alpha/M$ is the mass of the particle. We can immediately check that the last term in \eqref{eqn:hamiltonian-gravitomagnetism} gives rise to the expected hyperfine splitting,
\beq
\braket{n\ell m|H|n\ell m}=2\tilde aM^2m\Braket{n\ell m|\frac1{r^3}|n\ell m}=2\tilde aM^2m\,\frac{(\mu \alpha)^3}{n^3\ell(\ell+1/2)(\ell+1)}\,,
\eeq
which perfectly matches the last term in \eqref{eq:eigenenergy}.

\vskip 0pt
The orbital angular momentum $\vec L=\vec r\times\vec p$ evolves as
\beq
\frac{\dd \vec L}{\dd t}=i[H,\vec L]=\frac{2}{r^3}\vec J\times\vec L,
\eeq
which is the expected Lense-Thirring precession. Applying this equation to the cloud-binary system gives rise to two additional terms on the right-hand side of \eqref{eqn:Lz-balance} and \eqref{eqn:Lx-balance}, corresponding to the Lense-Thirring precession of the cloud (which vanishes in most cases, as $\vec S\ped{c}\parallel\vec J$ even during a transition, as we will see below) and of the binary. This precession is however \emph{parametrically small}. None of the other terms in \eqref{eqn:Lz-balance} and \eqref{eqn:Lx-balance} depends on the BH spin $\tilde a$, even in the case of hyperfine resonances, where the energy splitting is proportional to $\tilde a$. Not only for realistic parameters is this precession extremely slow, but it also does not disrupt the approach in the main text, as \eqref{eqn:Lx-balance} can be simply replaced by the analogous equation for the (precessing) equatorial projection of the angular momentum.

\vskip 0pt
Having justified the use of the conservation of total angular momentum, there is another potentially worrying aspect of the breaking of spherical symmetry, that has to do with the spin of the cloud when it is in a mixed state, for example during a transition. As long as the Hamiltonian is spherically symmetric, $\ket{n\ell m}$ are guaranteed to be eigenstates of the scalar field's orbital angular momentum $\vec L$. Its matrix elements are given by $L_z\ket{n\ell m}=m\ket{n\ell m}$ and, in the Condon– Shortley convention, $L_\pm\ket{n\ell m}=\sqrt{\ell(\ell+1)-m(m\pm1)}\ket{n\ell,m\pm1}$, where $L_\pm=L_x\pm iL_y$. If the cloud is in a mixed state of the form $\ket{\psi}=c_a\ket{n_a\ell_am_a}+c_b\ket{n_b\ell_bm_b}$, its $z$ component of the angular momentum is then $m_a\abs{c_a}^2+m_b\abs{c_b}^2$, while the equatorial components vanish unless $\ell_a=\ell_b$ and $\abs{m_a-m_b}=1$.

\vskip 0pt
Remarkably, all the previous results still hold for the Hamiltonian \eqref{eqn:hamiltonian-gravitomagnetism}. That is because the perturbation $\sim L_z/r^3$ is diagonal on the basis $\ket{\ell m}$, only mixing states with different $n$. Even though the spacetime is not spherically symmetric, the angular structure of the eigenstates is unchanged. The equations in the main text then do not need any modification, except for the case of hyperfine transitions with $\abs{\Delta m}=1$. For a hyperfine transition with $m_b=m_a-1$, careful computation (in the Schrödinger, not dressed, frame) of the equatorial components of $\vec S\ped{c}$ shows that equation \eqref{eqn:Lx-balance} would need to be corrected with a term
\beq
\frac{\dd S_{\text{c},x}}{\dd\tau}\sim\frac{B}{g}\sqrt{\ell(\ell+1)-m_a(m_a-1)}\sqrt{Z}(\abs{c_b}^2-\abs{c_a}^2)\sin(C\tau/3)\,.
\eeq
This is a fast oscillating term that averages to zero on timescales much shorter than the evolution of the orbital parameters and the duration of the resonance. We thus ignore it in the main text.

\section{General resonance breaking}
\label{sec:breaKING}

The phenomenon of resonance breaking was discussed in Section~\ref{sec:resonance-breaking} in the simplified scenarios where only one of the following quantities is allowed to vary at a time: the eccentricity $\varepsilon$, Landau-Zener parameter $Z$ and cloud's mass $M\ped{c}$. We derive here the result in the general case. Taking the time derivative of \eqref{eqn:dimensionless-omega-evolution-Gamma}, we find
\beq
\begin{split}
\frac{\dd^2\omega}{\dd\tau^2}={}&\frac{\dd f(\varepsilon)}{\dd\tau}+B\frac{\dd^2\abs{c_a}^2}{\dd\tau^2}=\frac{\dd f(\varepsilon)}{\dd\tau}+B\biggl(\frac{\dd^2c_a^*}{\dd\tau^2}c_a+2\frac{\dd c_a^*}{\dd\tau}\frac{\dd c_a}{\dd\tau}+c_a^*\frac{\dd^2c_a}{\dd\tau^2}\biggr)\\
={}&\frac{\dd f(\varepsilon)}{\dd\tau}-\frac{f(\varepsilon)}{2Z}\frac{\dd Z}{\dd\tau}+\Gamma-2ZB(\abs{c_a}^2-\abs{c_b}^2)\\
&+\biggl(\frac{1}{2Z}\frac{\dd Z}{\dd\tau}-\Gamma\biggr)\frac{\dd\omega}{\dd\tau}+\omega\sqrt ZB(c_a^*c_b+c_ac_b^*)\,,
\end{split}
\label{eqn:breaKING-harmonic-oscillator}
\eeq
where the second line is obtained by repeated use of the Schrödinger equation \eqref{eqn:schrodinger-Gamma} together with \eqref{eqn:dimensionless-omega-evolution-Gamma}. Under the assumption that all coefficients appearing above evolve slowly during a floating orbit, equation \eqref{eqn:breaKING-harmonic-oscillator} has the structure of a damped harmonic oscillator, with solution
\beq
\omega=\frac{\frac{\dd f(\varepsilon)}{\dd\tau}-\frac{f(\varepsilon)}{2Z}\frac{\dd Z}{\dd\tau}+\Gamma-2ZB(\abs{c_a}^2-\abs{c_b}^2)}{-(c_a^*c_b+c_ac_b^*)}+\text{damped oscillatory terms}\,.
\label{eqn:breaKING-harmonic-oscillator-solution}
\eeq
The resonance breaks whenever $c_b^*c_a+c_a^*c_b=0$. By direct application of the Schrödinger equation, we find
\beq
\sqrt Z\frac\dd{\dd\tau}(c_a^*c_b+c_ac_b^*)=-\omega\frac{\dd\abs{c_a}^2}{\dd\tau}-\Gamma\sqrt Z(c_a^*c_b+c_b^*c_a)\,.
\label{eqn:breaKING-dcacbcbcadt}
\eeq
By plugging in \eqref{eqn:breaKING-dcacbcbcadt} the non-oscillatory term of \eqref{eqn:breaKING-harmonic-oscillator-solution}, we arrive to an equation for the sole unknown $c_a^*c_b+c_b^*c_a$:
\beq
\frac{ZB}2\biggl(\frac\dd{\dd\tau}+2\Gamma\biggr)(c_a^*c_b+c_b^*c_a)^2=\biggl(\frac{\dd f(\varepsilon)}{\dd\tau}-\frac{f(\varepsilon)}{2Z}\frac{\dd Z}{\dd\tau}+\Gamma-2ZB(\abs{c_a}^2-\abs{c_b}^2)\biggr)\frac{\dd\abs{c_a}^2}{\dd\tau}\,.
\label{eqn:breaKING-master}
\eeq
Remarkably, the evolution of the eccentricity, the variation of the Landau-Zener parameter and the decay of the cloud contribute additively to \eqref{eqn:breaKING-master}, each with its own term. In realistic cases, $\Gamma$ is large enough to force the population of state $\ket{b}$ to reach a saturation value $\abs{c_b}^2=f(\varepsilon)/(2\Gamma B)$, which is usually small enough to be neglected in \eqref{eqn:breaKING-master}. The point of resonance breaking, then, only involves the population left in the initial state, $\abs{c_a}^2$.

\backmatter
\selectlanguage{english}

\summary{Summary / Samenvatting}

The existence of physics beyond the Standard Model is implied by both astronomical observations, through the dark matter paradigm, and by puzzles in particle physics, such as the strong CP problem. In these contexts, a generic prediction of many theoretical and phenomenological models is the existence of new bosons with ultralight masses and very weak couplings to the Standard Model particles. This property makes them extremely challenging to probe with particle colliders. The present thesis investigates a way to discover such ultralight bosons through observations of gravitational waves from binary black hole inspirals.

\vskip 0pt
The mechanism that makes this possible is known as black hole superradiance. In the spacetime of a rapidly spinning black hole, boson field perturbations are unstable, and therefore grow spontaneously until they extract enough mass and angular momentum from the black hole to shut down the instability. The result is a black hole surrounded by a cloud of bosons, a system known as a gravitational atom due to its similarity with the hydrogen atom. The cloud can occupy bound states with shapes and energies similar to electron orbitals, sustained by gravitational rather than electromagnetic forces. When the Compton wavelength of the boson is similar to the size of the black hole, the formation of the cloud is fast over astrophysical timescales. Bosons in the mass range $10^{-19}$ to $10^{-11}\,\si{eV}$, relevant for physics beyond the Standard Model, can thus be probed with astrophysical black holes of masses $10$ to $10^9M_\odot$.

\vskip 0pt
I focus here on a setup where a gravitational atom is orbited by a binary companion. The goal is to fully characterize the dynamics of the system and identify the signatures left by the boson cloud on the gravitational waves emitted by the binary. The predictions can be tested with current and future gravitational wave interferometers, such as LISA, LIGO, DECIGO, Einstein Telescope, and TianQin. The gravitational interaction between the cloud and the binary leads to a very rich phenomenology, most of which has analogues in atomic physics. Previous studies have demonstrated the existence of ``resonant'' orbits: when the binary's frequency matches the energy difference between two bound states, the cloud can make a transition between the states, while the binary undergoes a period of decelerated or accelerated inspiral.

\vskip 0pt
In this thesis, I introduce several new phenomena, achieve a self-consistent description of the dynamics of the system, and discover new striking observational signatures. The first new effect introduced here is the cloud's impact on the binary formation via dynamical capture. The interaction with the cloud dissipates energy and thus catalyzes the process. I demonstrate that the cross section for a successful binary formation after a close encounter can increase by a factor of $\mathcal O(10)$ compared to the vacuum case. The second and most prominent effect I examine in this thesis is the ionization of the cloud. In analogy with the photoelectric effect in atomic physics, the periodic perturbation from the binary can make the bosons transition into states that are gravitationally unbound from the black hole, thus ejecting away part of the cloud. The power lost by the binary through ionization can be orders of magnitude larger than that released in gravitational waves. Furthermore, the ionization power is a discontinuous function of the binary's frequency, leading to extremely distinctive features in the evolution of the system. The third and last new effect I study in this thesis is the accretion of the cloud on the binary companion if it is a black hole. This process alters the companion's mass and thus the dynamics of the system.

\vskip 0pt
After having introduced the array of new phenomena, I thoroughly investigate and generalize ionization and resonances, because they stand out as the effects with the most dramatic observational consequences. I demonstrate that ionization is the equivalent of dynamical friction in other astrophysical media, which behaves here in an unexpected way due to the specific energy spectrum of the cloud. I proceed to rigorously examine the characteristic features of ionization, understanding them with an analytical approach, and then generalize all results to orbits with nonzero eccentricity and inclination.

\vskip 0pt
In order to achieve a realistic and complete study of the evolution of the system, I also significantly generalize the previously available results on resonant orbits. This includes extending the framework to orbits with nonzero eccentricity and inclination, and taking into account self-consistently the backreaction of the resonances on the binary's orbit. These two aspects work together to change the resonance phenomenology considerably. I discover new analytical conditions for the resonances to start or end, and precisely quantify their impact on the orbital parameters.

\vskip 0pt
Making use of all these results, I systematically study the history of the system, from the binary formation to the black hole merger, with the goal of identifying the observational signatures. If the binary and the cloud rotate in near-opposite directions, then the cloud survives until the late stages of the inspiral, where it is able to directly affect the gravitational waveform. At this point, ionization takes over as the dominant force driving the inspiral, and its discontinuous features leave characteristic kinks in the frequency evolution of the gravitational wave. If instead the binary and the cloud do not rotate in near-opposite directions, the cloud is destroyed before it can directly affect the gravitational waveform. However, its destruction leaves the binary in an orbit with specific values of eccentricity and inclination, which can be searched for with a statistical analysis of a large number of gravitational wave signals.

\vskip 0pt
This thesis exhaustively describes the most important aspects of the phenomenology of gravitational atoms in binaries. The new results presented here pave the way for searches of ultralight bosons using gravitational waves, showing that they are a clear and very promising target due to their extremely sharp observational signatures.

\subsubsection*{Nederlandse Samenvatting}

Het bestaan van fysica buiten het Standaard Model wordt geïmpliceerd door zowel astronomische waarnemingen, via het donkere materie paradigma, als door raadsels in de deeltjesfysica, zoals het sterke CP-probleem. In deze contexten voorspellen vele theoretische en fenomenologische modellen het bestaan van een nieuw boson met een ultralichte massa en zeer zwakke koppelingen met deeltjes uit het Standaard Model. Deze eigenschappen maken het buitengewoon moeilijk om ze te ontdekken met deeltjesversnellers. Dit proefschrift onderzoekt een manier om dergelijke ultralichte bosonen te ontdekken via waarnemingen van zwaartekrachtsgolven van botsende zwarte gaten.

Het mechanisme dat dit mogelijk maakt, staat bekend als zwarte gat superradiantie. In de ruimtetijd van een snel draaiend zwart gat zijn verstoringen in een bosonveld instabiel, en dus groeien ze spontaan totdat ze voldoende massa en impulsmoment aan het zwarte gat onttrekken om de instabiliteit te stoppen. Het resultaat is een zwart gat omgeven door een wolk van bosonen, een systeem dat bekend staat als een zwaartekrachtsatoom vanwege de gelijkenis met het waterstofatoom. De wolk kan gebonden toestanden aannemen met vormen en energieën die vergelijkbaar zijn met elektronenbanen, maar in stand wordt gehouden door zwaartekracht in plaats van elektromagnetische krachten. Wanneer de Compton-golflengte van het boson vergelijkbaar is met de grootte van het zwarte gat, dan verloopt de vorming van de wolk snel op astrofysische tijdschalen. Bosonen in het massabereik van $10^{-19}$ tot $10^{-11}\,\si{eV}$ zijn relevant voor fysica buiten het Standaard Model en kunnen dus worden onderzocht met astrofysische zwarte gaten van massa’s $10$ tot $10^9M_\odot$.

Ik richt me hier op een situatie waarbij het zwaartekrachtsatoom omcirkeld wordt door een binaire metgezel. Het doel is om de dynamiek van het systeem volledig te karakteriseren en sporen te identificeren die de bosonenwolk achterlaat op de zwaartekrachtsgolven die worden uitgezonden door het binaire systeem. De voorspellingen kunnen zowel met huidige, als toekomstige zwaartekrachtgolf-interferometers worden getest, zoals LISA, LIGO, DECIGO, Einstein Telescope en TianQin. De zwaartekrachtsinteractie tussen de wolk en het binaire systeem leidt tot een zeer rijke fenomenologie, waarvan de meeste een analoog hebben in de atoomfysica. Eerdere studies hebben aangetoond dat er ``resonante’’ banen bestaan: wanneer de frequentie van het binaire systeem overeenkomt met het energieverschil tussen twee gebonden toestanden, kan de wolk een overgang maken tussen de toestanden, terwijl het binaire systeem een periode ondergaat van vertraagde of versnelde evolutie.

In dit proefschrift introduceer ik verschillende nieuwe fenomenen, bereik ik een zelf-consistente beschrijving van de dynamiek van het systeem en ontdek ik nieuwe, opvallende observationele kenmerken. Het eerste nieuwe effect dat hier wordt geïntroduceerd is de impact van de wolk op de formatie van het binaire systeem via dynamische vangst. De interactie met de wolk dissipeert energie en katalyseert zo het proces. Ik laat zien dat de doorsnede voor een succesvolle binaire formatie na een nauwe ontmoeting met een factor $\mathcal O(10)$ kan toenemen ten opzichte van het vacuümgeval. Het tweede en meest prominente effect dat ik in dit proefschrift onderzoek, is de ionisatie van de wolk. In analogie met het foto-elektrische effect in atoomfysica, kan de periodieke verstoring van het binaire systeem de bosonen laten overgaan in toestanden die gravitationeel ongebonden zijn aan het zwarte gat, waardoor een deel van de wolk wordt weggeslingerd. De kracht die het binaire systeem verliest door ionisatie kan ordes van grootte groter zijn dan de kracht die vrijkomt in zwaartekrachtsgolven. Bovendien is de ionisatiekracht een discontinue functie van de frequentie van het binaire systeem, wat leidt tot zeer kenmerkende eigenschappen in de evolutie van het systeem. Het derde en laatste nieuwe effect dat ik in dit proefschrift bestudeer, is de accretie van de wolk op de binaire metgezel als deze een zwart gat is. Dit proces verandert de massa van de metgezel en daarmee de dynamiek van het systeem.

Na de introductie van een scala aan nieuwe fenomenen onderzoek en generaliseer ik ionisatie en resonanties grondig, omdat zij zich onderscheiden als de effecten met de meest dramatische observationele gevolgen. Ik toon aan dat ionisatie het equivalent is van dynamische wrijving in andere astrofysische media, die zich hier op een onverwachte manier gedraagt vanwege het specifieke energiespectrum van de wolk. Ik vervolg met het grondig onderzoeken van de karakteristieke eigenschappen van ionisatie, waarbij ik ze begrijp met een analytische benadering en generaliseer vervolgens alle resultaten naar banen met een excentriciteit die niet nul is en een inclinatie.

Om een realistische en volledige studie van de evolutie van het systeem te verkrijgen, generaliseer ik ook aanzienlijk de eerder beschikbare resultaten over resonante banen. Dit omvat het uitbreiden van het raamwerk naar een excentriciteit die niet nul is en inclinatie, en daarbij op een consistente wijze rekening houdend met de terugkoppeling van de resonanties op de baan van het binaire systeem. Deze twee aspecten zorgen samen voor een aanzienlijke verandering in de fenomenologie van de resonantie. Ik ontdek nieuwe analytische voorwaarden voor het begin of einde van de resonanties en kwantificeer nauwkeurig hun invloed op de parameters van de baan.

Met behulp van al deze resultaten bestudeer ik systematisch de geschiedenis van het systeem, van de formatie van het binaire systeem tot de fusie van de zwarte gaten, met als doel het identificeren van de observationele kenmerken. Als het binaire systeem en de wolk in bijna tegengestelde richtingen roteren, overleeft de wolk tot de late stadia van de evolutie van het binaire systeem, waarbij deze direct invloed kan uitoefenen op de zwaartekrachtgolfvorm. Op dit punt neemt ionisatie het over als de dominante kracht die de evolutie aandrijft, en de discontinue kenmerken ervan laten karakteristieke knikken achter in de frequentie-evolutie van de zwaartekrachtsgolf. Als het binaire systeem en de wolk daarentegen niet in bijna tegengestelde richtingen roteren, wordt de wolk vernietigd voordat deze direct de zwaartekrachtsgolfvorm kan beïnvloeden. De vernietiging laat het binaire systeem echter achter in een baan met specifieke waarden van excentriciteit en inclinatie, waarvoor gezocht kan worden met een statistische analyse van een groot aantal zwaartekrachtgolfsignalen.

Dit proefschrift beschrijft grondig de belangrijkste aspecten van de fenomenologie van zwaartekrachtsatomen in binaire systemen. De nieuwe resultaten die hier worden gepresenteerd maken de weg vrij voor het zoeken naar ultralichte bosonen met behulp van zwaartekrachtsgolven, en laten zien dat ze een duidelijk en veelbelovend doelwit zijn vanwege hun extreem opvallende observationele kenmerken.

\hfill\emph{Vertaald door Thomas Spieksma.}

\acknowledgements{Acknowledgements}

I am lucky enough to have not one, but two amazing scientists as academic advisors. Their guidance has been complementary and crucial for my development.

Gianfranco turned me from a student into a researcher. When I began my PhD, all I could do was define and solve physics problems. It is largely due to his merit that I can now navigate the frontiers of research, understand how science is moving forward, establish contacts and collaborations with other scholars, make sure my work is recognized, and never forget the big picture of what I am doing. Without Gianfranco's contribution, I would not have had nearly as many opportunities to take a spot in international conferences. Being exposed to the world of research outside my bubble has been important not only for all the points mentioned above but also for improving my own work and ensuring I can continue to do physics at the highest level. Generally speaking, regular contact with Gianfranco has enriched me on every facet of the academic world.

Daniel completely transformed the way I communicate science. He made me improve so much that it would be hard to believe that paper drafts, presentations, and even notes from four years ago were written by the same person who authored this thesis. Daniel has also been influential on me for his ability to brainstorm and discuss new ideas, as well as an inspiration for the originality of his research. It is also impossible not to mention our unforgettable collaboration in teaching General Relativity, a true highlight of my PhD. Only the two of us know how much we enjoyed preparing that course and how glorious it felt when we came up with new problems for the exams. I am very grateful to Daniel for all of the above and, more broadly, for his overall contribution to my growth as a scientist.

Over the years, Thomas evolved from being my brilliant Master's student to my best collaborator. It is delightful that such a fruitful partnership was born spontaneously at such an early stage of our careers. I want to thank him for his contributions to the papers contained in this thesis and for his amazing attitude, which is always a motor to drive projects forward. John was also an important collaborator during the first half of my PhD. From him I learned the importance of questioning every step to truly understand what one is doing, and many tricks to create amazing Ti\emph{k}Z figures.

Among all the other scientists with whom I had stimulating physics discussions, those who stand out most prominently are Lam Hui, Rafael Porto, Enrico Trincherini, Rodrigo Vicente, Vitor Cardoso, and Ignacio Reyes. I also thank the entire Gianfranco group, especially Pippa Cole, for frequent meetings and fruitful exchanges. All the PhDs and postdocs I met at UvA, from GRAPPA, cosmology, astronomy, and string theory, contributed positively to my experience here. Two of my office mates became particularly good friends of mine: Noemi Anau Montel, whose PhD journey coincided with mine from the first day to the very last one, and Oleg Savchenko, with whom I share a similar scientific education and attitude. I must also thank Veronica Sacchi, whose brief stay in Amsterdam resulted in the two most restless yet memorable months I have experienced in this city.

Choosing to pursue my doctoral studies at the University of Amsterdam instead of Scuola Normale Superiore was a very tough decision. I am now certain it was the better choice because I was able to expand my horizons enormously, while not cutting ties with Normale. In fact, during my PhD I realized with my fellow normalisti a decade-long dream: organizing a team competition for the Physics Olympiads. I am very proud of this achievement, which might be the single most consequential project I have ever contributed to. I am also grateful to Gianfranco and Daniel for never complaining about all the time and energy I spent organizing this and other Olympiad-related events.

The past four years in Amsterdam have been an important step in my journey through science and life. I truly, deeply love physics and feel very grateful for being able to spend most of my time with it. The next step will bring me to yet another place, for a new adventure with physics I cannot wait to live.

\hfill \emph{Gimmy}

\hfill \emph{Amsterdam, 2024}

\newpage
\phantomsection
{\linespread{1.098}
\bibliographystyle{utphys}
\bibliography{main}
}
\end{document}